%% file: main.tex
\documentclass[twoside,openright,frontopenright]{ip3thesis}
\input{preamble}

\begin{document}
\setboolean{printBibInSubfiles}{false}

\title{Idealised Antenna Functions for Higher Order QCD Calculations}
\author{Oscar Ronald Torsten Braun-White}
\researchgroup{Institute for Particle Physics Phenomenology}
\maketitlepage*

\begin{abstract}
	In this thesis, the infrared structure of squared matrix elements in quantum chromodynamics (QCD) is scrutinised. Specifically, the triple-collinear splitting functions are decomposed and improvements to antenna subtraction are sought through the construction of idealised antenna functions. The antenna-subtraction technique has demonstrated remarkable effectiveness in handling next-to-next-to-leading order (NNLO) infrared divergences for a wide range of QCD processes relevant for colliders. However, since antenna functions were historically extracted from matrix elements, they did not have uniform properties, which made the generation of subtraction terms complex. Antenna subtraction up to NNLO is reviewed, including the role of antenna functions. A general algorithm is detailed in order to re-build antenna functions, with idealised features, directly from a specified list of unresolved limits, for any number of real and virtual emissions. Using this general algorithm, together with the decomposition of the triple-collinear splitting functions, all the antenna functions required for NNLO QCD calculations in the final-final configuration are constructed and it is demonstrated that they form a consistent NNLO subtraction scheme. The idealised antenna functions should simplify the generation of subtraction terms and minimise the introduction of spurious limits. Additionally, the general algorithm sets out an initial blueprint for next-to-NNLO (\NthreeLO) idealised antennae for use in \NthreeLO QCD calculations. 
\end{abstract}

\disableprotrusion
\tableofcontents*
\listoffigures
\listoftables
\enableprotrusion

\begin{declaration*}
	The work in this thesis is based on research carried out in the Department of
	Physics at Durham University. No part of this thesis has been
	submitted elsewhere for any degree or qualification. 
	Unless otherwise stated, all original work is my own in collaboration with my supervisor Professor Nigel Glover and Dr. Christian T Preuss. 
	Parts of this thesis are based on joint research, presented in the publications:
	\begin{itemize}
		\item O.~Braun-White and N.~Glover,
		``Decomposition of triple collinear splitting functions,''
		JHEP \textbf{09} (2022), 059
		doi:10.1007/JHEP09(2022)059
		[arXiv:2204.10755 [hep-ph]],
		\item O.~Braun-White, N.~Glover and C.~T.~Preuss,
		``A general algorithm to build real-radiation antenna functions for higher-order calculations,''
		JHEP \textbf{06} (2023), 065
		doi:10.1007/JHEP06(2023)065
		[arXiv:2302.12787 [hep-ph]],
		\item O.~Braun-White, N.~Glover and C.~T.~Preuss,
		``A general algorithm to build mixed real and virtual antenna functions for higher-order calculations,''
		JHEP \textbf{11} (2023), 179
		doi:10.1007/JHEP11(2023)179
		[arXiv:2307.14999 [hep-ph]].
	\end{itemize}
\end{declaration*}

\begin{acknowledgements*}
	Firstly, I would like to thank Nigel Glover for his excellent supervision. 
	Long parts of this PhD were undertaken during various forms of COVID lockdown. 
	While this was of course isolating (for everybody), Nigel was a valuable source of encouragement, motivation and friendship...not-to-mention his talent and ingenuity. 

	I would also like to thank my collaborator, Christian Preuss, for his support and enthusiasm. 
	Along with Nigel, we kept a regular correspondence since he first expressed interest in our work at the 2021 NNLOJET meeting. 
	This brings us to the NNLOJET group; it has been a pleasure to play a small part in the research group based around antenna functions. 
	The shared knowledge and dedication of the group has been very inspiring. 
	The time spent meeting with members of the group has been invaluable to my academic development, including the three-month stay in Thomas Gehrmann's group at the University of Zurich, for which I am very grateful. 

	I have also enjoyed being part of a thriving research community at the Institute for Particle Physics Phenomenology (IPPP) in Durham. 
	In particular, I would like to thank the following for helpful discussions and proofreading my thesis: Xuan Chen, Elliot Fox, Hitham Hassan, Matteo Marcoli, Peter Meinzinger and Joseph Walker. 

	My time at Durham would have been much less fulfilling without my church, my family and my wife. 
	The community of St. Margaret's Church has been a very loving second home. 
	I am grateful for the friendships developed there and for my growing relationship with God. 
	Thank you to my family for the love and support you give me. 
	For trying to understand my work and for giving me a distraction! 
	Finally, all my thanks and love to my wife, Georgia Braun-White. 
	We have grown up together since we met at 18 years old in Durham and I can't imagine my life without you. 
	Onwards to the next adventure! 
	%
%
\end{acknowledgements*}

\begin{epigraph*}
	People assume that time is a strict progression of cause to effect, but actually from a non-linear, non-subjective viewpoint — it’s more like a big ball of wibbly wobbly… timey wimey… stuff.
	\source{Blink, Doctor Who}{The Tenth Doctor}
\end{epigraph*}









%
%

\cleardoublepage

\chapter*{Preface} 
\label{chapter:intro}

Scientific research seeks to address the big questions. What is the nature of reality? How can we improve the lives of those around us? Where do we fit into it all? These questions cannot and should not be tackled alone. We must be able to communicate theories concisely and be able to dispute them effectively. Scientific research is a huge collaborative enterprise, more so than ever in the twenty-first century. It is a privilege to play a small part in this fascinating journey.

The research presented in this PhD thesis sits within the global efforts of the particle physics community to probe the fine structure of matter and its fundamental interactions. 
Our best theoretical description of this is the widely successful and sophisticated Standard Model of particle physics. 
The theories of the Standard Model are described in a framework seemingly beyond that of reality - we manipulate imaginary numbers, quantify infinities and calculate in a non-integer number of dimensions. 
Communicating in these terms has been our best way so far of explaining our real observations. 
However, there are still many open questions. 
How do we describe the fundamental nature of gravity? 
Why does our universe possess matter-antimatter asymmetry?
What is the source of neutrino masses? 
How do we address the hierarchy problem? 
And what is the nature of the huge portion of unexplained mass and energy, dubbed dark matter and dark energy? 

One way to address these questions is by rigorously testing the Standard Model against precise observations and by analysing any discrepancies. 
These may hold clues to the nature of New Physics. 
Since the correspondence between theories of the Standard Model and collider physics is a research field in its own right - particle physics phenomenology - the goal of precision testing requires significant progress and effort. 
In this thesis, we will focus on improvements to the antenna-subtraction formalism. 
This formalism facilitates the prediction of real observables to a high precision by using the underlying Standard Model theories, especially quantum chromodynamics (QCD). 
Antenna subtraction relies upon building blocks called antenna functions; it is the properties of these which we seek to idealise. 

The thesis is organised as follows. In Chapter~\ref{chapter:qcd}, we introduce the necessary theoretical framework in QCD for the main research. Chapter~\ref{chapter:pheno} justifies the role of precision QCD calculations in the field of particle physics phenomenology. In Chapter~\ref{chapter:paper1}, we present research into the singular structure of matrix elements, in the limit where three QCD particles are collinear. Chapter~\ref{chapter:antennasub} introduces the general antenna-subtraction formalism up to this point and motivates its necessary improvements. In Chapters~\ref{chapter:paper2} and~\ref{chapter:paper3}, we present the main research of this thesis: the idealisation of antenna functions for high precision calculations. In Chapter~\ref{chapter:conc}, we summarise the thesis and comment on possible future work. 

\chapter{Quantum Chromodynamics}
\label{chapter:qcd}

In this chapter, we will introduce and detail the concepts in QCD which are most relevant for the bulk of this thesis. 
We will start with a brief description of the QCD Lagrangian and its fields and our chosen matrix element formalism in Section~\ref{sec:QCDLang}. 
Next, we will introduce ultraviolet (UV) divergences in QCD and their handling via renormalisation in Section~\ref{sec:UVdivs}. 
Following on from that, we discuss the resultant running of the strong-coupling constant in Section~\ref{sec:gsrunning} and our use of perturbation theory in the strong-coupling in Section~\ref{sec:aspert}. 
We introduce infrared (IR) divergences in QCD and detail the universal infrared factorisation structures, which will be crucial for the remainder of this thesis, in Section~\ref{sec:IRdivs}. 
Finally, we summarise the chapter in Section~\ref{sec:QCDsummary}.

\section{QCD Lagrangian}
\label{sec:QCDLang}

QCD is a non-abelian field theory and the fundamental interactions of quarks and gluons are determined by the form of the QCD Lagrangian (density), 
\begin{equation}
\label{eq:QCDlag}
\mathcal{L}_{\text{QCD}} = \sum_q \overline{\psi}_q (i \slashed{D} - m_q) \psi_q - \frac{1}{4} F_{\mu \nu}^{a} F^{a,\mu \nu} .
\end{equation}
$\psi_q$ represents a quark field of flavour $q$ and transforms in the fundamental representation of the Lie group SU$(N_c)$, while $\overline{\psi}_q \equiv \psi_q^\dagger \gamma^0$.  
In the Standard Model, there are six flavours of quark ($u, d, s, c, b, t$), in increasing order of mass, $m_q$. 
The QCD Lagrangian is both Lorentz invariant and gauge invariant under local transformations in SU$(N_c)$, where $N_c=3$ is the number of colours. 
The quark field transforms as $\psi_q \to U \psi_q$, where $U$ is a matrix parametrised by local real rotations and the (hermitian) generators of the fundamental and irreducible representation of SU$(N_c)$, $t_a$, such that
\begin{equation}
U = \exp \bigg(-i g_s \theta^a (x) t^a \bigg),
\end{equation}
where $a=1,..., N_c^2 -1$ and $g_s$ is the strong-coupling constant. 
The generators $t^a$ obey the appropriate Lie algebra,
\begin{equation}
    [ t^a,t^b] = i f^{abc} t^c,
\end{equation}
with totally antisymmetric structure constants, $f^{abc}$, and the Gell-Mann matrices $\lambda^a/2$ satisfy this algebra. 
The covariant derivative $\slashed{D} = \gamma^\mu ( \partial_\mu + i g_s A_\mu^a t^a)$ is introduced such that $D_\mu \psi_q \to U D_\mu \psi_q$ under gauge transformations in the same way as $\psi_q$. 
This ensures that $\overline{\psi}_q \slashed{D} \psi_q $ is a gauge invariant term. 
As a result, gauge fields are introduced to the theory, in the form of vector fields $A_\mu^a$, which in QCD are called gluons. 
Gluons transform in the adjoint representation of SU$(N_c)$ and their interactions with quarks are encoded in the  $\overline{\psi}_q \slashed{D} \psi_q $ term. 
The quadratic terms in the quark fields in Eq.~\eqref{eq:QCDlag} give the masses.
If there were a mass term for gluons, it would look like,
\begin{equation}
    \frac{1}{2} m_A^2 A_\mu^a A^{\mu,a}. 
\end{equation}
Such a term is not gauge invariant, under the gauge transformation 
\begin{equation}
    A_\mu^a \to A_\mu^a - \frac{1}{g_s} \partial_\mu \theta^a (x) . 
\end{equation}
This means that mass terms for gluons are forbidden by gauge invariance and so gluons are massless.
However, the combination of fields in the final term of Eq.~\eqref{eq:QCDlag} is gauge invariant because of the form of the field-strength tensor,
\begin{equation}
    F_{\mu \nu}^a = \partial_\mu A_\nu^a - \partial_\nu A_\mu^a - g_s f^{abc} A_\mu^b A_\nu^c.
\end{equation}
This final term encodes the self interactions of gluons.

\begin{figure}[h]
    \epsfig{file=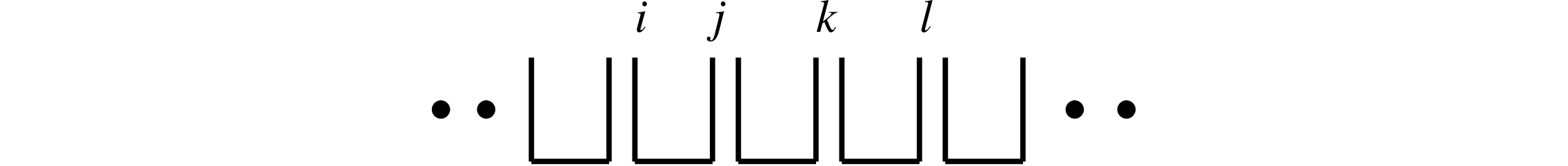,height=1.2cm}
    \caption{Diagram showing the colour-connections within a colour-ordered matrix element~\cite{Gehrmann-DeRidder:2005btv}.}
    \label{fig:connections}
    \end{figure}

Feynman rules can be deduced for the calculation of perturbative QCD matrix elements (amplitudes). 
The details of these rules have been well documented but are not necessary for explaining the work in this thesis, so we refer the reader to Ref.~\cite{Peskin:1995ev} and other standard textbooks on quantum field theory. 
Computations using the Feynman rules can become complicated very quickly with increasing numbers of scattered particles, so it is convenient to separate the colour from the kinematics in a calculation. 
In this thesis, we will be concerned with the unresolved limits of colour-ordered matrix elements (or partial amplitudes). 
Their calculation presents a number of challenges, which we will discuss later in this chapter.
Colour factors can be extracted by well documented rules in an automated process to all orders so they are not a barrier to calculations\cite{Sjodahl:2012nk,Sjodahl:2014opa,Gerwick:2014gya,Baberuxki:2019ifp}. 
The full matrix element can be reconstructed by a sum over all independent colour structures, for example the tree-level scattering amplitude, $\boldsymbol{\mathcal{M}}_{ng}^0$, for $n$ gluons is given by
\begin{equation}
    \boldsymbol{\mathcal{M}}_{ng}^0 = \sum_{\sigma \in S_n/Z_n} \text{Tr} (t^{a_{\sigma(1)}} ... t^{a_{\sigma(n)}}) \mathcal{M}_{ng}^0(\sigma(1) ... \sigma(n)), 
\end{equation}
where the sum is over the non-cyclic permutations of $n$ elements, $S_n/Z_n$, because of the cyclicity of the trace and $\mathcal{M}_{ng}^0$ is the colour-ordered matrix element. 
As the name suggests, the colour-ordered matrix element has specific colour connections. 
In the full matrix element, every gluon is equally connected to every other gluon but in the colour-ordered case, only neighbouring gluons are connected, which can be seen in Fig.~\ref{fig:connections}.
This means that, apart from simpler Feynman rules, the colour-ordered matrix element has a smaller set of collinear divergences (when two particles travel at such a similar angle as to be experimentally unresolved). 
From now on, we will simply refer to the colour-ordered matrix element as the matrix element.

\section{Ultraviolet Divergences}
\label{sec:UVdivs}

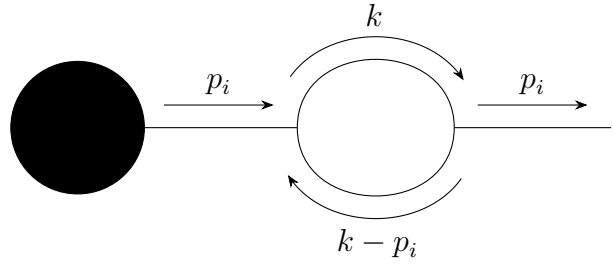
\begin{figure}[t]
    \centering
\begin{tikzpicture} 
        \begin{feynman}
        \vertex (a); 
        \vertex [right=2em of a] (a1);
        \vertex [right=7em of a] (b);
        \vertex [right=5em of b]  (c);
        \vertex [right=5em of c]  (d) ; 
        \filldraw (a) circle (25pt);
        \diagram* {
        (a1) -- [momentum=\( p_i  \)] (b),
        (b) -- [half left, momentum=\( k \)] (c), 
        (c) -- [half left, momentum=\( k- p_i \)] (b),
        (c) -- [momentum=\( p_i  \)] (d),
        };
        \end{feynman}
        \end{tikzpicture}
        \caption{Generic diagram, containing a loop with momentum $k$ on an external leg with momentum $p_i$.}
        \label{fig:loop}
\end{figure}

A recurring and prevalent theme of this thesis is the presence and handling of divergences (or infinities). 
In the context of quantum field theory, these appear in configurations with very large momentum or very small momentum structures, 
known respectively as ultraviolet and infrared. Infrared divergences are the main focus of this thesis and will be introduced later in this chapter. 
Ultraviolet divergences appear in the calculation of loop integrals, where the momentum, $k$, in the loop (the virtual particle)
must be allowed to take any value, including very large values. An example of this is shown in Fig.~\ref{fig:loop}. 
If we are to calculate real-world quantities, where there are no divergences, we must have a procedure for handling and removing these divergences; these are respectively regularisation and renormalisation. 

Regularisation consists of parametrising the divergences with a quantity which must be taken to a limit after performing the calculation. 
Therefore the divergences are not removed but merely exposed and it becomes possible to quantify them analytically. 
There are multiple regularisation schemes. 
The simplest is the momentum cut-off scheme, where a momentum cut off $\Lambda$ is introduced in any loop integral so that the integration region does not reach infinity. 
Then the divergence is expressed in terms of a function of 
$\Lambda$, which diverges when $\Lambda \to \infty$. 
The regularisation scheme which we use throughout this thesis is called conventional dimensional regularisation (CDR), where the number of space-time dimensions is deformed away from four to $d=4-2\e$ and a physical observable only exists in the $\e \to 0$ limit. 
This means that an integral over $d^4 k$, for loop-momentum $k$, is now over $d^d k$. 
The integrand is now integrated over a non-integer number of variables, rather than four, and a whole toolkit of mathematics has been introduced to handle this~\cite{tHooft:1972tcz}. 
As a result, dimensional regularisation is an extremely elegant method, which expresses UV (and some IR) divergences in terms of poles (inverse powers) of the small parameter $\e$, while also retaining Lorentz and gauge invariance. 
In particular, CDR treats all internal and external particles consistently within $d$ dimensions. 
In $d$ dimensions, the mass dimensions of the QCD Lagrangian is also $d$. 
In order to keep the strong-coupling constant, $g_s$, dimensionless, it must be rescaled by a regularisation scale, $g_s \to \mu^\e g_s$. The regularisation scale $\mu$ is arbitrary and in principle any observable should not depend on its value. 

As stated earlier, using a regularisation scheme allows for analytic handling of divergences. 
In the case of UV divergences in QCD we can handle them by renormalisation, since QCD is a renormalisable field theory; that is, there are only a finite number of matrix elements which superficially diverge in the UV regime. 
Since UV divergences respect gauge invariance, we can reabsorb them to all orders into the definitions of the fields, couplings and masses.
We do this by calculating multiplicative factors containing the UV divergences, 
which multiply the UV finite renormalised quantities to give the unrenormalised or bare quantities. For a generic quantity $\lambda$, this looks like
\begin{equation}
    \lambda_0 = Z_\lambda \lambda, 
\end{equation}
where $\lambda_0$ is the bare quantity, $\lambda$ the renormalised quantity and $Z_\lambda$ the divergent multiplicative factor. 
If we can expand $Z_\lambda$ in perturbation theory, $Z_\lambda \equiv 1 + \delta Z_\lambda = 1 + C g_s^2 /\e + \order{g_s^4}$. By rewriting the Lagrangian in terms of the various $\delta Z_\lambda$ and renormalised quantities, we find
\begin{equation}
    \mathcal{L_{\text{QCD}}} = \mathcal{L}_{\text{QCD,renorm}} + \delta \mathcal{L}_{\text{QCD}} 
\end{equation}
where the second term contains terms with $\delta Z_\lambda$ and are the renormalisation counterterms. 
By specifying renormalisation conditions based on the physical masses and the coupling constant, we can generate a new set of 
Feynman rules for the counterterms. 
If perturbation theory is appropriate for the coupling constant, we can calculate the contributions to $Z_\lambda$ order-by-order by identifying the UV divergent parts of the theory order-by-order. 
There is some ambiguity in how the divergent factors are determined. We can choose $\delta Z_\lambda$ to remove only explicit $\e$ poles - this is called the minimal subtraction scheme (MS). 
In practice, due to similar structures in loop integrals, it is convenient for the counterterms to absorb part of the finite terms for simpler book-keeping - this is called the $\overline{\text{MS}}$ scheme. Where $1/\e$ would be included in MS, we include $1/\bar{\e} = 1/\e - \gamma_E + \ln (4 \pi) + \order{\e}$ in $\overline{\text{MS}}$, where $\gamma_E$ is the Euler-Mascheroni constant.

\section{Running of the Strong-Coupling Constant}
\label{sec:gsrunning}

One of the renormalised quantities is the strong-coupling constant, $g_s$, given by
\begin{equation}
    g_{s,0} = \mu^\e Z_{g_s}(\mu) g_s(\mu), 
\end{equation}
where the renormalisation scale is $\mu$ and the scale-independence of the bare coupling is made explicit. The renormalisation is imposed at a particular scale when defining the counterterms and we set the regularisation scale equal to the renormalisation scale because any result is independent of the regularisation scale. We can think of the renormalisation scale as the typical energy scale for an experiment of interest. For us this is typically the high energy regime appropriate for colliders such as the Large Hadron Collider (LHC). 
This means that the renormalised measurable coupling $g_s (\mu)$, which we can extract from experiment, depends upon the scale $\mu$. We will now show that, given $g_s (\mu_0)$ at a particular scale $\mu_0$, we can predict the behaviour of the strong-coupling constant at other scales and what this means for QCD interactions. 

\begin{figure}[t]
    \centering
    \includegraphics[width=0.66\textwidth]{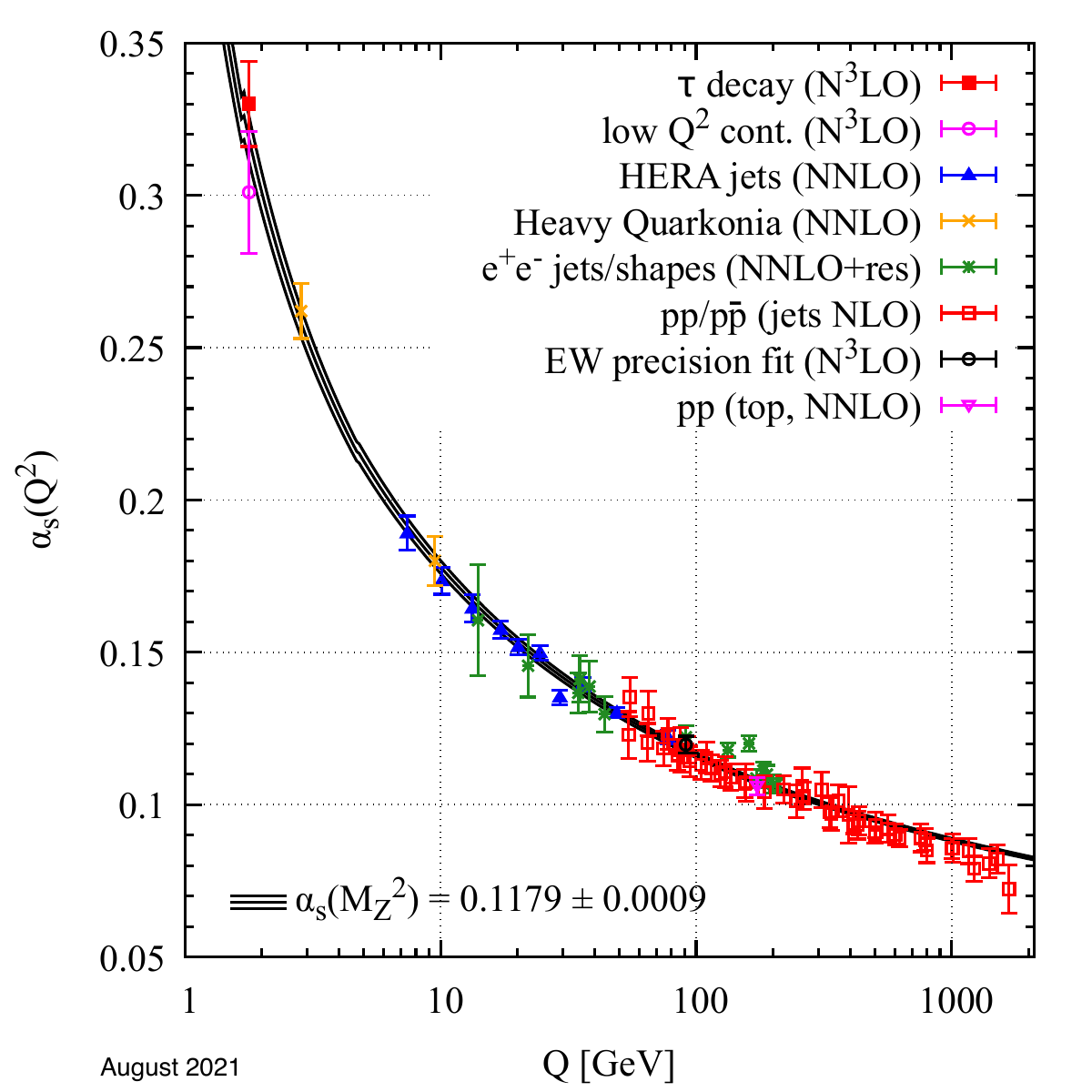}
    \caption{A collection of extractions of $\alpha_s$ at various scales consistent with the global average for $\alpha_s (M_Z^2) = 0.1179(9)$~\cite{PDG22}.}
    \label{fig:runningcoup}
\end{figure}

We can differentiate the renormalised coupling with respect to $\mu$ to give 
\begin{equation}
    \mu \frac{d g_s}{d \mu} = - \e g_s - \frac{\mu}{Z_{g_s}^2}\frac{d Z_{g_s}}{d \mu} \mu^{-\e} g_{s,0},
\end{equation}
and by the chain rule we have
\begin{equation}
    \frac{\mu}{g_s} \frac{d g_{s}}{d \mu} = \frac{ d \ln g_s}{d \ln \mu} = - \frac{\e}{1 + \frac{g_s}{Z_{g_{s}}} \frac{d {Z_{g_s}} }{d g_s}}. 
\end{equation}
Defining $\alpha_s \equiv g_s^2/(4 \pi)$, this equation defines the so-called QCD beta function, which when $\alpha_s$ is small can be written as
\begin{equation}
    \label{eq:betaeq}
    \beta (\alpha_s) \equiv \frac{d \ln \alpha_s}{d \ln \mu^2} = \frac{ d \ln g_s}{d \ln \mu} = - \sum_{n=1}^\infty \beta_{n-1} \left( \frac{\alpha_s}{2 \pi} \right)^n. 
\end{equation}
The coefficients of the QCD beta function in principle depend on the renormalisation scheme and have been computed up to five loops in $\overline{\text{MS}}$~\cite{Baikov:2016tgj,Luthe:2016ima,Luthe:2017ttc,Luthe:2017ttg,Chetyrkin:2017bjc}. The leading coefficient is scheme independent and is given by
\begin{equation}
\beta_0 = N_c b_0 + N_F b_{0,F},
\end{equation}
where $b_0 = 11/6$, $b_{0,F} = -1/3$ and $N_F$ is the number of active quark flavours accessible at the energy scale. 
Note that in the scenario where $N_c = 3$ and $N_F \le 16$, $\beta_0 >0$, which means that $\alpha_s$ decreases for higher energy scales. We can see this more clearly if we assume $\alpha_s$ is small, truncate the series in Eq.~\eqref{eq:betaeq} at $\beta_0$ and solve the equation to give
\begin{equation}
    \label{eq:alphas}
    \alpha_s ( \mu^2) = \frac{\alpha_s (\mu_0^2)}{1 + \beta_0 \frac{\alpha_s (\mu_0^2)}{2 \pi} \ln \left( \frac{\mu^2}{\mu_0^2} \right)},
\end{equation}
which can be used to find $\alpha_s$ at a scale $\mu$ if it is known at a scale $\mu_0$. Hence, $\alpha_s$ must be extracted from experiment but importantly this equation demonstrates the behaviour of $\alpha_s$ at different scales. This behaviour is visualised in Fig.~\ref{fig:runningcoup}, including experimental extractions of $\alpha_s$ at a remarkably large range of scales. 
This is distinct from conventional field theories, like quantum electrodynamics (QED), where the coupling constant increases with energy scale. 
In QCD we therefore have the feature that at high momentum transfer, quarks and gluons are free; this is called asymptotic freedom. 
The low energy regime can be described as below $\Lambda_{\text{QCD}} \approx 200 \text{ MeV}$, where $\alpha_s$ becomes large. In this regime, we cannot use perturbative QCD and we encounter colour confinement, 
where quarks and gluons are bound in colour-neutral hadrons (baryons and mesons). Therefore we can only indirectly measure the properties of free quarks and gluons. 

At large energy scales, where the strong-coupling constant, $\alpha_s$, is small, we can use perturbation theory to approximate calculations which are too complex to undertake to all orders. We will now detail the perturbative framework used in high energy QCD calculations.

\section{Perturbation Theory in QCD}
\label{sec:aspert}

\begin{figure}[t]
    \centering
    \includegraphics[width=0.32\textwidth]{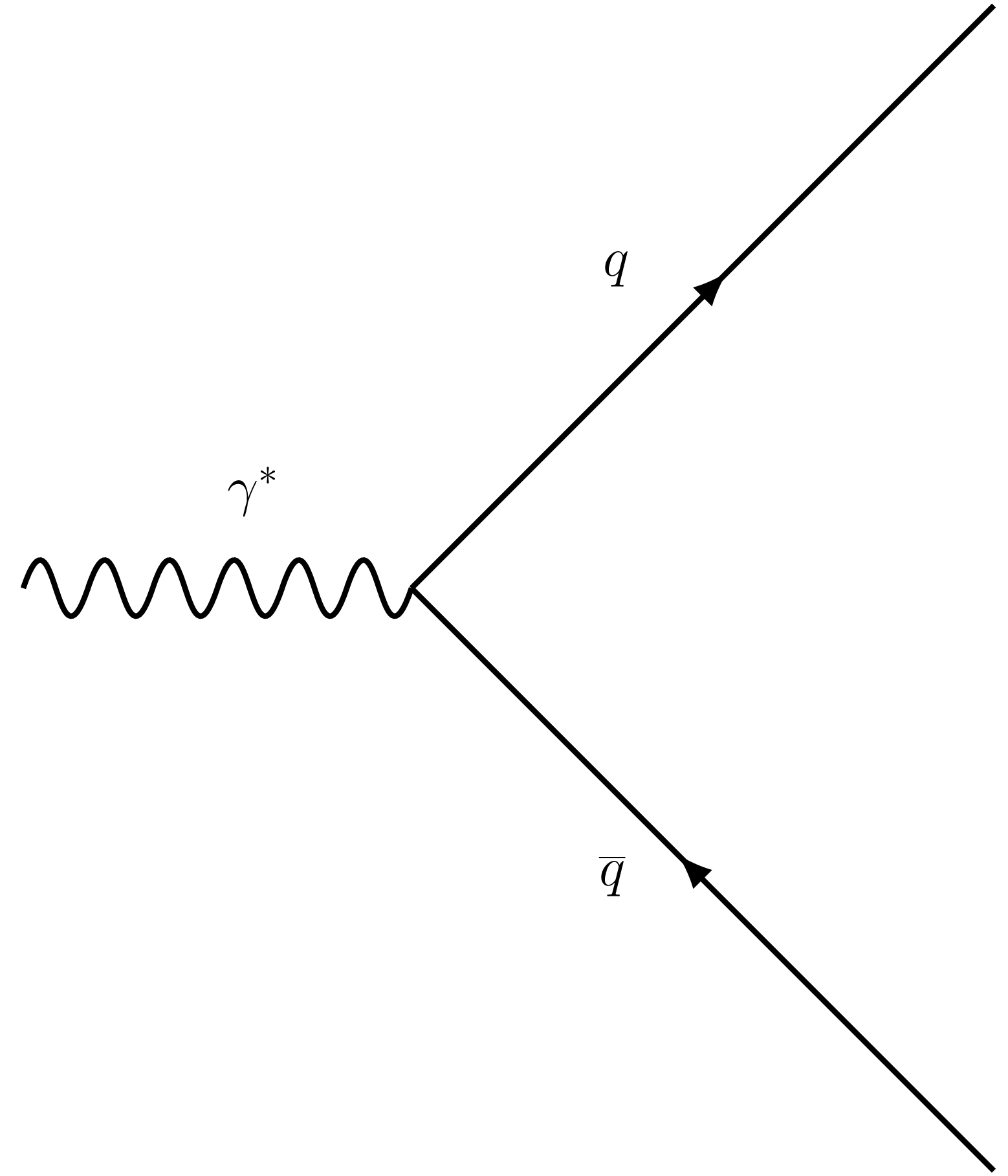}
    \includegraphics[width=0.32\textwidth]{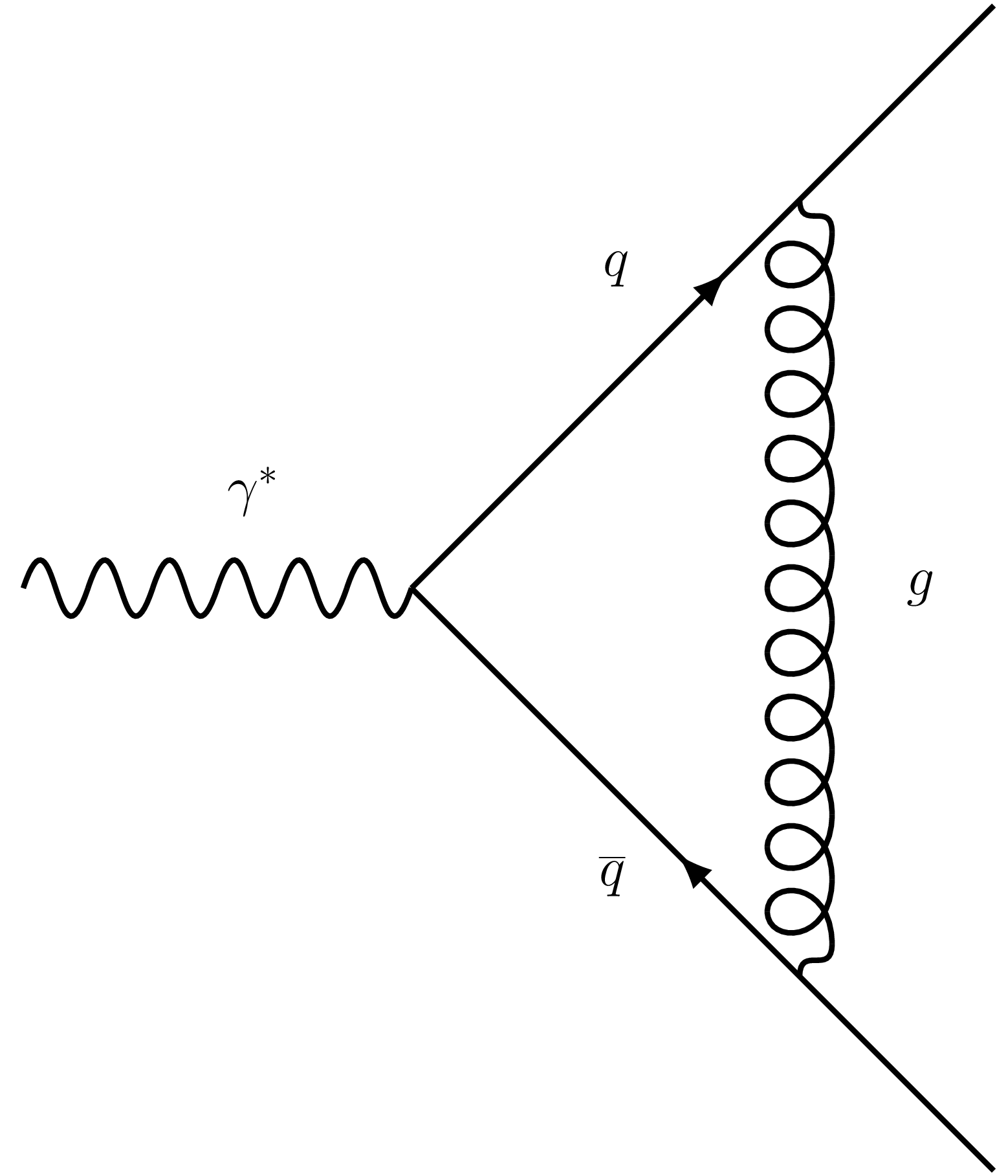}
    \includegraphics[width=0.32\textwidth]{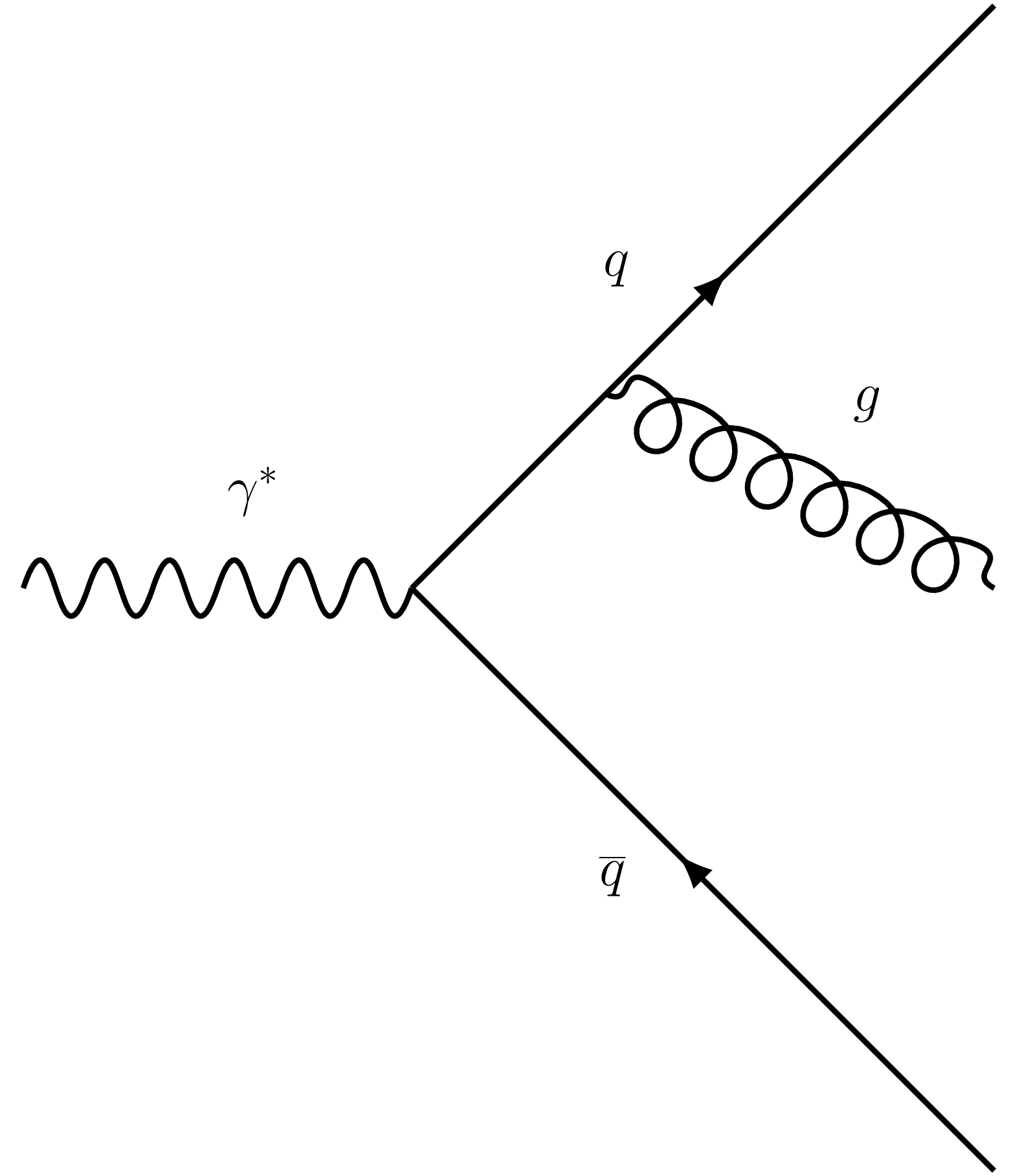}
    \caption{A small number of contributing diagrams to $ e^+e^- \to 2 \text{ jets}$. From left to right: the Born/LO diagram, the NLO loop diagram, one of the NLO unresolved emission diagrams. }
    \label{fig:NLOdiags}
\end{figure}

If we consider $e^+e^-$ colliders, like LEP, at large centre of mass energies, we can calculate approximate predictions for collider observables for different processes. For example, $ e^+e^- \to 2 \text{ jets}$. Jets can be defined as collimated beams of hadrons, for which colliders can make observations, having classified them according to jet algorithms~\cite{Catani:1993hr,Ellis:1993tq,Dokshitzer:1997in,Wobisch:1998wt,Blazey:2000qt,Cacciari:2008gp}. 
A calculation we may wish to perform is the cross section, $\sigma$, for this process. Experimentally, a cross section is defined as the rate of collisions $N$ for the given process, normalised by the particle luminosity $L$, such that,
\begin{equation}
    \sigma = \frac{d N}{dt} \frac{1}{L}.
\end{equation}
Theoretically, a cross section is extracted via a phase space integration over the matrix elements, averaged over initial states and summed over final states. In calculating the cross section we must include contributions from any Feynman diagrams which would result in the measurement of two jets, that is two resolvable coloured particles. 
Transitioning from the high energy collision scale to the low energy measurement scale, the coloured particles (or partons) will spread their energy through a parton shower and hadronisation process. They split into collinear groups and can emit soft (zero energy) radiation, until they form bound states (hadrons) at energy scales where $\alpha_s$ is large. Focussing again on the hard cross section, the tree-level diagrams are those which result in two jets with the smallest power of $\alpha_s$ - leading order (LO). At higher orders in $\alpha_s$, we can calculate corrections to the LO cross section, which should be progressively smaller because $\alpha_s$ is small at high energy scales. As such, we can form a perturbative expansion in $\alpha_s$, given by
\begin{eqnarray}
    {\sigma}(e^+e^- \to 2 \text{ jets}) &=& \left(\frac{\alpha_s}{2\pi}\right)^m {\sigma}^\text{LO} + \left(\frac{\alpha_s}{2\pi}\right)^{m+1} {\sigma}^\text{NLO}  \\
     &+ & \left(\frac{\alpha_s}{2\pi}\right)^{m+2} {\sigma}^\text{NNLO} 
    + \left(\frac{\alpha_s}{2\pi}\right)^{m+3} {\sigma}^\text{\NthreeLO} + \mathcal{O} (\alpha_s^{m+4}), \nonumber
\end{eqnarray}
where $m$ is determined by the vertices of the lowest-order contribution. 
The left-hand side of Fig.~\ref{fig:NLOdiags} shows the Born/LO diagram for the process, $ e^+e^- \to 2 \text{ jets}$. At next-to-leading order (NLO), there are two types of correction: virtual and real. 
Neglecting corrections in quantum electrodynamics (QED), the only diagram with one additional loop (virtual contribution) is the gluon-loop diagram in the centre of Fig.~\ref{fig:NLOdiags}. 
The gluon is an internal unobserved particle and therefore this diagram contributes to the cross section of interest. 
Again, if we suppress QED, the only diagrams with one additional unresolved real emission, are the diagram on the right-hand side of Fig.~\ref{fig:NLOdiags} and one where the gluon is emitted from the anti-quark. 
These real corrections must be included because the gluon can be unresolved; unresolved means the gluon either is soft (zero energy) or is collinear (same direction) as either the quark or the anti-quark. 
In both cases, experimentally, this will appear as two jets and so we must include this correction. 
The next-to-next-to-leading order (NNLO) correction consists in general of three contributions: double-virtual (VV), real-virtual (RV) and double-real (RR). At next-to-next-to-next-to-leading order (\NthreeLO), we have four contributions (VVV, RVV, RRV, RRR) etc. for higher orders. At higher orders in $\alpha_s$, we encounter greatly increasing complexity. There are two reasons for this. The first is that powerful methods for performing multi-loop calculations are needed at higher orders, especially when the LO diagram is loop-induced. 
The second is the presence of IR divergences of increasing complexity at higher orders. 
IR divergences due to loop integrals appear explicitly as poles in the dimensional regulator $\e$, while those due to unresolved real radiation appear implicitly in poles of vanishing momentum structures. 

\begin{figure}[t]
    \centering
    \includegraphics[width=0.66\textwidth]{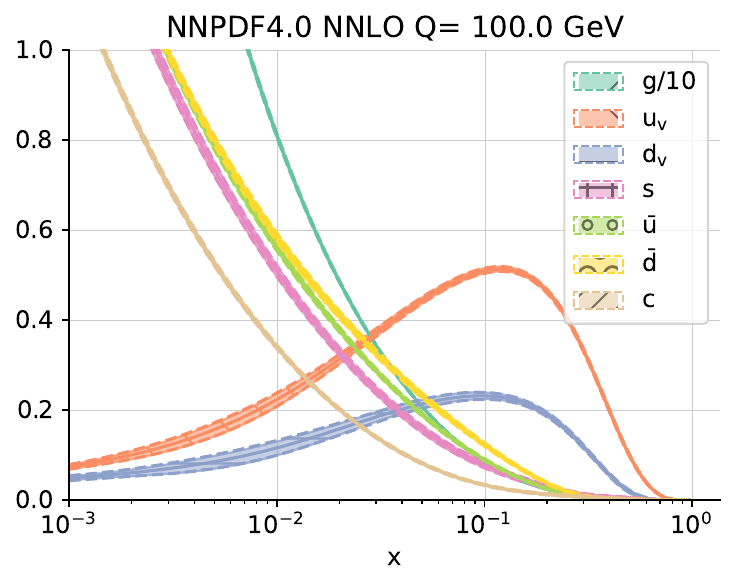}
    \caption{The NNPDF4.0 NNLO proton PDF at $Q=100.0$ GeV~\cite{NNPDF:2021njg}. Different colours represent the PDFs for different flavours of quark, anti-quark and gluon as a function of parton momentum fraction $x$, with error bars.}
    \label{fig:NNPDF}
\end{figure}

Before continuing this discussion, it is important to introduce the theory behind the other important classes of colliders: electron-proton ($ep$) and proton-proton ($pp$) colliders. 
$ep$ colliders, like HERA at DESY, and $pp$ colliders, like the LHC, have been highly successful at testing the theories of the Standard Model; we will explore this in Section~\ref{chapter:pheno}. 
However, to discuss these types of collisions, we need to introduce the parton model. 
The parton model is based on the assumption that when a hadron is accelerated to high energies, the constituents behave like free point-like particles; so quarks and gluons are the partons. 
In a scattering experiment, like deep inelastic scattering in an $ep$ collider, we assume that the electron scatters off a single parton and so the hard interaction is between the electron and the single parton. 
In order to relate the hard partonic cross section to the full hadronic cross section we need information about the binding of partons in the hadron. 
This information is formulated into parton distribution functions (PDFs), which give the number density for a parton of flavour $a$ to carry a momentum fraction $x_a$ of its parent hadron. 
PDFs are non-perturbative in nature because they are determined by low energy scale QCD phenomena. 
As such, PDFs are extracted by experiment, or by lattice QCD\footnote{Although we note that they are universal quantities.}. 
A specific extraction of a proton PDF is shown in Fig.~\ref{fig:NNPDF}, where both the valence quarks ($uud$) and the sea quarks and gluons are indicated. 
In calculations, we work with renormalised PDFs, which depend on the factorisation scale, $\mu_F$, and absorb IR divergences due to initial state configurations. 
The factorisation scale is arbitrary and separates the long distance physics, encoded in the PDF, from the short distance physics, encoded in the partonic cross section. 
Due to the particular running of the strong-coupling constant, the parton model is a good approximation for hadrons in high energy collisions and we can write QCD factorisation formulae. 
For the case of an $ep$ collider, we have 
\begin{equation}
\sigma_{eA} = \sum_a \int_0^1 d x_a f_{a|A} (x_a, \mu_F^2) \hat{\sigma}_{ea} (x_a, \mu_F^2) \left( 1 + \order{\Lambda_{\mathrm{QCD}}/Q} \right),
\end{equation}
where $A$ is the hadron, the sum runs over all parton flavours $a$, the integral is over the momentum fraction in the hadron, $f_{a|A} (x_a, \mu_F^2)$ is the PDF, $\hat{\sigma}_{ea} (x_a)$ is the partonic hard cross section and $Q$ is the momentum transfer from the electron to the parton. 
Note that it is the partonic scattering cross section, for an electron and parton $a$, which we calculate in perturbation theory. 
This collision does not happen at a well defined centre of mass energy, like in the $e^+e^-$ collider, because only a fraction of the hadron momentum takes part in the collision. 
As such, $ep$ (and $pp$) colliders can scan a range of partonic collision energies. 
This is also because we do not know what type of parton will take part in the collision. For the case of a $pp$ collider, we have that
\begin{equation}
    \sigma_{AB} = \sum_{a,b} \int_0^1 d x_a \int_0^1 d x_b f_{a|A} (x_a, \mu_F^2) f_{b|B} (x_b, \mu_F^2) \hat{\sigma}_{ab} (x_a,x_b, \mu_F^2) \left( 1 + \order{\Lambda_{QCD}/Q} \right),
\end{equation}
where $A$ and $B$ are the hadrons, partons $a$ and $b$ have momentum $x_a$ and $x_b$, in $A$ and $B$ respectively and they enter the partonic cross section $\hat{\sigma}_{ab} (x_a,x_b)$. 
We have given expressions here for full cross sections but similar relations hold for various differential observables. 

\section{Infrared Divergences}
\label{sec:IRdivs}

Finally we can discuss the finer details of IR divergences in QCD calculations. As mentioned earlier, IR divergences present themselves differently in virtual and real corrections. 
In virtual corrections, IR divergences appear due to the low energy limit of loop integrals and in CDR take the form of poles in $\e$. 
We sometimes call these {\em explicit} IR divergences. In real corrections, IR divergences appear due to the unresolved configurations and take the form of vanishing momentum structures (in the unresolved limits). We sometimes call these {\em implicit} IR divergences. 
This is because any suitable observable (we will tighten this definition later) will require integration over the phase space of the final state particles; when the real correction is integrated over the unresolved phase space, the implicit IR divergences become explicit in $\e$ poles. 
These IR divergences would be a barrier to numerical calculations and automation, if we could not demonstrate cancellation of the IR divergences in the overall result for an observable. 
Fortunately, in QCD, we have the Kinoshita, Lee, Nauenberg (KLN) theorem, which states that, for IR-safe observables, we have cancellation between real and virtual IR divergences~\cite{KLN1,KLN2}. 
This cancellation can be demonstrated at every order. 
IR-safe observables are functions of particle momenta, before phase space integration, and are inclusive with respect to soft emissions and collinear splittings. 
This means that for an IR-safe observable, $\mathcal{F}^{(n+1)} (\left\{ p_i\right\})$ depending on $(n+1)$ final state momenta, we have two types of feature:
\begin{eqnarray}
    &(\text{a})& \hspace{1cm} \mathcal{F}^{(n+1)} (...,p_i,...) \stackrel{i~{\rm soft}}{\longrightarrow} \mathcal{F}^{(n)} (...,p_{i-1},p_{i+1},...) ,\\
    &(\text{b})& \hspace{1cm} \mathcal{F}^{(n+1)} (...,p_i,p_j,...) \stackrel{i j~{\rm collinear}}{\longrightarrow} \mathcal{F}^{(n)} (...,(p_{i}+p_{j}),...) .
\end{eqnarray}
Total cross sections are of course IR-safe but less inclusive observables can also be IR-safe, including event selection cuts, detector geometry and differential cross sections.
Additionally, event shapes like thrust and number of jets are IR-safe. 

Unfortunately, demonstrating the cancellation of IR divergences is non-trivial, especially beyond NLO, because the different corrections need integrating over different multiplicity phase spaces. 
This is handled by a subtraction or slicing scheme, such as antenna subtraction, which will be introduced in Chapter~\ref{chapter:antennasub}. 
Subtraction schemes in general require knowledge of the universal IR singularity structure of matrix elements, which is the focus of the rest of this chapter. 

The structure of solely-virtual IR divergences is well documented at one and two loops, in terms of Catani's IR singularity operators~\cite{Catani:1998bh}. 
In the language of colour-ordered matrix elements, which we use throughout, it is convenient to recast Catani's structures in terms of integrated dipoles in colour space~\cite{Chen:2022ktf,Gehrmann:2023dxm}:
\begin{eqnarray}
    \label{eq:Jcoldef}
\Jcol{\ell} (\e) &=& \sum_{(i,j)}{\mathcal{J}}^{(\ell)}_2 (i,j) \textbf{T}_i\cdot\textbf{T}_j , \\ 
\Jcolb{2} (\e) &=&\sum_{(q,\bar{q})}\sum_{g}\overline{\mathcal{J}}^{(2)}_2\left(q,\bar{q}\right)\,\left(\textbf{T}_q+\textbf{T}_{\bar{q}}\right)\cdot\textbf{T}_g,
\end{eqnarray}
where $l=0,1$ indicates the number of loops and $i,j \in \{ q,\bar{q},g \}$. The operators, $\textbf{T}_i = (T_i)^a_{bc}$, are the generators $(t^a)_{bc}$ for quarks and the structure constants $f_{abc}$ for gluons. 
We further divide the ${\mathcal{J}}^{(\ell)}_2 (i,j)$ according to their colour-structures,
\begin{eqnarray}
\mathcal{J}^{(1)}_2\left(q,\bar{q}\right)&=&N_c\,\J{1}\left(q,\bar{q}\right), \\ 
\mathcal{J}^{(1)}_2\left(q,g\right)&=&N_c\J{1}\left(q,g\right)+N_F\,\Jh{1}\left(q,g\right), \\ 
\mathcal{J}^{(1)}_2\left(g,g\right)&=&N_c\J{1}\left(g,g\right)+N_F\,\Jh{1}\left(g,g\right),  
\end{eqnarray}
and
\begin{eqnarray}
\mathcal{J}^{(2)}_2\left(q,\bar{q}\right)&=&N_c^2\,\J{2}\left(q,\bar{q}\right)-\Jt{2}\left(q,\bar{q}\right)+N_c N_F\Jh{2}\left(q,\bar{q}\right), \\ 
\mathcal{J}^{(2)}_2\left(q,g\right)&=&N_c^2\,\J{2}\left(q,g\right)+N_c N_F\,\Jh{2}\left(q,g\right)-\dfrac{N_F}{N_c}\Jht{2}\left(q,g\right)+N_F^2\,\Jhh{2}\left(q,g\right), \hspace{1.5cm} \\ 
\mathcal{J}^{(2)}_2\left(g,g\right)&=&N_c^2\,\J{2}\left(g,g\right)+N_c N_F\,\Jh{2}\left(g,g\right)-\dfrac{N_F}{N_c}\Jht{2}\left(g,g\right)+N_F^2\,\Jhh{2}\left(g,g\right), \hspace{1.5cm} \\ 
\overline{\mathcal{J}}^{(2)}_2\left(q,\bar{q}\right)&=&\dfrac{N_c^2}{2}\,\Jb{2}\left(q,\bar{q}\right)-\dfrac{1}{2}\,\Jt{2}\left(q,\bar{q}\right) .
\end{eqnarray}

\begin{figure}[th!]
    \centering
\begin{tikzpicture} 
        \begin{feynman}
        \vertex (a); 
        \vertex [right=2em of a] (a0);
        \vertex [right=10em of a] (b);
        \vertex [below right=7em of b]  (f3);
        \vertex [above right=7em of b]  (d) ; 
        \filldraw (a) circle (25pt);
        \diagram* {
        (a0) -- [momentum=\( p_i + p_j \)] (b),
        (b) -- [momentum=\( p_i \)] (d), 
        (b) -- [momentum=\( p_j \)] (f3),
        };
        \end{feynman}
        \end{tikzpicture}
        \caption{Generic diagram, where massless final states $i$ and $j$ result from an internal propagator with momentum $(p_i + p_j)$. }
        \label{fig:prop}
\end{figure}
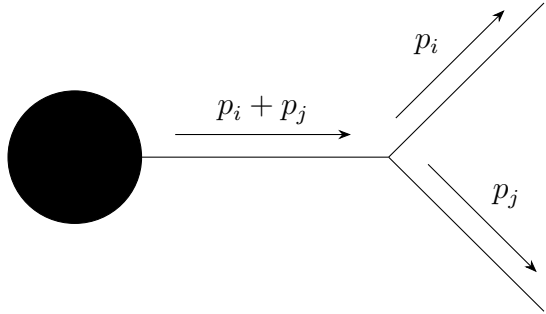

We will now detail the structure of IR divergences in matrix elements when there is at least one real emission and we focus on final-state (time-like) limits. We will also present formulae in massless QCD for the remainder of this thesis. This is where there are $N_F$ massless quarks, (naturally) massless gluons and $(6-N_F)$ heavy quarks, which are inaccessible at the relevant energy scales. 
We define Lorentz invariant momentum quantities,
\begin{equation}
	s_{i,\ldots,n} \equiv (p_{i}+...+p_{n})^2.
\end{equation}
In the massless limit, $s_{ij} = 2p_i \cdot p_j = 2E_iE_j(1-\cos\theta_{ij})$, where $E_i$, $E_j$ are the energies of particles $i$, $j$ and $\theta_{ij}$ is the angle between them. 
Consider a typical propagator on a Feynman diagram with $i$ and $j$ connected to an internal propagator with momentum $(p_i + p_j)$, as illustrated in Fig.~\ref{fig:prop}.
This propagator provides a denominator,
\begin{equation}
\frac{1}{(p_i + p_j)^2 + i \phi} = \frac{1}{s_{ij} + i\phi} = \frac{1}{2 E_i E_j (1- \cos \theta_{ij}) + i\phi},
\end{equation}
which diverges if either particle is soft or they are collinear. 
In both of these limits, $s_{ij} \to 0$, so $s_{ij}$ is a very convenient structure for describing real IR singularities. 
We now have the tools to detail the singular factors of squared colour ordered matrix elements in the relevant unresolved limits listed below. 
The factorisation formula for a tree-level matrix element with $n$ final state partons in an $m$-unresolved limit is, in general, given by
\begin{equation}
    M_n^0 \equiv |\mathcal{M}_n^0|^2  \to V_m^0 M_{n-m}^0,
\end{equation}
where $m$ is the number of colour-connected unresolved partons, $V_m^0$ is the IR singular factor (listed below) and $\mathcal{M}_{n-m}^0$ is the reduced matrix element, with $(n-m)$ final state partons and a relevant reparametrisation of the momenta. If the unresolved partons are not colour-connected, the unconnected limits can be taken sequentially and the overall singular factor is given by a product of the colour-connected factors. 

The universal factorisation properties of multi-particle matrix elements are important for generating counterterms that can be used to isolate the infrared singularities that are produced in particular regions of phase space, when one or more particles are unresolved\footnote{See Ref.~\cite{Agarwal:2021ais} for a review. These factorisation properties are also key in quantifying the accuracy of parton branching algorithms in event generators and how these algorithms can eventually be extended to increase their logarithmic accuracy, see for example Refs.~\cite{Li:2016yez,Hoche:2017iem,Dulat:2018bfe,Dulat:2018vuy,Dasgupta:2020fwr,Loschner:2021keu,Gellersen:2021eci}}. 
Most well studied are the single unresolved limits, where either one particle is soft, or two are collinear, which are relevant for NLO calculations.  
At NNLO, one is concerned with the double unresolved limits of tree-level matrix elements~\cite{campbell,Catani:1998nv,Catani:1999ss,Kosower:2002su}, as well as the single unresolved limits of one-loop matrix elements~\cite{Bern:1994zx,Bern:1998sc,Kosower:1999rx,Bern:1999ry}.  

\subsection{Tree-level Single Unresolved Limits}
\label{sec:tree-single-lims}

Our notation is that $i,j,k$ are momenta for three colour-connected partons of particle type $a,b,c$. The tree-level soft factors are given by the following for a soft particle with momentum $j$, radiated between two hard radiators with momenta $i$ and $k$,
\begin{align}
    \label{eq:Sg}
    \Sg(i^h,j_g,k^h) &= \Sp(i^h,j_{\gamma},k^h) =   \frac{2s_{ik}}{s_{ij}s_{jk}},  \\
    \Sq(i^h,j_q,k^h) &= \Sqb(i^h,j_{\bar{q}},k^h) =0.
\end{align}
The superscript $h$ indicates when the particle is hard, while the subscript on the $j$ indicates its particle type.  

The tree-level splitting functions $P_{ab}^{(0)}(i^h,j)$ only contain singular configurations consistent with $b$ being unresolved, ie. $j$ soft and $ij$ collinear. 
They are related to the usual spin-averaged splitting functions, cf.~\cite{Altarelli:1977zs,Dokshitzer:1977sg}, by
\begin{align} 
\label{eqn:Pqg}
\Pqgzero(i^h,j) &= \frac{1}{s_{ij}} \Pqg(\xj), \\
\Pqgzero(i,j^h) &= 0,\\
\label{eqn:Pqq}
\Pqqzero(i^h,j) &= \frac{1}{s_{ij}} \Pqq(\xj),\\
\Pqqzero(i,j^h) &= \frac{1}{s_{ij}} \Pqq\omxj,\\
\label{eqn:Pgg}
\Pggzero(i^h,j) &= \frac{1}{s_{ij}} \PggS(\xj),\hfill\\
\label{eqn:Pggalt}
\Pggzero(i,j^h) &= \frac{1}{s_{ij}}\PggS\omxj ,
\end{align}
with
\begin{eqnarray}
\label{eq:Pqg}
\Pqg(\xj) &=& \left(\frac{2\omxj}{\xj} + \ome \xj \right) , \\
\label{eq:Pqq}
\Pqq(\xj) &=& \left( 1 -\frac{2\omxj\xj}{\ome} \right) = \Pqq\omxj , \\
\label{eq:PggS}
\PggS(\xj)&=& \left( \frac{2\omxj}{\xj} + \xj \omxj \right) , 
\end{eqnarray}
and
\begin{equation}
\PggS(\xj) + \PggS\omxj \equiv \Pgg(\xj).
\end{equation}
Here, $\xj$ is the momentum fraction of parton $j$ in the $ij$ collinear pair. 

\subsection{Tree-level Double Unresolved Limits}
\label{sec:notation}

First, we consider the singular factors for the colour-ordered partons $i^h,j,k,l^h$, where $j,k$ are soft. The colour-connected double-soft limit is defined as the kinematic regime where the invariants $s_{ij},$ $s_{jk}$, $s_{ik}$, $s_{jl}$, $s_{kl}$, $s_{ijk}$, $s_{jkl}$ become small. 
The tree-level colour connected double-soft factors are given by 
\begin{eqnarray}
    \label{eq:Sggtwo}
    \Sgg(i^h,j,k,l^h) &=&
    \frac{2 s_{ik}}{s_{ij}s_{jk}} 
    \frac{2 s_{il}}{s_{ijk}s_{jkl}}
    +
    \frac{2 s_{jl}}{s_{kl}s_{jk}} 
    \frac{2 s_{il}}{s_{ijk}s_{jkl}}
    \nonumber \\
    &&+ \frac{2s_{il} \Tr{i}{l}{j}{k} }{s_{ij}s_{jk}s_{kl}s_{ijk}s_{jkl}}
    + \frac{2\ome 
    \left(s_{ij}s_{kl}-s_{ik}s_{jl}\right)^2}
    {s_{jk}^2s_{ijk}^2s_{jkl}^2}, \\
    \label{eq:Spp}    
    \Spp(i^h,j,k,l^h) &=& \frac{2 s_{il}}{s_{ij}s_{jl}}
    \frac{2 s_{il}}{s_{ik}s_{kl}}, \\
    \label{eq:Sqq}
    \Sqq(i^h,j,k,l^h) &=&
    \frac{2 s_{il}}{s_{jk}s_{ijk}s_{jkl}} 
    -\frac{2
    \left(s_{ij}s_{kl}-s_{ik}s_{jl}\right)^2}{s_{jk}^2s_{ijk}^2s_{jkl}^2},
\end{eqnarray}
and zero otherwise.
Here we write the two-gluon soft factor in a suggestive manner such that the first two terms are iterations of the single-soft-gluon eikonal factor and
\begin{equation}
    \Tr{i}{l}{j}{k} = s_{il}s_{jk}-s_{ij}s_{kl}+s_{ik}s_{jl} \, .
\end{equation}

We also present the time-like triple-collinear limits of colour-connected particles that were first discussed in Ref.~\cite{campbell}. 
Note that in this thesis we consider spin-averaged collinear limits, which are directly obtained by taking the collinear limit of colour-ordered squared matrix elements. 
One could equivalently work in colour space and retain information about the spin of the parton formed from the merger of the collinear particles, as was done in Refs.~\cite{Catani:1998nv,Catani:1999ss}. 
The spin-unaveraged splitting functions contain additional azimuthal correlations when the parent parton is a gluon.
These reflect different orientations of the final state particles with respect to the gluon polarisation (and effectively with respect to other particles not involved in the triple-collinear limit). 
These azimuthal correlations are not present in the case where the parent parton is a quark, since the splitting function is proportional to the identity operator in spin space. 

We consider three colour-connected partons of particle type $a,b,c$ with four-momenta $i,j,k$, which become collinear in a process involving four or more partons. 
There are also configurations in which particles that are not colour-connected can usefully be thought of as colour-connected. 
This happens when there is more than one colour-string - there is an antiquark at the end of one colour-string and a like flavour quark at the beginning of another. For example, the matrix element
\begin{equation}
    M_n^0 (...,\bar{q}|q,...)
\end{equation}
represents a situation where there are two colour-strings, one terminated by the fundamental colour index of the $\bar{q}$ and another initiated by the fundamental colour index of the $q$. In this case, when the quark-antiquark pair are collinear, they combine to form a gluon, which then connects, or pinches together, the two colour-strings,
\begin{equation}
\label{eq:TC2}
	M_n^0(...,i,j|k...) \rightarrow P_{a\bar{q}q}^{(0)}(i,j,k) M_{n-2}^0(..,P,...).
\end{equation}
In the triple-collinear limit, the collinear cluster has momentum 
\begin{equation}
p_i^{\mu} + p_j^{\mu}+ p_k^{\mu} = p_P^\mu.
\end{equation}
The triple-collinear limit is defined as the kinematic regime where the invariants $s_{ij},$ $s_{jk}$, $s_{ik}$, $s_{ijk}$ become small and therefore $p_P^2 \sim 0$. In this limit, we can write $p_i = x_i p_P$, $p_j = x_j p_P$ and $p_k = x_k p_P$ with $x_i+x_j+x_k = 1$.  
The particle $P$ retains the quantum numbers of the collinear partons and there are seven possible clusterings: $\nameggg$, $\nameqgg$, $\nameqpp$, $\namegqbq$, $\namepqbq$, $\nameqQQ$ and $\nameqqq$, where we have distinguished quarks of different flavour by $q$ and $Q$.  
The presence of the photon, $\gamma$, is to indicate the `QED-like' triple collinear limits in sub-leading colour squared matrix elements. One can consider the photons to be abelianised gluons or colour-unconnected gluons. In the case of $\nameqpp$, each $\gamma$ is equally connected to $q$ and not to the other $\gamma$. 
The triple-collinear splitting functions depend on the momentum fractions and the small invariants.  However, for brevity we will suppress these arguments and use a shorthand notation,
\begin{equation}
\label{eq:Pabcdecomp}
P_{abc}^{(0)} (i,j,k) \equiv P_{abc}^{(0)}(x_i, x_j, x_k; s_{ij},s_{ik},s_{jk},s_{ijk}).
\end{equation} 
Triple-collinear splitting functions and their structure are the focus of Chapter~\ref{chapter:paper1}, thus we will not list their expressions here. 
Raw expressions for the spin-averaged tree-level time-like triple-collinear splitting functions were first detailed in Ref.~\cite{campbell} and unaveraged in Ref.~\cite{Catani:1998nv}.

The final types of double unresolved limits are iterated single unresolved limits. 
If the particles involved in the two single unresolved limits are colour-unconnected, then the singular factor is the product of the two single unresolved factors. 
In the colour-connected double-collinear limit, the singular factor also takes a simple form. 
That is, in the limit when two pairs of colour-connected particles become collinear, the singular factor is a product of two simple-collinear splitting functions. 
The final limit is the colour-connected soft-collinear limit, see Refs.~\cite{campbell,Catani:1999ss}, more specifically the $j$ soft and $kl$ collinear limit of a colour-ordered matrix element with ordering, $...,i,j,k,l,...$. 
In this limit the singular factor is given by 
\begin{equation}
    S_{i;jkl}^{(0)} (x_k;i,j,k,l) \frac{1}{s_{kl}} P_{kl}^{(0)} (1-x_k),
\end{equation}
where $x_k$ is the momentum fraction of $k$ in the $kl$ collinear pair and 
\newpage
\begin{eqnarray}
    S_{i;jkl}^{(0)} (x_k;i,j,k,l) &=& \frac{(s_{ik} + s_{il})}{s_{ij} s_{jk}} \left(x_k + \frac{s_{jk} + x_k s_{kl}}{s_{jkl}} \right) \nonumber \\
    &=& \frac{(s_{ik} + s_{il})}{s_{ij}} \left\{ \frac{2}{s_{jk} + s_{jl}} + \left(1 + \frac{s_{kl}}{s_{jkl}} \right) \left[ \frac{x_k}{s_{jk}} - \frac{1}{s_{jk} + s_{jl}} \right]\right\} \nonumber \\
    &\approx& \frac{2 (s_{ik} + s_{il})}{s_{ij} (s_{jk} + s_{jl})} = \Sgzero (i^h,j,(k+l)^h),
\end{eqnarray}
where last line is due to the squared bracket vanishing in the $kl$ collinear limit and we encounter the expected soft-collinear factor,
\begin{equation}
    \Sgzero (i^h,j,(k+l)^h) \frac{1}{s_{kl}} P_{kl}^{(0)} (1-x_k). 
\end{equation}

\subsection{One-loop Single Unresolved Limits}
\label{sec:limits}

The factorisation formula for a one-loop matrix element with $n$ final state partons, in a single unresolved limit, takes a different form to the tree-level case,
\begin{equation}
    \label{eq:1loopfact}
    M_n^1 \equiv \langle \mathcal{M}_n^1 | \mathcal{M}_n^0 \rangle + \langle \mathcal{M}_n^0 | \mathcal{M}_n^1 \rangle \to V_1^1 M_{n-1}^0 + V_1^0 M_{n-1}^1 ,
\end{equation}
where ${M}_{n-1}^\ell$ is the reduced squared matrix element with $\ell$ loops, $(n-1)$ final state partons and a relevant reparametrisation of the momenta. Note that the matrix element, $|\mathcal{M}_n^1 \rangle$, is a renormalised colour-ordered matrix element. $V_1^0$ takes the form of the tree-level single unresolved factors given above, while $V_1^1$ are the one-loop single unresolved factors given below. Note that we will use the convenient definition for ``squared matrix elements'', which include all interference terms with $\ell$ loops, for example,
\begin{equation}
    \label{eq:2loopstruc}
    M_n^2 \equiv \langle \mathcal{M}_n^2 | \mathcal{M}_n^0 \rangle + \langle \mathcal{M}_n^0 | \mathcal{M}_n^2 \rangle + \langle \mathcal{M}_n^1 | \mathcal{M}_n^1 \rangle,
\end{equation}
where $|\mathcal{M}_n^2 \rangle$ is also a renormalised colour-ordered matrix element. 

The universal soft and collinear factorisation properties of multi-particle real-virtual matrix elements have been well studied in the literature~\cite{Bern:1994zx,Bern:1998sc,Kosower:1999rx,Bern:1999ry}.
In this section, we list the unrenormalised, colour-ordered single unresolved factors at one loop in CDR, which are consistent with the formulations in Refs.~\cite{Kosower:1999rx,Catani:2000pi,Weinzierl:2003ra,Badger:2004uk,Gehrmann-DeRidder:2005btv,Bern:1999ry,Duhr:2013msa}. 
The overall factor $\Af$ is given by
\begin{equation}
\label{eq:Afdef}
\Af = e^{\e \gamma_E} \frac{\Gamma^2(1-\e) \Gamma(1+\e)}{\Gamma(1-2\e)} \text{Re} \left[ (-1)^{-\e} \right] ,
\end{equation}
depending on the traditional gamma functions, $\Gamma (x)$, defined in Appendix~\ref{app:integration}. 
We also use the convenient notation, 
\begin{equation}
    S_{ij} = \left(\frac{s_{ij}}{\mu^2}\right)^{-\e}.
\end{equation}

The full-colour (unrenormalised) one-loop soft factor is given by 
\begin{equation}
    \mathbf{S}_g^{(1)} (i^h,j,k^h) = N_c \Sgone(i^h,j,k^h) - \frac{1}{N_c} \Sgtone(i^h,j,k^h) + N_F \Sghone(i^h,j,k^h),
\end{equation}
where
\begin{align}
\label{eq:Sg1}
    \Sgone(i^h,j,k^h) &= - \Af\frac{\Gamma(1-\e)\Gamma(1+\e)}{\e^2} \frac{S_{ij}S_{jk}}{S_{ik}}  \Sgzero (i^h,j,k^h)  \, , \\
    \Sgtone(i^h,j,k^h) &= 0 \, , \\
    \Sghone(i^h,j,k^h) &= 0 \, ,    
\end{align}
and formally we define any soft factor, where the particle type of $j$ is not a gluon, as zero. The tree-level single unresolved limits are given in Eqs.~\eqref{eq:Sg}-\eqref{eqn:Pggalt}.

In general, the full-colour (unrenormalised) one-loop splitting function is colour-decomposed according to
\begin{equation}
    \mathbf{P}_{ab}^{(1)}(i,j) = N_c P_{ab}^{(1)}(i,j) - \frac{1}{N_c} \widetilde{P}_{ab}^{(1)}(i,j) + N_F \widehat{P}_{ab}^{(1)}(i,j) \, .
\end{equation}
As at tree-level, we organise the splitting functions by which particle is a hard-radiator.  
This means that $\mathbf{P}_{ab}^{(1)}(i^h, j)$ only contains singular configurations consistent with $b$ being unresolved ($j$ soft and $ij$ collinear) and is directly related to the usual spin-averaged one-loop splitting functions given in terms of the momentum fraction carried by particle $j$ ($\xj$).

The $q \to q g$ one-loop splitting functions for $i$ being the hard radiator are given by
\begin{align}
    \Pqgone(i^h,j)  &= \frac{1}{s_{ij}} \Pqgone(\xj)\, ,  \\
    \Pqgtone(i^h,j) &= \frac{1}{s_{ij}} \Pqgtone(\xj)\, ,  \\
    \Pqghone(i^h,j) &= 0 \, ,
\end{align}
where
\begin{align}
    \Pqgone(\xj) &= S_{ij} \frac{\Af}{\e^2} \bigg[ - \Gamma(1-\e) \Gamma(1+\e) \left(\frac{1-x_j}{x_j}\right)^\e + G\left(\frac{x_j}{1-x_j},\e\right) \bigg] \Pqg(\xj) \nonumber \\
    &\hspace{0.5cm} +  S_{ij} \Af\frac{(1-x_j \e )}{2(1-2\e)}
    \, , \label{eq:Pqg1} \\
    \Pqgtone(\xj) &= 
     - S_{ij} \frac{\Af}{\e^2}  G\left(\frac{x_j}{1-x_j},-\e\right) \Pqg(\xj) -  S_{ij} \Af \frac{(1-x_j \e )}{2(1-2\e)}
    \, .
\end{align}
Here $G(w,\e)$ is defined as a class of hypergeometric functions for brevity,
\begin{eqnarray}
\label{eq:Ghyperdef}
G(w,\e) &=&  {}_2F_1\left(1,\e,1+\e,-w\right) -1  , \nonumber \\
&=& - \sum_{n=1}^{\infty} (-\e)^n \Li_{n}\left(-w\right), \nonumber \\
&\equiv& (1+w)^{-\e} {}_2F_1\left(\e,\e,1+\e,\frac{w}{1+w}\right) -1 ,
\end{eqnarray}
where we have given $G(w,\e)$ in three useful forms, including in terms of polylogarithms, $\Li_n (-w)$. 
The definitions for the functions ${}_2F_1$ and $\Li_n (-w)$ are given in Appendix.~\ref{app:integration}. 
Note that in the $w \to 0$ limit, $G(w,\e)$ vanishes. It has the $\e$-expansion,
\begin{equation}
    G\left(\frac{x_j}{1-x_j},\e\right) = \e \ln\omxj -\e^2 \Li_2\left(\frac{-\xj}{1-\xj}\right) + \order{\e^3}.
\end{equation}

The tree-level splitting function $\Pqg$ is given in Eq.~\eqref{eq:Pqg}.
In the complementary case of $j$ being the hard radiator, all splitting functions vanish identically,
\begin{equation}
    \Pqgone(i,j^h) = 0 \, , \quad
    \Pqgtone(i,j^h) = 0 \, , \quad
    \Pqghone(i,j^h) = 0 \, . 
\end{equation}

The $g \to q \qb$ one-loop splitting functions are 
\begin{align}
    \Pqqone(i^h,j)  &= \frac{1}{s_{ij}} \Pqqone(\xj)\, , \\
    \Pqqtone(i^h,j) &= \frac{1}{s_{ij}} \Pqqtone(\xj) \, , \\
    \Pqqhone(i^h,j) &= \frac{1}{s_{ij}} \Pqqhone(\xj) \, ,
\end{align}
and
\begin{align}
    \Pqqone(i,j^h)  &= \frac{1}{s_{ij}} \Pqqone\omxj , \\ 
    \Pqqtone(i,j^h) &= \frac{1}{s_{ij}} \Pqqtone\omxj \, , \\
    \Pqqhone(i,j^h) &= \frac{1}{s_{ij}} \Pqqhone\omxj \, ,
\end{align}
where
\begin{align}
    \Pqqone(\xj) &=  S_{ij} \frac{\Af}{ \e ^2} \bigg[ - \Gamma(1-\e) \Gamma(1+\e) \left( \frac{1-x_j}{x_j} \right)^\e + 1 + G\left(\frac{x_j}{1-x_j},\e\right) \nonumber \\ 
    & \hspace{1.75cm} - G\left(\frac{x_j}{1-x_j},-\e\right)  + \frac{\e (13 - 8 \e)}{2(3-2\e)(1-2\e)} \bigg] \Pqq (\xj) \, , \\    
    \Pqqtone(\xj) &= - S_{ij} \Af \bigg[   \frac{1}{ \e^2} + \frac{3+2\e}{2 \e (1-2 \e)}   \bigg] \Pqq (\xj) \, , \\
    \Pqqhone(\xj) &=  S_{ij} \Af \bigg[ 
    - \frac{2(1-\e)}{\e (3-2 \e)(1-2\e)} \bigg] \Pqq (\xj)  \, ,
\end{align}
with $\Pqq$ defined in Eq.~\eqref{eq:Pqq}. Note that the symmetry of $\Pqqone$ is preserved;
\begin{equation}
     \Pqqone\omxj \equiv  \Pqqone(\xj).
\end{equation}

Finally, the $g \to g g$ one-loop splitting functions for $i$ being the hard radiator are given by
\begin{align}
    \Pggone(i^h,j) &= \frac{1}{s_{ij}} \Pggonesub (\xj) \, ,  \\
    \Pggtone(i^h,j) &= 0 \, , \\
    \Pgghone(i^h,j) &= \frac{1}{s_{ij}} \Pgghonesub(\xj), \, 
\end{align}
while when $j$ is the hard radiator,
\begin{align}
    \Pggone(i,j^h) &= \frac{1}{s_{ij}} \Pggonesub\omxj \, , \\
    \Pggtone(i,j^h) &= 0 \, , \\
    \Pgghone(i,j^h) &= \frac{1}{s_{ij}} \Pgghonesub\omxj \, .
\end{align}
Here, the one-loop splitting functions are given in terms of the tree-level splitting function $\PggS$, defined in Eq.~\eqref{eq:PggS},
\begin{align}
    \Pggonesub(\xj) &= S_{ij} \frac{\Af}{\e^2} \bigg[ - \Gamma(1-\e) \Gamma(1+\e) \left(\frac{1-x_j}{x_j}\right)^\e + G\left(\frac{x_j}{1-x_j},\e\right) 
\nonumber \\
    &\hspace{1.5cm}    - G\left(\frac{x_j}{1-x_j},-\e\right) \bigg] \PggSzero (\xj) \nonumber \\
    &\hspace{0.5cm} + S_{ij} \Af \frac{(1-2 x_j (1-x_j) \e)}{2(1-\e)(1-2\e)(3-2\e)}
    \, , \\
    \Pgghonesub(\xj) &= S_{ij} \Af \left(\frac{-(1-2 x_j (1-x_j) \e)}{2(1-\e)^2(1-2\e)(3-2\e)}\right) \, , 
\end{align}
and they satisfy
\begin{align}
   \Pggone(\xj) &= \Pggonesub(\xj) +  \Pggonesub\omxj \, , \\
   \Pgghone(\xj) &= \Pgghonesub(\xj) +  \Pgghonesub\omxj \, .
\end{align}

\subsection{\NthreeLO Unresolved Limits}

At \NthreeLO, one encounters the triple unresolved limits of tree-level matrix elements~\cite{Catani:2019nqv,DelDuca:1999iql,DelDuca:2019ggv,DelDuca:2020vst,DelDuca:2022noh}, the double unresolved limits of one-loop matrix elements~\cite{Catani:2003vu,Sborlini:2014mpa,Badger:2015cxa,Zhu:2020ftr,Catani:2021kcy,Czakon:2022fqi} and the single unresolved limits of two-loop matrix elements~\cite{Bern:2004cz,Badger:2004uk,Duhr:2014nda,Li:2013lsa,Duhr:2013msa}. 
These expressions are very large and complex and we do not need to reproduce them for the purposes of this thesis. 

\section{Summary}
\label{sec:QCDsummary}

The theoretical exploration of QCD stems from the definition of the fields, masses and couplings in the QCD Lagrangian. Quarks and gluons receive their couplings and self-couplings from these terms, from which we can deduce Feynman rules for the calculation of QCD matrix elements. 
We explained that we decompose full matrix elements into sums of products of colour structures and colour-ordered matrix elements. 
For the purposes of this thesis, we are often concerned with the IR divergent structure of these colour-ordered matrix elements. 
UV divergences were introduced as the result of high energy limits in loop integrals and regularisation as a way of parametrising the divergences. 
We also explained how renormalisation absorbs the UV divergences into Lagrangian counterterms, leaving us with physical UV-finite renormalised fields.
As part of renormalisation, we encountered the running of the strong-coupling constant, which explained asymptotic freedom and colour confinement in QCD.
At large energy scales, we showed that we can calculate observables approximately by calculating a fixed number of leading terms in a perturbative expansion in the strong-coupling. 
We also introduced the concepts of real and virtual corrections to the leading order contribution to an observable. 
This led to a discussion of the differing physics in $ep$ and $pp$ colliders compared to $e^+ e^-$ colliders, including the definition of hadron PDFs. 
Finally we described how IR divergences appear explicitly in $\e$ poles in virtual corrections and implicitly in $s_{ij}$ poles in real corrections.
By the KLN theorem, these IR divergences cancel order-by-order in the strong-coupling expansion for IR-safe observables. 
We introduced notation for the IR structure at one and two loops and also detailed the unresolved limits for use in NLO and NNLO calculations. 
This chapter gives us the necessary toolkit to describe the research presented in this thesis. However, we will spend Chapter~\ref{chapter:pheno} motivating the research as part of greater efforts in particle physics phenomenology. 

\chapter{Phenomenology}
\label{chapter:pheno}

In this chapter, we explain why higher order perturbative calculations in QCD are important. We do this by placing the calculations in their phenomenological context. 
In Section~\ref{sec:higherorders}, we overview the theoretical benefits of higher order calculations and the experimental motivation for predictions with greater precision. 
Next we review the state-of-the-art in calculations for direct phenomenological comparison in Section~\ref{sec:phenocomp}. 
We also remark, in this section, on the excellent agreement between experiment and Standard Model predictions. 
In Section~\ref{sec:errors}, we introduce and evaluate the estimation of uncertainties in fixed-order calculations. 
We also compare this to other sources of error in calculating a prediction for an observable. 
Finally, we summarise the chapter in Section~\ref{sec:phenosummary}. 

\section{Why Higher Orders?}
\label{sec:higherorders}

We start with a quote from the European Strategy for Particle Physics (2020,\cite{CERN-ESU-015}):
\begin{quote}
    Theoretical physics is an essential driver of particle physics that opens new, daring lines of research, motivates experimental searches and provides the tools needed to fully exploit experimental results ... The success of the
    field depends on dedicated theoretical work and intense collaboration between the theoretical and experimental communities. 
\end{quote}
Theoretical research in particle physics encapsulates a large and varied field, including fixed-order calculations, parton distribution functions, parton showers and the modelling of non-perturbative effects, as well as investigating models for new physics. 
Precise theoretical predictions that are adapted to the specific experimental observables and that match the accuracy of the experimental measurements are needed to extract fundamental Standard Model parameters.  
Typically, theoretical predictions are obtained using perturbation theory as an expansion in the coupling.  
The precision of the theoretical predictions is generally limited by a dependence on unphysical renormalisation and factorisation scales, or through the modelling of complicated final states with relatively few final state particles. 
This can be systematically improved by including higher-order corrections. 
The LO prediction captures the gross features of an observable. 
For example, an LO $2 \to 2$ calculation cannot result in final states with transverse momentum and so higher orders are needed to approximate the transverse momenta of final states. 
Inclusion of NLO corrections is also required to estimate the normalisation of the predictions. 
Even higher orders (NNLO, \NthreeLO, \ldots) are needed to describe detailed event properties, improve sensitivity to experimental cuts, or to achieve the goal of percent level precision. 
Additionally, theoretical predictions at higher orders in QCD are useful for improving the splittings modelled in the parton shower. 
This enables improved matching of jet algorithms between theory and experiment. 

Through precise experimental measurements, we can directly investigate the fundamental interactions of elementary particles at short distances, pushing the boundaries of our knowledge and providing valuable insights into the fundamental interactions that govern the universe. 
However, we can only learn from many experiments, in no small part due to colour confinement, by comparison against precise theoretical predictions. 
The exploration of particle physics, particularly in the absence of new particle discoveries, holds immense significance. 
By scrutinising experimental data with high precision, even the slightest deviations from the predictions of the Standard Model can have profound implications for our understanding of the natural world. 
Such small deviations in measurements have the potential to revolutionise our knowledge and guide us towards physics beyond the Standard Model. 
Hence, precision phenomenology emerges as a crucial component in the quest for new physics. 

In the next section we will discuss the experimental and theoretical frontiers and their delicate interplay.

\section{Phenomenological Comparisons}
\label{sec:phenocomp}

On the theoretical side of particle physics, there have been a wide range of achievements, particularly driven by parallel experimental investigations. 
For multi-particle final states (corresponding to $2 \to 4,5,\ldots$ kinematics), the state-of-the-art are NLO perturbative corrections.  
Automated programmes exist for calculating tree and one-loop matrix elements together with the necessary infrared subtraction terms. 
These are encapsulated in a number of multi-purpose event generator programs~\cite{herwig:2015jjp,Sherpa:2019gpd,powheg:2010xd,madgraph:2011uj}, enabling NLO-accurate predictions for essentially any relevant collider process\footnote{See Ref.~\cite{snowmass:2022qmc} for a summary of available tools.}. 

At NNLO, calculations are mostly limited to $2\to 2$ kinematics, such as di-jet production~\cite{Currie:2016bfm,Czakon:2019tmo}, vector-boson-plus-jet production~\cite{Boughezal:2015dva,Gehrmann-DeRidder:2015wbt, Boughezal:2015ded}, photon-plus-jet-production~\cite{Campbell:2016lzl,Chen:2019zmr} or top quark pair production~\cite{Czakon:2015owf,Catani:2019iny}.  
Recent progress in the derivation of two-loop $2\to 3$ matrix elements has led to calculations for three-photon production~\cite{Chawdhry:2019bji}, diphoton-plus-jet production~\cite{Chawdhry:2021hkp} and three-jet production~\cite{Czakon:2021mjy}. 
Several infrared subtraction methods have been developed for NNLO calculations; see Ref.~\cite{TorresBobadilla:2020ekr} for a review. 
We will also discuss these in Chapter~\ref{chapter:antennasub}. 
Implementations using these methods are largely made on a process-by-process basis; most methods scale either poorly or not at all to higher multiplicities. 

At \NthreeLO, inclusive~\cite{Anastasiou:2015vya,Anastasiou:2016cez,Mistlberger:2018etf,Dreyer:2016oyx,Duhr:2019kwi,Duhr:2020kzd,Chen:2019lzz,Currie:2018fgr,Dreyer:2018qbw,Duhr:2020sdp,Duhr:2020seh} as well as more differential calculations have started to emerge~\cite{Dulat:2017prg,Dulat:2018bfe,Cieri:2018oms,Chen:2021isd,Chen:2021vtu,Billis:2021ecs,Chen:2022cgv,Neumann:2022lft,Camarda:2021ict,Chen:2022lwc,Baglio:2022wzu,Jakubcik:2022zdi}, the latter mainly for $2 \to 1$ processes via the use of the Projection-to-Born method~\cite{Cacciari:2015jma} or transverse-momentum-slicing techniques ($q_T$-slicing)~\cite{Catani:2007vq} to promote known NNLO subtraction schemes to \NthreeLO.
Calculations for higher multiplicities are currently hindered by the lack of process-independent local \NthreeLO subtraction schemes.
The need for NNLO and \NthreeLO predictions for phenomenologically relevant high-multiplicity processes highlights the importance of developing a more systematic and structured infrared subtraction formalism.  

NLO matching schemes such as MC@NLO~\cite{Frixione:2002ik} and POWHEG~\cite{Nason:2004rx,Frixione:2007vw} have been developed which systematically combine NLO fixed-order calculations with all-order parton-shower resummation at leading logarithmic accuracy. 
These innovations laid the foundations for the state-of-the-art multi-purpose event generators~\cite{powheg:2010xd,madgraph:2011uj,Bellm:2019zci,Sherpa:2019gpd,Bierlich:2022pfr}, see Ref.~\cite{Campbell:2022qmc} for a review. 
Similarly, fully-differential NNLO matching techniques are so far still in their infancy, as they require higher-order corrections to be exponentiated in the shower algorithm \cite{Campbell:2021svd}. 
Progress on including such corrections in parton showers has been reported in Refs.~\cite{Jadach:2011kc,Jadach:2013dfd,Hartgring:2013jma,Li:2016yez,Hoche:2017hno,Hoche:2017iem,Dulat:2018vuy,Gellersen:2021eci,Loschner:2021keu}. 
Also, note that the concept of jets requires a correspondence between the coloured final-state partons in a theoretical calculation and the collimated beams of hadrons observed in an experiment. 
Identifying groups of detected hadrons into jets, by an appropriate jet algorithm, is the IR-safe way of relating the measurements to a hard partonic QCD cross section. Examples of IR-safe jet algorithms are the anti-$k_t$~\cite{Cacciari:2008gp}, $k_t$~\cite{Catani:1993hr} and Cambridge-Aachen~\cite{Dokshitzer:1997in} algorithms. 
Then a matching scheme bridges the gap between energy scales by a parton shower and hadronisation process.  

\begin{figure}[t]
    \centering
    \includegraphics[width=0.95\textwidth]{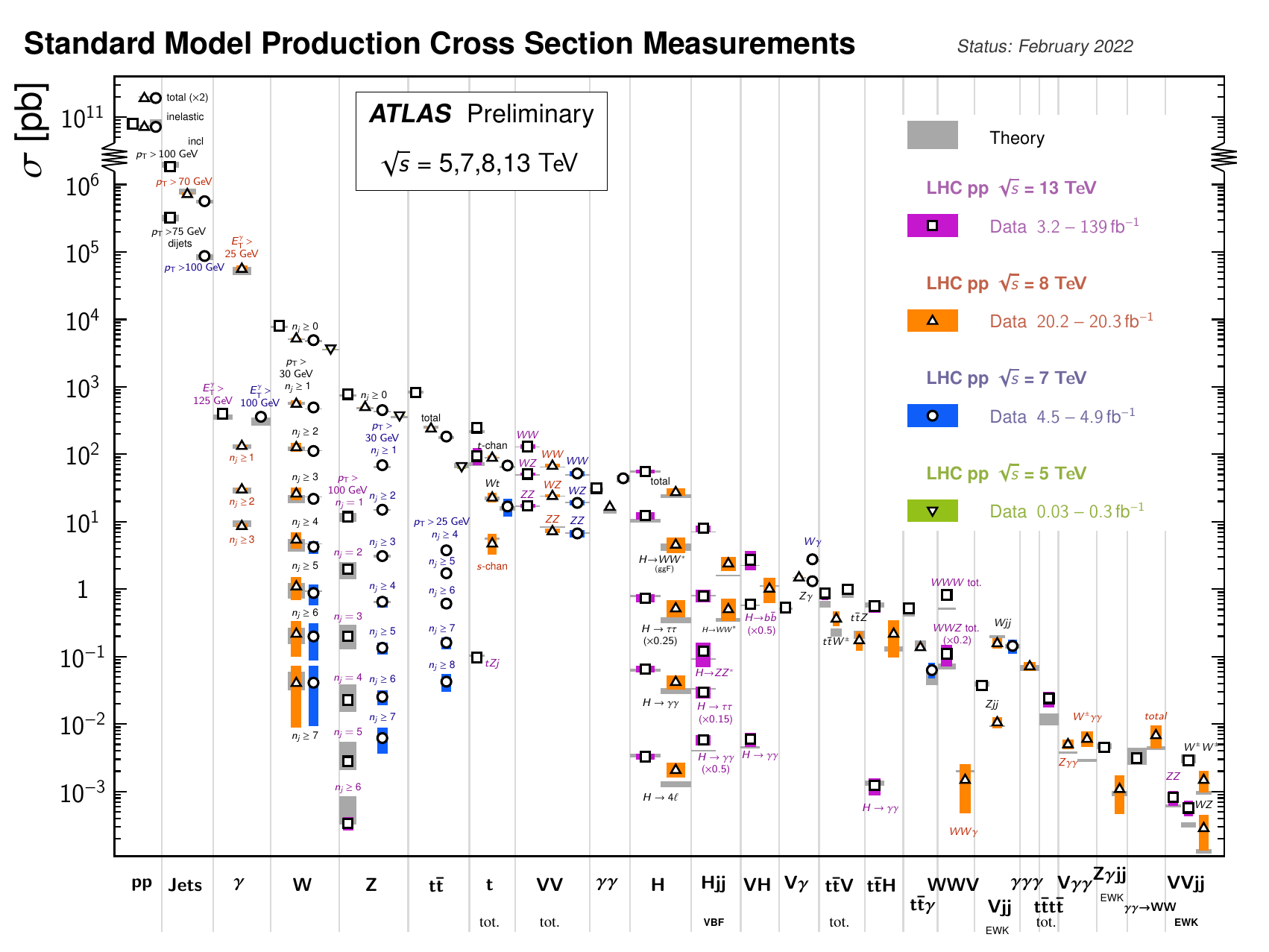}
    \caption{Plot of cross section measurements by ATLAS at various energies according to process and respective cross section predictions~\cite{ATLAS:2022djm}.}
    \label{fig:crosssec}
\end{figure}

The calculations and tools described above have generally delivered very good agreement between Standard Model predictions and collider experiments. In particular, we can highlight the discovery of the Higgs boson at the LHC in 2012~\cite{ATLAS:2012yve,CMS:2012qbp}. 
Of course, the LHC is just one of many colliders worldwide, which are all designed to probe particular areas of particle physics. 
Colliders can be categorised according to their shape (circular or linear) and according to the colliding particles (hadrons, electrons or various ions). 
Focussing on the LHC, we can see remarkable agreement between theoretical and experimental cross sections across a large number of processes and many orders of magnitude in Fig.~\ref{fig:crosssec}.
As can be seen in Fig.~\ref{fig:crosssec}, the LHC enables us to scrutinise a wide range of observables involving Higgs bosons, electroweak bosons, top quarks and hadronic jets with high accuracy. 
With the anticipated dataset from the High-Luminosity LHC, the statistical uncertainties on many observables will be negligible and percent-level accuracy is likely to be achieved experimentally. 
In the realm of perturbative QCD, reaching the desired level of refinement typically involves extending fixed-order calculations to at least NNLO in the strong-coupling expansion. 
Equally, we can expect future colliders (eg. Future Circular Collider, Circular Electron-Positron Collider, Super Proton-Proton Collider) to both increase the energy of collisions (thus allowing more accessible states) and increase luminosity (more collisions and better statistics). 
As such, development of the tools required for high-multiplicity NNLO calculations and for \NthreeLO calculations is necessary to infer physics from the improved experimental tools.

\section{Uncertainties in Calculations}
\label{sec:errors}


In performing fixed-order calculations, we accept that our prediction is not exact but how do we estimate the uncertainty without calculating the next order? The commonly accepted answer is to use scale variation. The idea can be demonstrated in the running of the strong coupling, Eq.~\eqref{eq:alphas}, Taylor-expanded to give, 
\begin{equation}
    \alpha_s ( \mu^2) = \alpha_s (\mu_0^2) - \frac{\beta_0}{2\pi} \alpha_s^2 (\mu_0^2) \ln \left( \frac{\mu^2}{\mu_0^2} \right) +\order{\alpha_s^3}.
\end{equation}
Thus by varying the renormalisation scale from $\mu$ to $\mu_0$, we are including terms of higher order in the perturbation, where $\alpha_s (\mu_0^2)$ is now the small parameter. 
Similarly, we can vary the factorisation scale, which affects the renormalised PDFs according to the DGLAP evolution equations. The evolution of a PDF depends on the all-orders splitting functions but the NLO correction due to factorisation scale variation depends only on the tree-level splitting functions and is given by,
\begin{equation}
    f_{a|A} (x_a, \mu^2) = f_{a|A} (x_a, \mu_0^2) - \frac{\alpha_s (\mu_0^2)}{2 \pi} \log \left( \frac{\mu_0^2}{\mu^2} \right) \sum_b \int_{x_a}^1 \frac{ d x_b}{x_b} P_{ab}^{(0)} \left( \frac{x_a}{x_b} \right) f_{b|A} (x_b, \mu_0^2) + \order{\alpha_s^2}.
\end{equation}
Note that the sum is over all flavours of quarks and the gluons and we have chosen to set the renormalisation scale equal to the factorisation scale here. We observe that by varying the factorisation scale, we include additional terms at higher orders in $\alpha_s (\mu_0^2)$.
Of course, scale variation is no substitute for calculating higher orders, where additional channels and features can be modelled. Typically, a calculation is performed by varying both the renormalisation and factorisation scales up and down by a factor of 2 from some typical energy scale determined by the experiment. 

\begin{figure}[t]
    \centering
    \includegraphics[width=0.66\textwidth]{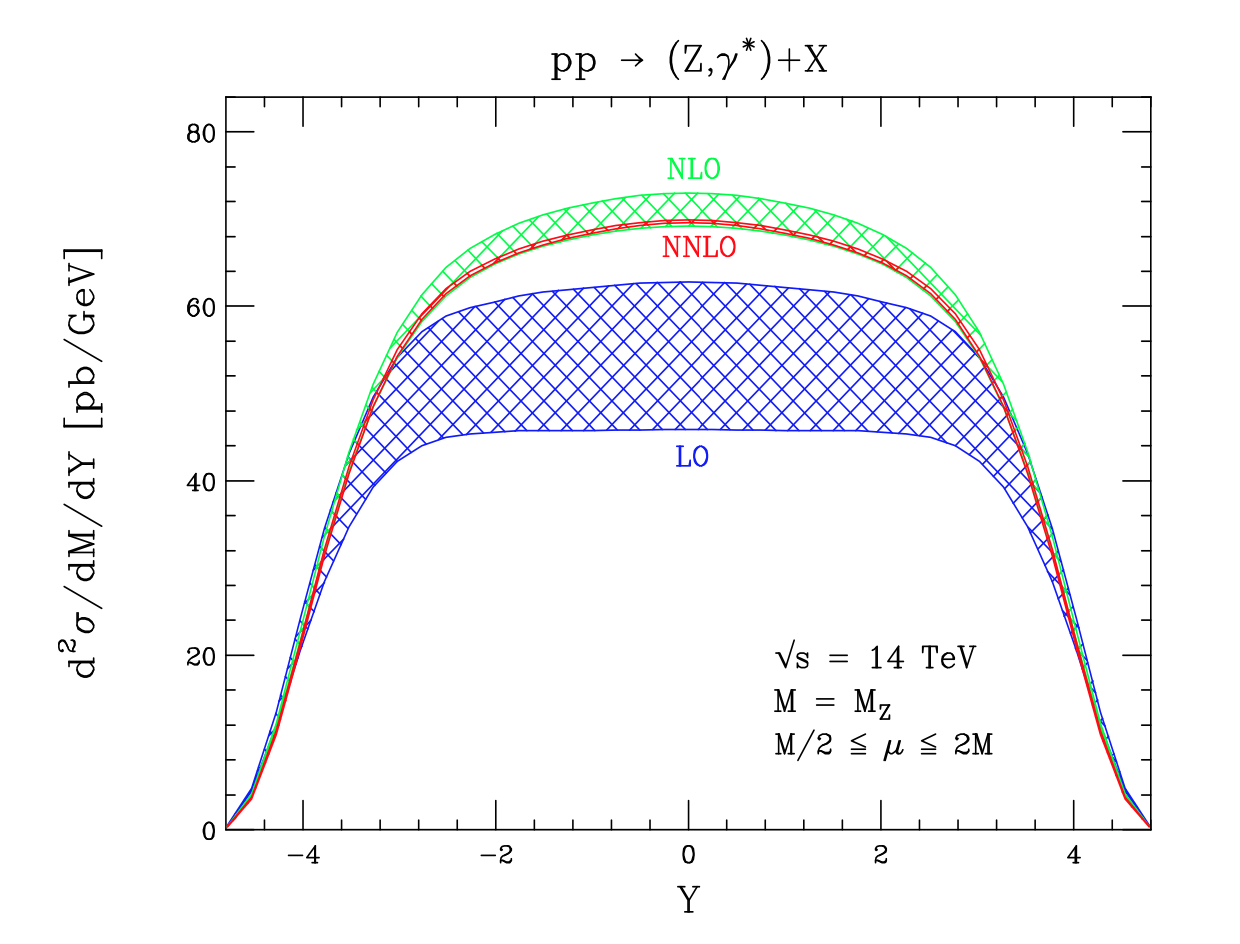}
    \caption{Theory predictions of the rapidity distribution for Z boson production in proton-proton collisions at centre-of-mass energy $14$ TeV~\cite{Anastasiou:2003ds}. The error bands show the scale uncertainties, where the renormalisation and factorisation scales are varied in the range $M_Z/2 \le \mu \le 2 M_Z$. }
    \label{fig:scaleerrors}
\end{figure}

An example of a calculation, showing the scale uncertainties at various orders, is shown in Fig.~\ref{fig:scaleerrors}. 
We can make a few observations. 
Firstly, it is clear that the dependence on the unphysical scales falls dramatically with higher orders, motivating the need for higher orders. 
Secondly the ratio of the NLO to the LO result is particularly large and the NLO result falls outside of the scale uncertainty bounds for the LO result. 
This is because of particular structures appearing in the NLO expressions and also because a new channel, $q g \to Z q$, only contributes from NLO. 
Since the gluon distribution is large, this creates a large ratio order-to-order. This illustrates how scale uncertainty can only account for a portion of unknown higher-order effects. 
Thirdly, the NNLO result does fall within the scale uncertainty bounds for the NLO result, suggesting that the scale variations give a reliable estimate for the NLO theoretical uncertainty. It also shows signs of perturbative convergence at NNLO. 
Finally, due to the increased availability of channels and increased variability of momentum configurations at NLO and NNLO, we see a different shape for the distribution (not just normalisation), compared to LO, where it is approximately flat for small rapidity. 

\begin{figure}[t]
    \centering
    \includegraphics[width=0.66\textwidth]{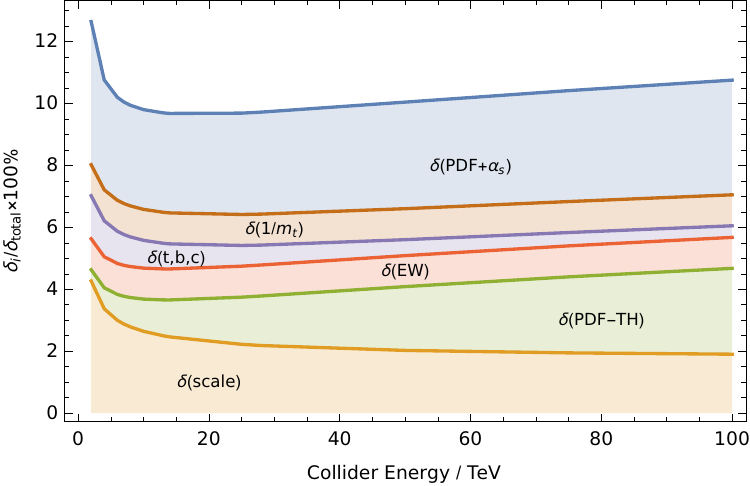}
    \caption{The sum of the sources of relative uncertainty as a function of collider energy in an \NthreeLO calculation of inclusive Higgs boson production~\cite{Cepeda:2019klc}. Each band shows the size of the labelled source of uncertainty. } 
    \label{fig:error-sources}
\end{figure}

Of course, the uncertainty in truncating a perturbative series is not the only source of uncertainty in the full calculation of an observable. 
While those other parts of a calculation are not the focus of this thesis, it is instructive to see the relative sources of uncertainty. 
Fig.~\ref{fig:error-sources} shows the various sources of error for an inclusive Higgs boson production cross section. 
The scale uncertainty gives an estimate of the missing higher orders beyond the \NthreeLO calculation. 
The errors associated with neglecting quark masses beyond NLO are given as $\delta(1/m_t)$ and $\delta(t,b,c)$. 
The error associated with missing higher electroweak corrections is given by $\delta(\text{EW})$. $\delta(\text{PDF}+\alpha_s)$ denotes the uncertainty in our knowledge of PDFs and $\alpha_s$, while $\delta(\text{PDF}+\text{TH})$ is due to using PDFs evaluated at NNLO but evaluating cross-sections at \NthreeLO. 
The PDF and the scale sources of uncertainty are the greatest sources in this case, both motivating study of QCD at higher orders in $\alpha_s$. 
In fact, when scale uncertainty can often be regarded as too small an estimate, calculations up to at least NNLO seem crucial. 
Even in the case of Fig.~\ref{fig:error-sources}, where the calculation has been performed to \NthreeLO, the scale uncertainty is of similar size to other uncertainty contributions, justifying the calculation up to this order. 

\section{Summary and Outlook}
\label{sec:phenosummary}

With the increase in experimental precision, phenomenological tools need to be refined to improve bounds on fundamental parameters and to detect potential deviations from theory. 
For many observables, percent-level accuracy cannot be reached without improvements in fixed-order calculations, parton distribution functions, parton showers and modelling of non-perturbative effects. 
Improvements are being made in all these areas.  
In the field of perturbative QCD, this typically requires fixed-order calculations to at least NNLO. 
While there are efficient automated programs for calculating tree-level and one-loop matrix elements and for complete IR subtraction at NLO, this is not the case at NNLO. 
At NNLO, calculations are limited to lower multiplicity states, particularly due to the complex patterns of IR cancellation. 
At \NthreeLO, calculations are mainly limited to $2 \to1$ processes, where NNLO subtraction terms with one extra jet can be promoted to \NthreeLO. 
Thanks to the developments in both experimental and theoretical particle physics, we observe excellent agreement across a wide range of observables. 
However, the search for discrepancies and new physics necessitates reduction of theory uncertainties, including by calculating higher orders and by developing the tools required for them. 
Scale variation is a useful tool for estimating uncertainty in fixed-order calculations, despite its limitations because it cannot account for channels and features unavailable at a certain order. 
Developments in fixed order calculations are also important as they can be a substantial contributor to the overall uncertainty in Standard Model predictions. 

In general, there are two obstacles to calculation at higher orders.  
First, knowledge of the relevant tree-level and loop multi-particle matrix elements. 
There is factorial growth in the number of Feynman diagrams and therefore the number of loop structures which need integrating with increasing perturbative order. 
In the framework of dimensional regularisation, gauge-theory loop matrix elements contain explicit infrared poles in the regulator $\e$ of up to two powers per loop. 
The computation of such matrix elements is sufficiently complicated that it is a field in its own right. 
Second, a scheme to extract the implicit infrared divergences.  
These are produced by integration of matrix elements with fewer loops and more external particles over the unresolved or infrared-singular regions of the phase space, as discussed in Chapter~\ref{chapter:qcd}. 
The explicit poles and implicit poles are cancelled in IR-safe observables, thereby enabling the numerical evaluation over the whole of phase space. 
Subtraction schemes are currently regarded as the most elegant solution to address these complexities and they will be discussed in Chapter~\ref{chapter:antennasub}.

\chapter{Decomposition of Triple-Collinear Splitting Functions}
\label{chapter:paper1}

\maketitle
\flushbottom

\section{Introduction \label{sec:intro}}

One of the complications immediately evident at NNLO is the overlap between iterated single unresolved and double unresolved limits. For example, the limit in which three particles $i$, $j$ and $k$ become collinear (studied in Refs.~\cite{campbell,Catani:1998nv,Catani:1999ss}) is obtained when invariants in the set $\{s_{ij}$, $s_{jk}$, $s_{ik}$, $s_{ijk} \}$ are small and there are two inverse powers of them. This limit contains both single and double unresolved limits - an iterated collinear contribution (which overlaps with soft and collinear limits), as well as a genuinely double unresolved contribution.  In this chapter we decompose the triple-collinear splitting functions into products of two-particle splitting functions and a remainder that is explicitly finite when any two of $\{i,j,k\}$ are collinear.
 
To help with the discussion of the singularities present in the collinear limits of real radiation matrix elements, we introduce the notion of {\bf internal} and {\bf external} singularities. Internal singularities are associated with small invariants amongst the set of collinear particles. External singularities involve other (spectator) particles involved in the scattering through the definition of the momentum fraction. For example, when two particles, $i$ and $j$ are collinear we find the well known collinear limit proportional to the two-particle splitting function, 
$$
\frac{1}{s_{ij}} P_{ab}^{(0)}(x_i).
$$
The limit as $s_{ij} \to 0$ references only particles in the collinear set and is therefore an internal singularity. External singularities are both present in the splitting function $P_{ab}^{(0)}(x_i)$ and associated with the momentum fraction limits $x_i \to 0$ or $x_i \to 1$. These external singularities correspond to situations where one of $\{i,j\}$ is collinear with a spectator particle, or where one of the particles is soft.
As in Ref.~\cite{campbell}, we work with colour-ordered matrix elements.

This chapter is organised as follows. We discuss the infrared singularity structure of the triple-collinear splitting functions in Section~\ref{sec:singstructure}. In Section~\ref{sec:genstructure} we discuss the general structure of the triple-collinear limit and explain how to restructure it such that the strongly-ordered limit is explicit and the remaining terms are manifestly finite when any two of $\{i,j,k\}$ are collinear. 
Results for the triple-collinear splitting function for all of the various parton configurations are collected in Section~\ref{sec:results}.  We also analyse all of the internal and external single unresolved singularities of each of the splitting functions. Finally, we summarise our findings in Section~\ref{sec:summary}.

\section{Singularity Structure of the Triple-Collinear Splitting Function} \label{sec:singstructure}

The primary aim of this chapter is to rewrite the $P_{abc}^{(0)}$ splitting function in a way that exposes its singularity structure. In particular, we aim to isolate the strongly-ordered iterated contributions.  In other words, we aim to rewrite the spin-averaged and colour-ordered three-particle splitting function as,
\begin{equation}
\label{eq:Psplit}
P_{abc}^{(0)} (i,j,k) = \sum_{\text{perms}} 
\frac{1}{s_{ijk}} P_{(ab)c}^{(0)}\left(x_k\right)
\frac{1}{s_{ij}} P_{ab}^{(0)}\left(\frac{x_j}{1-x_k}\right) 
+ \frac{1}{s_{ijk}^2}R_{abc}^{(0)} (i,j,k),
\end{equation}
where $P_{ab}^{(0)}$ are the usual spin-averaged two-particle splitting functions (listed in Section~\ref{sec:tree-single-lims}) and the remainder $R_{abc}^{(0)} (i,j,k)$ depends on the momentum fractions and small invariants. 

\begin{figure}[th!]
    \centering

\begin{tikzpicture} 
\begin{feynman}
\vertex (a); 
\vertex [right=4em of a] (b);
\vertex [above right=6em of b]  (d) ; 
\vertex [above right=of d]      (f1) {\(x_i\)};
\vertex [below right=of d]      (f2) {\(x_j\)};
\vertex [below right=7em of b]  (f3) {\(x_k\)};

\diagram* {
(a) -- [fermion, edge label=\( P \)] (b),
(b) -- [fermion, edge label=\( (ab) \)] (d), 
(d) -- [fermion, edge label=\( a \)] (f1), 
(d) -- [fermion, edge label=\( b \)] (f2),
(b) -- [fermion, edge label=\( c \)] (f3),
};
\end{feynman}
\end{tikzpicture}

    \caption{The iterated simple-collinear contribution to the triple-collinear splitting function.  }
    \label{fig:iterated}
\end{figure}
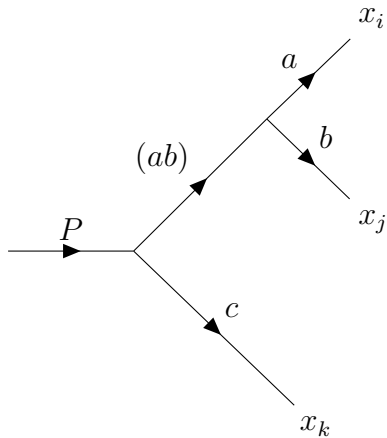

\begin{figure}[th!]
    \centering
\begin{tikzpicture} 
\begin{feynman}
\vertex (a); 
\vertex [right=4em of a] (b);
\vertex [above right=7em of b]  (f1) {\(x_i\)};
\vertex [right=6em of b]        (f2) {\(x_j\)};
\vertex [below right=7em of b]  (f3) {\(x_k\)};

\diagram* {
(a) -- [fermion, edge label=\( P \) ] (b),
(b) -- [fermion, edge label=\( a \)] (f1),
(b) -- [fermion, edge label=\( b \)] (f2),
(b) -- [fermion, edge label=\( c \)] (f3), 
};
\end{feynman}
\end{tikzpicture}            

    \caption{The remainder function $R_{abc}^{(0)}$ contains the parts of the triple-collinear splitting function that are not contained in the strongly-ordered iterated contribution.}
    \label{fig:Rijk}
\end{figure}
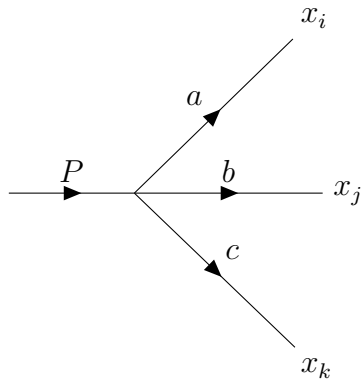

An iterated (or strongly-ordered) contribution is obtained through the product of leading-order splitting functions, $P \times P$, as illustrated in Fig.~\ref{fig:iterated} and is given by terms of the type,
\begin{equation}
    \frac{P_{(ab)c}^{(0)}(x_k)}{s_{ijk}} \times
     \frac{P_{ab}^{(0)}(y_j)}{s_{ij}} ,
\end{equation}  
where $y_j$ is the momentum fraction of the second splitting, 
\begin{equation}
y_j = \frac{x_j}{x_i+x_j} = \frac{x_j}{1-x_k}.
\end{equation}
The invariants in the denominator are simply those corresponding to the two- and three-particle invariants, $s_{ij}$ and $s_{ijk}$\footnote{Note that one could have chosen to define the strongly-ordered limit in which $s_{ijk}$ is replaced by $s_{ik}+s_{jk}$.}. The remainder (or uniterated) $1\rightarrow 3$ splitting function $R_{abc}^{(0)}$ is illustrated in Fig.~\ref{fig:Rijk}. 

The triple-collinear splitting functions contain both single and double unresolved limits:
\begin{enumerate}
\item simple-collinear limits when two of $\{i,j,k\}$ are collinear
or one of $\{i,j,k\}$ is collinear with a spectator particle, 
\item single-soft limits when one particle is soft and is colour-connected to either the other two particles, or a spectator particle,
\item double-collinear limits when two of $\{i,j,k\}$ are collinear and the third is collinear with a spectator particle,
\item soft-collinear limits when one particle is soft and the other two are collinear or one is collinear with a spectator particle,
\item double-soft limits when two particles are soft,
\item triple-collinear limits when two of $\{i,j,k\}$ are collinear with a spectator particle (this occurs only in the double-soft limits).
 
\end{enumerate}

In order to explain what types of single unresolved singularities appear in the iterated contributions, $P\times P$, and the uniterated splitting function, $R_{abc}^{(0)}$, we must first discuss the different types of single unresolved singularities present in $P_{abc}^{(0)}$. 
 
There are two types of simple-collinear singularities present in a triple-collinear splitting function. 
First, internal simple-collinear singularities like $1/s_{ij}$, where $i,j$ are collinear. 
Internal simple-collinear singularities appear only in the iterated two-particle splitting contributions. 
Second, external simple-collinear singularities like $1/x_i$, which indicate potential collinear factors with the spectator particles used to define the momentum fractions. 
External simple-collinear singularities which are present in $P_{abc}^{(0)}$ appear only in the iterated two-particle splitting contributions. 
Although, if $P_{abc}^{(0)}$ does not contain external simple-collinear singularities, there could be a cancellation between external simple-collinear singularities in the $P\times P$ contribution and those in the remainder $R_{abc}^{(0)}$. 

There are also two types of single-soft singularities. 
First, internal soft $j$ singularities encoded through typical eikonal factors like $s_{ik}/(s_{ij} s_{jk})$. 
This type of singularity is explicitly embedded in a triple-collinear splitting function and only appears in the remainder $R_{abc}^{(0)}$. 
This makes sense because internal single soft singularities are inherently uniterated - this type of eikonal factor contributes the full weight of a triple-collinear term without an $s_{ijk}$ pole. 
Second, there are external soft $j$ singularities that appear in `hidden' eikonal factors like $x_i/(x_j s_{ij})$. 
This type of singularity is produced when the soft particle is colour-connected to a spectator particle. 
If present, it only appears in the iterated two-particle splitting contributions.  

We can summarise these important features as follows:
\begin{itemize}
    \item[-] {\bf Internal} simple-collinear singularities like $1/s_{ij}$ appear only in $P \times P$ terms (the iterated two-particle splitting contributions). 
    \item[-] When {\bf external} simple-collinear singularities like $1/x_i$ appear in $P_{abc}^{(0)}$, they are all contained in $P \times P$ terms.
    \item[-] When {\bf external} simple-collinear singularities like $1/x_i$ {\bf do not} appear in $P_{abc}^{(0)}$, there could be terms proportional to $1/x_i$ in $P \times P$ and $R_{abc}^{(0)}$ which cancel. 
    \item[-] {\bf Internal} single soft singularities like $s_{ik}/(s_{ij} s_{jk})$ appear only in $R_{abc}^{(0)}$.
    \item[-] {\bf External} single soft singularities like $x_i/(x_j s_{ij})$ appear only in the iterated  $P \times P$ terms.
    \end{itemize}

\section{General Structure of the Triple-Collinear Splitting Function} \label{sec:genstructure}

As mentioned earlier, the triple-collinear limit is defined as the kinematic regime where the invariants $s_{ij},$ $s_{jk}$, $s_{ik}$ and $s_{ijk}$ all become small. 
In this region, the singular factor has at most two net inverse powers of the small invariants. 
Additionally, since the splitting functions are limits of squared matrix elements, there is the additional physics constraint that there are at most two inverse powers of double invariants ($s_{IJ}$ where $I,J \in \{i,j,k\}$) and at most two inverse powers of the triple invariant $s_{ijk}$. 
Therefore, any triple-collinear splitting function can be represented by coefficients $\beta_i (x_i,x_j,x_k,\epsilon)$ of 37 invariant pole structures: 
\begin{equation} \label{eqn:spoles}
	\begin{split} 
P_{abc}^{(0)}(i,j,k) =	& \color{black} \frac{\beta_1}{s_{jk}s_{ijk}} + \frac{\beta_2}{s_{ij}s_{ijk}}+ \frac{\beta_3}{s_{ik}s_{ijk}}  \color{black} + \frac{\beta_4}{s_{jk}^2}+ \frac{\beta_5}{s_{ij}^2}+ \frac{\beta_6}{s_{ik}^2} \\
	\color{black} +& \color{black} \frac{\beta_7}{s_{jk} s_{ij}} + \frac{\beta_8}{s_{jk}s_{ik}}+ \frac{\beta_9}{s_{ij}s_{ik}} + \frac{\beta_{10} s_{ij}}{s_{jk}^2 s_{ijk}} + \frac{\beta_{11} s_{ij}}{s_{ik}^2 s_{ijk}} + \frac{\beta_{12} s_{jk}}{s_{ij}^2 s_{ijk}}  \\
	\color{black} + & \color{black} \frac{\beta_{13}}{s_{ijk}^2}+  \frac{\beta_{14} s_{ij}}{s_{jk} s_{ijk}^2} + \frac{\beta_{15} s_{ij}}{s_{ik} s_{ijk}^2} + \frac{\beta_{16} s_{jk}}{s_{ij} s_{ijk}^2} 
	+ \frac{\beta_{17} s_{ij}^2}{s_{jk}^2 s_{ijk}^2}+ \frac{\beta_{18} s_{ij}^2}{s_{ik}^2 s_{ijk}^2} + \frac{\beta_{19} s_{jk}^2}{s_{ij}^2 s_{ijk}^2}\\
	\color{black} + & \color{black} \bigg[ \frac{\beta_{20} s_{jk}}{s_{ik}^2 s_{ijk}} + \frac{\beta_{21} s_{jk}}{s_{ik} s_{ijk}^2} + \frac{\beta_{22} s_{jk}^2}{s_{ik}^2 s_{ijk}^2} +  \frac{\beta_{23} s_{ij}^2}{s_{ik} s_{jk} s_{ijk}^2} 
	+ \frac{\beta_{24} s_{ij}}{s_{ik} s_{jk} s_{ijk}} + \frac{\beta_{25} s_{jk}}{s_{ik} s_{jk} s_{ijk}} \\
    \color{black} + & \color{black} \frac{\beta_{26} s_{ik}}{s_{ij}^2 s_{ijk}} +  \frac{\beta_{27} s_{ik}}{s_{jk}^2 s_{ijk}} +  \frac{\beta_{28} s_{ik}}{s_{ij} s_{ijk}^2} + \frac{\beta_{29} s_{ik}}{s_{jk} s_{ijk}^2} + \frac{\beta_{30} s_{ik}^2}{s_{ij}^2 s_{ijk}^2} + \frac{\beta_{31} s_{ik}^2}{s_{jk}^2 s_{ijk}^2} + \frac{\beta_{32} s_{ik}}{s_{ij} s_{jk} s_{ijk}} \\
    \color{black} +& \color{black} \frac{\beta_{33} s_{ik}^2}{s_{ij} s_{jk} s_{ijk}^2}  
    +\frac{\beta_{34} s_{jk}^2}{s_{ij} s_{ik} s_{ijk}^2}  + \frac{\beta_{35} s_{ij} s_{ik}}{s_{jk}^2 s_{ijk}^2} + \frac{\beta_{36} s_{ij} s_{jk}}{s_{ik}^2 s_{ijk}^2}+ \frac{\beta_{37} s_{ik} s_{jk}}{s_{ij}^2 s_{ijk}^2} \bigg].
\end{split}
\end{equation}
Using momentum conservation, any triple-collinear splitting function can be expressed in the basis of the first three lines of Eq. (\ref{eqn:spoles}) (ie. $\beta_1$ --- $\beta_{19}$, the non-square bracketed terms). 

The factorisation properties of squared matrix elements impose six additional relationships amongst the $\beta_i$. 
Of these there are two relationships between $\beta_4$, $\beta_{10}$, $\beta_{17}$, which are due to the absence of $jk$ simple-collinear contributions of the form $1/s_{jk}^2$. 
Similar relationships hold for the coefficients of $1/s_{ij}^2$ and $1/s_{ik}^2$. 
Therefore, we propose an alternative basis in terms of 13 $\alpha_i (x_i,x_j,x_k,\epsilon)$ invariant structures that make the physical constraints more evident,
\begin{eqnarray}
\label{eq:abasis}
P_{abc}^{(0)}(i,j,k) &= &\phantom{+}
\frac{\alpha_{12}}{s_{jk}s_{ijk}}+ 
\frac{\alpha_{13}}{s_{ij}s_{ijk}}+
\frac{\alpha_{14}}{s_{ik}s_{ijk}} \nonumber \\
& & + \frac{\alpha_1}{s_{ijk}^2} 
+ \frac{\alpha_2 \Tr{j}{k}{i}{\ell}}{ s_{jk} s_{ijk}^2} 
+ \frac{\alpha_3 \Tr{i}{j}{k}{\ell}}{ s_{ij} s_{ijk}^2} 
+ \frac{\alpha_4 \Tr{k}{i}{j}{\ell}}{ s_{ik} s_{ijk}^2} \nonumber \\
& &
+ \frac{\alpha_{23} \Tr{i}{j}{k}{\ell}}{ s_{ij}  s_{jk} s_{ijk} } 
+ \frac{\alpha_{24} \Tr{j}{k}{i}{\ell}}{ s_{jk}  s_{ik} s_{ijk}} 
+ \frac{\alpha_{34} \Tr{k}{i}{j}{\ell}}{ s_{ik}  s_{ij} s_{ijk}}  \nonumber \\
& &
+ \frac{\alpha_{22} W_{jk}}{ s_{jk}^2 s_{ijk}^2} 
+ \frac{\alpha_{33} W_{ij}}{ s_{ij}^2 s_{ijk}^2} 
+ \frac{\alpha_{44} W_{ik}}{ s_{ik}^2 s_{ijk}^2} .
\end{eqnarray}
Here $\ell$ is a suitably normalised spectator momentum such that,
\begin{equation}
\Tr{i}{j}{k}{\ell} = x_k s_{ij} -x_j s_{ik} + x_i s_{jk} ,    
\end{equation}
while the quantity $W_{ij}$ is defined as
\begin{equation}
\label{eq:Wdef}
        W_{ij} = (x_i s_{jk} - x_j s_{ik})^2 - 
        \frac{ 2} {\ome}
        \frac{ x_i x_j x_k} {(1-x_k)}
        s_{ij} s_{ijk}.
\end{equation}
In Eq.~\eqref{eq:abasis}, the first three coefficients ($\alpha_{12}$, $\alpha_{13}$, $\alpha_{14}$) display any strongly-ordered structure present, like in Eq.~(\ref{eq:Psplit}).  
The remaining structures are written in combinations that are designed to be less singular in the simple-collinear limits. 
For example, in the $s_{ij} \to 0$ limit,
\begin{equation} \label{eq:Trsingularity}
\Tr{i}{j}{k}{\ell} = \mathcal{O} (\sqrt{s_{ij}}),
\end{equation}
so that there is no singular contribution in the $ij$ collinear limit from the $\alpha_3$, $\alpha_{23}$ or $\alpha_{34}$ terms\footnote{Note that an alternative basis to the $\alpha$ basis could be chosen with somewhat different structures to the trace structure used here. We choose the trace structure for its `natural' interpretation and see that it reflects the colour-ordering in the results. Another suitable basis would require properties which follow equations similar to Eq.~\eqref{eq:Trsingularity}.}.
This can be shown explicitly by using the Sudakov momenta parametrisation for two collinear partons~\cite{Catani:1996vz,parton}, 
\begin{eqnarray}
    p_i^\mu &=& x_i p^\mu + k_T^\mu - \frac{k_T^2}{2 x_i p \cdot n} n^\mu, \\
    p_j^\mu &=& (1-x_i) p^\mu - k_T^\mu - \frac{k_T^2}{2(1-x_i) p \cdot n} n^\mu, 
\end{eqnarray}
where $p$ is the total longitudinal momentum of the $ij$ pair, $k_T$ is the transverse momentum (so $k_T \cdot p = 0$), which approaches zero in the collinear limit and $n$ is an arbitrary light-like vector, such that
\begin{equation}
    n^2 = 0, \hspace{2 cm} n \cdot k_T = 0.
\end{equation}
The parametrisation also ensures the on-shell conditions:
\begin{equation}
    p_i^2 = p_j^2 = p^2 = 0.
\end{equation}
Using this parametrisation, we note that 
\begin{equation}
    \Tr{i}{j}{k}{\ell} = 4 ( p_\ell \cdot k_T) (p_k \cdot p) - 4(p_k \cdot k_T) (p_\ell \cdot p) + \order{k_T^2},
\end{equation}
and 
\begin{equation}
    s_{ij} = - \frac{k_T^2}{x_i (1-x_i)}, 
\end{equation}
so therefore Eq.~\eqref{eq:Trsingularity} holds and both $k_T$ and $s_{ij}$ approach zero in the $ij$ collinear limit, as expected.  

Similarly, the $\alpha_{33}$ term also has no contribution in the $s_{ij} \rightarrow 0$ limit. 
$W_{ij}$ has been constructed from terms that appear in the triple-collinear limit and a second term that is added to $\alpha_{13}$ (and subtracted from $\alpha_{33}$ in order to have the full spin-averaged splitting functions in the strongly-ordered contributions).  
Both terms in $W_{ij}$ are individually $\mathcal{O} (s_{ij})$ when expanded but have opposite signs so that $W_{ij}/s_{ij}^2 = \mathcal{O} (1/\sqrt{s_{ij}})$. 
This is an integrable singularity that vanishes upon azimuthal integration (in $d$-dimensions). 
To make this clear, strictly in the collinear $ij$ limit, we can interpret $W_{ij}$ in terms of the azimuthal angle with respect to the $(ij)$ direction. 
Following Ref.~\cite{Dulat:2018vuy}, we find that we can write
\begin{equation}
\label{eq:azim}
(x_i s_{jk} - x_j s_{ik})^2 = \frac{4 x_i x_j x_k}{\omxk} s_{ij} s_{ijk} \cos^2 \phi_{ij,kl},
\end{equation}
such that $W_{ij}$ has the form,
\begin{equation}
    W_{ij} = \frac{4 x_i x_j x_k}{\omxk} s_{ij} s_{ijk} \left( \cos^2 \phi_{ij,kl} - \frac{1}{2 \ome} \right).
\end{equation}

\section{Results \label{sec:results} }

In this section, we summarise our results for the triple-collinear splitting functions.  In each case, we find that the remainder $R_{abc}^{(0)}$ can be expressed in terms of a single trace (rather than three in general). The $\alpha_1$ term from Eq.~\eqref{eq:abasis} is always some combination of two auxiliary functions and they are a feature of the $\alpha$ basis: 
\begin{eqnarray}
\label{eq:A0def}
    A_0(x,y) &=& 1 - \frac{(1-x)}{(1-y)}, \\
\label{eq:B0def}
    B_0(x,y) &=& 1 + \frac{2x(x-2)}{(1-y)^2} + \frac{4x}{(1-y)}.
\end{eqnarray}

It is also convenient to divide each splitting function according to structures where one parton can be described as hard, such that
\begin{align}
    P_{abc}^{(0)}(i,j,k) &= P_{abc}^{(0)}(i^h,j,k)
    +P_{abc}^{(0)}(i,j^h,k)
    +P_{abc}^{(0)}(i,j,k^h).
\end{align}

\subsection{Three Collinear Gluons}

We consider the case where gluons $i,j,k$ are in a particular colour-ordering. In other words, the outer gluons $i$ and $k$ play a different role to the inner gluon $j$. We find that, 
\begin{eqnarray}
    \label{eq:Pggg}
    \Pggg(i^h,j,k) &=& 
    \frac{\PggS(x_k)}{s_{ijk}}  
    \frac{\PggS\left(\frac{x_j}{1-x_k} \right)}{s_{ij}} +
    \frac{\PggS(1-x_i)}{s_{ijk}}  
    \frac{\Pgg\left(\frac{x_j}{1-x_i} \right)}{s_{jk}} \nonumber \\
    &&  + 
    \frac{1}{s_{ijk}^2} \Rgggsub(i,j,k) ,\\
    \Pggg(i,j^h,k) &=& 
    \frac{\PggS(x_k)}{s_{ijk}}  
    \frac{\PggS\left(\frac{x_i}{1-x_k} \right)}{s_{ij}}
    +
    \frac{\PggS(x_i)}{s_{ijk}}  
    \frac{\PggS\left(\frac{x_k}{1-x_i} \right)}{s_{jk}} , \hspace{1cm} \label{eq:Pgpp}\\
    \Pggg(i,j,k^h) &=& \Pggg(k^h,j,i),
\end{eqnarray}
%
where
\begin{eqnarray}
\label{eq:Rggg}
\Rgggsub(i,j,k) &=&  
\frac{2 \ome W_{jk}}{(1-x_i)^2 s_{jk}^2} 
+ \frac{4 \ome x_k}{(1-x_i)^2} \frac{ \Tr{i}{j}{k}{\ell} }{ s_{jk} } \nonumber \\
&& + \fa_0(x_i,x_j,x_k) 
 +  \fa(x_i,x_j,x_k) \frac{s_{ijk}\Tr{i}{j}{k}{\ell}}{s_{ij} s_{jk} } ,
\end{eqnarray}
and
\begin{eqnarray}
\label{eq:fa0}
\fa_0(x_i,x_j,x_k) &=& \ome B_0(x_k,x_i)  ,\\
\label{eq:fa}
\fa(x_i,x_j,x_k) &=& -\frac{x_k \Pgg(x_k)}{x_j(1-x_i)} 
- \frac{\Pgg(x_j)}{x_k} \nonumber \\  
&& + \frac{2}{x_j(1-x_k)}  -1 - \frac{1}{(1-x_i)(1-x_k)} .
\end{eqnarray}

We note that $\fa$ contains poles in $x_i$, $x_j$ and $x_k$. Therefore, we write $\fa$ in a manner that exposes the residue of these poles, in terms of two-particle splitting functions. Eqs.~\eqref{eq:Pggg}--\eqref{eq:fa} are equivalent to Eq.~(5.4) in Ref.~\cite{campbell} up to a normalisation of a factor of 4. 

As expected, there are no internal simple-collinear limits (i.e. relating to any of the simple-collinear limits ($s_{ij} \to 0$, $s_{jk} \to 0$ or $s_{jk} \to 0$) present in Eq.~\eqref{eq:Rggg}. All of the internal simple-collinear limits are contained in the iterated contribution.
However, there are possible external and internal singularities when 
\begin{itemize}
\item[(i)] gluon $I$ (for $I \in \{i,j,k\}$) is collinear with the spectator particle $\ell$, indicated when there is one singular power of $x_I$,
\item[(ii)] gluon $I$ is soft, indicated when there are two singular factors in the set $\{s_{IJ}, s_{IK}, x_I\}$.
\end{itemize}
These collinear and/or soft singularities can be present in the $\Pgg \times \Pgg$ contribution and/or in the remainders.  Within the remainders, they are produced entirely by the final term in Eq. (\ref{eq:Rggg}) when,
\begin{equation}
    \fa(x_i,x_j,x_k) \propto \frac{1}{x_I}.
\end{equation}
Given that 
\begin{equation}
\Tr{i}{j}{k}{\ell} = x_k s_{ij} - x_j s_{ik} + x_i s_{jk},
\end{equation}
there are two types of contribution in the remainders. Let us consider the two cases in turn:
\begin{itemize}
    \item $I = k$ (or $I = i$)
\begin{equation}
    \frac{1}{x_k} \frac{s_{ijk}\Tr{i}{j}{k}{\ell}}{s_{ij} s_{jk} } \longrightarrow 
      \frac{x_i}{x_k} \frac{s_{ijk}}{s_{ij}} 
    - \frac{x_j}{x_k} \frac{s_{ijk} s_{ik}}{s_{ij} s_{jk}},
\end{equation}
    
    \item $I = j$
\begin{equation}
\label{eq:Jsoft}
    \frac{1}{x_j} \frac{s_{ijk}\Tr{i}{j}{k}{\ell}}{s_{ij} s_{jk} } \longrightarrow 
    \frac{x_k}{x_j} \frac{s_{ijk}}{s_{jk}} 
    + \frac{x_i}{x_j} \frac{s_{ijk}}{s_{ij}} 
    - \frac{s_{ijk}s_{ik}}{s_{ij}s_{jk}}.
\end{equation}

\end{itemize}

\begin{itemize}

\item[(i)] Gluon $I$ is collinear with the spectator particle $\ell$ - external collinearity.

Let us first consider the external limits where the particle with small momentum fraction is collinear to the spectator particle, $s_{I\ell} = x_I \to 0$. These singular structures are tabulated in Table~\ref{table:ggg}.

The $x_i \to 0$ and $x_k \to 0$ contributions are shown in the first and third rows of Table~\ref{table:ggg}.  These limits are related by the $i \leftrightarrow k$ symmetry, so let us focus on the $x_k \to 0$ limit in the third row.  All contributions are proportional to $\Pgg(x_i)$. They originate in the iterated two-particle splitting and the second term of Eq.~\eqref{eq:fa}. Note that by construction, there are no contributions from $\Rgggsub(k,j,i)$. 

In the $x_j \to 0$ limit, there are contributions from the iterated two-particle splitting and the double unresolved $\Rgggsub$ splittings.  However, these contributions cancel and the $\Pggg$ splitting function does not exhibit a singularity in this limit. This is as expected, since gluon $j$ is only colour-connected to gluons $i$ and $k$ so that there should be no collinear limit for gluon $j$ with any spectator. 

\item[(ii)] Gluon $I$ is soft.

The soft $I$ limit is obtained when those in the set $\{s_{IJ}, s_{IK}, x_I\}$ are small and there are two inverse powers of them. The external soft contributions of the form $1/(s_{IJ} x_I)$ can be read off from Table~\ref{table:ggg}. However there are also internal soft $j$ contributions coming from the third term in Eq.~\eqref{eq:Jsoft}.
  
When gluon $k$ is soft, we recover the expected limit describing collinear gluons $i$ and $j$ with the soft gluon $k$ radiated between the colour-connected partners $j$ and $\ell$,
\begin{equation}
    \Pggg(i,j,k) \stackrel{k~{\rm soft}}{\longrightarrow} \frac{2 x_j}{s_{jk}x_k} \frac{1}{s_{ij}}\Pgg(x_i).
\end{equation}
This limit comes entirely from the iterated two-particle splitting.
The soft $i$ limit is obtained by $k \leftrightarrow i$ symmetry.  

In the soft $j$ limit, the $1/(x_j s_{ij})$ and $1/(x_j s_{jk})$ terms cancel between the $\Pgg \times \Pgg$ and $\Rgggsub$ contributions, such that
\begin{eqnarray}
    \frac{\Rgggsub(i,j,k)}{s_{ijk}^2} &
    \stackrel{j~{\rm soft}}{\longrightarrow}& \left(
    -\frac{x_k}{x_j s_{jk}} 
    -\frac{x_i}{x_j s_{ij}} 
    +\frac{s_{ik}}{s_{ij}s_{jk}}
    \right)
    \frac{2}{s_{ik}} \PggS(x_k), \hspace{1cm} \\    
    \Pggg(i,j,k) &
    \stackrel{j~{\rm soft}}{\longrightarrow}& \frac{2 s_{ik}}{s_{ij}s_{jk}}\frac{1}{s_{ik}}\Pgg(x_k).
\end{eqnarray}
This is precisely as expected for the emission of a soft gluon between the hard (and collinear) radiators $i$ and $k$.
\end{itemize}

The limit where both $j$ and $k$ are soft encodes $x_j \to 0$, $x_k \to 0$ and therefore $x_i \to 1$.  There are two types of contribution. First, there are iterated soft singularities in $\Pgg \times \Pgg$,
\begin{eqnarray} 
\label{eq:DsoftinPxPggg}
&& \frac{\Pgg(x_i)}{s_{ijk}}  \frac{\Pgg\left(\frac{x_k}{1-x_i} \right)}{s_{jk}} + \frac{\Pgg(x_k)}{s_{ijk}}  \frac{\Pgg\left(\frac{x_i}{1-x_k} \right)}{s_{ij}}
\stackrel{j,k~{\rm soft}}{\longrightarrow} \nonumber \\ 
&& \hspace{3.5cm} \frac{2}{\omxi s_{ijk} s_{jk}} \Pgg\left(\frac{x_k}{1-x_i}\right) 
 + \frac{4}{x_j x_k s_{ijk} s_{ij}}.
\end{eqnarray}
Second, there are double soft contributions in $\Rgggsub(i,j,k)$,
\begin{equation}
\label{eq:DsoftinRggg}
\frac{\Rgggsub(i,j,k) }{s_{ijk}^2}
\stackrel{j,k~{\rm soft}}{\longrightarrow}
 \frac{2 \ome W_{jk}}{(1-x_i)^2 s_{jk}^2 s_{ijk}^2} 
  - \left( 
 \frac{2}{x_k (1-x_i)} + \frac{4}{x_j\omxi}
 \right) \frac{\Tr{i}{j}{k}{\ell}}{s_{ij} s_{jk} s_{ijk}}.
\end{equation}
The second term in Eq.~\eqref{eq:DsoftinRggg} is produced by $\fa(x_i,x_j,x_k)$.

The double soft singularities, when gluons $i,j$ are soft, are obtained by the $i \leftrightarrow k$ interchange in Eqs.~\eqref{eq:DsoftinPxPggg} and \eqref{eq:DsoftinRggg}. 

Finally, there are also double soft singularities when gluons $i,k$ are soft, however, because they are not colour-adjacent, they only appear in the $\Pgg \times \Pgg$ contributions as a product of two eikonal factors.

We note that projecting the splitting function onto the $\alpha$-basis of Eq.~\eqref{eq:abasis} forces a link between the trace-like structures and the $B_0$ terms that appear in $\fa_0$, which is evident in the $x_i \to 1$ limit. This corresponds to the $x_j \to 0$, $x_k \to 0$ limit because the three momentum fractions sum to unity. We see that the second and third terms of Eq.~\eqref{eq:Rggg} are separately singular in this limit, 
\begin{eqnarray}
\frac{4\ome x_k}{\omxi^2}\frac{\Tr{i}{j}{k}{\ell}}{s_{jk}} &\to& \phantom{-}\frac{4\ome x_k}{\omxi^2}, \\
\fa_0(x_i,x_j,x_k) =  \ome B_0(x_k,x_i) &\to& - \frac{4\ome x_k}{\omxi^2},
\end{eqnarray}
and that the singular behaviour cancels when the terms are combined.
The link between the trace-like structures and the $B_0$ terms (including $A_0$ terms in generality) is a feature of the $\alpha$-basis of Eq.~\eqref{eq:abasis} and is repeated in all of the triple-collinear splitting functions. 

\begin{landscape}
\begin{table}[p]
\centering
\begin{center}
\begin{tabular}{|c|| c |c| c||c|} 
 \hline
 $\nameggg$ & 
 $\frac{\Pgg(x_i)}{s_{ijk}}  \frac{\Pgg\left(\frac{x_k}{1-x_i} \right)}{s_{jk}} + (i \leftrightarrow k) $ 
 & $\frac{1}{s_{ijk}^2} \Rgggsub(i,j,k)$ 
 & $\frac{1}{s_{ijk}^2} \Rgggsub(k,j,i)$ 
 & $\frac{1}{s_{ijk}^2} \Pggg(i,j,k)$ \\  
 \hline\hline
 $x_i \rightarrow 0$ 
 &\makecell[l]{   \\ $+\frac{1}{s_{ij}s_{ijk}} \frac{ x_j}{x_i} \bigg[2 \Pgg(x_k) \bigg] $ \\$+ \frac{1}{s_{jk}s_{ijk}} \frac{1}{x_i} \bigg[2 \Pgg(x_k) \bigg]$}
 & \makecell[c]{  \\ 0 \\} 
 & \makecell[l]{ $+\frac{1}{s_{ij}s_{jk}} \frac{x_j}{x_i} \bigg[ \Pgg(x_k) \bigg] $ \\ + $\frac{1}{s_{ij}s_{ijk}} \frac{x_j}{x_i} \bigg[ -\Pgg(x_k) \bigg] $ \\ +$\frac{1}{s_{jk}s_{ijk}} \frac{1}{x_i} \bigg[ -\Pgg(x_k)\bigg]$} 
 & \makecell[l]{ $+\frac{1}{s_{ij}s_{jk}} \frac{ x_j}{x_i} \bigg[\Pgg(x_k) \bigg]$ \\ $+ \frac{1}{s_{ij}s_{ijk}} \frac{ x_j}{x_i} \bigg[\Pgg(x_k) \bigg]$ \\$+ \frac{1}{s_{jk}s_{ijk}} \frac{ 1}{x_i} \bigg[ \Pgg(x_k) \bigg]$} \\ 
 \hline
 $x_j \rightarrow 0$ 
 & \makecell[l]{  \\  $+ \frac{1}{s_{ij}s_{ijk}} \frac{ x_i}{x_j} \bigg[ 2 \Pgg(x_k) \bigg]  $  \\ $+ \frac{1}{s_{jk}s_{ijk}} \frac{ x_k}{x_j} \bigg[ 2 \Pgg(x_k) \bigg]$} 
 & \makecell[l]{  \\ $+ \frac{1}{s_{ij}s_{ijk}} \frac{ x_i}{x_j} \bigg[ -2 \PggS(x_k) \bigg]  $ \\$+ \frac{1}{s_{jk}s_{ijk}} \frac{ x_k}{x_j} \bigg[ -2 \PggS(x_k) \bigg]$} 
 & \makecell[l]{  \\ $+ \frac{1}{s_{ij}s_{ijk}} \frac{ x_i}{x_j} \bigg[ -2 \PggS(1-x_k) \bigg]  $ \\$+ \frac{1}{s_{jk}s_{ijk}} \frac{ x_k}{x_j} \bigg[ -2\PggS(1-x_k) \bigg]$} 
 & \makecell[c]{  \\ 0 \\} \\
 \hline
 $x_k \rightarrow 0$ 
 & \makecell[l]{   \\ $+\frac{1}{s_{ij}s_{ijk}} \frac{ 1}{x_k} \bigg[2 \Pgg(x_i) \bigg] $ \\$+ \frac{1}{s_{jk}s_{ijk}} \frac{x_j}{x_k} \bigg[2 \Pgg(x_i) \bigg]$ } 
 & \makecell[l]{  $+\frac{1}{s_{ij}s_{jk}} \frac{x_j}{x_k} \bigg[ \Pgg(x_i) \bigg] $ \\ + $\frac{1}{s_{ij}s_{ijk}} \frac{1}{x_k} \bigg[ -\Pgg(x_i)  \bigg] $ \\ +$\frac{1}{s_{jk}s_{ijk}} \frac{x_j}{x_k} \bigg[ -\Pgg(x_i) \bigg]$} 
 & \makecell[c]{  \\ 0 \\}  
 & \makecell[l]{ $+\frac{1}{s_{ij}s_{jk}} \frac{ x_j}{x_k} \bigg[\Pgg(x_i) \bigg]$ \\ $+ \frac{1}{s_{ij}s_{ijk}} \frac{1}{x_k} \bigg[\Pgg(x_i) \bigg]$ \\$+ \frac{1}{s_{jk}s_{ijk}} \frac{ x_j}{x_k} \bigg[ \Pgg(x_i) \bigg]$} \\
 \hline
\end{tabular}
\end{center}
\caption{Singular behaviour of the $\Pggg$ triple-collinear splitting function in the limit where individual momentum fractions are small. The contributions from the iterated two-particle splittings are shown in column 2, while the contributions from the two permutations of $\Rgggsub$ are shown in columns 3 and 4 and the contributions for the entire splitting function $\Pggg$ is shown in column 5. Each row shows the singular limit for a different momentum fraction tending to zero. The vertical displacement within each cell is organised by $\{s_{ij},s_{jk},s_{ik},s_{ijk}\}$. }
\label{table:ggg}
\end{table}
\end{landscape}

\subsection{Two Gluons with a Collinear Quark or Antiquark}

There are two distinct splitting functions representing the clustering of two gluons and a quark which depend on whether or not the gluons are symmetrised over.

\vspace{3mm}\noindent (a) 
In the case where gluon $j$ is colour-connected to quark $i$ and gluon $k$, we find that,
\begin{align}
    \Pqgg(i^h,j,k) &= \frac{\Pqg(x_k)}{s_{ijk}} \frac{\Pqg\left(\frac{x_j}{1-x_k}\right)}{s_{ij}}
    + \frac{\Pqg(1-x_i)}{s_{ijk}}   \frac{\Pgg\left( \frac{x_j}{1-x_i} \right)}{s_{jk}} 
    \nonumber \\
    & \qquad + \frac{1}{s_{ijk}^2} \Rqgg(i,j,k) \, , \\
    \Pqgg(i,j^h,k) &= 0 \, , \\
    \Pqgg(i,j,k^h) &= 0 \, , 
    \label{eq:Pqgg}
    \end{align}

where
\begin{eqnarray}
\label{eq:Rqgg}
\Rqgg (i,j,k) &=& 
\frac{2 \ome}{(1-x_i)^2}  \frac{W_{jk}}{s_{jk}^2} 
+ \frac{4 \ome x_k }{(1-x_i)^{2}} \frac{\Tr{i}{j}{k}{\ell}}{s_{jk}} + \frac{\ome^2}{(1-x_k)}  \frac{\Tr{i}{j}{k}{\ell}}{s_{ij}}
\nonumber \\
&&
+ \fb_0 (x_i,x_j,x_k) 
+  \fb(x_i,x_j,x_k) \frac{s_{ijk} \Tr{i}{j}{k}{\ell}}{s_{ij}s_{jk}}, 
\end{eqnarray}
and
\begin{eqnarray}
\label{eq:fb0}
\fb_0 (x_i,x_j,x_k) &=& 
\ome \left(B_0(x_k,x_i) -1 + \ome A_0(x_i,x_k)\right),\\
\label{eq:fb}
\fb(x_i,x_j,x_k) &=&- \frac{x_j \Pqg(x_j)}{x_k (1-x_i)} - \frac{2 x_k \Pqg(x_k)}{x_j (1-x_i)} + \frac{4}{(1-x_i)} - 3 \ome.
\end{eqnarray}
Eqs.~\eqref{eq:Pqgg}--\eqref{eq:fb} are equivalent to Eq.~(5.5) in Ref.~\cite{campbell} up to a normalisation of a factor of 4. By charge conjugation, we also have,
\begin{equation}
P_{ \bar q g g}^{(0)} (i,j,k) = \Pqgg (i,j,k).
\end{equation}

We observe that $\fb$ contains inverse powers of $x_j$ and $x_k$. The 
behaviour of the $\Pqgg$ triple-collinear splitting function in the limit where individual momentum fractions are small is tabulated in Table~\ref{table:qgg}.
We see that there is no singular behaviour as $x_i \to 0$.  This reflects the fact that there is no singularity when the quark and spectator momentum are collinear and that there is no soft quark singularity.  When $x_j \to 0$, we see that there are contributions from both the strongly-ordered contribution and from $\Rqgg$ which cancel in the full $\Pqgg$ splitting function, 
\begin{equation}
    \Pqgg(i,j,k) \stackrel{x_j \to 0}{\longrightarrow} 0.
\end{equation}
When $x_k \to 0$, we see that the contributions from the strongly-ordered contribution and from $\Rqgg$ do not cancel in full $\Pqgg$ splitting function.

\begin{table}[t]
\centering
\begin{center}
\begin{tabular}{|c|| c |c|| c|} 
 \hline
 $\nameqgg$ & \makecell[l]{
 $\phantom{+} \frac{\Pqg(x_k)}{s_{ijk}} \frac{\Pqg\left(\frac{x_j}{1-x_k}\right)}{s_{ij}}$
 \\
 + $\frac{\Pqg(1-x_i)}{s_{ijk}}   
 \frac{\Pgg\left( \frac{x_j}{1-x_i} \right)}{s_{jk}}$ 
 } 
 & $\frac{1}{s_{ijk}^2} \Rqgg(i,j,k)$ 
 & $\frac{1}{s_{ijk}^2} \Pqgg(i,j,k)$ \\  
 \hline\hline
 $x_i \rightarrow 0$ 
 &\makecell[c]{0}
 & \makecell[c]{0}
 & \makecell[c]{0} \\ 
 \hline
 $x_j \rightarrow 0$ 
 & \makecell[l]{  \\  $+ \frac{1}{s_{ij}s_{ijk}} \frac{ x_i}{x_j} \bigg[ 2 \Pqg(x_k) \bigg]  $  \\ $+ \frac{1}{s_{jk}s_{ijk}} \frac{ x_k}{x_j} \bigg[ 2 \Pqg(x_k) \bigg]$} 
 & \makecell[l]{  \\  $+ \frac{1}{s_{ij}s_{ijk}} \frac{ x_i}{x_j} \bigg[ -2 \Pqg(x_k) \bigg]  $  \\ $+ \frac{1}{s_{jk}s_{ijk}} \frac{ x_k}{x_j} \bigg[ -2 \Pqg(x_k) \bigg]$} 
 & \makecell[c]{  \\ 0 \\} \\
 \hline
 $x_k \rightarrow 0$ 
 & \makecell[l]{   \\ $+\frac{1}{s_{ij}s_{ijk}} \frac{ 1}{x_k} \bigg[2 \Pqg(x_j) \bigg] $ \\$+ \frac{1}{s_{jk}s_{ijk}} \frac{x_j}{x_k} \bigg[2 \Pqg(x_j) \bigg]$ } 
 & \makecell[l]{ $+ \frac{1}{s_{ij}s_{jk}} \frac{x_j}{x_k} \bigg[ \Pqg(x_j) \bigg] $ \\ $+\frac{1}{s_{ij}s_{ijk}} \frac{ 1}{x_k} \bigg[- \Pqg(x_j) \bigg] $ \\$+ \frac{1}{s_{jk}s_{ijk}} \frac{x_j}{x_k} \bigg[-\Pqg(x_j) \bigg]$} 
 & \makecell[l]{$+ \frac{1}{s_{ij}s_{jk}} \frac{x_j}{x_k} \bigg[ \Pqg(x_j) \bigg] $ \\ $+\frac{1}{s_{ij}s_{ijk}} \frac{ 1}{x_k} \bigg[\Pqg(x_j) \bigg] $   \\$+ \frac{1}{s_{jk}s_{ijk}} \frac{x_j}{x_k} \bigg[\Pqg(x_j) \bigg]$} \\
 \hline
\end{tabular}
\end{center}
\caption{Singular behaviour of the $\Pqgg$ triple-collinear splitting function in the limit where individual momentum fractions are small. }
\label{table:qgg}
\end{table}

In the soft $k$ limit, only the strongly-ordered term contributes and we recover the expected limit describing collinear partons $i$ and $j$ with the soft gluon $k$ radiated between the colour-connected partners $j$ and $\ell$,
\begin{equation}
    \Pqgg(i,j,k) \stackrel{k~{\rm soft}}{\longrightarrow} \frac{2x_j}{s_{jk}x_k} \frac{1}{s_{ij}}\Pqg(x_j).
\end{equation}

However, in the soft $j$ limit the $1/(x_j s_{ij})$ and $1/(x_j s_{jk})$ terms cancel between the $P \times P$ and $\Rqgg$ contributions, such that
\begin{eqnarray}
    \frac{1}{s_{ijk}^2} \Rqgg(i,j,k) &
    \stackrel{j~{\rm soft}}{\longrightarrow}& \left(
    -\frac{2x_i}{x_j s_{ij}} 
    -\frac{2x_k}{x_j s_{jk}} 
    +\frac{2s_{ik}}{s_{ij}s_{jk}}
    \right)
    \frac{1}{s_{ik}}\Pqg(x_k),\\    
    \Pqgg(i,j,k) &
    \stackrel{j~{\rm soft}}{\longrightarrow}& \frac{2 s_{ik}}{s_{ij}s_{jk}}\frac{1}{s_{ik}}\Pqg(x_k).
\end{eqnarray}
This is precisely as expected for the emission of a soft gluon between the hard (and collinear) radiators $i$ and $k$.

As in the three gluon splitting function, there are double soft singularities when gluons $j,k$ are soft. These are contained iteratively in the $P \times P$ contributions and in $\Rqgg (i,j,k)$, and are identical to Eqs.~\eqref{eq:DsoftinPxPggg} and~\eqref{eq:DsoftinRggg},
\begin{eqnarray} 
\label{eq:DsoftinPxPqgg}
\frac{\Pqg(1-x_i)}{s_{ijk}}  \frac{\Pgg\left(\frac{x_j}{1-x_i} \right)}{s_{jk}}  &+& \frac{\Pqg(x_k)}{s_{ijk}}  \frac{\Pqg\left(\frac{x_j}{1-x_k} \right)}{s_{ij}}\nonumber \\
\stackrel{j,k~{\rm soft}}{\longrightarrow}&&
\frac{2}{\omxi s_{ijk} s_{jk}} \Pgg\left(\frac{x_j}{1-x_i}\right) + \frac{4}{x_j x_k s_{ijk} s_{ij}},\\
\label{eq:DsoftinRqgg}
\frac{1}{s_{ijk}^2}
\Rqgg(i,j,k) \stackrel{j,k~{\rm soft}}{\longrightarrow}&&
 \frac{2 \ome W_{jk}}{(1-x_i)^2 s_{jk}^2 s_{ijk}^2}
 - \left( 
\frac{2}{x_k (1-x_i)} + \frac{4}{x_j\omxi}
 \right) \frac{\Tr{i}{j}{k}{\ell}}{s_{ij} s_{jk} s_{ijk}}. \nonumber \\
\end{eqnarray}
There are no other double soft singularities.

As noted earlier, Eq.~\eqref{eq:Rqgg} also appears to have spurious singularities in both the $x_i \to 1$ and $x_k \to 1$ limits. As in the three gluon splitting function, the singular $x_i \to 1$ behaviour present in the second term in Eq.~\eqref{eq:Rqgg} cancels against the $B_0(x_k,x_i)$ term in $\fb_0$.  The singularity as $x_k \to 1$ in the third term cancels against a similar singularity produced by the $A_0(x_i,x_k)$ term in $\fb_0$.

\vspace{3mm}\noindent (b)  
In the case where the gluons are abelianised ($\tilde{g}$) or two photons are collinear to the quark, then the splitting function is symmetric under the exchange of the two bosons ($j,k$).  We find,
\begin{align}
    \label{eq:Pqpp}
    \Pqpp(i^h,j,k) &=
            \frac{\Pqg(x_k)}{s_{ijk}}  
            \frac{\Pqg\left(\frac{x_j}{1-x_k}\right)}{s_{ij}} 
            + \frac{\Pqg(x_j)}{s_{ijk}}  
            \frac{\Pqg\left(\frac{x_k}{1-x_j}\right)}{s_{ik}} \nonumber \\
            &\qquad	+ \frac{1}{s_{ijk}^2} \Rqpp (i,j,k) \, , \\ 
    \Pqpp(i,j^h,k) &= 0 \, , \\
    \Pqpp(i,j,k^h) &= 0 \, ,
\end{align}

where
\begin{eqnarray}
\label{eq:Rqpp}
\Rqpp (i,j,k) &=&
  -  \frac{\ome^2}{(1-x_k)} \frac{\Tr{j}{i}{k}{\ell}}{s_{ij}} 
  -  \frac{\ome^2}{(1-x_j)} \frac{\Tr{j}{i}{k}{\ell}}{s_{ik}} \nonumber \\
&& +  \fc_0(x_i,x_j,x_k)
		+ \fc(x_i,x_j,x_k) 
		\frac{s_{ijk} \Tr{j}{i}{k}{\ell}}{s_{ij}s_{ik}},
\end{eqnarray}
and
\begin{eqnarray}
\label{eq:fc0}
    \fc_0(x_i,x_j,x_k) &=& \ome\left(2 - \ome A_0(x_j,x_k) -\ome A_0(x_k,x_j)\right), \\
\label{eq:fc}
	\fc(x_i,x_j,x_k) &=& - \frac{x_j \Pqg(x_j)}{x_k (1-x_i)}  - \frac{x_k \Pqg(x_k)}{x_j (1-x_i)} \nonumber \\ 
    &&+ \frac{4}{(1-x_i)} - 4 \ome + \ome^2.
\end{eqnarray}
Eqs.~\eqref{eq:Pqpp}--\eqref{eq:fc} are equivalent to Eq.~(5.6) in Ref.~\cite{campbell} up to a normalisation of a factor of 4. By charge conjugation, we have 
\begin{equation}
P_{\bar q \gamma \gamma }^{(0)} (i,j,k) = \Pqpp (i,j,k).
\end{equation}

The 
behaviour of the $\Pqpp$ triple-collinear splitting function in the limit where individual momentum fractions are small is tabulated in Table~\ref{table:qpp}. As in the previous case, there is no singular behaviour as $x_i \to 0$ reflecting the fact that there is no singularity when the quark and spectator momentum are collinear and that there is no soft quark singularity.
We also see that there are contributions from both the strongly-ordered contribution and from $\Rqpp$ when $x_j \to 0$ and $x_k \to 0$ that do not cancel in the full $\Pqpp$ splitting function. 
However, only the strongly-ordered term contributes in the soft $j$ or soft $k$ limits, 
\begin{eqnarray}
    \Pqpp(i,j,k) &\stackrel{j~{\rm soft}}{\longrightarrow}& \frac{2x_i}{s_{ij}x_j} \frac{1}{s_{ik}}\Pqg(x_k),\\
   \Pqpp(i,j,k) &\stackrel{k~{\rm soft}}{\longrightarrow}& \frac{2x_i}{s_{ik}x_k} \frac{1}{s_{ij}}\Pqg(x_j).   
\end{eqnarray}
It can be seen that the strongly-ordered terms contribute the full double soft $j,k$ limit (a product of two eikonal factors) and there are no contributions from $\fc (x_i,x_j,x_k)$. There are no other double soft singularities.

\begin{table}[H]
\centering
\begin{center}
\begin{tabular}{|c|| c |c|| c|} 
 \hline
 $\nameqpp$ & 
 \makecell[l]{ $\phantom{+}\frac{\Pqg(x_k)}{s_{ijk}} \frac{\Pqg\left(\frac{x_j}{1-x_k}\right)}{s_{ij}}$
\\ + $ (j \leftrightarrow k) $}
 & $\frac{1}{s_{ijk}^2} \Rqpp(i,j,k)$ 
 & $\frac{1}{s_{ijk}^2} \Pqpp(i,j,k)$ \\  
 \hline\hline
 $x_i \rightarrow 0$ 
 &\makecell[c]{0}
 & \makecell[c]{0}
 & \makecell[c]{0} \\ 
 \hline
 $x_j \rightarrow 0$ 
 & \makecell[l]{  
 $\phantom{\bigg[}$ \\ 
 $+\frac{1}{s_{ij}s_{ijk}} \frac{ x_i}{x_j} \bigg[2 \Pqg(x_k) \bigg] $ \\
 $+ \frac{1}{s_{ik}s_{ijk}} \frac{1}{x_j} \bigg[2 \Pqg(x_k) \bigg]$ } 
 & \makecell[l]{ $+ \frac{1}{s_{ij}s_{ik}} \frac{x_i}{x_j} \bigg[\Pqg(x_k) \bigg]$ \\ 
 $+ \frac{1}{s_{ij}s_{ijk}} \frac{x_i}{x_j} \bigg[-\Pqg(x_k) \bigg]$ \\
 $+ \frac{1}{s_{ik}s_{ijk}} \frac{1}{x_j} \bigg[-\Pqg(x_k) \bigg]$} 
 & \makecell[l]{ $+ \frac{1}{s_{ij}s_{ik}} \frac{x_i}{x_j} \bigg[\Pqg(x_k) \bigg]$ \\ 
 $+ \frac{1}{s_{ij}s_{ijk}} \frac{x_i}{x_j} \bigg[\Pqg(x_k) \bigg]$ \\
 $+ \frac{1}{s_{ik}s_{ijk}} \frac{1}{x_j} \bigg[\Pqg(x_k) \bigg]$} \\
 \hline
 $x_k \rightarrow 0$ 
 & \makecell[l]{   $\phantom{\bigg[}$ \\ 
 $+\frac{1}{s_{ij}s_{ijk}} \frac{ 1}{x_k} \bigg[2 \Pqg(x_j) \bigg] $ \\
 $+ \frac{1}{s_{ik}s_{ijk}} \frac{x_i}{x_k} \bigg[2 \Pqg(x_j) \bigg]$ } 
 & \makecell[l]{ $+ \frac{1}{s_{ij}s_{ik}} \frac{x_i}{x_k} \bigg[\Pqg(x_j) \bigg]$ \\ 
 $+ \frac{1}{s_{ij}s_{ijk}} \frac{1}{x_k} \bigg[-\Pqg(x_j) \bigg]$ \\
 $+ \frac{1}{s_{ik}s_{ijk}} \frac{x_i}{x_k} \bigg[-\Pqg(x_j) \bigg]$} 
 & \makecell[l]{$+ \frac{1}{s_{ij}s_{ik}} \frac{x_i}{x_k} \bigg[\Pqg(x_j) \bigg]$ \\
  $+ \frac{1}{s_{ij}s_{ijk}} \frac{1}{x_k} \bigg[\Pqg(x_j) \bigg]$ \\
  $+ \frac{1}{s_{ik}s_{ijk}} \frac{x_i}{x_k} \bigg[\Pqg(x_j) \bigg]$} \\
 \hline
\end{tabular}
\end{center}
\caption{Singular behaviour of the $\Pqpp$ triple-collinear splitting function in the limit where individual momentum fractions are small.  }
\label{table:qpp}
\end{table}

\subsection{Quark-Antiquark Pair with a Collinear Gluon}

There are also two distinct splitting functions representing the clustering of a gluon with a quark-antiquark pair into a parent gluon.

\vspace{3mm}\noindent (a)  
When the gluon is colour-connected to the antiquark, we find that, 
\begin{align}
    \label{eq:Pgqbq}
    \Pgqbq(i^h,j,k) &=
    \frac{\PggS(1-x_i)}{s_{ijk}}
    \frac{\Pqq\left(\frac{x_k}{1-x_i}\right)}{s_{jk}}
    + \frac{1}{s_{ijk}^2} \Rgqbq(i,j,k) \, , \\
    \Pgqbq(i,j^h,k) &= 0 \, , \\
    \Pgqbq(i,j,k^h) &= \frac{\Pqq(x_k)}{s_{ijk}} 
    \frac{\Pqg\left(\frac{x_i}{1-x_k}\right)}{s_{ij}} 
    + \frac{\PggS(x_i)}{s_{ijk}} 
    \frac{\Pqq\left(\frac{x_k}{1-x_i}\right)}{s_{jk}} \, ,
\end{align}
%
where
\begin{eqnarray}
\label{eq:Rgqbq}
\Rgqbq(i,j,k) &= & 
- \frac{2}{(1-x_i)^2} \frac{W_{jk}}{s_{jk}^2} 
- \frac{\ome}{(1-x_k)} \frac{\Tr{i}{j}{k}{\ell}}{s_{ij}} 
- \frac{4 x_k}{(1-x_i)^2} \frac{ \Tr{i}{j}{k}{\ell}}{s_{jk}} \nonumber \\
&&
+ \fd_0 (x_i,x_j,x_k) 
+ \fd (x_i,x_j,x_k) \frac{s_{ijk} \Tr{i}{j}{k}{\ell}}{s_{ij}s_{jk}}, 
\end{eqnarray}
and
\begin{eqnarray}
\label{eq:fd0}
    \fd_0 (x_i,x_j,x_k) &=&- B_0(x_k,x_i) + 1 - \ome A_0(x_i,x_k) 
    , \\
\label{eq:fd}    
    \fd (x_i,x_j,x_k) &=& 
    - \frac{\Pqq (x_k)}{x_i (1-x_i)} + \frac{2}{(1-x_i)} \nonumber \\ 
    &&+ 1 - 2 x_i + \frac{2( x_j -x_k - 2x_j x_k)}{\ome (1-x_i)}. 
\end{eqnarray}
Eqs.~\eqref{eq:Pgqbq}--\eqref{eq:fd} are equivalent to Eq.~(5.8) in Ref.~\cite{campbell} up to a normalisation of a factor of 4.  By charge conjugation, we have that, 
\begin{equation}
\Pgqbq (i,j,k) = P_{g q \bar q}^{(0)} (i,j,k).
\end{equation}

\begin{table}[H]
\centering
\begin{center}
\begin{tabular}{|c|| c |c|| c|} 
 \hline
 $\namegqbq$ & 
\makecell[l]{$\phantom{+}\frac{\Pqq(x_k)}{s_{ijk}} 
\frac{\Pqg(\frac{x_i}{1-x_k})}{s_{ij}}$ 
\\
+
$\frac{\Pgg(x_i)}{s_{ijk}} 
\frac{\Pqq(\frac{x_k}{1-x_i})}{s_{jk}}$}
 & $\frac{1}{s_{ijk}^2} \Rgqbq(i,j,k)$ 
 & $\frac{1}{s_{ijk}^2} \Pgqbq(i,j,k)$ \\  
 \hline\hline
 $x_i \rightarrow 0$ 
 &\makecell[l]{  \phantom{\bigg[} \\ $+\frac{1}{s_{ij}s_{ijk}} \frac{ x_j}{x_i} \bigg[2 \Pqq(x_k) \bigg] $ \\$+ \frac{1}{s_{jk}s_{ijk}} \frac{1}{x_i} \bigg[2 \Pqq(x_k) \bigg]$ }
 & \makecell[l]{ $+ \frac{1}{s_{ij}s_{jk}} \frac{x_j}{x_i} \bigg[ \Pqq(x_k) \bigg]$  \\ $+\frac{1}{s_{ij}s_{ijk}} \frac{ x_j}{x_i} \bigg[-\Pqq(x_k) \bigg] $ \\$+ \frac{1}{s_{jk}s_{ijk}} \frac{1}{x_i} \bigg[-\Pqq(x_k) \bigg]$ }
 & \makecell[l]{$+ \frac{1}{s_{ij}s_{jk}} \frac{x_j}{x_i} \bigg[ \Pqq(x_k) \bigg]$  \\ $+\frac{1}{s_{ij}s_{ijk}} \frac{ x_j}{x_i} \bigg[\Pqq(x_k) \bigg] $ \\$+ \frac{1}{s_{jk}s_{ijk}} \frac{1}{x_i} \bigg[\Pqq(x_k) \bigg]$} \\ 
 \hline
 $x_j \rightarrow 0$ 
 & \makecell[c]{0}
 & \makecell[c]{0}
 & \makecell[c]{0} \\
 \hline
 $x_k \rightarrow 0$ 
 & \makecell[c]{0}
 & \makecell[c]{0}
 & \makecell[c]{0} \\
 \hline
\end{tabular}
\end{center}
\caption{
Singular Behaviour of the $\Pgqbq$ triple-collinear splitting function in the limit where individual momentum fractions are small. }
\label{table:gqbq}
\end{table}

The behaviour of the $\Pgqbq$  triple-collinear splitting function in the limit where individual momentum fractions are small is tabulated in Table~\ref{table:gqbq}. 
There are no collinear limits between the quark/antiquark and the spectator. There is singular behaviour as $x_i \to 0$.  
In the soft $i$ limit only the strongly-ordered term contributes, 
\begin{eqnarray}
    \Rgqbq(i,j,k) &\stackrel{i~{\rm soft}}{\longrightarrow}& 0,\\
    \Pgqbq(i,j,k) &\stackrel{i~{\rm soft}}{\longrightarrow}& \frac{2x_j}{s_{ij}x_i} \frac{1}{s_{jk}}\Pqq(x_k).
\end{eqnarray}

There are double soft singularities when the $q \bar q$ pair are both soft. These are contained iteratively in the $\Pgg \times \Pqq$ contribution and in $\Rgqbq (i,j,k)$,
\begin{eqnarray} 
\label{eq:DsoftinPxPgqbq}
\frac{\Pgg(x_i)}{s_{ijk}}  \frac{\Pqq\left(\frac{x_k}{1-x_i} \right)}{s_{jk}} 
&\stackrel{j,k~{\rm soft}}{\longrightarrow}&
\frac{2}{\omxi s_{ijk} s_{jk}} \Pqq\left(\frac{x_k}{1-x_i}\right),\\
\label{eq:DsoftinRgqbq}
\frac{1}{s_{ijk}^2}
\Rgqbq(i,j,k) &\stackrel{j,k~{\rm soft}}{\longrightarrow}&
 -\frac{2 W_{jk}}{(1-x_i)^2 s_{jk}^2 s_{ijk}^2}.
\end{eqnarray}
We identify the double soft terms in Eq.~\eqref{eq:DsoftinRgqbq} as uniquely double unresolved. 
There are no other double soft singularities.

\vspace{3mm}\noindent (b) 
The QED-like splitting, where the gluon, quark and antiquark form a photon-like colour singlet is given by, 
\begin{align}
    \label{eq:Ppqbq}
    \Ppqbq(i^h,j,k) &=
    \frac{\Pqq(1-x_k)}{s_{ijk}} 
    \frac{\Pqg\left(\frac{x_j}{1-x_k}\right)}{s_{ij}} 
    +
    \frac{\Pqq(1-x_i)}{s_{ijk}} 
    \frac{\Pqg\left(\frac{x_j}{1-x_i}\right)}{s_{jk}} \nonumber \\ 
    & \qquad + \frac{1}{s_{ijk}^2} \Rpqbq(i,j,k) \, , \\
    \Ppqbq(i,j^h,k) &= 0 \, , \\
    \Ppqbq(i,j,k^h) &= \frac{\Pqq(1-x_k)}{s_{ijk}} 
    \frac{\Pqg\left(\frac{x_j}{1-x_k}\right)}{s_{ij}} 
    +
    \frac{\Pqq(1-x_i)}{s_{ijk}} 
    \frac{\Pqg\left(\frac{x_j}{1-x_i}\right)}{s_{jk}} \nonumber \\ 
    & \qquad + \frac{1}{s_{ijk}^2} \Rpqbq(i,j,k) \, , 
\end{align}
%
where
\begin{eqnarray}
\label{eq:Rpqbq}
\Rpqbq(i,j,k) &= & 
 \frac{\ome}{(1-x_k)} \frac{\Tr{i}{j}{k}{\ell}}{s_{ij}} 
+ \frac{\ome}{(1-x_i)} \frac{\Tr{i}{j}{k}{\ell}}{s_{jk}} \nonumber \\
&&
+ \fe_0 (x_i,x_j,x_k) 
+ \fe (x_i,x_j,x_k) \frac{s_{ijk} \Tr{i}{j}{k}{\ell}}{s_{ij}s_{jk}}, 
\end{eqnarray}
and
\begin{eqnarray}
\label{eq:fe0}
    \fe_0 (x_i,x_j,x_k) &=& 
    -2 + \ome A_0(x_i,x_k) + \ome A_0(x_k,x_i)
    , \\
\label{eq:fe}
    \fe (x_i,x_j,x_k) &=& 
    -\frac{\Pqq(x_i)}{x_j}-\frac{\Pqq(x_k)}{x_j} + \frac{2 \epsilon}{\ome} x_j.
\end{eqnarray}
Eqs.~\eqref{eq:Ppqbq}--\eqref{eq:fe} are equivalent to Eq.~(5.10) in Ref.~\cite{campbell} up to a normalisation of a factor of 4. 
Note that because of charge conjugation this splitting function is symmetric under the exchange of the quark and antiquark $i,k$. 

\begin{table}[t]
\centering
\begin{center}
\begin{tabular}{|c|| c |c|| c|} 
 \hline
$\namepqbq$ & 
  \makecell[l]{
  $\frac{\Pqq(1-x_k)}{s_{ijk}} 
\frac{\Pqg(\frac{x_j}{1-x_k})}{s_{ij}}$ \\
+ $(i\leftrightarrow k)$}
 & $\frac{1}{s_{ijk}^2} \Rpqbq(i,j,k)$ 
 & $\frac{1}{s_{ijk}^2} \Ppqbq(i,j,k)$ \\  
 \hline\hline
 $x_i \rightarrow 0$ 
 & \makecell[c]{0}
 & \makecell[c]{0}
 & \makecell[c]{0} \\
 \hline
 $x_j \rightarrow 0$ 
 &\makecell[l]{ $+\frac{1}{s_{ij}s_{ijk}} \frac{ x_i}{x_j} \bigg[2 \Pqq(x_k) \bigg] $ \\$+ \frac{1}{s_{jk}s_{ijk}} \frac{x_k}{x_j} \bigg[2 \Pqq(x_k) \bigg]$ }
 & \makecell[l]{ $+\frac{1}{s_{ij}s_{ijk}} \frac{ x_i}{x_j} \bigg[-2 \Pqq(x_k) \bigg] $ \\$+ \frac{1}{s_{jk}s_{ijk}} \frac{x_k}{x_j} \bigg[-2 \Pqq(x_k) \bigg]$ }
 & \makecell[c]{0} \\ 
 \hline
 $x_k \rightarrow 0$ 
 & \makecell[c]{0}
 & \makecell[c]{0}
 & \makecell[c]{0} \\
 \hline
\end{tabular}
\end{center}
\caption{Singular behaviour of the $\Ppqbq$ triple-collinear splitting function in the limit where individual momentum fractions are small.}
\label{table:pqqb}
\end{table}

The 
behaviour of the $\Ppqbq$  triple-collinear splitting function in the limit where individual momentum fractions are small is tabulated in Table~\ref{table:pqqb}. 
There are no collinear limits between the quark/antiquark and the spectator.  
In the $x_j \to 0$ limit, the contributions from the strongly-ordered terms and $\Rpqbq$ cancel.
However, in the soft $j$ limit the $1/(x_j s_{ij})$ and $1/(x_j s_{jk})$ terms cancel between the $P \times P$ and $\Rpqbq$ contributions, such that
\begin{eqnarray}
    \frac{1}{s_{ijk}^2} \Rpqbq(i,j,k) &\stackrel{j~{\rm soft}}{\longrightarrow}& 
  \left(
  -\frac{2x_i}{x_j s_{ij}}
    -\frac{2x_k}{x_j s_{jk}}
    +\frac{2s_{ik}}{s_{ij}s_{jk}}\right) \frac{1}{s_{ik}}\Pqq(x_k),\\
    \Ppqbq(i,j,k) &\stackrel{j~{\rm soft}}{\longrightarrow}& \frac{2s_{ik}}{s_{ij}s_{jk}} \frac{1}{s_{ik}}\Pqq(x_k).
\end{eqnarray}
There are no double soft singularities.

\subsection{Quark-Antiquark Pair  with a Collinear Quark or Antiquark}

Finally, we consider the clustering of a quark-antiquark pair ($Q\bar{Q}$) and a quark $q$ to form a parent quark with the same flavour as $q$. 
There are two splitting functions, one where the quark flavours are different and one where the quarks have the same flavour.  

\vspace{3mm}\noindent (a) 
For distinct quarks, we have
\begin{align}
    \label{eq:PqQQ}
    \PqQQ(i^h,j,k) &=
    \frac{\Pqg(1-x_i)}{s_{ijk}} \frac{\Pqq\left( \frac{x_j}{1-x_i} \right)}{s_{jk}} 
    + \frac{1}{s_{ijk}^2} \RqQQ(i,j,k) \, , \\
    \PqQQ(i,j^h,k) &= 0 \, ,\\
    \PqQQ(i,j,k^h) &= 0 \, ,
\end{align}
%
where
\begin{eqnarray}
\label{eq:RqQQ}
\RqQQ(i,j,k) &=&  - \frac{2}{\omxi^2} \frac{W_{jk}}{s_{jk}^2} 
 - \frac{2 x_k}{\omxi^2} \frac{\Tr{i}{j}{k}{\ell}}{ s_{jk}} - \frac{2 x_j}{\omxi^2} \frac{\Tr{i}{k}{j}{\ell}}{ s_{jk}}  \nonumber\\
&& +\ff_0 (x_i,x_j,x_k)  , 
\end{eqnarray}
and
\begin{equation}
\label{eq:ff0}
\ff_0 (x_i,x_j,x_k) = - \frac{1}{2} ( B_0(x_j,x_i) + B_0(x_k,x_i)) + 1 +\epsilon .
\end{equation}

Eqs.~\eqref{eq:PqQQ}--\eqref{eq:ff0} are equivalent to Eq.~(5.12) in Ref.~\cite{campbell}, up to a normalisation of a factor of 4. 
Note that this splitting function is symmetric under the exchange of the quark and antiquark $j,k$ of the same flavour and by charge conjugation we have that, 
\begin{equation}
\PqQQ (i,j,k) = P_{q Q \bar Q}^{(0)} (i,j,k).
\end{equation}
Note that we have chosen to make this symmetry explicit in the trace structures.
There are no collinear limits between the quark/antiquark and the spectator, and no soft limits.

The double soft singularities when the $Q \bar Q$ pair are both soft are contained iteratively in the $\Pqg \times \Pqq$ contribution and in the $W_{jk}$ term in $\RqQQ (i,j,k)$, and are equivalent to those given in Eqs.~(\ref{eq:DsoftinPxPgqbq},\ref{eq:DsoftinRgqbq}),
\begin{eqnarray} 
\label{eq:DsoftinPxPqQQ}
\frac{\Pqg(1-x_i)}{s_{ijk}}  \frac{\Pqq\left(\frac{x_j}{1-x_i} \right)}{s_{jk}} 
&\stackrel{j,k~{\rm soft}}{\longrightarrow}&
\frac{2}{\omxi s_{ijk} s_{jk}} \Pqq\left(\frac{x_j}{1-x_i}\right),\\
\label{eq:DsoftinRqQQ}
\frac{1}{s_{ijk}^2}
\RqQQ(i,j,k) &\stackrel{j,k~{\rm soft}}{\longrightarrow}&
 -\frac{2 W_{jk}}{(1-x_i)^2 s_{jk}^2 s_{ijk}^2}.
\end{eqnarray}
There are no other double soft singularities.

\vspace{3mm}\noindent (b) 
For identical quarks, we have 
\begin{align}
    \label{eq:Pqqq}
    \Pqqq(i^h,j,k) &= \frac{1}{s_{ijk}^2} \Rqqq(i,j,k) \, , \\
    \Pqqq(i,j^h,k) &= 0 \, , \\
    \Pqqq(i,j,k^h) &= \frac{1}{s_{ijk}^2} \Rqqq(i,j,k) \, , 
\end{align}
%
where
\begin{eqnarray}
\label{eq:Rqqq}
\Rqqq(i,j,k) &=& 
-  \frac{2\ome}{\omxi} \frac{\Tr{i}{j}{k}{\ell}}{s_{jk}} 
-  \frac{2\ome}{\omxk} \frac{\Tr{i}{j}{k}{\ell}}{s_{ij}} 
\nonumber \\
&& + \fg_0(x_i,x_j,x_k) + \fg(x_i,x_j,x_k) 
\frac{s_{ijk}\Tr{i}{j}{k}{\ell}}{s_{ij} s_{jk}},
\end{eqnarray}
and
\begin{eqnarray}
\label{eq:fg0}
\fg_0 (x_i,x_j,x_k) &=&  
 -2 \ome (\epsilon + A_0(x_i,x_k) + A_0(x_k,x_i)), \\
\label{eq:fg}
    \fg(x_i,x_j,x_k)  &=& 
    -\frac{2x_j}{\omxi\omxk} \nonumber \\ 
    && +\ome \left(\frac{\omxk}{\omxi} + \frac{\omxi}{\omxk}
    +2+\epsilon \right).
\end{eqnarray}
Eqs.~\eqref{eq:Pqqq}--\eqref{eq:fg} are equivalent to Eq.~(5.14) in Ref.~\cite{campbell}, up to a normalisation of a factor of 4. 
Because of charge conjugation,
\begin{equation}
P_{\bar q q \bar q }^{(0)}  (i,j,k) = \Pqqq(i,j,k).
\end{equation}
There are no collinear limits between the quark/antiquark and the spectator, and no soft limits. 
There are no double soft singularities.

\subsection{N=1 SUSY Identity}
\label{sec:SUSY}

The two particle and three particle splitting functions are related by an $N=1$ supersymmetry (SUSY) identity that relates the mass of the spin-1 gluon to the spin-1/2 gluino. 
The gluino can be identified as a quark in this scenario. At one loop, the two particle cuts of the one-loop self energy are equal, leading to the identity~\cite{Antoniadis:1981zv}
\begin{equation} 
\label{eq:SUSY2}
	\Pgg (x)  + \Pqq (x) = \Pqg (x) + \Pgq (x) ,
\end{equation}
which only holds in the $d=4, \epsilon = 0$ limit. This is because in dimensional regularisation the number of degrees of freedom of the gluon and gluino are not equal and SUSY is broken.
The left-hand side of Eq.~\eqref{eq:SUSY2} are the two particle cuts of the one-loop gluonic self energy - i.e. the splitting functions which split from a gluon,  while on the right are 
those which split from a quark (gluino). 

Similarly the triple-collinear splitting functions are related by the three-particle cuts of the two-loop self energies leading to the $N=1$ SUSY identity~\cite{campbell} (where $\epsilon=0$), 
\begin{eqnarray}
\label{eq:SUSY3}
\lefteqn{
	(\Pggg + 2 \Pgqbq + \Ppqbq) (i,j,k)+ (\text{5 perms.}) }\nonumber \\ 
&=&	(2 \Pqgg + \Pqpp + 2 \PqQQ +  \Pqqq ) (i,j,k) + (\text{5 perms.}).
\end{eqnarray}
The strongly-ordered contributions automatically satisfy Eq.~\eqref{eq:SUSY3} through repeated use of Eq.~\eqref{eq:SUSY2}.  The remaining contributions satisfy, 
\begin{eqnarray}
\label{eq:SUSYR3}
\lefteqn{
	(\Rggg + 2 \Rgqbq + \Rpqbq) (i,j,k)+ (\text{5 perms.}) }\nonumber \\ 
&=&	(2 \Rqgg + \Rqpp + 2 \RqQQ +  \Rqqq ) (i,j,k) + (\text{5 perms.}).
\end{eqnarray}
In Eq.~\eqref{eq:SUSYR3}, the terms proportional to each possible kinematic pole structure, $1/s_{ijk}^2$, $1/(s_{ijk}s_{ij})$,  $1/(s_{ij}s_{jk})$, $1/s_{ij}^2$ and cyclic permutations separately cancel. 
This leads to relations amongst the coefficients of the trace-structures and amongst the functions multiplying $1/s_{ijk}^2$.
Additionally, by analysing terms proportional to $1/(s_{ij}s_{jk})$, the following relationship holds (where $\epsilon =0$):
\begin{eqnarray}
    && 2 \fa(x_i,x_j,x_k) + 2 \fd(x_i,x_j,x_k) + \fe(x_i,x_j,x_k) + (i \leftrightarrow k) \nonumber \\
     && \hspace{1cm} =  2 \fb(x_i,x_j,x_k) + \fc(x_j,x_i,x_k) + \fg(x_i,x_j,x_k) + (i \leftrightarrow k) . 
\end{eqnarray}

\section{Summary} \label{sec:summary} 

In this chapter, we have rewritten the triple-collinear splitting functions $P_{abc}^{(0)}$ in a way that exposes the single and double unresolved limits.  
In particular, we have isolated the strongly-ordered iterated contributions as products of the usual spin-averaged two-particle splitting functions (generically $P \times P$) and a remainder function $R_{abc}^{(0)} (i,j,k)$ that is finite when any pair of $\{i,j,k\}$ are collinear.  

To help with the discussion of the unresolved limits, we introduced the notion of internal and external singularities.   
Internal singularities are only associated with small invariants in the set $\{ s_{ij}, s_{ik}, s_{jk}, s_{ijk}\}$ and correspond to simple-collinear, single soft or triple-collinear $i,j,k$ contributions. By construction, 
\begin{itemize}
    \item[-] {\bf Internal} simple-collinear singularities between a pair of $\{i,j,k\}$ lead to a factor of $1/s_{ij}$ and are captured by the iterated two-particle splitting contributions ($P \times P$). We write $R_{abc}^{(0)}$ in a way that makes it visibly finite in each of these simple-collinear limits.
    \item[-] {\bf Internal} single soft singularities, when the soft particle is colour-connected to the other two collinear particles, produce terms like $s_{ik}/(s_{ij}s_{jk})$ and appear only in $R_{abc}^{(0)}$. 
\end{itemize}

External singularities reference other particles involved in the scattering - for example, the spectator particles used to define the momentum fractions of the three collinear particles.  This includes external simple-collinear singularities involving one of the collinear particles and a spectator particle, soft radiation where one of the spectator particles is colour-connected to the collinear particle or other external double unresolved singularities. These show up in the following way,
\begin{itemize}
    \item[-] When {\bf external} simple-collinear singularities like $1/x_i$ are present in the full $P_{abc}^{(0)}$ splitting function, they are all contained in $P \times P$ terms.
    \item[-] When {\bf external} simple-collinear singularities like $1/x_i$ {\bf do not} appear in the full $P_{abc}^{(0)}$ splitting function, then any terms proportional to $1/x_i$ in $P \times P$ will cancel with analogous terms coming from $R_{abc}^{(0)}$. 
    \item[-] {\bf External} single soft singularities where the soft particle is colour-connected to a spectator particle produce terms like $x_i/(x_j s_{ij})$ and appear only in the iterated  $P \times P$ terms. 
\end{itemize}
In the triple-collinear splitting function, there are two inverse powers of the small invariants.  Double collinear (two pairs of collinear particles), soft-collinear, double soft or other triple-collinear limits than $i,j,k$, all depend on singularities involving one or more of the momentum fractions and are all therefore external singularities.  
In particular, the double soft limit requires at least one singular factor involving the momentum fractions and is classed as an external singularity. These singularities appear in both the iterated $P \times P$ terms and in $R_{abc}^{(0)}$. Double soft singularities are always the overlap between triple-collinear $\{i,j,k\}$ and external triple-collinear singularities.

We find it useful to think of a hard radiator particle that emits possibly unresolved radiation, together with a spectator particle. Therefore, we have split each splitting function into at most three pieces according to this principle. 

Decomposing the triple-collinear splitting functions, as presented in this chapter, will prove crucial in developing idealised antenna functions for an improved NNLO antenna-subtraction scheme. We will explore the idealisation of antenna functions in Chapters~\ref{chapter:paper2} and~\ref{chapter:paper3}, after introducing the antenna-subtraction architecture in Chapter~\ref{chapter:antennasub}. 

\chapter{Antenna Subtraction}
\label{chapter:antennasub}

This chapter is intended to illustrate the antenna subtraction method up to NNLO. 
We will set the context in Section~\ref{sec:slicsub} by introducing the theory for both slicing and subtraction schemes and examples of them at various orders. 
After that, we will summarise the concept of antenna functions and the successes of antenna subtraction at NNLO in Section~\ref{sec:introant}. 
We will introduce all the antenna functions necessary for NNLO calculations, including their integrals, and the momentum mappings necessary for building subtraction terms. 
Next, we will explore the structure of antenna subtraction at NLO in Section~\ref{sec:antNLO} before engaging in an extended study of NNLO antenna subtraction in Section~\ref{sec:antNNLO}. 
This includes the definition of subtraction terms for matrix elements with two real emissions, one real emission with one virtual particle and two virtual particles. 
We end by summarising the chapter in Section~\ref{sec:antsummary}.

\section{Slicing and Subtraction}
\label{sec:slicsub}

As discussed in Chapters~\ref{chapter:qcd} and~\ref{chapter:pheno}, when performing higher-order calculations, a particular challenge is the handling and cancellation of IR divergences across different multiplicity phase spaces. 
In general, there are two classes of scheme for extracting and handling the IR divergences: slicing and subtraction. 

Slicing schemes work by identifying a parameter which separates the IR-finite from the IR-divergent regions of the phase space. 
For a generic differential $m$-jet cross section, we have the full $m+1$ particle phase space $d \Phi_{m+1}$, split into the IR-divergent region $d \Psi_{m+1}$ and the IR-finite region $d \Phi_{m+1} \setminus d \Psi_{m+1}$. 
The idea at NLO is presented schematically by
\begin{eqnarray}
    d \hat{\sigma}^{\text{NLO}} &=& \int_{d \Phi_{m}} V + \int_{d \Phi_{m+1}} R  ,\\
        &\approx&  \int_{d \Phi_{m}} V + \int_{d \Phi_{m+1} \setminus d \Psi_{m+1}} R + \int_{d \Psi_{m+1}} F ,
\end{eqnarray}
where $V$ is the virtual correction and $R$ is the real correction to the differential cross section $d \hat{\sigma}$. $F$ will be introduced below. 
In the IR-finite regions, the phase-space integration can be calculated numerically in 4 dimensions. In the IR-divergent regions, slicing schemes utilise the universal factorisation properties of matrix elements.
They use these properties to approximate the real corrections in the IR-divergent regions with a relevant function, $F$ in the example above, which can be analytically integrated in $d$ dimensions. 
After integration, one can verify the cancellation of IR-divergences expressed as $\e$-poles between the integrated analytic functions and the virtual contributions. 
Complications arise in ensuring that the result is independent of the slicing parameter, which only occurs if the approximation used is applied deep enough within the singular regions. 
This cancellation of the slicing parameter must be performed numerically, to a high enough accuracy, between the integrated parts of the real corrections. 
However, the deeper we slice into the IR-divergent region, the more unstable and computationally costly the numerical integration becomes. 
Therefore, the final result is necessarily approximate because the functions introduced in the IR-divergent regions are only completely valid when in the exact IR limits and $F$ neglects sub-leading terms. 

Slicing schemes demonstrate non-local cancellation of the IR divergences because they do not cancel point-by-point in phase space but instead cancel after integration. 
Examples of slicing schemes include initial developments in the 1990s~\cite{PhysRevD.46.1980,Giele:1993dj} and more recent schemes such as $q_T$-slicing~\cite{Catani:2007vq,Bozzi:2005wk}, which slices on the transverse momentum of a colour singlet, and $N$-jettiness~\cite{Boughezal:2015eha,Gaunt:2015pea}, which slices on an event shape parameter. 

Subtraction schemes work by the introduction of well chosen counterterms which cancel the IR divergences point-by-point (locally) in the phase space. 
The combination of all the counterterms must contribute 0 to the final result. 
At NLO, this is described schematically by
\begin{eqnarray}
    \label{eq:NLOsubtract}
    d \hat{\sigma}^{\text{NLO}} &=& \bigg[ \int_{d \Phi_{m}} V + \int_{d \Phi_{m+1}} S \bigg] 
    + \int_{d \Phi_{m+1}} ( R  - S ) \nonumber \\
    &=& \int_{d \Phi_{m}} \bigg[ V + \int_{d \Psi} S \bigg] 
    + \int_{d \Phi_{m+1}} ( R  - S ),
\end{eqnarray}
where $S$ is the real subtraction term, which ensures that $(R  - S)$ is IR-finite at every point in phase space. 
Since this combination is IR-finite, it can be integrated numerically, utilising modern Monte Carlo sampling techniques. 
In order to demonstrate full cancellation of IR divergences, a subtraction scheme must verify the cancellation of $\e$-poles due to IR divergences in the first two terms of Eq.~\eqref{eq:NLOsubtract}. 
This requires the subtraction term, $S$, to be analytically integrable over a well-defined single unresolved phase space, $d \Psi$, which is defined according to
\begin{equation}
d \Phi_{m+1} = d \Psi d \Phi_m .   
\end{equation}
If this is the case, the implicit divergences in $S$, once integrated over $d \Psi$, become explicit as $\e$-poles and cancel against the integrand $V$ under the $d \Phi_m$ integral. 

At NLO, fully-differential calculations have been automated thanks to two general subtraction schemes known as Catani-Seymour dipole subtraction~\cite{Catani:1996vz} and FKS subtraction~\cite{Frixione:1995ms}, which were developed in the late-1990's. 
The generality of these subtraction schemes has facilitated two fully-differential NLO matching schemes, known as MC@NLO \cite{Frixione:2002ik} and POWHEG \cite{Nason:2004rx,Frixione:2007vw} which systematically combine NLO fixed-order calculations with all-order parton-shower resummation. 
Together with automated one-loop matrix-element generators~\cite{madgraph:2011uj,Cascioli:2011va}, these schemes are used for fully-differential high-multiplicity processes. 
These NLO subtraction schemes form the backbone of state-of-the-art multi-purpose event generators~\cite{powheg:2010xd,Alwall:2014hca,Bellm:2019zci,Sherpa:2019gpd,Bierlich:2022pfr}, see Ref.~\cite{snowmass:2022qmc} for a recent summary. Furthermore, newer methods at NLO have also been developed~\cite{Prisco:2020kyb,Bertolotti:2022ohq,Giachino:2023loc}. 

At NNLO, subtraction can be described schematically by
\begin{eqnarray}
    \label{eq:NNLOsubtract}
    d\hat\sigma^{\text{NNLO}} &=& \int_{d\Phi_{m}} VV  \nonumber \\
    &+& \int_{d \Phi_{m+1}} ( RV - S_{RV} ) + \int_{d \Phi_{m+1}} S_{RV} \nonumber \\
    &+& \int_{d\Phi_{m+2}} ( RR - S_{RR}) + \int_{d\Phi_{m+2}} S_{RR} .
\end{eqnarray}
There are three corrections at NNLO to the differential cross section $d\hat\sigma$: double-real ($RR$), real-virtual ($RV$) and double-virtual ($VV$). 
$S_{RR}$ is the double-real subtraction term, which ensures that $(RR-S_{RR})$ is IR-finite at every point in phase space; in other words, $S_{RR}$ matches the implicit singularity structure of $RR$. 
The real-virtual correction contains both types of IR divergence: implicit and explicit. $S_{RV}$ is chosen such that the implicit divergences of $RV$ are cancelled. 
While parts of $S_{RV}$ may contain explicit divergences, the full $S_{RV}$ must be $\e$-finite. The $\e$-poles in $RV$ are subtracted by a different contribution, as we will see.  
The double-real correction contains both single and double unresolved divergences which must be subtracted by $S_{RR}$. 
As such, we can rewrite $S_{RR}$ as
\begin{equation}
    S_{RR} = S_{RR,1} + S_{RR,2},
\end{equation}
to separate the single and double unresolved parts of $S_{RR}$ respectively. 
Terms in $S_{RR,1}$ must then be integrated over a single unresolved phase space ($d \Psi_1$, similarly to NLO), generating $\e$-poles (up to $\e^{-2}$), which cancel against those in $RV$ under the $d{\Phi}_{m+1}$ integral, due to the KLN theorem. 
Terms in $S_{RR,2}$ must be integrated over a double unresolved phase space, $d\Psi_2$, generating $\e$-poles (up to $\e^{-4}$). 
These single and double unresolved phase spaces must satisfy, with unspecified momentum mappings,
\begin{equation}
    d \Phi_{m+2} = d \Psi_i d \Phi_{m+2-i} ,
\end{equation}
for $i \in \{1,2\}$. Using this we can express the finite blocks, in squared brackets, at each level of the NNLO subtraction,
\begin{eqnarray}
    d\hat\sigma^{\text{NNLO}} &=& \int_{d\Phi_{m}} \bigg[ VV + \int_{d \Psi_{1}} S_{RV} + \int_{d \Psi_2} S_{RR,2} \bigg]   \nonumber \\
    &+& \int_{d \Phi_{m+1}} \bigg[ RV - S_{RV} + \int_{d \Psi_{1}} S_{RR,1} \bigg] \nonumber \\
    &+& \int_{d\Phi_{m+2}} \bigg[ RR - S_{RR} \bigg]  .
\end{eqnarray}
The $S_{RV}$ term contains only single unresolved divergences and requires integration over a single unresolved phase space ($d \Psi_1$, similarly to NLO), generating up to 2 further $\e$-poles in addition to those already present in $S_{RV}$. 
The KLN theorem ensures that, at the double-virtual level, all the explicit divergences ($\e$-poles) in $VV$ are cancelled under the $d \Phi_m$ integral by $S_{RV}$ integrated over a single unresolved phase space and $S_{RR,2}$ integrated over a double unresolved phase space. 

Subtraction at NNLO therefore contains multiple complications not present at NLO: we require cancellations to be apparent for many overlapping unresolved configurations; at the real-virtual level we require cancellation of both implicit and explicit singularities simultaneously; and we require more complex phase-space integrations. 
There have been multiple approaches taken to create subtraction terms which minimise these complications, each representing a different subtraction scheme. 
At NNLO, there are the antenna subtraction scheme~\cite{Gehrmann-DeRidder:2005btv}, projection-to-Born~\cite{Cacciari:2015jma}, CoLoRFulNNLO~\cite{DelDuca:2016ily}, nested soft-collinear subtraction~\cite{Caola:2017dug}, local analytic sector subtraction~\cite{Magnea:2018hab}, geometric subtraction~\cite{Herzog:2018ily} and sector-improved residue subtraction~\cite{Czakon:2010td}; see Ref.~\cite{TorresBobadilla:2020ekr} for a review. 
The implementation of these methods is currently done one process at a time and they do not straightforwardly scale to higher multiplicities. 

\newpage
At \NthreeLO, subtraction can be described schematically by
\begin{eqnarray}
    \label{eq:N3LOsubtract}
    d\hat\sigma^{\text{\NthreeLO}} &=& \int_{d\Phi_{m}} VVV \nonumber \\
    &+& \int_{d \Phi_{m+1}} ( RVV - S_{RVV} ) + \int_{d \Phi_{m+1}} S_{RVV}  \nonumber \\
    &+& \int_{d \Phi_{m+2}} ( RRV - S_{RRV} ) + \int_{d\Phi_{m+2}} S_{RRV} \nonumber \\
    &+& \int_{d\Phi_{m+3}} ( RRR - S_{RRR}) + \int_{d\Phi_{m+3}} S_{RRR} .
\end{eqnarray}
There are four corrections at \NthreeLO to the differential cross section $d\hat\sigma$: $RRR$, $RRV$, $RVV$ and $VVV$. $S_{RRR}$ is the triple-real subtraction term, which ensures that $(RRR-S_{RRR})$ is IR-finite at every point in phase space. 
The double-real-virtual correction ($RRV$) contains both types of IR divergence: implicit and explicit. 
$S_{RRV}$ is chosen such that the implicit divergences of $RRV$ are cancelled. 
The same comments for $RRV$ can be applied to the real-double-virtual correction ($RVV$) and its subtraction term. 
The triple-real correction ($RRR$) contains single, double and triple unresolved divergences which must be subtracted by $S_{RRR}$. 
We can rewrite $S_{RRR}$ into single, double and triple unresolved parts respectively,
\begin{equation}
    S_{RRR} = S_{RRR,1} + S_{RRR,2} + S_{RRR,3}.
\end{equation}
Terms in $S_{RRR,1}$ must then be integrated over a single unresolved phase space ($d \Psi_1$, similarly to NLO), generating $\e$-poles (up to $\e^{-2}$), which cancel against those in $RRV$ under the $d{\Phi}_{m+2}$ integral, by the KLN theorem. 
Terms in $S_{RRR,2}$ must be integrated over a double unresolved phase space ($d \Psi_2$, similarly to NNLO), generating $\e$-poles (up to $\e^{-4}$). 
Terms in $S_{RRR,3}$ must be integrated over a triple unresolved phase space, $d \Psi_3$, generating $\e$-poles (up to $\e^{-6}$). The $S_{RRV}$ term contains single and double unresolved parts, 
\begin{equation}
    S_{RRV} = S_{RRV,1} + S_{RRV,2},
\end{equation}
which require integration over a single ($d \Psi_1$) and double ($d \Psi_2$) unresolved phase space respectively. 
The $S_{RVV}$ term contains single unresolved limits which require integrating over a single unresolved phase space, $d \Psi_1$. 
The KLN theorem ensures that, at every level, all the explicit divergences ($\e$-poles) in each correction are cancelled by integrated parts of various subtraction terms. 
Using the decomposed structure of the subtraction terms we can present the finite blocks, in squared brackets, at each level of the \NthreeLO calculation,
\begin{eqnarray}
    d\hat\sigma^{\text{\NthreeLO}} &=& \int_{d\Phi_{m}} \bigg[ VVV + \int_{d \Psi_1} S_{RVV} + \int_{d \Psi_2} S_{RRV,2} + \int_{d \Psi_3} S_{RRR,3} \bigg] \nonumber \\
    &+& \int_{d \Phi_{m+1}} \bigg[ RVV - S_{RVV} + \int_{d \Psi_1} S_{RRV,1} + \int_{d \Psi_{2}} S_{RRR,2} \bigg] \nonumber \\
    &+& \int_{d \Phi_{m+2}} \bigg[ RRV - S_{RRV} + \int_{d \Psi_1} S_{RRR,1} \bigg]  \nonumber \\
    &+& \int_{d\Phi_{m+3}} \bigg[ RRR - S_{RRR}\bigg] .
\end{eqnarray}

Unfortunately, due to the great complexity only hinted at here, a process-independent subtraction scheme is not available. 
Whether an \NthreeLO calculation is phenomenologically justified depends on the particular process but nonetheless there have been a range of \NthreeLO calculations to date. 
At \NthreeLO, inclusive~\cite{Anastasiou:2015vya,Anastasiou:2016cez,Mistlberger:2018etf,Dreyer:2016oyx,Duhr:2019kwi,Duhr:2020kzd,Chen:2019lzz,Currie:2018fgr,Dreyer:2018qbw,Duhr:2020sdp,Duhr:2020seh} and differential calculations have started to emerge~\cite{Dulat:2017prg,Dulat:2018bfe,Cieri:2018oms,Chen:2021isd,Chen:2021vtu,Billis:2021ecs,Chen:2022cgv,Neumann:2022lft,Camarda:2021ict,Chen:2022lwc,Baglio:2022wzu}, the latter mainly for $2 \to 1$ processes via the use of the Projection-to-Born method~\cite{Cacciari:2015jma} or $q_T$-slicing techniques~\cite{Catani:2007vq} to promote established NNLO calculations to \NthreeLO. 
This is done by using fully differential NNLO calculations for production of $X +$jet for an \NthreeLO calculation of $X$ production, where the jet has been taken unresolved. 
We note that the first steps towards an \NthreeLO antenna-subtraction scheme have been taken in Refs.~\cite{Jakubcik:2022zdi,Chen:2023fba,Chen:2023egx}. 
Nevertheless, calculations for higher multiplicities are currently hindered by the lack of process-independent local \NthreeLO subtraction schemes, since good progress is being made on three-loop matrix elements, see Refs.~\cite{Caola:2020dfu,Caola:2021rqz,Caola:2021izf,Bargiela:2021wuy,Gehrmann:2023jyv}. 

\section{Introduction to Antenna Subtraction}
\label{sec:introant}

The antenna subtraction scheme is one of the most successful methods for fully-differential NNLO calculations in QCD.
It was first proposed for perturbative QCD calculations with massless partons in electron-positron annihilation in Refs.~\cite{Gehrmann-DeRidder:2005btv,Gehrmann-DeRidder:2005alt,Gehrmann-DeRidder:2005svg}. 
It allowed the calculation of the NNLO corrections to 3-jet production and related event-shape observables in electron-positron annihilation \cite{Catani:2007vq,Gehrmann-DeRidder:2007foh,Gehrmann-DeRidder:2007nzq,Gehrmann-DeRidder:2007vsv,Gehrmann-DeRidder:2008qsl}. 
The extension of the scheme to the treatment of initial-state radiation relevant to processes with initial-state hadrons was established at NLO in Ref.~\cite{Daleo:2006xa} and extended to NNLO in Refs.~\cite{Daleo:2009yj,Pires:2010jv,Boughezal:2010mc,Gehrmann:2011wi,Gehrmann-DeRidder:2012too,Currie:2013vh}. 
A cornerstone of the antenna-subtraction framework is that all of the integrals relevant for processes at NNLO with massless quarks are known analytically \cite{Daleo:2009yj,Boughezal:2010mc,Gehrmann:2011wi,Gehrmann-DeRidder:2012too}. 
The scheme has now been applied to a range of LHC processes through the parton-level NNLOJET Monte Carlo event generator. 
The extension of antenna subtraction for the production of heavy particles at hadron colliders has been studied in Refs. \cite{Gehrmann-DeRidder:2009lyc,Abelof:2011ap,Bernreuther:2011jt,Abelof:2011jv,Abelof:2012bga,Abelof:2012rv,Bernreuther:2013uma,Dekkers:2014hna}. 
Besides its application in fixed-order calculations, the antenna framework has been utilised in antenna-shower algorithms \cite{Gustafson:1987rq,Lonnblad:1992tz,Giele:2007di,Giele:2011cb,Fischer:2016vfv,Brooks:2020upa}, where it enabled proof-of-concept frameworks for higher-order corrections \cite{Li:2016yez} and fully-differential NNLO matching \cite{Campbell:2021svd}.

Antenna subtraction is based on a simple idea: the singular behaviour of a matrix element is determined by universal factorisation properties, so they are also present in the lowest-multiplicity matrix elements. 
That is, any subtraction term can be built out of universal building blocks, antenna functions, extracted from ratios of simpler matrix elements. 
This has the benefit that the integration of each subtraction term is usually less challenging than the state-of-the-art loop integrals. 
Additionally, antenna functions smoothly interpolate between natural groups of unresolved singularities. 
In particular, an antenna function should encode all the colour-ordered unresolved singularities between two hard partons. 
An example of an antenna subtraction term for a matrix element $M_m^L$ is given by
\begin{equation} \label{eqn:subterm}
    X_{n+2}^\ell(i_1^h,i_3,\ldots,i_{n+2},i_2^h) 
    M_{m-n}^{L-\ell} (\ldots,I_1,I_2,\ldots) \, ,
\end{equation}
where $X_{n+2}^\ell$ represents an $\ell$-loop, $(n+2)$-particle antenna,
$i_1^h$ and $i_2^h$ represent the hard radiators, and $i_3$ to $i_{n+2}$ denote the $n$ unresolved particles. Particle $i$ carries a four-momentum $p_i^{\mu}$.
As the hard radiators may either be in the initial or in the final state, final-final (FF), initial-final (IF) and initial-initial (II) configurations need to be considered in general. $M_{m-n}^{L-\ell}$ is the reduced matrix element, with $n$ fewer particles, $\ell$ fewer loops and where $I_1^h$ and $I_2^h$ represent the particles obtained through an appropriate mapping,
\begin{align}
\{ p_{i_1},p_{i_3},\ldots,p_{i_{n+2}},p_{i_2} \} \mapsto \{ p_{I_1}, p_{I_2} \}.  
\end{align}

At NLO, the number of loops $\ell$ is equal to 0 and the number of unresolved particles is equal to 1; we only have the $X_3^0 (i^h,j, k^h)$-type antennae. 
These contain the unresolved singularities for $i^h j$ collinear, $j$ soft and $k^h j$ collinear. A suitable $3 \to 2$ mapping, for NLO FF configurations, requires 
\begin{eqnarray}
    p_I + p_K = p_i + p_j + p_k, &&\hspace{1cm} p_I^2 = p_K^2 = 0, \hspace{1cm} \text{always}, \nonumber \\
    p_I = p_i + p_j, &&\hspace{1cm} p_K = p_k, \hspace{1cm} \text{ in }i j \text{ collinear limit}, \nonumber \\
    p_K = p_k + p_j, &&\hspace{1cm} p_I = p_i, \hspace{1cm} \text{ in }k j \text{ collinear limit}, \nonumber \\
    p_I = p_i, &&\hspace{1cm} p_K = p_k, \hspace{1cm} \text{ in } j \text{ soft limit}. \nonumber 
\end{eqnarray}
A general on-shell momentum mapping which satisfies these properties is the antenna mapping, first presented in Ref.~\cite{Kosower:2002su}. It is given by,
\begin{eqnarray}
    \label{eq:antmommap}
p_I &=& x p_i + r p_j + z p_k, \\
\label{eq:antmommap2}
p_K &=& (1-x)p_i + (1-r)p_j + (1-z) p_k ,
\end{eqnarray} 
where
\begin{eqnarray}
    x &=& \frac{1}{2(s_{ik}  + s_{ij})} \left[ (1+\rho_1)s_{ijk} - 2 r s_{jk} \right], \\
    r &=& \frac{s_{jk}}{s_{ij} + s_{jk}} , \\
    z &=& \frac{1}{2 (s_{ik} + s_{jk})} \left[ (1-\rho_1)s_{ijk} - 2 r s_{ij} \right], \\
    \rho_1^2 &=& 1 + 4 r (1-r) \frac{s_{ij} s_{jk}}{s_{ijk} s_{ik}} . 
\end{eqnarray}
In the antenna framework, kinematic mappings are agnostic to the roles of the parent radiator partons and the transverse recoil of any additional emission is shared between them. Note that this is different to dipole-like kinematics, in which one of the parents is identified as the emitter and the other as the recoiler, whose role is solely to absorb the transverse recoil. Similar mappings are available for the IF and II configurations, which differ in how the recoil is redistributed~\cite{Daleo:2006xa}. 

At NNLO there are two additional cases: $\X$ corresponding to $\ell=0$ loops and $n=2$ unresolved particles; and $X_3^1$ corresponding to $\ell=1$ loop and $n=1$ unresolved particle. For subtraction terms at NNLO containing products of $X_3^0$ functions, the $3 \to 2 $ mapping can be iterated between the two terms. For subtraction terms containing $X_3^1$, the $3 \to 2$ mapping can be used. However, for subtraction terms at NNLO containing $\X$ antenna functions, used for subtracting colour-connected double unresolved limits, a new $4 \to 2 $ mapping is required. 

At NNLO there is a FF double unresolved antenna momentum mapping to describe the coalescence of four particles into two, $(i,j,k,l) \to (I,L)$, which satisfies similar properties to its NLO counterpart~\cite{Kosower:2002su}. The properties it must obey are given by
\begin{eqnarray}
    \label{eq:4to2mapprops}
    p_I + p_L = p_i + p_j + p_k+p_l, &&\hspace{0.5cm} p_I^2 = p_L^2 = 0, \hspace{0.5cm} \text{always}, \nonumber \\
    p_I = p_i + p_j, &&\hspace{0.5cm} p_L = p_k+p_l, \hspace{0.5cm} \text{ in }i j \text{ and } kl \text{ collinear limit}, \nonumber \\
    p_I = p_i , &&\hspace{0.5cm} p_L = p_k + p_l, \hspace{0.5cm} \text{ in } j \text{ soft and } k l \text{ collinear limit}, \nonumber \\
    p_L = p_l , &&\hspace{0.5cm} p_I = p_i + p_j, \hspace{0.5cm} \text{ in } k \text{ soft and } i j \text{ collinear limit}, \nonumber \\
    p_I = p_i + p_j + p_k , &&\hspace{0.5cm} p_L = p_l, \hspace{0.5cm} \text{ in } i j k \text{ collinear limit}, \nonumber \\
    p_L = p_j + p_k + p_l , &&\hspace{0.5cm} p_I = p_i, \hspace{0.5cm} \text{ in } j k l \text{ collinear limit}, \nonumber \\
    p_I = p_i, &&\hspace{0.5cm} p_L = p_l, \hspace{0.5cm} \text{ in } j k \text{ soft limit}. 
\end{eqnarray}
Therefore, it is suitable for four-parton antenna functions with divergent behaviour in these limits only, where $i$ and $l$ are hard and $j$ and $k$ are unresolved. 
The FF $4 \to 2$ mapping is given in full by
\begin{eqnarray}
p_I &=& x_1 p_i + x_2 p_j +x_3 p_k+ x_4 p_l, \\
p_L &=& (1-x_1)p_i + (1-x_2)p_j + (1-x_3) p_k + (1-x_4) p_l ,
\end{eqnarray}
where 
\begin{eqnarray}
    x_1 &=& \frac{1}{2(s_{ij} + s_{ik} + s_{il})} \bigg[ (1+\rho_2)s_{ijkl} -x_2 (s_{jk} + 2 s_{jl}) - x_3 (s_{jk} + 2 s_{kl}) \nonumber \\
    && + (x_2 - x_3) \left( \frac{s_{ij} s_{kl} - s_{ik} s_{jl}}{s_{il}} \right)\bigg], \\
    x_2 &=& \frac{s_{jk}+s_{jl}}{s_{ij} + s_{jk}+s_{jl}} , \\
    x_3 &=& \frac{s_{kl}}{s_{ik} + s_{jk}+s_{kl}} , \\
    x_4 &=& \frac{1}{2 (s_{il} + s_{jl} + s_{kl})} \bigg[ (1-\rho_2)s_{ijkl} -x_2 (s_{jk} + 2 s_{ij}) - x_3 (s_{jk} + 2 s_{ik}) \nonumber \\
    && - (x_2 - x_3) \left( \frac{s_{ij} s_{kl} - s_{ik} s_{jl}}{s_{il}} \right)  \bigg]. 
\end{eqnarray}
$\rho_2$ is defined by
\begin{eqnarray}
    \rho_2^2 &=& 1 + \frac{(x_2 - x_3)^2}{s_{il}^2 s_{ijkl}^2} \lambda (s_{ij} s_{kl}, s_{il} s_{jk}, s_{ik} s_{jl}) \nonumber \\
    &&+ \frac{1}{s_{il}s_{ijkl}} \bigg( 2(x_2(1-x_3) + x_3(1-x_2)) (s_{ij}s_{kl} + s_{ik} s_{jl} - s_{jk} s_{il}) \nonumber \\
    && + 4x_2 (1-x_2) s_{ij} s_{jl} + 4 x_3 (1-x_3) s_{ik} s_{kl} \bigg),
\end{eqnarray}
while the Källen function is given by
\begin{equation}
    \lambda (x,y,z) = x^2 + y^2 + z^2 - 2(xy + xz + yz). 
\end{equation}
This mapping has the useful property that it reduces to the $3 \to 2$ antenna mapping in any single unresolved limit. 
This proves crucial so that the single unresolved divergences can be removed from $\X M^0_{m-2}$ by products like $X_3^0 X_3^0 M^0_{m-2}$ with two iterated $3 \to 2$ mappings. 
We will see this in more detail in Section~\ref{sec:RRsub}. 

At \NthreeLO there are three additional cases: $X_5^0$ corresponding to $\ell=0$ loops and $n=3$ unresolved particles; $X_4^1$ corresponding to $\ell=1$ loop and $n=2$ unresolved particles; and $X_3^2$ corresponding to $\ell=2$ loops and $n=1$ unresolved particle.  

\section{NLO Antenna Subtraction}
\label{sec:antNLO}

\begin{figure}[h!]
    \epsfig{file=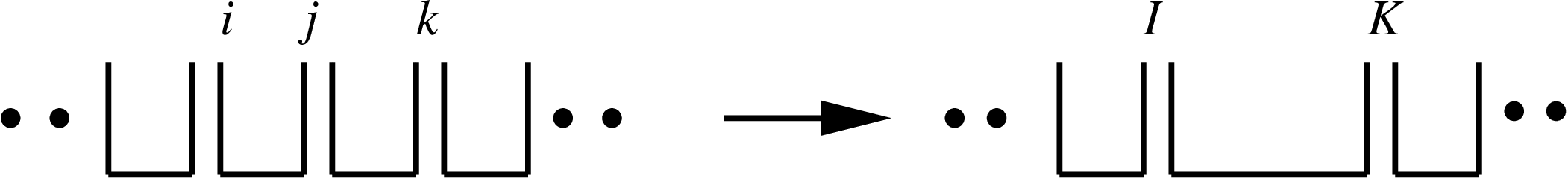,height=1.2cm}
    \caption{Diagram showing the colour-connections within the matrix element before and after the $3 \to 2$ antenna momentum mapping~\cite{Gehrmann-DeRidder:2005btv}.}
    \label{fig:combine}
    \end{figure}

While NNLO antenna subtraction will form the bulk of the discussion, it is instructive to detail antenna subtraction at NLO, as it is considerably simpler. 
We can begin with the generic NLO subtraction formula in Eq.~\eqref{eq:NLOsubtract}, where we now specify the content of the subtraction term in the FF case for the leading-colour calculation of an $m$-jet cross section, correct to an overall factor,
\begin{equation}
    \label{eq:NLOantsub}
    S \sim \sum_{\text{perms}} d \Phi_{m+1} \sum_j X_3^0 (i^h,j,k^h) {M}_m^0 (...,I,K,...) J_m^{(m)}.
\end{equation}
The sum over permutations is the sum over all contributing colour-orderings, while the sum over the unresolved particle $j$ is the sum over all final state particles, so that the subtraction term captures unresolved limits from any of the particles in a particular colour-ordering. The antenna function $X_3^0$ is different depending on the particle content of $i,j,k,$ denoted by a different letter. ${M}_m^0 (...,I^h,K^h,...)$ is the reduced squared matrix element relative to the real squared matrix element correction, 
\begin{equation}
    R \sim \sum_{\text{perms}} {M}_{m+1}^0 (...,i,j,k,...) J_m^{(m+1)},
\end{equation}
that $S$ is subtracting against. The jet function $J_m^{(m+n)}$ applies the appropriate jet algorithm for the experimental comparison, which ensures there are $m$ jets from $m+n$ partons, where we have suppressed its $m+n$ arguments. The arguments of the reduced matrix element are related to those of the real matrix element by the antenna momentum mapping for $(i,j,k) \to (I,K)$ in Eqs.~\eqref{eq:antmommap} and~\eqref{eq:antmommap2} and the colour-connections can be seen in Fig.~\ref{fig:combine}. 
We specify that $d \Phi_m$ denotes the $2 \to m$ particle phase space given in its unfactorised form by,
\begin{eqnarray}
    && d \Phi_m (p_1,...,p_m;  k_1,k_2) = \nonumber \\
    && \hspace{1cm} \frac{d^{d-1} p_1}{2 E_1 (2 \pi)^{d-1}}  \ldots \frac{d^{d-1} p_m}{2 E_m (2 \pi)^{d-1}} (2 \pi)^d \delta^d (p_1 + \ldots + p_m - k_1 - k_2), 
\end{eqnarray}
where $\{p_i\}$ are the $m$ final-state momenta, $E_i$ is the energy component of the four-momentum $p_i$ and $\{k_1,k_2\}$ are the initial-state momenta. The subtraction terms for a sub-leading colour calculation include more colour-connections because the full correction cannot be written as a sum of squared colour-ordered matrix elements, see Refs.~\cite{Currie:2013vh,Gehrmann:2023dxm}.  

The success of antenna subtraction depends upon the analytic integrability of the antenna functions. In the FF case at NLO which we focus on here, we integrate antennae over the single unresolved antenna phase space, $d \Phi_{X_{ijk}}$. In order to compare the integrated antennae to the virtual matrix element, the antenna phase space must satisfy
\begin{eqnarray}
    && d \Phi_{m+1} (p_1,...,p_i, p_j, p_k,...,p_{m+1}; k_1,k_2) = \nonumber \\
    && \hspace{1cm} d \Phi_{X_{ijk}} (p_i,p_j,p_k; p_I, p_K) d \Phi_m (p_1,...,p_I, p_K, ...,p_{m+1}; k_1,k_2), 
\end{eqnarray}
which fixes $d \Phi_{X_{ijk}}$ at $m=2$ because $d \Phi_2 = P_2$, a constant, we have
\begin{equation}
    d \Phi_{X_{ijk}} = \frac{d \Phi_3}{P_2}.
\end{equation}
Full details are given in Ref.~\cite{Gehrmann-DeRidder:2003pne} and $d \Phi_{X_{ijk}}$ in its factorised form is given in Appendix~\ref{app:integration}. We define the integrated antennae as 
\begin{equation}
    \mathcal{X}_3^0 (s_{ijk})= 8 \pi^2 (4 \pi)^{-\e} e^{\e \gamma_E} \int d \Phi_{X_{ijk}} X_3^0 (i^h,j,k^h),
\end{equation}
where we have included a normalisation factor to account for powers of the strong-coupling constant. 
By integrating the antennae, the subtraction term which cancels against the virtual matrix element correction,
\begin{equation}
    V \sim \sum_{\text{perms}} {M}_{m}^{1} (...,i,j,k,...) J_m^{(m)},
\end{equation}
can be denoted schematically by
\begin{equation}
    \int d \Phi_{X_{ijk}} S \sim \sum_{\text{perms}} \sum_{i,j} \mathcal{X}_3^0 (s_{ij}) {M}_m^0 (...,i,j,...) J_m^{(m)},
\end{equation}
where the permutations are over those colour-orderings which contribute and the sum is over all colour-adjacent dipole pairs, $(i,j)$. 
Technically, there can be multiple distinct antennae with the same-type parents, so the integrated antennae are grouped into the integrated antenna dipoles, $\J{1}\left(i,j \right)$, introduced in Section~\ref{sec:IRdivs}. Since the $\J{1}$ have the same pole structure as Catani's IR singularity operators~\cite{Catani:1998bh,Chen:2022ktf}, the virtual subtraction term is guaranteed to cancel the one-loop IR singularities in the virtual correction. 
Additionally, the KLN theorem guarantees this cancellation, since each term in the virtual subtraction term is an integrated version of a term in the real subtraction term. 

In Chapters~\ref{chapter:paper2} and~\ref{chapter:paper3} we propose an alternative formulation of antenna functions, so we label the antenna functions extracted using the original method as $X_{n+2}^{\ell, \text{OLD}}$. The original method for extracting the $X_3^{0,\text{OLD}}$ is by fixing them according to a ratio of simple tree-level squared matrix elements, 
\begin{equation}
    X_3^{0, \text{OLD}} (i,j,k)= \mathcal{S}_{ijk/IK} \frac{{M}_3^0 (i,j,k)}{{M}_2^0 (I,K)}, 
\end{equation}
which contain the desired single unresolved limits. The two-particle squared matrix element depends only on $s_{IK} = s_{ijk}$. 
Here we focus only on the final-final configuration. 
Antenna functions can be classified into three categories: quark-antiquark ($q \bar{q}$), quark-gluon ($qg$) and gluon-gluon ($gg$). These represent the two parents in the antenna's momentum mapping. That is, the particle-type of $I_1^h$ and $I_2^h$ for an antenna mapping $(i_1^h,i_3,\ldots,i_{n+2},i_2^h) \to (I_1^h, I_2^h)$. Note that this is often, but not always, the same as the particle-type of the antenna's hard radiators, $i_1^h$ and $i_2^h$. 
\begin{itemize}
    \item $q \bar{q}$ $X_3^{0,\text{OLD}}$ are obtained from matrix elements for the decay of a virtual photon into three and two partons~\cite{Gehrmann-DeRidder:2005btv}. There is only one three-parton squared matrix element for this case, $qg \bar{q}$, so one antenna function, $A_3^{0,\text{OLD}}$. 
    \item $qg$ $X_3^{0,\text{OLD}}$ are obtained from matrix elements for the decay of a heavy neutralino, in the minimally supersymmetric standard model (MSSM)~\cite{Gehrmann-DeRidder:2005svg}. The gluino takes the role of the quark in this scenario. For the case of $qg \sim \tilde{g} g$ parents, there are two three-parton squared matrix elements, $\tilde{g}gg$ and $\tilde{g} \bar{Q} Q$, so there are two antenna functions, $D_3^{0,\text{OLD}}$ and $E_3^{0,\text{OLD}}$. 
    \item $gg$ $X_3^{0,\text{OLD}}$ are obtained from matrix elements for the decay of a Higgs boson into three and two partons~\cite{Gehrmann-DeRidder:2005alt}. For the case of $gg$ parents, there are two three-parton squared matrix elements, $ggg$ and $g \bar{Q} Q$, so there are two antenna functions, $F_3^{0,\text{OLD}}$ and $G_3^{0,\text{OLD}}$.
\end{itemize}

The definitions and integrals of these $X_3^{0,\text{OLD}}$ are sufficient for a functioning NLO subtraction scheme to handle IR divergences. 
The $X_3^{0,\text{OLD}}$ antennae are considerably simpler objects than those at NNLO, although we can see the traces of the complexity present at NNLO. 
Let us consider the unresolved limits of two antennae, $A_3^{0,\text{OLD}} (i_q,j_g,k_{\bar{q}})$ and $D_3^{0,\text{OLD}}(i_q,j_g,k_g)$.  $A_3^{0,\text{OLD}} (i_q,j_g,k_{\bar{q}})$  contains the singular limits,
\begin{eqnarray}
    A_3^{0,\text{OLD}} (i_q,j_g,k_{\bar{q}}) &\to& \Pqgzero(i^h,j) \hspace{1cm} \text{ in }i j \text{ collinear limit}, \\
    A_3^{0,\text{OLD}} (i_q,j_g,k_{\bar{q}}) &\to& \Pqgzero(k^h,j) \hspace{1cm} \text{ in }j k \text{ collinear limit}, \\
    A_3^{0,\text{OLD}} (i_q,j_g,k_{\bar{q}}) &\to& \Sg(i^h,j_g,k^h) \hspace{1cm} \text{ in } j \text{ soft limit}, 
\end{eqnarray}
and nothing else. 
Note that $A_3^{0,\text{OLD}} (i_q,j_g,k_{\bar{q}})$ does exactly as expected by containing the three NLO unresolved limits between a hard quark-antiquark pair. 
$D_3^{0,\text{OLD}} (i_q,j_g,k_{g})$ however, contains the singular limits, 
\begin{eqnarray}
    D_3^{0,\text{OLD}} (i_q,j_g,k_{g}) &\to& \Pqgzero(i^h,j) \hspace{1cm} \text{ in }i j \text{ collinear limit}, \\
    D_3^{0,\text{OLD}} (i_q,j_g,k_{g}) &\to& \Pggzero(k^h,j) + \Pggzero(j^h,k) \text{ in }j k \text{ collinear limit}, \\
    D_3^{0,\text{OLD}} (i_q,j_g,k_{g}) &\to& \Pqgzero(i^h,k) \hspace{1cm} \text{ in }i k \text{ collinear limit}, \\
    D_3^{0,\text{OLD}} (i_q,j_g,k_{g}) &\to& \Sg(i^h,j_g,k^h) \hspace{1cm} \text{ in }j  \text{ soft limit}, \\
    D_3^{0,\text{OLD}} (i_q,j_g,k_{g}) &\to& \Sg(i^h,k_g,j^h) \hspace{1cm} \text{ in }k  \text{ soft limit}, 
\end{eqnarray}
and nothing else. 
There are extra limits present, compared to $A_3^{0,\text{OLD}}$, because the gluon $k_g$ is connected to the quark $i_q$ `around the back' in the defining matrix element ${M}_3^0 (i_q,j_g,k_g)$, and as such $D_3^{0,\text{OLD}} (i_q,j_g,k_{g})$ contains a $j \leftrightarrow k$ symmetry. 
This means that $D_3^{0,\text{OLD}} (i_q,j_g,k_{g})$ contains more unresolved limits than the NLO unresolved limits between a hard quark-gluon pair. 
This is potentially problematic for creating subtraction terms like Eq.~\eqref{eq:NLOantsub}, where each term in the $\sum_j$ should account for $j$ unresolved singularities only; that is $i^h j$ collinear, $j$ soft and $j k^h$ collinear. 
If one were to use $D_3^{0,\text{OLD}} $ directly in Eq.~\eqref{eq:NLOantsub}, unresolved singularities would be double-counted. 
Additionally, the antenna momentum mapping requires that two of the particles in the antenna can be identified as hard and the other unresolved. 
In practice there is a simple decomposition, 
\begin{equation}
    D_3^{0,\text{OLD}}(i,j,k) = \Xold{d} (i,j,k) + \Xold{d} (i,k,j),
\end{equation}
where $\Xold{d}$ has the required antenna properties.
Similarly, 
\begin{equation}
    F_3^{0,\text{OLD}}(i,j,k)  = \Xold{f} (i,j,k) + \Xold{f} (j,k,i) + \Xold{f} (k,i,j),
\end{equation}
where $\Xold{f}(i,j,k)$ has the required antenna properties. Both $\Xold{d}$ and $\Xold{f}$ are simple to integrate. 
Solving the same issue is not as convenient at NNLO, due to the huge increase in complexity in the matrix element expressions, as we will see later. 

\section{NNLO Antenna Subtraction}
\label{sec:antNNLO}

At NNLO, we encounter a number of additional features not present at NLO. 
The most obvious feature is that we can form subtraction terms out of three types of antenna functions: $\X$, $X_3^1$ and $X_3^0$. 
This means that we have many types of subtraction term, designed to subtract different IR limits. 
Before discussing the subtraction terms, we will introduce the new antenna functions at NNLO and how to manipulate them. 

\subsection{$\X$}

$\X (i^h,j,k,l^h)$ are used to encapsulate all the single and double unresolved divergences between two hard radiators, $i^h$, $l^h$, at tree-level. 
They are functions of the four momenta (in the full phase space) and smoothly interpolate between the following limits: double-soft ($j,k$), triple-collinear ($i^hjk$ or $l^hkj$), double-collinear (two pairs of collinear particles simultaneously), soft-collinear (one particle soft and another pair collinear), single-soft ($j$ or $k$) and simple-collinear. 
There are different antenna functions depending on particle content. 
Since we also decompose squared matrix elements according to colour structures, we have both leading-colour $\X$ and sub-leading-colour $\Xt$, with specific letters to indicate the particle content of $i,j,k,l$.
Their original extraction is similar to that of the $X_3^{0,\text{OLD}}$ and is given by
\begin{equation}
        X_4^{0, \text{OLD}} (i,j,k,l)= \mathcal{S}_{ijkl/IL} \frac{{M}_4^0 (i,j,k,l)}{{M}_2^0 (I,L)}, 
\end{equation}
in terms of a ratio of colour-ordered squared matrix elements, where the two-particle squared matrix element is a function of $s_{IL} = s_{ijkl}$ only. 
Like the $X_3^0$, they can be categorised according to the parent partons, $q \bar{q}$, $qg$ or $gg$. 
Exactly as at NLO, the matrix elements used are decays of a virtual photon, a heavy neutralino in the MSSM and a Higgs boson respectively. 
We will not discuss the specific $\X$ here because we will explore their full details in Chapter~\ref{chapter:paper2}. 

Once we have the appropriate $ 4 \to 2$ mapping, we can introduce the four-parton antenna phase space for integrating $\X$, which satisfies the factorisation formulae,
\begin{eqnarray}
    && d \Phi_{m+2} (p_1,...,p_i, p_j, p_k,p_l,...,p_{m+2}; k_1,k_2) = \nonumber \\
    && \hspace{1cm} d \Phi_{X_{ijkl}} (p_i,p_j,p_k,p_l; p_I, p_L) d \Phi_m (p_1,...,p_I, p_L, ...,p_{m+2}; k_1,k_2). 
\end{eqnarray}
The definition of the four-parton antenna phase space can be taken by setting $m=2$, to give
\begin{equation}
    d \Phi_{X_{ijkl}} = \frac{d \Phi_4}{P_2}.
\end{equation}
We then define the integrated antennae as 
\begin{equation}
    \mathcal{X}_4^0 (s_{ijkl})= \left[8 \pi^2 (4 \pi)^{-\e} e^{\e \gamma_E} \right]^2 \int d \Phi_{X_{ijkl}} X_4^0 (i^h,j,k,l^h),
\end{equation}
with an appropriate normalisation. 
The method for performing the phase-space integration consists of a few stages. 
Firstly, the four-particle phase space is rewritten in terms of a tripole phase space, where momentum invariants are normalised to be between 0 and 1. 
Then reverse unitarity-Cutkosky rules~\cite{Cutkosky:1960sp,Anastasiou:2002yz} are exploited to express the integrals in terms of cut multi-loop diagrams. 
This facilitates the use of multi-loop tools, which utilise integration-by-parts and other methods to express the cut diagrams in terms of known scalar master integrals. 
The method is detailed in full for the FF case in Ref.~\cite{Gehrmann-DeRidder:2003pne}, the IF case in Ref.~\cite{Daleo:2009yj} and the II case in Ref.~\cite{Boughezal:2010mc}. 

The double unresolved antenna momentum mapping requires the identification of two hard particles $i^h,l^h$ and two unresolved particles $j,k$, and only has the desired properties for certain colour-ordered unresolved limits. 
Unfortunately, the $X_4^{0, \text{OLD}}$ extracted from four-particle matrix elements may have one or both of the following problems. 
Firstly, the $X_4^{0, \text{OLD}}$ are not always compatible with the mapping. 
This happens when the antenna contains unresolved configurations for which the mapping does not have favourable properties, like those four-parton limits absent in Eq.~\eqref{eq:4to2mapprops}. 
Secondly, the $X_4^{0, \text{OLD}}$ may contain spurious limits that an antenna should not typically contain. 
In the original formulation of antenna subtraction, certain antenna functions were decomposed into sub-antennae, for which two particles can be identified as hard and two unresolved, and which were compatible with the $4\to 2 $ mapping. 
This was done at the cost of very long expressions (longer than the full $X_4^{0, \text{OLD}}$) and non-integrability of the sub-antennae. 
These sub-antennae are extracted using complex combinations of $N=1$ supersymmetry identities and partial fractioning, which result in families of integrals that are outside the standard techniques available. 
In turn, we gain the restriction that sub-antennae must appear in subtraction terms in conjunction with certain other sub-antennae, such that they can be recombined at the integrated level to give the full integrated $\mathcal{X}_4^0$. 
The ideal design features of the $\X$ and how to maximise them will be addressed in Chapter~\ref{chapter:paper2}. 

\subsection{$X_3^1$}

$X_3^1 (i^h,j,k^h)$ are used to encapsulate the single unresolved divergences between two hard radiators, $i^h$, $k^h$, at one-loop. 
We can recall from Section~\ref{sec:limits} that one-loop squared matrix elements, ${M}_n^1$, display a different type of IR factorisation compared to tree-level, as in Eq.~\eqref{eq:1loopfact}. 
There is a term corresponding to the unresolved limits of $\mathcal{M}_n^0$, which are encapsulated by the $X_3^0$ (we call this (tree$\times$loop)) and a term corresponding to the unresolved limits of $\mathcal{M}_n^1$ (loop$\times$tree), which are truly one-loop limits. 
For this second term we define the $X_3^1$, which contains the one-loop single unresolved limits set out in Section~\ref{sec:limits}. 
The extraction of the $X_3^1$ in the original formulation uses the ratio of matrix elements~\cite{Gehrmann-DeRidder:2005btv},
\begin{equation}
    X_3^{1,\text{OLD}} (i,j,k) = \mathcal{S}_{ijk/IK} \frac{ {M}_3^1 (i,j,k)}{{M}_2^0 (I,K)} - X_3^{0,
    \text{OLD}} (i,j,k) \frac{ {M}_2^1  (I,K)}{{M}_2^0 (I,K)},
\end{equation}
where we can see the (tree$\times$loop) contribution to the one-loop three-parton matrix element has been removed using the $X_3^0$ antennae and the two-particle matrix elements depend only on $s_{IK} = s_{ijk}$. 
The same momentum mapping, $(i,j,k) \to (I,K)$, as for the $X_3^0$ in Eqs.~\eqref{eq:antmommap} and~\eqref{eq:antmommap2} can be used for the $X_3^1$. 
We expect different singular factors, depending on particle content and colour structure. 
Since we decompose matrix elements according to colour structures, we have both leading-colour $X_3^1$, sub-leading-colour $\widetilde{X}_3^1$ and closed-quark-loop $\widehat{X}_3^1$, with specific letters to indicate the particle content of $i,j,k$. 
The $X_3^1$ are integrated over the single unresolved antenna phase space, $d \Phi_{X_{ijk}}$, the same as at NLO, so we define the integrated antennae as
\begin{equation}
    \mathcal{X}_3^1 (s_{ijk})= 8 \pi^2 (4 \pi)^{-\e} e^{\e \gamma_E} \int d \Phi_{X_{ijk}} X_3^1 (i^h,j,k^h). 
\end{equation}
Unlike the tree-level antennae, the $X_3^1$ contain both implicit and explicit IR divergences. 
We will see that this adds another layer of complication to NNLO subtraction. 
This is because the $X_3^1$ are used to match and subtract certain implicit IR divergences from the real-virtual matrix element, while their explicit IR divergences will not, in general, match those of the real-virtual matrix element. 
The $X_3^1$, in particular, will be further explored in Chapter~\ref{chapter:paper3}. 

\subsection{An Overview of the Subtraction Terms}

\begin{figure}[t]
    \centering
    \includegraphics[width=0.95\textwidth]{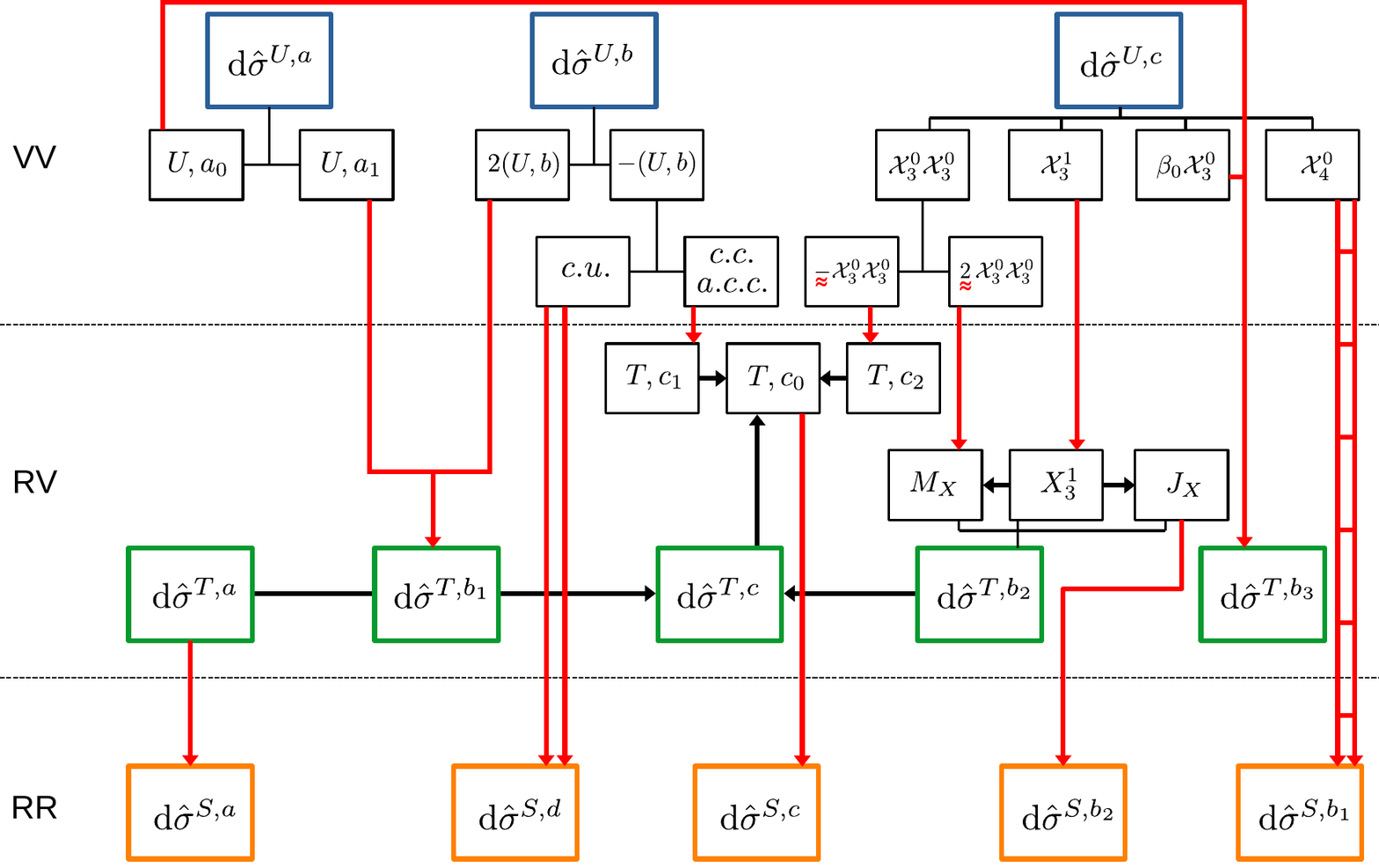}
    \caption{Structure of the antenna subtraction terms at NNLO~\cite{Gehrmann:2023dxm}. Red arrows link together subtraction terms at one level with an integrated counterpart at a higher level. A single arrow represents (un)integration over a single unresolved phase space, two parallel disconnected arrows represent (un)integration over two disconnected single unresolved phase spaces and two parallel connected arrows represent (un)integration over a double unresolved phase space. Each contribution will be discussed in the remainder of this chapter. }
    \label{fig:NNLOdiag}
\end{figure}

The handling and cancellation of IR divergences across three different multiplicity phase spaces requires the definition of many types of subtraction term. 
As can be seen in the generic NNLO subtraction formula in Eq.~\eqref{eq:NNLOsubtract}, every subtraction term appears in two places, so as to not change the overall result. 
It is useful to rewrite Eq.~\eqref{eq:NNLOsubtract} in terms of three integrals over different phase spaces,
\begin{eqnarray}
    \label{eq:NNLOantsubtract}
    d\hat\sigma^{\text{NNLO}} &=& \int_{d\Phi_{m}} ( VV - d \hat{\sigma}^U)  \nonumber \\
    &+& \int_{d \Phi_{m+1}} ( RV - d \hat{\sigma}^T)  \nonumber \\
    &+& \int_{d\Phi_{m+2}} ( RR - d \hat{\sigma}^S) ,
\end{eqnarray}
where 
\begin{equation}
    \int_{d\Phi_{m}} d \hat{\sigma}^U + \int_{d\Phi_{m+1}} d \hat{\sigma}^T + \int_{d\Phi_{m+2}} d \hat{\sigma}^S = 0.
\end{equation}
Every term in $d \hat{\sigma}^S$ has an integrated counterpart in either $d \hat{\sigma}^T$ or $d \hat{\sigma}^U$. 
Every term in $d \hat{\sigma}^T$ is either an integrated counterpart of $d \hat{\sigma}^S$ or has an integrated counterpart in $d \hat{\sigma}^U$. 
Every term in $d \hat{\sigma}^U$ is an integrated counterpart of certain terms in $d \hat{\sigma}^T$ or $d \hat{\sigma}^S$. 
The intricate structure of antenna subtraction can be seen in Fig.~\ref{fig:NNLOdiag}, across the three levels. 
We will now give an overview of the purpose of each subtraction term but more detailed discussions can be found in Refs.~\cite{Currie:2013vh,Gehrmann-DeRidder:2005btv,Gehrmann:2023dxm}.
The discussion here aims to explain the roles of the antenna functions in NNLO antenna subtraction and to motivate the idealisation of antenna functions, described in Chapters~\ref{chapter:paper2} and~\ref{chapter:paper3}. 
In those chapters we only focus on the construction of antennae for FF calculations, so we will not describe subtraction terms which are only present in IF and II configurations, including mass factorisation terms. 
We will also frame the discussion around creating subtraction terms for one particular colour-ordering. 
This is appropriate for a calculation at leading-colour. 
In the case of sub-leading colour, the subtraction terms include more colour connections. This is because the full corrections cannot be written as a sum of squared colour-ordered matrix elements but rather in terms of interferences of matrix elements with two colour-orderings. 

\subsection{Double-Real Subtraction Terms}
\label{sec:RRsub}

At the double-real level, there are five contributions to the subtraction term, as depicted in Fig.~\ref{fig:NNLOdiag}:
\begin{equation}
    d \hat{\sigma}^S = d \hat{\sigma}^{S,a} + d \hat{\sigma}^{S,b_1} + d \hat{\sigma}^{S,b_2} + d \hat{\sigma}^{S,c} + d \hat{\sigma}^{S,d}. 
\end{equation}
The overall subtraction term, $d \hat{\sigma}^S$, must capture all the single and double unresolved divergences in the double-real correction,
\begin{equation}
    RR \sim \sum_{\text{perms}} {M}_{m+2}^0 (...,i,j,k,l,...) J_m^{(m+2)}.
\end{equation}
The following formulae should be taken as schematic, not accounting for symmetry factors and possible over-counting of singularities when using unidealised antenna functions. 

The first contribution to the double-real subtraction term is given by
\begin{equation}
    d \hat{\sigma}^{S,a} \sim \sum_{\text{perms}} d \Phi_{m+2} \sum_j X_3^0 (i^h,j,k^h) {M}_{m+1}^0 (...,I,K,...) J_m^{(m+1)}. 
\end{equation}
This contribution looks very similar to the NLO subtraction term in Eq.~\eqref{eq:NLOantsub} because it subtracts all the single unresolved divergences in the double-real correction, via the sum over $j$, one unresolved parton at a time. The NLO antenna mapping is also used for $(i,j,k) \to (I,K)$, with the colour-connections demonstrated in Fig.~\ref{fig:combine}. Note that the jet function, $J_m^{(m+1)}$, ensures that the $m+1$ partons of the reduced matrix element form the $m$ jets of the Born matrix element. 
Unfortunately, this results in the inclusion of double unresolved divergences. 
This is because one of the partons in the reduced matrix element is allowed to be unresolved, in addition to the unresolved parton $j$, handled by the $X_3^0$ antenna. 
Some of these double unresolved divergences match those in the double-real correction, namely the colour-connected iterated single unresolved divergences. 
Others, however, are spurious (do not match those in the double-real correction) and are counter-subtracted by other contributions, which are introduced fix these limits. 
Since this contribution handles single unresolved divergences, it has a counterpart, $d \hat{\sigma}^{T,a},$ at the $RV$ level, which is built out of integrated antennae, $\mathcal{X}_3^0$. 
This can be seen on the bottom-left of Fig.~\ref{fig:NNLOdiag}. 

\begin{figure}[h!]
    \epsfig{file=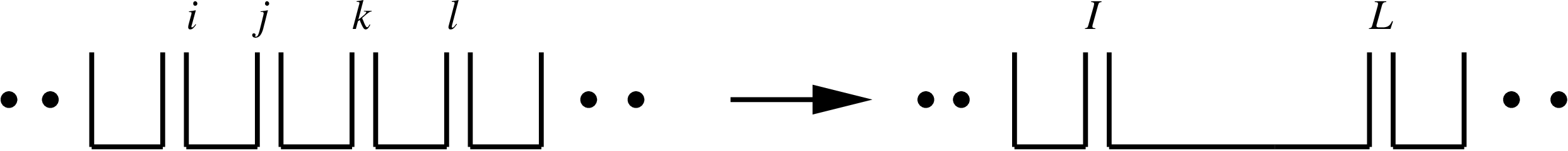,height=1.2cm}
    \caption{Diagram showing the colour-connections within the matrix element before and after the $4 \to 2$ antenna momentum mapping~\cite{Gehrmann-DeRidder:2005btv}.}
    \label{fig:double}
\end{figure}

The second contribution to $d \hat{\sigma}^S$ is given by
\begin{equation}
    d \hat{\sigma}^{S,b_1} \sim \sum_{\text{perms}} d \Phi_{m+2} \sum_{j,k} \X(i^h,j,k,l^h) {M}_{m}^0 (...,I,L,...) J_m^{(m)}, 
\end{equation}
where $j,k$ are colour adjacent and we use the double unresolved antenna momentum mapping, $(i,j,k,l) \to (I,L)$. The colour-connections for both full and reduced matrix elements are shown in Fig.~\ref{fig:double}. 
This contribution is introduced to match the colour-connected and uniterated double unresolved divergences of the double-real correction. 
Each term in $d \hat{\sigma}^{S,b_1}$ subtracts the double- and single unresolved divergences of $j,k$, sandwiched between $i^h,l^h$, according to the antenna, $\X(i^h,j,k,l^h)$. 
For the case when certain $X_4^{0, \text{OLD}}$ are used, one cannot identify two hard radiators and two unresolved partons. 
In this case, combinations of sub-antennae are used, with different mappings on each, and spurious unresolved divergences may be introduced. 
This in turn requires the introduction of other antenna functions to counter-subtract the spurious divergences. 
The arguments of these additional antenna functions may not match the particle type or colour-ordering of the partons in the double-real correction. 

Since the single unresolved limits are handled by the $d \hat{\sigma}^{S,a}$ contribution, the $d \hat{\sigma}^{S,b_1}$ must be combined with the contribution $d \hat{\sigma}^{S,b_2}$, which subtracts the iterated single unresolved limits from each term in $d \hat{\sigma}^{S,b_1}$. 
This gives the formula,
\begin{equation}
    d \hat{\sigma}^{S,b_2} \sim - \sum_{\text{perms}} d \Phi_{m+2} \sum_{j} X_3^0 (i^h,j,k^h) X_3^0 (I^h,K,l^h) {M}_{m}^0 (...,\widetilde{IK},\widetilde{Kl},...) J_m^{(m)}, 
\end{equation}
where we encounter products of two $X_3^0$ antennae to account for iterated single unresolved limits. 
The iterated NLO antenna momentum mappings, $(i,j,k) \to (I,K)$ and $(I,K,l) \to (\widetilde{IK}, \widetilde{Kl})$, ensure that the iterated single unresolved limits are colour-connected and ordered. 
The combination of $d \hat{\sigma}^{S,b_1}$ and $d \hat{\sigma}^{S,b_2}$, along with $d \hat{\sigma}^{S,a}$, ensures the cancellation of all colour-connected double unresolved divergences. 
Since the $d \hat{\sigma}^{S,b_1}$ contribution contains $\X$ antennae, it has a counterpart on the double-virtual level, containing the integrated antennae, $\calX$. The $d \hat{\sigma}^{S,b_2}$ contribution has a counterpart on the real-virtual level, containing products of $\mathcal{X}_3^0$ and $X_3^0$. 
These relationships can both be seen in Fig.~\ref{fig:NNLOdiag}.  

The first three contributions are enough to subtract all the IR divergences from the double-real correction, when considering processes with up to four partons in the double-real matrix elements. 
This is because all four partons are colour-connected and no spurious limits are introduced in $d \hat{\sigma}^{S,a}$. 
Examples of Born processes which generate up to four partons in the double-real correction include di-jet production in $e^+ e^-$ colliders (FF), single-jet production in $ep$ colliders (IF and FF), Drell-Yan production (II and IF) and inclusive Higgs production (II and IF).

If we want to consider more complicated QCD processes with at least five partons at the double-real level, we need to consider almost-colour-connected double unresolved limits. 
For processes with at least six partons, we need the colour-unconnected double unresolved limits. 
These are fixed by $d \hat{\sigma}^{S,c}$ and $d \hat{\sigma}^{S,d}$ respectively. 

\begin{figure}[h!]
    \epsfig{file=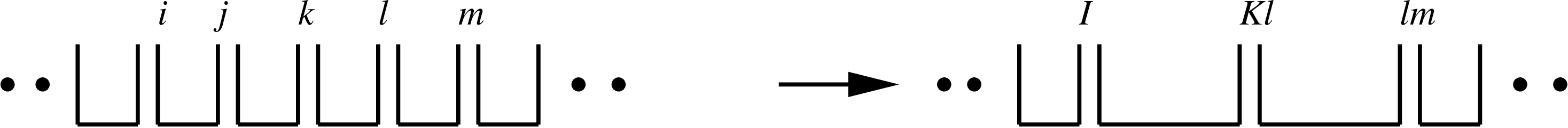,height=1.2cm}
    \caption{Diagram showing the colour-connections within the matrix element before and after two $3 \to 2$ antenna momentum mappings, where the two antennae are adjacent~\cite{Gehrmann-DeRidder:2005btv}.}
    \label{fig:overlap}
\end{figure}

Let us consider a double-real matrix element with the colour-ordering
\begin{equation}
    ...,i,j,k,l,m,n,...
\end{equation} 
Since we are concerned with almost-colour-connected limits, we will focus on the limits when $j$ and $l$ are unresolved. In the double-real matrix element, we would expect some limits corresponding to $j$ and $l$ unresolved at the same time. 
In general, we may subtract divergences in this limit by terms in $d \hat{\sigma}^{S,a}$ and $d \hat{\sigma}^{S,b}$. In $d \hat{\sigma}^{S,a}$, there are pairs of terms like
\begin{eqnarray}
    \label{eq:sigSaover}
    &&X_3^0 (i^h,j,k^h) {M}_{m+1}^0 (...,I,K,l,m,n,...) J_m^{(m+1)} \nonumber \\
    + &&X_3^0 (k^h,l,m^h) {M}_{m+1}^0 (...,i,j,K',M,n,...) J_m^{(m+1)},
\end{eqnarray}
which over-count by a factor of two the limits when $j$ and $l$ are both unresolved. This is because in the first term $l$ can be unresolved in the reduced matrix element (in addition to $j$ in the antenna) and vice versa for the second term.
This requires correcting in $d \hat{\sigma}^{S,c}$ by terms like
\begin{eqnarray}
    \label{eq:Sceg1}
   &&- \frac{1}{2} X_3^0 (i^h,j,k^h) X_3^0 (K^h,l,m^h) {M}_{m}^0 (...,I,\widetilde{Kl},\widetilde{lm},n,...) J_m^{(m)} \nonumber \\
    &&- \frac{1}{2} X_3^0 (k^h,l,m^h) X_3^0 (i^h,j,{K'}^h) {M}_{m}^0 (...,\widetilde{ij},\widetilde{jK'},M,n,...) J_m^{(m)}.
\end{eqnarray}
The colour connections before and after mappings for the first term are displayed in Fig.~\ref{fig:overlap}.

If the presence of a $\Xt(k^h,j,l,m^h)$ term in $d \hat{\sigma}^{S,b_1}$ is required, then the respective terms in $d \hat{\sigma}^{S,b_2}$, which should remove the single unresolved limits from $\Xt(k^h,j,l,m^h)$, are given by
\begin{equation}
    \label{eq:sigSb2over}
    - X_3^0 (k^h,j,m^h) X_3^0 (K^h,l,M^h) {M}_{m}^0 (...,i,\widetilde{Kl},\widetilde{lM},n,...) J_m^{(m)}  + (j\leftrightarrow l).
\end{equation}
In fact, these subtraction terms over-count divergences in $\Xt(k^h,j,l,m^h)$, so they need correcting by additional terms in $d \hat{\sigma}^{S,c}$, since these types of limit are not present in the double-real correction. 
The respective terms in $d \hat{\sigma}^{S,c}$ to fix the over-counting in Eq.~\eqref{eq:sigSb2over} are given by
\begin{equation}
    \label{eq:Sceg2}
    + \frac{1}{2} X_3^0 (k^h,j,m^h) X_3^0 (K^h,l,M^h) {M}_{m}^0 (...,i,\widetilde{Kl},\widetilde{lM},n,...) J_m^{(m)}  + (j\leftrightarrow l).
\end{equation}
A subtraction term including $\Xt(k^h,j,l,m^h)$ may be required when considering the sum of multiple colour-orderings, where we can recover Eq.~(3.14) of Ref.~\cite{Currie:2013vh}.
Unfortunately, the structures introduced in $d \hat{\sigma}^{S,c}$ include wide-angle soft divergences. 
Additional terms are then added to $d \hat{\sigma}^{S,c}$, built out of products of single-soft factors and $X_3^0$ antennae, in order to remove these spurious divergences~\cite{Gehrmann-DeRidder:2007foh,Pires:2010jv}.
For example, the following is added to Eq.~\eqref{eq:Sceg2}, in order to remove the wide-angle soft divergences,
\begin{eqnarray}
    &-&\frac{1}{2} \left(  \Sg (K^h,j,M^h) - \Sg (\widetilde{Kl}^h,j,\widetilde{lM}^h) \right) X_3^0 (K,l,M) M_m^0 (...,i,\widetilde{Kl},\widetilde{lM},n,...) J_m^{(m)} \nonumber \\
    && \hspace{2cm} + (j\leftrightarrow l).  
\end{eqnarray}
The structure of the $d \hat{\sigma}^{S,c}$ contribution, as a whole, is inherited from the structure of the more fundamental subtraction terms, $d \hat{\sigma}^{S,a}$, $d \hat{\sigma}^{S,b_1}$ and $d \hat{\sigma}^{S,b_2}$, ensuring that only the divergences present in the double-real matrix element are subtracted. 
In practice this is done process-by-process, where cancellation may only be apparent after summing over multiple colour-orderings. 
The $d \hat{\sigma}^{S,c}$ contribution has a counterpart on the real-virtual level, containing products of $\mathcal{X}_3^0$ and $X_3^0$ and also products of integrated eikonal factors and $X_3^0$. This correspondence can be seen in Fig.~\ref{fig:NNLOdiag}.  

\begin{figure}[h!]
    \epsfig{file=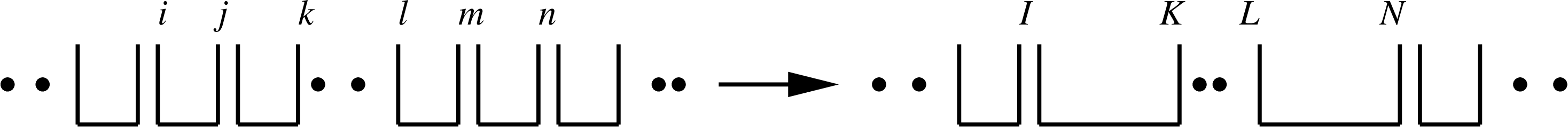,height=1.2cm}
    \caption{Diagram showing the colour-connections within the matrix element before and after two $3 \to 2$ antenna momentum mappings, where the two antennae are disconnected~\cite{Gehrmann-DeRidder:2005btv}.}
    \label{fig:nonoverlap}
\end{figure}

The final term to be introduced is only present for processes with at least six partons at the double-real level. 
It is included to correct for the colour-unconnected double unresolved limits introduced in $d \hat{\sigma}^{S,a}$. 
In a similar way to Eq.~\eqref{eq:sigSaover}, there are pairs of terms in $d \hat{\sigma}^{S,a}$ which over-count these limits by factor of two. 
For example the colour-unconnected unresolved limits of $j$ and $m$ are over-counted in
\begin{eqnarray}
    && X_3^0 (i^h,j,k^h) {M}_{m+1}^0 (...,I,K,l,m,n,...) J_m^{(m+1)} \nonumber \\
    + && X_3^0 (l^h,m,n^h) {M}_{m+1}^0 (...,i,j,k,L,N,...) J_m^{(m+1)},
\end{eqnarray}
since the other unconnected parton can become unresolved in the reduced matrix element. This means that the $d \hat{\sigma}^{S,d}$ contribution takes the form,
\begin{equation}
    d \hat{\sigma}^{S,d} \sim - \sum_{\text{perms}} d \Phi_{m+2} \sum_{j,m} X_3^0 (i^h,j,k^h) X_3^0 (l^h,m,n^h) {M}_{m}^0 (...,I,K,L,N,...) J_m^{(m)}, 
\end{equation}
where the sum is over all colour-unconnected pairs $(j,m)$. Note that $(i,k)$ are the neighbours of $j$, while $(l,n)$ are the neighbours of $m$, although it is not necessary that $k$ is colour-adjacent to $l$. The colour connections before and after the two $3 \to 2$ mappings are displayed in Fig.~\ref{fig:nonoverlap}.
The contribution $d \hat{\sigma}^{S,d}$ is suitable for integration over two disconnected antenna phase spaces, $d \Phi_{X_{ijk}}$ and $d \Phi_{X_{lmn}}$, which means that it has a counterpart at the double-virtual level, as can be seen in Fig.~\ref{fig:NNLOdiag}. 

\subsection{Real-Virtual Subtraction Terms}

At the real-virtual level, there are five contributions to the subtraction term, as depicted in Fig.~\ref{fig:NNLOdiag}:
\begin{equation}
    d \hat{\sigma}^T = d \hat{\sigma}^{T,a} + d \hat{\sigma}^{T,b_1} + d \hat{\sigma}^{T,b_2} + d \hat{\sigma}^{T,b_3} + d \hat{\sigma}^{T,c}. 
\end{equation}
The overall subtraction term, $d \hat{\sigma}^T$, must cancel all the single unresolved divergences and the explicit $\e$-poles in the real-virtual correction,
\begin{equation}
    RV \sim \sum_{\text{perms}} {M}_{m+1}^{1} (...,i,j,k,...) J_m^{(m+1)}.
\end{equation}
The following formulae should be taken as schematic, not accounting for symmetry factors and possible over-counting of singularities. 


The first contribution to the real-virtual subtraction term is very similar to the single-virtual subtraction term at NLO and is given by
\begin{equation}
    \label{eq:Ta}
    d \hat{\sigma}^{T,a} \sim \sum_{\text{perms}} d \Phi_{m+1} \sum_{i,j} \mathcal{X}_3^0 (i,j) {M}_{m+1}^0 (...,i,j,...) J_m^{(m+1)},
\end{equation}
where the permutations are over those colour-orderings which contribute and the sum is over all colour-adjacent dipole pairs, $(i,j)$. 
A more general expression would replace the integrated antenna with the integrated antenna dipole, $\J{1} (i,j)$, since in principle, there can be multiple distinct antennae with the same-type parents. 
Although it is the case that each $\J{1} (i,j)$ depends on one integrated antenna, this is not the case for the $\J{2} (i,j)$, which show more complex dependences. 
We will return to the $\J{1}$ and $\J{2}$ in Chapter~\ref{chapter:paper3}. 
Similarly to the virtual subtraction term at NLO, since the $\J{1}$ are directly related to Catani's IR singularity operators~\cite{Catani:1998bh,Chen:2022ktf}, this contribution is guaranteed to cancel the explicit one-loop IR singularities ($\e$-poles) in the real-virtual correction. Each term in $d \hat{\sigma}^{T,a}$ corresponds to an integrated version of a term in $d \hat{\sigma}^{S,a}$. 
Note that the jet function, $J_m^{(m+1)}$, ensures that the $m+1$ partons of the tree-level matrix element form the $m$ jets of the Born matrix element. 
This results in spurious single unresolved limits being introduced in this contribution, which may not match those in the real-virtual correction. 
These spurious singularities are handled by the $d \hat{\sigma}^{T,c}$ contribution. 

Since one-loop matrix elements obey Eq.~\eqref{eq:1loopfact}, when in a single unresolved limit, there are two contributions to handle the single unresolved divergences in the real-virtual correction. $d \hat{\sigma}^{T,b_1}$ and $d \hat{\sigma}^{T,b_2}$ handle the two pieces in 
\begin{equation}
    \text{one-loop} \to (\text{tree} \times \text{loop}) + (\text{loop} \times \text{tree}),
\end{equation}
respectively. 
The first case can be interpreted as requiring subtraction terms where the unresolved factors are at tree-level (therefore $X_3^0$) and the loop dependence is held within the reduced matrix element. 
The second case can be interpreted as requiring subtraction terms where the unresolved factors are at one-loop (therefore $X_3^1$) and the reduced matrix element is then tree-level. 

A candidate for $d \hat{\sigma}^{T,b_1}$ could be
\begin{equation}
    \sum_{\text{perms}} d \Phi_{m+1}  \sum_{j} {X}_3^0 (i^h,j,k^h) {M}_{m}^1 (...,I,K,...) J_m^{(m)},
\end{equation}
where the sum is over $j$ unresolved between its colour-neighbours ($i$ and $k$) and the reduced matrix element (depending on $(i,j,k)\to(I,K)$) is at one-loop. 
Unfortunately, this is not adequate, since the reduced one-loop matrix element itself contains explicit IR divergences. 
These have already been handled at the real-virtual level by $d \hat{\sigma}^{T,a}$. 
Therefore we also need to add a counter term, which makes the contribution $\e$-finite and can be constructed using the $\J{1}$ operators. 
This gives us the $\e$-finite contribution,
\begin{eqnarray}
    d \hat{\sigma}^{T,b_1} \sim \sum_{\text{perms}}  d \Phi_{m+1} \sum_{j} {X}_3^0 (i^h,j,k^h) &\bigg[& {M}_{m}^1 (...,I,K,...) \\
    && - \sum_{I,K} \J{1} (I,K) M_m^0 (...,I,K,...) \bigg] J_m^{(m)}, \nonumber
\end{eqnarray}
where the additional $\sum_{I,K}$ is the sum over all colour-adjacent pairs within the reduced matrix element $M_m^0 (...,I,K,...)$, including the mapped momenta and the unmapped momenta. 
The downside of making this contribution $\e$-finite is that the second term itself contains single unresolved divergences, due to the $X_3^0$ prefactor, which must also be handled by the $d \hat{\sigma}^{T,c}$ contribution. 
Both terms in $d \hat{\sigma}^{T,b_1}$ are introduced at the real-virtual level and the presence of the $X_3^0$ antenna makes them appropriate for integration over the single unresolved phase space $d \Phi_{X_{ijk}}$. 
In practice, the first and second terms relate to different parts of the double-virtual subtraction terms after integration, $d \hat{\sigma}^{U,a_1}$ and $d \hat{\sigma}^{U,b}$ respectively.

A candidate for $d \hat{\sigma}^{T,b_2}$ could be
\begin{equation}
    \sum_{\text{perms}} d \Phi_{m+1} \sum_{j} {X}_3^1 (i^h,j,k^h) {M}_{m}^0 (...,I,K,...) J_m^{(m)},
\end{equation}
where the sum is over $j$ unresolved, with the same mapping as the first term in $d \hat{\sigma}^{T,b_1}$. This candidate correctly cancels the $(\text{loop} \times \text{tree})$ unresolved limits in the real-virtual correction. The combination of $RV$, the first term of $d \hat{\sigma}^{T,b_1}$ and this candidate is non-divergent in any unresolved limit. 
However, similarly to $d \hat{\sigma}^{T,b_1}$, this candidate is not $\e$-finite, so the full contribution is given by
\begin{equation}
    \label{eq:Tb2}
    d \hat{\sigma}^{T,b_2} \sim \sum_{\text{perms}} d \Phi_{m+1} \sum_{j} \bigg[ {X}_3^1 (i^h,j,k^h) - \bar{\boldsymbol{J}}_3^1 (i,j,k) X_3^0 (i^h,j,k^h) \bigg] {M}_{m}^0 (...,I,K,...) J_m^{(m)},
\end{equation}
where the term $\bar{\boldsymbol{J}}_3^1 X_3^0$ is defined such that it cancels the $\e$-poles present in the $X_3^1$. 
This term assumes that the $X_3^1$ have an $\e$-pole structure which is proportional to an $X_3^0$. 
This feature is guaranteed when the $X_3^1$ are extracted from matrix elements. 
In Chapter~\ref{chapter:paper3}, when we rebuild the $X_3^1$ from the unresolved limits, we impose this $\e$-pole structure on the $X_3^1$ for internal consistency. 
Similarly to $d \hat{\sigma}^{T,b_1}$, the second term in $d \hat{\sigma}^{T,b_2}$ introduces spurious unresolved limits at the real-virtual level. 
These must be handled by the $d \hat{\sigma}^{T,c}$ contribution. 
The first term in $d \hat{\sigma}^{T,b_2}$ is introduced at the real-virtual level and is suitable for integration over the single unresolved phase space $d \Phi_{X_{ijk}}$. 
This is the $X_3^1$ part of $d \hat{\sigma}^{T,b_2}$ in Fig.~\ref{fig:NNLOdiag}, which corresponds to the integrated antenna, $\mathcal{X}_3^1$, part of $d \hat{\sigma}^{U,c}$. 
The relationship between the second term of $d \hat{\sigma}^{T,b_2}$ and the other levels depends on the objects present in the pole-structure of $X_3^1$ and therefore $\bar{\boldsymbol{J}}_3^1$. 
In generic terms, one can write this integrated antenna string as a linear combination of $\J{1}$ with different arguments,
\begin{equation}
    \bar{\boldsymbol{J}}_3^1(i,j,k) = J_X \left(  a \J{1} (s_{ij}) + b \J{1} (s_{jk}) + c \J{1} (s_{ik}) \right) + M_X \J{1} (s_{ijk})  .   
\end{equation}
Each $\J{1}$ can be related directly to one $\mathcal{X}_3^0$. If an $X_3^0 (i',j',k')$ were integrated over the relevant single unresolved phase space, the integrated counterpart would depend on two of the momenta in the real-virtual matrix element; this corresponds to the $J_X$ terms. 
See $d \hat{\sigma}^{T,a}$ in Eq.~\eqref{eq:Ta} for a similar integrated antenna dependence. 
Therefore the $J_X$ terms in $d \hat{\sigma}^{T,b_2}$ are integrated counterparts of the $d \hat{\sigma}^{S,b_2}$ terms at the double-real level, as can be seen in Fig.~\ref{fig:NNLOdiag}. 
Recall that the terms in $d \hat{\sigma}^{S,b_2}$ contained iterated products of two $X_3^0$ antennae. 
The integration of $d \hat{\sigma}^{S,b_2}$ to the $J_X$ terms can be sketched out as
\begin{eqnarray}
    \label{eq:Sb2int}
    && \int d \Phi_{X_{i'j'k'}} X_3^0 (i',j',k') X_3^0 ((\widetilde{i'j'}),(\widetilde{j'k'}),l') M_m^0 (...,(\widetilde{(\widetilde{i'j'}) (\widetilde{j'k'})}),(\widetilde{(\widetilde{j'k'}) l'}),...) \nonumber \\
    && \hspace{2 cm} \sim \mathcal{X}_3^0 ((\widetilde{i'j'}),(\widetilde{j'k'})) X_3^0 ((\widetilde{i'j'}),(\widetilde{j'k'}),l') M_m^0 (...,(\widetilde{(\widetilde{i'j'}) (\widetilde{j'k'})}),(\widetilde{(\widetilde{j'k'}) l'}),...) \nonumber \\
    && \hspace{2 cm} \sim \J{1} (s_{ij}) X_3^0 (i,j,k) M_m^0 (...,I,K,...).
\end{eqnarray}
The first equality shows the integration over the phase-space of the primary antenna. 
The second equality uses a relabelling of mapped momenta to the momenta of the real-virtual matrix element and the correspondence of one $\J{1}$ to one $\mathcal{X}_3^0$, which are correct to normalisation factors. 
The final form of the term in Eq.~\eqref{eq:Sb2int} can be identified with the $J_X$ terms in $d \hat{\sigma}^{T,b_2}$.
The terms with the $M_X$ prefactor cannot be identified as the integrated counterpart of any term at the double-real level, so they must be introduced at the real-virtual level. 
As such, there is an integrated counterpart to the $M_X$ terms at the double-virtual level, as can be seen in Fig.~\ref{fig:NNLOdiag}. 
After integration over $d \Phi_{X_{ijk}}$, the $X_3^0$ in Eq.~\eqref{eq:Tb2} becomes an $\mathcal{X}_3^0$ and we find terms which are an iterated product of two integrated antennae, $\mathcal{X}_3^0 \otimes \mathcal{X}_3^0$. 

The next contribution is introduced to match the renormalisation scale of the $X_3^1$ to the renormalisation scale of the real-virtual matrix element $\mu^2$, rather than the scale of the antenna $s_{ijk}$. 
Matching the renormalisation scale consists of redefining the $X_3^1$ in $d \hat{\sigma}^{T,b_2}$ according to 
\begin{equation}
    X_3^1 (i^h,j,k^h) \to X_3^1 (i^h,j,k^h) + \frac{\beta_0}{\e} \frac{(4 \pi)^\e e^{-\e \gamma_E}}{8 \pi^2} X_3^0 (i^h,j,k^h) \left[ \left(\frac{s_{ijk}}{\mu^2} \right)^{-\e} -1 \right].
\end{equation}
Note that this redefinition does not change the explicit $\e$-poles due to the form of the squared bracket. $\beta_0$ takes the form of $b_0$ for a leading-colour $X_3^1$, $b_{0,F}$ for a quark-loop $\widehat{X}_3^1$ and $0$ for a sub-leading-colour $\widetilde{X}_3^1$. This contribution takes the form,
\begin{equation}
    \label{eq:Tb3}
    d \hat{\sigma}^{T,b_3} \sim \sum_{\text{perms}} d \Phi_{m+1} \frac{\beta_0}{\e} \frac{(4 \pi)^\e e^{-\e \gamma_E}}{8 \pi^2} \sum_{j} X_3^0 (i^h,j,k^h) \left[ \left(\frac{s_{ijk}}{\mu^2} \right)^{-\e} -1 \right] {M}_{m}^0 (...,I,K,...) J_m^{(m)},
\end{equation}
where the sums are the same as in Eq.~\eqref{eq:Tb2}. 
This contribution ensures that the $(\text{loop} \times \text{tree})$ unresolved divergences are exactly cancelled. 
Since this contribution must be introduced at this level, like the $M_X$ terms, these terms have integrated counterparts at the double-virtual level, as shown in Fig.~\ref{fig:NNLOdiag}. 
The first term in the square bracket in Eq.~\eqref{eq:Tb3}, once integrated, corresponds to the term labelled $\beta_0 \mathcal{X}_3^0$ in $d \hat{\sigma}^{U,c}$. 
The second term in the square bracket, once integrated, corresponds to the term labelled $d \hat{\sigma}^{U,a_0}$. 

The final contribution at the real-virtual level, $d \hat{\sigma}^{T,c}$, is introduced to cancel the spurious unresolved divergences introduced by terms in $d \hat{\sigma}^{T,a}$, $d \hat{\sigma}^{T,b_1}$ and $d \hat{\sigma}^{T,b_2}$. 
By this stage, all the $\e$-poles of $RV$ have been cancelled by $d \hat{\sigma}^{T,a}$. Also, $d \hat{\sigma}^{T,b_1}$, $d \hat{\sigma}^{T,b_2}$ and $d \hat{\sigma}^{T,b_3}$ are individually $\e$-finite. 
This means that the contribution $d \hat{\sigma}^{T,c}$ must also be $\e$-finite. 
We have still not accounted for the integrated counterparts to the double-real contribution $d \hat{\sigma}^{S,c}$, which form $d \hat{\sigma}^{T,c_0}$, as in Fig.~\ref{fig:NNLOdiag}. 
Since we have only sketched-out the form of $d \hat{\sigma}^{S,c}$, we can only engage in a similar discussion for $d \hat{\sigma}^{T,c}$. $d \hat{\sigma}^{T,c_0}$ contains products of either integrated antennae, $\mathcal{X}_3^0$ and $X_3^0$, or integrated eikonal factors and $X_3^0$. 
The terms like Eq.~\eqref{eq:Sceg1} correspond to terms in $d \hat{\sigma}^{T,c_0}$ like
\begin{eqnarray}
    &&\frac{1}{2} \J{1} (s_{ij}) X_3^0 (j^h,k,l^h) M_m^0 (...,i,J,L,...) J_m^{(m)} \nonumber \\
    + && \frac{1}{2} \J{1} (s_{kl}) X_3^0 (i^h,j,k^h) M_m^0 (...,I,K,l,...) J_m^{(m)} ,
\end{eqnarray}
with mappings $(j,k,l) \to (J,L)$ and $(i,j,k) \to (I,K)$ for the two terms respectively. 
These types of terms are needed to cancel some of the spurious unresolved divergences in $d \hat{\sigma}^{T,a}$ and the $J_X$ part of $d \hat{\sigma}^{T,b_2}$.
In order to make $d \hat{\sigma}^{T,c}$ $\e$-finite, terms like the following would be added,
\begin{eqnarray}
    - &&\frac{1}{2} \J{1} (s_{iJ}) X_3^0 (j^h,k,l^h) M_m^0 (...,i,J,L,...) J_m^{(m)} \nonumber \\
    - && \frac{1}{2} \J{1} (s_{Kl}) X_3^0 (i^h,j,k^h) M_m^0 (...,I,K,l,...) J_m^{(m)},
\end{eqnarray}
where the mapped momenta $J$ and $K$ enter the $\J{1}$ dependence. 
These types of terms are needed to cancel some of the spurious unresolved divergences in $d \hat{\sigma}^{T,b_1}$. 
However, we note that the combination of the previous two expressions is not yet $\e$-finite, this is only true for the full $d \hat{\sigma}^{T,c}$. 
These types of terms are introduced at the real-virtual level and make up $d \hat{\sigma}^{T,c_1}$. 
They have integrated counterparts within $d \hat{\sigma}^{U,b}$, as shown in Fig.~\ref{fig:NNLOdiag}. 
Other terms originating at the double-real level, like Eq.~\eqref{eq:Sceg2}, correspond to terms in $d \hat{\sigma}^{T,c_0}$ like
\begin{equation}
     - \frac{1}{2} \J{1} (s_{jl}) X_3^0 (j^h,k,l^h) {M}_{m}^0 (...,J,L,...) J_m^{(m)},
\end{equation}
with the mapping $(j,k,l) \to (J,L)$. These terms are needed to cancel some of the spurious unresolved divergences in the $J_X$ part of $d \hat{\sigma}^{T,b_2}$. 
In order to make $d \hat{\sigma}^{T,c}$ $\e$-finite, terms like the following would be added,
\begin{equation}
    \label{eq:Tc2}
    + \frac{1}{2} \J{1} (s_{jkl}) X_3^0 (j^h,k,l^h) {M}_{m}^0 (...,J,L,...) J_m^{(m)}.
\end{equation}
These terms are needed to cancel some of the spurious unresolved divergences in $d \hat{\sigma}^{T,b_1}$ and the $M_X$ part of $d \hat{\sigma}^{T,b_2}$. 
Terms like those in Eq.~\eqref{eq:Tc2} are introduced at the real-virtual level and make up $d \hat{\sigma}^{T,c_2}$. 
They have integrated counterparts within $d \hat{\sigma}^{U,c}$, as shown in Fig.~\ref{fig:NNLOdiag}. 
Only after the full combination of these types of terms and the terms with integrated eikonal factors do we recover an $\e$-finite grouping for  $d \hat{\sigma}^{T,c}$, which cancels all spurious unresolved divergences introduced in the other subtraction terms. 

\subsection{Double-Virtual Subtraction Terms}

The IR structure of the two-loop squared matrix elements, with no unresolved partons, mimics their defining structure in Eq.~\eqref{eq:2loopstruc}. 
These IR divergences can be written in terms of Catani's dipole operators, from Ref.~\cite{Catani:1998bh}, inserted between reduced matrix elements (reduced by either one or two loops). 
Equivalently, we can write them in terms of the integrated dipoles in colour space from Ref.~\cite{Chen:2022ktf}, which we have been using throughout. 
This gives us a complete and general formula,
\begin{equation}
    \label{eq:Ustruc}
    d \hat{\sigma}^{U} = d \hat{\sigma}^{U,a_1} + d \hat{\sigma}^{U,a_0} + d \hat{\sigma}^{U,b} + d \hat{\sigma}^{U,c}, 
\end{equation}
which is also given by
\begin{eqnarray}
    \label{eq:Ufull}
    d \hat{\sigma}^{U} & \sim& \sum_{\text{perms}} d \Phi_m 2 \bigg\{ \langle \mathcal{M}_m^0 | \Jcol{1} (\e) | \mathcal{M}_m^1 \rangle + \langle \mathcal{M}_m^1 | \Jcol{1} (\e) | \mathcal{M}_m^0 \rangle \nonumber \\
    && - \frac{\beta_0}{\e} \langle \mathcal{M}_m^0 | \Jcol{1} (\e) | \mathcal{M}_m^0 \rangle \nonumber \\ 
    &&- \langle \mathcal{M}_m^0 | \Jcol{1} (\e) \otimes \Jcol{1} (\e) | \mathcal{M}_m^0  \rangle \nonumber \\
    && + \langle \mathcal{M}_m^0 | \Jcol{2} (\e) | \mathcal{M}_m^0 \rangle - \langle \mathcal{M}_m^0 | \Jcolb{2} (\e) | \mathcal{M}_m^0 \rangle \bigg\} J_m^{(m)} ,
\end{eqnarray}
in terms of the full-colour structures defined in Eq.~\eqref{eq:Jcoldef}, where each term in Eq.~\eqref{eq:Ustruc} is the respective line in Eq.~\eqref{eq:Ufull}. 
The integrated counterparts of the subtraction terms, unaccounted for at the double-real and real-virtual levels, are guaranteed to correctly cancel the IR singularities of the double-virtual correction by the KLN theorem. 
The presence of the $\Jcolb{2}$ term is necessary to cancel unphysical $\e$-poles in the quark-gluon integrated dipoles. 
This is due to to the extraction of quark-gluon antennae from heavy neutralino decays in the MSSM. 
Such extractions include a small number of unresolved limits which are not present in QCD matrix elements. 
These spurious limits are cancelled by combinations of other antenna functions, which filter to the double-virtual level in $\Jcolb{2}$~\cite{Chen:2022clm,Gehrmann:2023dxm}. 

$d \hat{\sigma}^{U,a_1}$ encapsulates the one-loop insertions in the one-loop squared matrix elements. 
It can also be written as
\begin{equation}
    d \hat{\sigma}^{U,a_1} \sim \sum_{\text{perms}} d \Phi_m \sum_{i,j}\mathcal{X}_3^0 (i,j) M_m^1 (...,i,j,...) J_m^{(m)},
\end{equation}
where the sum is over all colour-adjacent dipole pairs, $(i,j)$. 
This formula bears a close resemblance to Eq.~\eqref{eq:Ta}, except with a different squared matrix element. 
We could also write this term with $\mathcal{X}_3^0 (i,j)$ replaced by $\J{1} (i,j)$ (and a constant). 
As can be seen in Fig.~\ref{fig:NNLOdiag}, this term is the integrated counterpart of the first term in $d \hat{\sigma}^{T,b_1}$, integrated over the phase space of the $X_3^0$ antenna. 

$d \hat{\sigma}^{U,a_0}$ is present to match the renormalisation scale in all elements of the subtraction term. It can be written as
\begin{equation}
    d \hat{\sigma}^{U,a_0} \sim \sum_{\text{perms}} d \Phi_m \frac{\beta_0}{\e} \frac{(4 \pi)^\e e^{-\e \gamma_E}}{8 \pi^2} \sum_{i,j} \mathcal{X}_3^0 (i,j) M_m^0 (...,i,j,...) J_m^{(m)},
\end{equation}
which is the integrated counterpart of one of the terms in $d \hat{\sigma}^{T,b_3}$ and we could equally replace $\mathcal{X}_3^0 (i,j)$ with $\J{1} (i,j)$. 

$d \hat{\sigma}^{U,b}$ encapsulates the case where there are two separate one-loop insertions within the tree-level squared matrix element. It can also be written as
\begin{equation}
    d \hat{\sigma}^{U,b} \sim \sum_{\text{perms}} d \Phi_m  \sum_{i,j} \sum_{k,l} \J{1} (i,j) \J{1} (k,l) M_m^0 (...,i,j,...,k,l,...) J_m^{(m)},
\end{equation}
where the two sums over the two integrated dipoles are made explicit and run over all contributing dipole pairs. 
This contribution can be usefully split into three categories, as in Refs.~\cite{Gehrmann-DeRidder:2005btv,Currie:2013vh}: the colour-connected terms (c.c.), where the two integrated dipoles share the same hard radiators, as shown in
\begin{equation}
    d \hat{\sigma}^{U,b,c.c.} \sim \sum_{\text{perms}} d \Phi_m \sum_{i,j} \J{1} (i,j) \J{1} (i,j) M_m^0 (...,i,j,...) J_m^{(m)};
\end{equation}
the almost-colour-connected terms (a.c.c), where the integrated dipoles share one hard radiator, as shown in
\begin{equation}
    d \hat{\sigma}^{U,b,a.c.c.} \sim \sum_{\text{perms}} d \Phi_m \sum_{i,j} \sum_{k} \J{1} (i,j) \J{1} (j,k) M_m^0 (...,i,j,k,...) J_m^{(m)};
\end{equation}
and the colour-unconnected terms (c.u.), where the integrated dipoles depend on distinct hard radiators, as shown in 
\begin{equation}
    d \hat{\sigma}^{U,b,c.u.} \sim \sum_{\text{perms}}  d \Phi_m \sum_{i,j} \sum_{k,l \ne i,j} \J{1} (i,j) \J{1} (k,l) M_m^0 (...,i,j,...,k,l,...) J_m^{(m)}.
\end{equation}
All three categories are present in the integrated counterparts of the second term of $d \hat{\sigma}^{T,b_1}$. 
The colour-connected and almost-colour-connected terms are present in the integrated counterparts of $d \hat{\sigma}^{T,c_1}$. 
The integrated counterparts (over two single unresolved phase spaces) of $d \hat{\sigma}^{S,d}$ fall into the colour-unconnected category of $d \hat{\sigma}^{U,b}$, as should be expected. 
These relationships are all evident in Fig.~\ref{fig:NNLOdiag}. 

$d \hat{\sigma}^{U,c}$ encapsulates the two-loop insertion in the tree-level squared matrix element. 
It can be written as
\begin{equation}
    d \hat{\sigma}^{U,c} \sim \sum_{\text{perms}} d \Phi_m \sum_{i,j} \J{2} (i,j) M_m^0 (...,i,j,...) J_m^{(m)},
\end{equation}
where the sum is over all colour-adjacent dipole pairs, $(i,j)$. For FF configurations, the two-loop integrated dipoles can be written as a linear combination of the following integrated antenna structures:
\begin{equation}
    \J{2} (i,j) = c_1 (\mathcal{X}_3^0 \otimes \mathcal{X}_3^0) (i,j) + c_2 \mathcal{X}_3^1 (i,j) + c_3 \frac{\beta_0}{\e} \left( \frac{s_{ij}}{\mu^2} \right)^{-\e} \mathcal{X}_3^0 (i,j) + c_4 \mathcal{X}_4^0 (i,j).
\end{equation}
These structures correspond to the pieces of $d \hat{\sigma}^{U,c}$ as displayed in Fig.~\ref{fig:NNLOdiag}. 
$(\mathcal{X}_3^0 \otimes \mathcal{X}_3^0) (i,j)$ refers to the convolution of two integrated $X_3^0$. 
Terms like this are the integrated counterparts of the $d \hat{\sigma}^{T,c_2}$ terms and the $M_X$ terms in $d \hat{\sigma}^{T,b_2}$. 
Terms including $\mathcal{X}_3^1$ are naturally the integrated counterparts of the $X_3^1$ terms in $d \hat{\sigma}^{T,b_2}$. 
The terms with $\beta_0$ are the other terms necessary at the double-virtual level to match the renormalisation scale of all elements to $\mu^2$. 
They are the integrated counterparts of part of $d \hat{\sigma}^{T,b_3}$. 
The final category of $d \hat{\sigma}^{U,c}$ includes $\mathcal{X}_4^0$ and these terms are the integrated counterparts (over a double unresolved phase space) of $d \hat{\sigma}^{S,b_1}$.  

This concludes the discussion of the general structure of antenna subtraction at NNLO. 
We have focussed on the FF configuration, suitable for predictions for $e^+ e^-$ colliders but the same discussion generalises to the IF and II configurations, after adding suitable mass factorisation terms and integrations over the momentum fractions of initial parton splittings. 

We have so far given the traditional discussion of antenna subtraction, starting at double-real level, then the real-virtual level and then the double-virtual level. 
However, we note that there has been recent work in building the subtraction terms in the opposite order. 
This is known as the colourful antenna subtraction method detailed in Refs.~\cite{Gehrmann:2023dxm,Chen:2022ktf}. 
In this method, the first subtraction term to be deduced is Eq.~\eqref{eq:Ufull}, which is completely general. 
The subtraction term is separated into the contributions discussed here. 
The deduction of a corresponding subtraction term, at the real-virtual or double-real level, relies upon the appropriate insertion of one or two unresolved partons into the subtraction term, thus `unintegrating' the term. 
The insertion demands a variety of replacements and remapping. 
These include replacing an integrated antenna with an unintegrated antenna, and remapping the momenta in the reduced matrix element and jet function appropriate to the unintegrated antenna function. 
The benefit of this procedure is that it is completely predictable and general, so the procedure can be automated. 
Additionally, the colourful method works in full-colour since the inference of subtraction terms begins from Catani's formulae for IR singularities at one and two loops in full-colour. 
This means sub-leading colour is not significantly more complicated to calculate, as the double-virtual subtraction terms determine the colour-correlations across the rest of the calculation. 
The colourful antenna subtraction method complements the work presented in this thesis; they both aim to make the generation of subtraction terms at NNLO simpler and more automatable.

\section{Summary}
\label{sec:antsummary}

In this chapter, we began by introducing the two main classes of method for the handling of IR divergences in higher order calculations. 
Both classes handle implicit and explicit IR divergences. 
Slicing schemes demonstrate non-local cancellation, while subtraction schemes demonstrate local cancellation. 
Subtraction schemes are generally preferable in the context of NNLO calculations, partly due to the need to minimise CPU time in calculations. 
Since more elements of subtraction are handled analytically and the subtraction is performed locally, the CPU load is often lower than if slicing is used. 
On the other hand, slicing techniques are playing a crucial role in the rise of \NthreeLO calculations. 
This is because there is currently no process-independent method for generating subtraction terms at \NthreeLO. 
Instead, at \NthreeLO, slicing schemes benefit from the well-known IR structures at NNLO and promote them by one order. 

Antenna subtraction has shown to be highly successful at NNLO, as well as facilitating \NthreeLO calculations. 
This is partly because of the rich tapestry of subtraction terms which can be constructed at NNLO using the antenna functions; these can handle the IR divergences for any generic process, including at high-multiplicity. 
The primary reason for the success of antenna subtraction is that the integrals of antenna functions (and therefore subtraction terms) are known. 
This feature is often absent or complex in other NNLO subtraction schemes and facilitates the construction of simple subtraction terms. 
In the original formulation, antenna functions were extracted from low-multiplicity matrix elements, ensuring their simple-integrability. 
This had the cost that antenna functions can include limits `around the back', which in turn complicates the creation of subtraction terms due to the inclusion of spurious limits. 
We leave the fuller discussion of this topic and how it can be addressed to Chapter~\ref{chapter:paper2}. 

We summarised the simple and elegant format of antenna subtraction at NLO first and introduced the NLO antenna mapping suitable for FF configurations. 
We considered the $2 \to m$ phase space and its factorisation into a $2 \to m-1$ phase space multiplied by the single unresolved antenna phase space used for integrating $X_3^0$. 
We also described the fundamental features of the $X_3^0$, before moving onto those of the $X_4^0$ and the $X_3^1$. 
The remainder of the chapter consists in the definition of the subtraction terms at NNLO for a generic IR-safe observable in the FF configuration. 
Here we demonstrated the relationships between pairs of terms at two levels in the subtraction and the intricate handling of both explicit and implicit IR divergences and both double and single unresolved divergences. 

However, the construction of subtraction terms is more complex than represented here in three situations. 
Firstly, when one or two of the hard radiators is in the initial state. 
In this situation, we require mass factorisation terms and the integrated dipoles depend on the initial splittings. 
Secondly, when performing a sub-leading colour calculation. 
In this situation the colour-orderings in the product of two matrix elements is not necessarily the same, so the subtraction terms must cancel differently-connected limits. 
Thirdly, when the antenna function needed for a particular term contains limits other than those between two hard radiators. 
In this situation, we cannot use the antenna mapping and we would also be including spurious limits. 
The current solution is to utilise sub-antennae but this sacrifices the feature of integrability of antenna functions. 
In the following chapters, we will address the third element mentioned here, since the first two have been solved elsewhere. 

\chapter{Constructing Idealised Real-Radiation \\ Antenna Functions}
\label{chapter:paper2}

\section{Introduction}

This chapter's focus is on improvements to the antenna-subtraction scheme at NNLO. 
In particular, we describe an algorithm to re-build real-radiation antenna functions directly from the divergences we want them to contain, thus creating idealised antenna functions. 
We note that some of the techniques employed in this chapter have been used in other NNLO subtraction schemes such as Refs.~\cite{Caola:2017dug,Czakon:2010td,Magnea:2018hab}. 
While the antenna-subtraction formalism successfully enabled the calculation of many processes at NNLO and even at \NthreeLO, the complexity of the subtraction terms becomes increasingly difficult with growing particle multiplicity.
This is mainly due to two reasons. Firstly, double-real radiation antenna functions derived from matrix elements do not always identify which particles are the hard radiators. 
This is particularly the case for gluons. To get around this, so-called sub-antenna functions are introduced. 
The construction of sub-antenna functions at NNLO is extremely cumbersome and typically introduces unphysical denominators that make analytic integration difficult. 
Often analytic integrals are known only for the full antenna functions. 
This means that antenna-subtraction terms have to be assembled in such a way that sub-antenna functions recombine to full antenna functions before integration. 
Secondly, NNLO antenna functions can contain spurious singularities that have to be removed by explicit counterterms, which in turn can introduce further spurious singularities. 
In general, this can trigger an intricate chain of cross-dependent subtraction terms that have no relation to the actual singularity structure of the process at hand.

As a first step towards a refined antenna-subtraction scheme at NNLO, we construct a full set of idealised single-real and double-real antenna functions. 
Instead of building these from physical matrix elements, as done originally \cite{Gehrmann-DeRidder:2005alt,Gehrmann-DeRidder:2005svg,Gehrmann-DeRidder:2005btv}, we build antenna functions directly from the relevant limits properly accounting for the overlap between different limits. 
The universal factorisation properties of multi-particle matrix elements when one or more particles are unresolved have been well studied in the literature and serve as an input to the algorithm.  
The single unresolved limits of tree-level matrix elements, where either one particle is soft or two are collinear, are used to construct  NLO antennae. 
While for the real-radiation NNLO antennae, the double unresolved limits of tree-level matrix elements, with up to two soft particles, or three collinear particles~\cite{campbell,Catani:1998nv,Catani:1999ss,Kosower:2002su} are needed.   
We note that the triple-unresolved limits of tree-level matrix elements are available for the construction of real radiation \NthreeLO antennae~\cite{Catani:2019nqv,DelDuca:1999iql,DelDuca:2019ggv,DelDuca:2020vst,DelDuca:2022noh}. 
However, they may not be in a form to be directly useful for \NthreeLO antennae at this time. 
 
The chapter is structured as follows. 
We outline the design principles for idealised real-radiation antenna functions in Section~\ref{sec:design-principles}, before discussing the general construction algorithm in Section~\ref{sec:algorithm1}.
To illustrate the algorithm, we explicitly construct a full set of single-real and double-real antenna functions for use in NLO and NNLO antenna subtraction in Sections~\ref{sec:single-real-antennae} and \ref{sec:double-real-antennae}, respectively.
We conclude and give an outlook on further work in Section~\ref{sec:outlook1}. 

\section{Design Principles}
\label{sec:design-principles}
Within the antenna framework, subtraction terms are constructed from antenna functions which describe all unresolved partonic radiation (soft and collinear) between a hard pair of radiator partons.
In general, an antenna-subtraction term requires:
\begin{itemize}
\item antennae composed of two hard radiators that accurately reflect the infrared singularities of the $n$ unresolved partons radiated ``between'' them;
\item an on-shell momentum mapping $\mathcal{F}_{n+2\mapsto 2}$, clustering $n+2$ particles into $2$, while preserving the invariant mass of the radiators used to define the ``reduced'' matrix element; and
\item a process- and antenna-dependent colour factor.
\end{itemize}
The calculation of colour factors strictly follows the non-abelian structure of QCD and, while in principle cumbersome, can be automated to all orders \cite{Sjodahl:2012nk,Sjodahl:2014opa,Gerwick:2014gya,Baberuxki:2019ifp}.
A general on-shell momentum mapping for multi-particle emissions in the antenna language has been derived in Ref.~\cite{Kosower:2002su} for massless particles and is given in Section~\ref{sec:introant}.  

Historically, antenna functions, $X_{n+2}^\ell$, have been constructed directly from matrix elements which have the desired singularities. 
However, we note that these ``natural'' antenna functions do not share the same design principles we are describing -- for example, many natural antenna functions do not have two identified hard radiators. 
All quark-antiquark antennae have two identified hard radiators but this is not the case for quark-gluon or gluon-gluon antennae. 
This is because the matrix elements used for extraction will inevitably have a divergent limit when one of the gluonic radiators becomes soft.  
This makes the construction of the subtraction terms more complex and correspondingly less automatable. 
Additionally, antenna momentum mappings require a clear identification of two hard radiators so that the mapping is appropriate for the limits the antenna is intended to describe. 
In order to counter these issues, so-called sub-antenna functions have been created by the complicated use of supersymmetry relations, other antenna functions and partial fractioning \cite{NigelGlover:2010kwr,Gehrmann-DeRidder:2007foh}. 
The momentum mapping is then different for each sub-antenna. 
These sub-antenna functions are in general difficult to integrate but can be combined such that only the full antenna requires integration. 
In some cases, sub-antennae map onto different types of matrix elements and multiple subtraction terms cannot be combined easily. 
This means that a direct integration of the sub-antenna functions is required, which is not feasible. 
In these cases, intricate process-dependent combinations of other antennae have to be used to correct for over-subtraction of spurious limits. 

We aim to design idealised antenna functions directly from their desired properties, with a uniform template, in a way that simplifies the construction of subtraction terms in general, while being straightforwardly integrable.
Specifically, we impose the following requirements on the idealised antenna functions:
\begin{enumerate}
\item each antenna function has exactly two hard particles (``radiators'') which cannot become unresolved;
\item each antenna function captures all (multi-)soft limits of its unresolved particles;
\item where appropriate (multi-)collinear and mixed soft and collinear limits are decomposed over ``neighbouring'' antennae;
\item antenna functions do not contain any spurious (unphysical) limits;
\item antenna functions only contain singular factors corresponding to physical propagators; and
\item where appropriate, antenna functions obey physical symmetry relations (such as line reversal).
\end{enumerate}
We wish to emphasise again that the original NNLO antenna functions derived in Refs.~\cite{Gehrmann-DeRidder:2005svg,Gehrmann-DeRidder:2005alt,Gehrmann-DeRidder:2005btv} do not obey these requirements, as they typically violate (some of) these principles, as alluded to above.
A subtlety is connected to (multi-)collinear and soft-collinear limits. 
For example, any gluon can become soft in a multi-gluon matrix element, so a prescription has to be defined for distributing limits between antennae that identify each of the gluons as being hard in turn. 

In the following sections we will describe a general algorithm to construct real-radiation antenna functions, following strictly these design principles. 
We will apply it to the case of single-real and double-real radiation, required for NLO and NNLO calculations. 

\section{The Algorithm}
\label{sec:algorithm1}
The main goal of the algorithm is to construct (multiple-)real-radiation antenna functions containing singular limits pertaining to exactly two hard radiators plus an (in principle) arbitrary number of additional particles that are allowed to become unresolved. 
To this end, each limit is defined by a ``target function'', which in the following we will denote by $L_i$. 
In general, there will be $N$ such limits and we denote the ordered set of limits by $\{L_j\}$. 
The target functions have to capture the behaviour of the colour-ordered matrix element squared in the given unresolved limit and are taken as input to the algorithm. 
While the target functions may include process-dependent azimuthal terms, for the purposes of the present chapter we will limit ourselves to azimuthally-averaged functions.

\begin{figure}[t]
    \centering
    \begin{tikzpicture}
    \node (x1) [fullspace] {Full Phase Space \\ $X^0_{n;i-1}$};
    \node (x2) [subspace, below=2cm of x1] {Subspace \\ $L_i$};
    \node (center) [below=1cm of x1] {};
    \node (AR) [right=0.001cm of x1] {};
    \node (AL) [left=0.001cm of x1] {};
    \node (BR) [right=0.001cm of x2] {};
    \node (BL) [left=0.001cm of x2] {};
    \node (a1) [left=2cm of center] {$\mathrm{\textbf{P}}^\downarrow_i$};
    \node (a2) [right=2cm of center] {$\mathrm{\textbf{P}}^\uparrow_i$};
    
    \draw [arrow] (AL) arc[start angle= 90,end angle=270,x radius=1,y radius=1.80];
    \draw [arrow] (BR) arc[start angle=270,end angle=450,x radius=1,y radius=1.80];
    \end{tikzpicture}
    \caption{Visual representation of the down- and up-projectors translating between the full phase space, on which antenna functions are defined, and the subspaces, on which the target functions are defined. Here $X^0_{n;i-1}$ represents the accumulated antenna function having taken into account the limits $L_1, \ldots L_{i-1}$, which is projected into the subspace relevant for limit $L_i$ by $\PPdown_i$, subtracted from $L_i$ and the remainder projected back into the full phase space by $\PPup_i$. }
    \label{fig:spaces}
\end{figure}
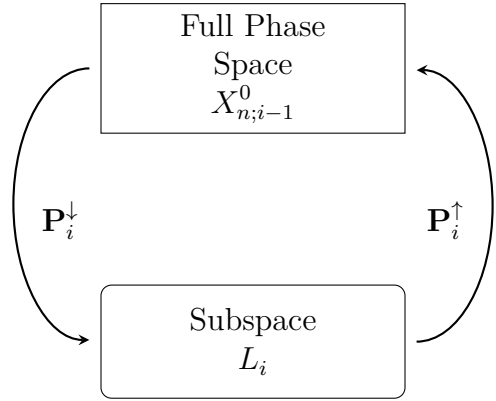

Antenna functions are defined over the full phase space appropriate to the respective antenna, whereas each singular limit lives on a restricted part of phase space, with one or more of the momenta being soft or collinear.
We relate the two by a ``down-projector'' $\PPdown$ that maps the invariants of the full phase space into the relevant subspace. 
An associated ``up-projector'' $\PPup$ restores the full phase space. That is, it re-expresses all variables valid in the subspace in terms of invariants valid in the full phase space. This is illustrated in Fig.~\ref{fig:spaces}. 
It is to be emphasised that down-projectors $\PPdown$ and up-projectors $\PPup$ are typically not inverse to each other, as down-projectors destroy information about less-singular and finite pieces.

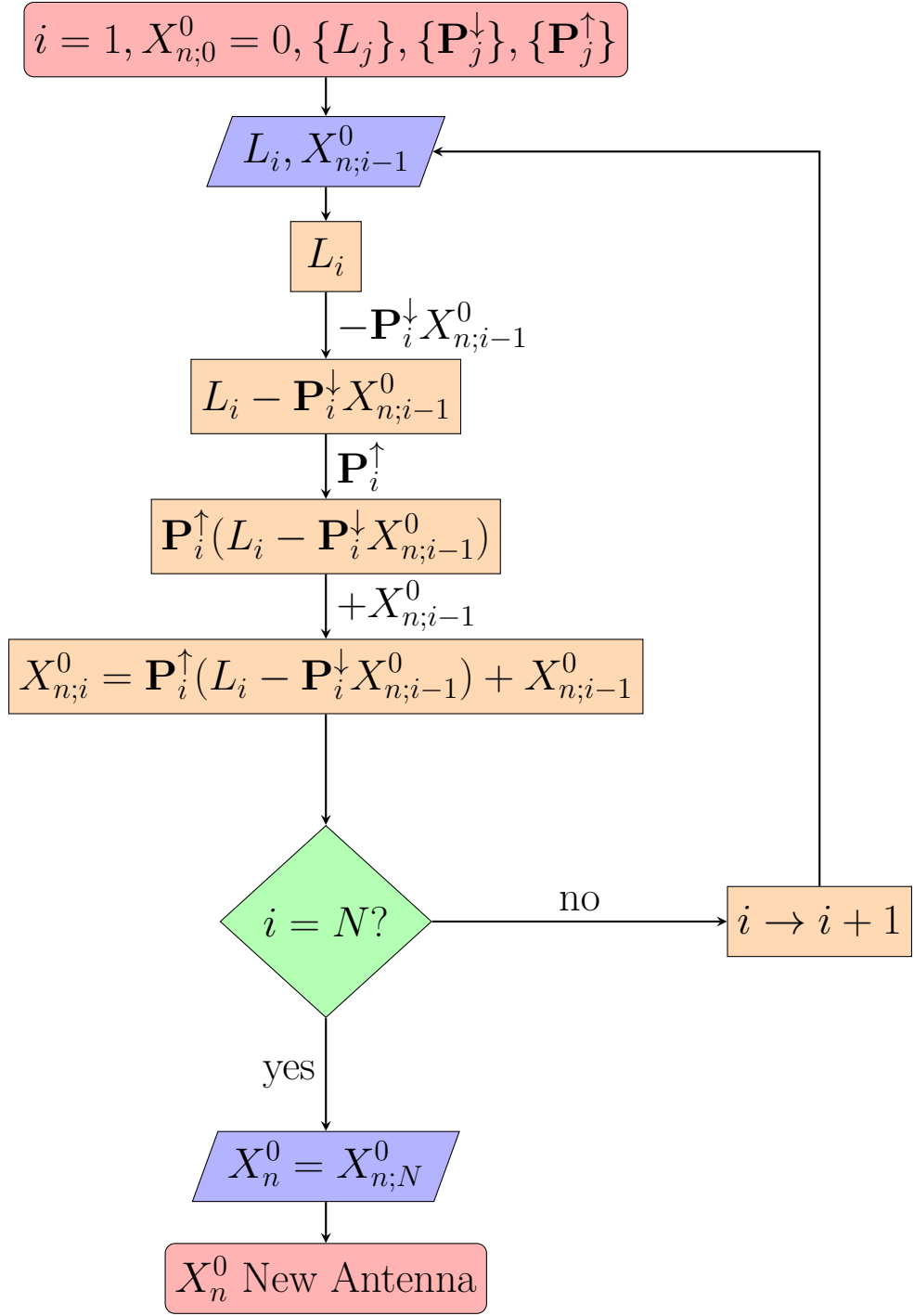
\begin{figure}[p]
\centering
\begin{tikzpicture}[node distance=2.0cm]

\node (start) [startstop] {\Large $i=1, X_{n;0}^0 = 0, \{L_j\}, \{\PPdown_j\},\{\PPup_j\}$};
\node (Li) [io, below of=start, yshift=0.4cm] {\Large $L_i, X_{n;i-1}^0$};
\node (pro1) [process, below of=Li, yshift=0.5cm] {\Large $L_i$};
\node (pro2) [process, below of=pro1] {\Large $L_i -\PPdown_i X_{n;i-1}^0$};
\node (pro3) [process, below of=pro2] {\Large $\PPup_i (L_i -\PPdown_i X_{n;i-1}^0)$};
\node (pro4) [process, below of=pro3] {\Large $X_{n;i}^0 = \PPup_i (L_i -\PPdown_i X_{n;i-1}^0)+X_{n;i-1}^0$};
\node (dec1) [decision, below of=pro4, yshift=-1.5cm] {\Large $i=N$?};

\node (pro2b) [process, right of=dec1, xshift=5cm] {\Large $i\rightarrow i+1$};

\node (stop) [io, below of=dec1, yshift=-1.5cm] {\Large $X_n^0 = X_{n;N}^0 $};
\node (end) [startstop, below of=stop, yshift=0.4cm] {\Large $X_n^0$ New Antenna};

\draw [arrow] (Li) -- (pro1);
\draw [arrow] (start) -- (Li);
\draw [arrow] (pro1) -- node[anchor=west] {\Large $-\PPdown_i X_{n;i-1}^0$} (pro2);
\draw [arrow] (pro2) -- node[anchor=west] {\Large $\PPup_i$} (pro3);
\draw [arrow] (pro3) -- node[anchor=west] {\Large $+X_{n;i-1}^0$} (pro4);
\draw [arrow] (pro4) -- (dec1);
\draw [arrow] (dec1) -- node[anchor=east] {\Large yes} (stop);
\draw [arrow] (dec1) -- node[anchor=south] {\Large no} (pro2b);
\draw [arrow] (pro2b) |- (Li);
\draw [arrow] (stop) -- (end);

\end{tikzpicture}
\caption{Flowchart representing the algorithm to construct a tree-level $n$-particle antenna $X_n^0$ using an ordered set of $N$ limits $\{L_j\}$ with associated down- and up-projectors. $X_{n;i}^0$ represents the accumulation of the contributions from the first $i$ limits.  The full antenna is obtained when all $N$ limits have been satisfied, $X_{n}^0 = X_{n;N}^0$.}
\label{fig:algorithm}
\end{figure}

We construct antenna functions using an iterative process, which requires as input the full set of $N$ unresolved limits we aim to capture (defined in terms of target functions $L_i$, with $i$ running from 1 to $N$), together with the appropriate set of ``down-'' and ``up-projectors'' pertaining to each of these limits.

The algorithm starts with the contribution corresponding to the deepest pole on the integrated level and terminates upon reaching the level of finite corrections, i.e. terms that integrate to corrections of $\order{\e^0}$.
In each step of the iteration, we remove the overlap of the target function with all previously considered limits and accumulate the remainder.
To this end, we subtract the projection of the accumulated antenna function into the subspace relevant to the target limit, $\PPdown_iX^0_{n;i-1}$, from the target function. 
The remainder $(L_i -\PPdown_i X^0_{n;i-1})$ is then restored to the full phase space via the associated up-projector $\PPup_i$ before adding it to the accumulated antenna function.
Schematically, this procedure is shown in Fig.~\ref{fig:algorithm} for $N$ target functions $L_i$ and can be written as 
\begin{equation}
\label{eq:algorithm}
  \begin{split}
    X^0_{n;1} &= \PPup_1 L_1 \, , \\
    X^0_{n;2} &= X^0_{n;1} + \PPup_2 (L_2 - \PPdown_2 X^0_{n;1}) \, ,\\
    & \vdots \\
    X^0_{n;N} &= X^0_{n;N-1} + \PPup_N (L_N-\PPdown_N X^0_{n;N-1}) \, ,
  \end{split}
\end{equation}
where $X^0_n \equiv X^0_{n;N}$. 
In particular, restoring the kinematics to the full phase space in a judicious way ensures that the antenna function can be expressed solely in terms of invariants corresponding to physical propagators. 
This specifically guarantees that the full antenna function can be integrated easily over its Lorentz-invariant antenna phase space, just like the original antenna functions.

The algorithm resembles the ones in Refs.~\cite{Czakon:2010td,Caola:2017dug,Magnea:2018hab} in spirit, in the sense that singular limits are considered subsequently, ordered according to the depth of the associated explicit pole. 
However, there are two important differences to the algorithms described in Refs.~\cite{Czakon:2010td,Caola:2017dug,Magnea:2018hab}. 
Firstly, we do not use our algorithm to construct process-dependent subtraction terms (even though the construction of the counterterms is in principle process independent) but to construct universal antenna functions that are only assembled to process-dependent subtraction terms in a separate (automatable) step, see Refs.~\cite{Gehrmann-DeRidder:2005btv,Currie:2013vh,Chen:2022ktf,Gehrmann:2023dxm}. 
Secondly, we always reconstruct the kinematics of the subtraction term to the full phase space via up-projectors, a step that is not strictly necessary for the sake of the subtraction but is vital to build a function in terms of multi-particle invariants that is valid in the full phase space.

The set of target functions unambiguously defines the behaviour of the real-radiation antenna function in all unresolved limits pertaining to the antenna at hand. 
In any unresolved limit, the full antenna function has to approach the target function in order to capture the correct singular behaviour of the respective squared matrix element.
In particular, the real-radiation antenna function has to be finite in all limits not described explicitly by a target function.
This ensures that no spurious singularities enter, a feature not shared by antenna functions constructed directly from physical matrix elements.
As alluded to above, an important aspect to consider in the construction of antenna functions pertains to the presence of certain multi-collinear and soft-collinear limits, which are shared by ``neighbouring'' antennae in the antenna formalism.
In these cases, the correct multi-collinear or soft-collinear limit is only recovered in the sum over multiple antenna functions, each containing one of the involved partons as a hard radiator.
This means that each multi-collinear/soft-collinear splitting function has to be decomposed over all possibilities to identify one of the partons as the hard radiator before it can be used as a target function for constructing idealised antenna functions.
For simple-collinear splitting functions, this decomposition is simple and has been identified already in Refs.~\cite{Altarelli:1977zs,Dokshitzer:1977sg}. A generalisation for triple-collinear splitting functions has been derived in Ref.~\cite{paper1} and was detailed in Chapter~\ref{chapter:paper1}.
As a by-product of this procedure, soft-collinear limits do not have to be entered as explicit inputs into the algorithm at NNLO.

A core part of our algorithm is the definition of down-projectors into singular limits with corresponding up-projectors into the full phase space.
In each step of the construction, down-projectors are needed to identify the overlap of the so-far constructed antenna function with the target function of the respective unresolved limit, whereas up-projectors are required to re-express the subtracted target function in terms of antenna invariants.
In this way, the full (accumulated) antenna function can be expressed solely in terms of $n$-particle invariants and is therefore valid in the full phase space. 
By choosing the up-projectors judiciously, the antenna function can furthermore be expressed exclusively in terms of invariant structures corresponding to physical propagators.
As alluded to above, down-projectors $\PPdown$ and up-projectors $\PPup$ are not required to be inverse to each other.

The number of projectors depends directly on the perturbative order.
At NLO, only two types of down-projectors and their up-projectors into the full phase space are needed,
\begin{equation}
    \PSdown, \PCdown, \text{ and } \PSup, \PCup,
\end{equation}
corresponding to the single-soft and simple-collinear limits. 
These will be discussed in detail in Section~\ref{sec:single-real-antennae}.
At NNLO, three additional types of down-projectors are needed,
\begin{equation}
    \PDSdown, \PTCdown, \PDCdown,
\end{equation}
corresponding to the double-soft, triple-collinear and double-collinear limits. The up-projectors into the full phase space are given by
\begin{equation}
    \PDSup, \PTCup, \text{ and } \PDCup,
\end{equation}
respectively. While not relevant to the construction of the antenna functions, for the sake of validating the correct singular behaviour of the constructed antenna functions, one can further define a down-projector related to the soft-collinear limit, $\PSCdown$.
Both down-projectors and up-projectors at NNLO will be discussed in Section~\ref{sec:double-real-antennae}.

Up to NNLO, we have automated our algorithm in computer code based on MAPLE and FORM~\cite{Vermaseren:2000nd,Kuipers:2012rf}. 
To make the construction explicit, we work through the construction of all idealised single-real antenna functions as an example in Section~\ref{sec:single-real-antennae} before constructing a full set of idealised double-real antenna functions in Section~\ref{sec:double-real-antennae}.
This constitutes a first step towards a refined antenna-subtraction formalism at NNLO and beyond.

\section{Single-Real Radiation Antennae}
\label{sec:single-real-antennae}
At NLO, we want to construct three-particle antenna functions $X_3^0(i^h_a,j_b,k^h_c)$, where the particle types are denoted by $a$, $b$ and $c$, which carry four-momenta $i$, $j$ and $k$ respectively. 
Particles $a$ and $c$ should be hard, and the antenna functions must have the correct limits when particle $b$ is unresolved. 
Frequently, we drop explicit reference to the particle labels in favour of a specific choice of $X$ according to Table~\ref{tab:X30}.

\begin{table}[t]
\centering
\begin{tabular}{ccc}
\underline{Quark-antiquark} & & \\
$qg\bar{q}$ & $X_3^0(i_q^h,j_g,k_{\bar{q}}^h)$ & $A_3^0(i^h,j,k^h)$ \\
\underline{Quark-gluon} & &  \\
$qgg$ & $X_3^0(i_q^h,j_g,k_g^h)$ & $D_3^0(i^h,j,k^h)$  \\
$q\bar{Q}Q$ & $X_3^0(i_q^h,j_{\bar{Q}},k_Q^h)$  & $E_3^0(i^h,j,k^h)$ \\ 
\underline{Gluon-gluon} & & \\
$ggg$ & $X_3^0(i_g^h,j_g,k_g^h)$ & $F_3^0(i^h,j,k^h)$  \\
$g\bar{Q}Q$ & $X_3^0(i_g^h,j_{\bar{Q}},k_Q^h)$ & $G_3^0(i^h,j,k^h)$  \\
\end{tabular}
\caption{Identification of $X_3^0$ antennae according to particle type. These antennae only contain singular limits when particle $b$ (or equivalently momentum $j$) is unresolved. Antennae are classified as quark-antiquark, quark-gluon and gluon-gluon according to the particle type of the parents (i.e. after the antenna mapping). }
\label{tab:X30}
\end{table}

We systematically start from the most singular limit and build the list of target functions from single-soft and simple-collinear limits.  
For the particles of $X_3^0(i^h_a,j_b,k^h_c)$ there are three such limits, corresponding to particle $b$ becoming soft and particle $b$ becoming collinear to either $a$ or $c$, expressed as
\begin{equation}
\begin{split}
    L_1(i^h,j,k^h) &= \Sb(i^h,j,k^h) \, , \\
    L_2(i^h,j;k) &= P_{ab}^{(0)}(i^h,j) \, , \\
    L_3(k^h,j;i) &= P_{cb}^{(0)}(k^h,j) \, .
\end{split}
\label{eq:NLOL}
\end{equation}
The tree-level soft factor $\Sb$ is given by the eikonal factor for particle $b$ radiated between two hard radiators in Eq.~\eqref{eq:Sg} if $b$ is a gluon and $0$ if $b$ is a quark.
The splitting functions $P_{ab}^{(0)}(i^h,j)$ are given in Eqs.~\eqref{eqn:Pqg}-\eqref{eq:PggS}.
Here, the momentum fraction $\xj$ is defined with reference to the third particle in the antenna, $\xj = s_{jk}/(s_{ik}+s_{jk})$.

We note that we choose to use spin-averaged splitting functions, so there are no azimuthal correlations in the constructed $X_3^0$. 
We will now take a short detour to consider the handling of azimuthal correlations in antenna subtraction. 

Constructing antenna functions with no azimuthal correlations is the correct choice for antenna subtraction because one set of azimuthal correlations in an antenna function will not in general match those in the process-specific matrix elements. 
Instead, the azimuthal terms in the matrix elements are effectively averaged-out by summing multiple pairs of correlated phase-space points during the subtraction procedure. 
We can do this because azimuthal terms cancel against those terms related by a phase space rotation~\cite{Gehrmann-DeRidder:2007foh, Weinzierl:2006wi}. 
The azimuthal terms can be shown to be proportional to $\cos (2 \phi + \alpha)$, where $\phi$ is the azimuthal angle around the collinear direction. 
This means that if we sum the contributions from phase space points with azimuthal angles $\phi$ and $\phi + \pi/2$, the azimuthal terms cancel. 
In this sense, antenna subtraction is not truly local because cancellation is only demonstrated in the sum of phase-space points, not point-by-point. 

\begin{figure}[t]
    \centering
    \includegraphics[width=0.49\textwidth]{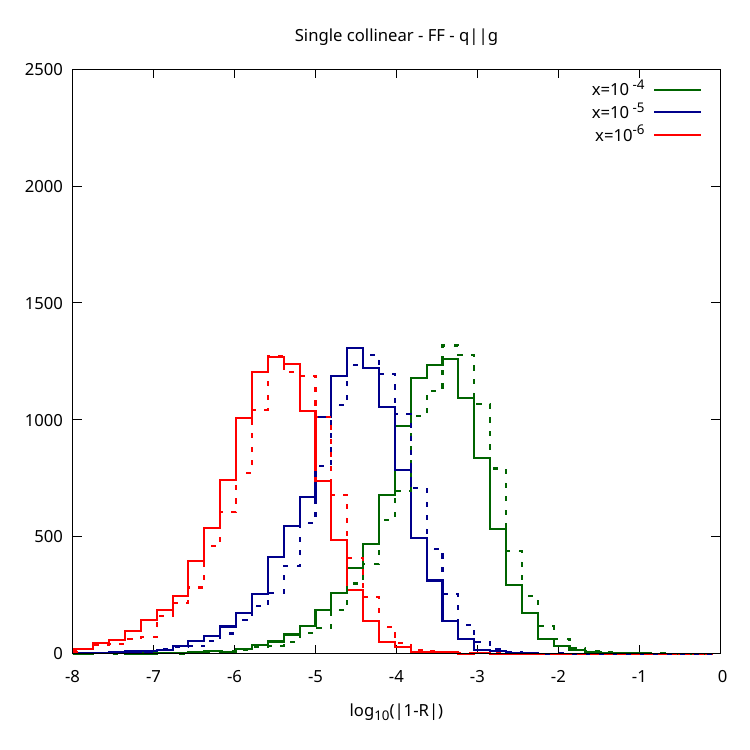}
    \includegraphics[width=0.49\textwidth]{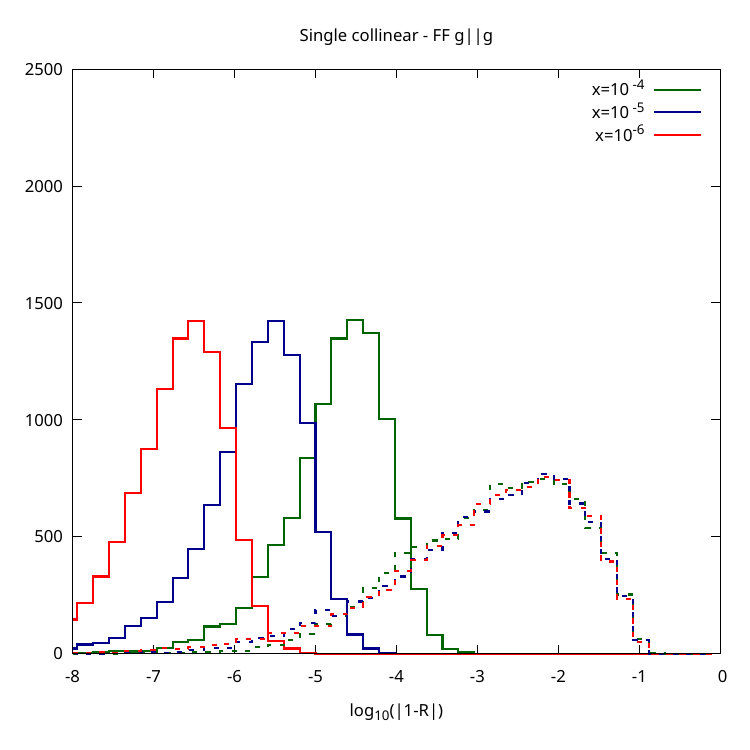}
    \caption{Numerical tests of the NLO final-final real subtraction terms in the limit where a quark-gluon pair are collinear (left frame) and where two gluons are collinear (right frame). For three different values of the scaling parameter $x$, the relative agreement of the ratio $R$ with $1$ is shown on a logarithmic axis. The dashed lines show the trajectory into the limit point-by-point. The solid lines combine phase-space points related by an azimuthal rotation of $\pi/2$ around the collinear direction~\cite{Fox:2023bma}.}
    \label{fig:azimuthal}
\end{figure}

In a recent paper, Ref.~\cite{Fox:2023bma}, the authors use the idealised $X_3^0$ antenna functions in a first application to NLO subtraction of $pp \to 3$ jets. They focus on the subtraction terms for the real correction matrix elements, $gg \to gggg$ and $qg \to qggg$. 
Following Ref.~\cite{Pires:2010jv}, trajectories are built into unresolved limits by scaling the relevant invariants by a fraction $x$ relative to the antenna invariant mass, $s_{ij\ldots} = x s_{ijkl}$. 
Due to the absence of azimuthal terms in our antenna functions, phase-space points are combined that are correlated by angular rotations about the collinear direction in every (multi-)collinear and soft-collinear limit in the full antenna-subtraction formalism.
Each histogram in Fig.~\ref{fig:azimuthal} shows the relative agreement of the real subtraction terms with the real matrix elements in digits, $\log_{10}\left(\left\vert 1-R\right\vert\right)$, in two separate limits where $R$ is the ratio of the real correction and its subtraction term. 
The ratio, $R$, should approach $1$ in all unresolved limits and so $(1-R)$ should approach $0$. 
We display the plots for the final-final configuration only. 
Note that a smaller $x$ indicates being numerically closer to the limit. 
The dashed lines show the trajectory into the limit point-by-point. 
The solid lines combine phase-space points related by an azimuthal rotation of $\pi/2$ around the collinear direction. 
Since there are no azimuthal terms in quark-initiated splittings, both the azimuthally averaged and unaveraged trajectories show increasing agreement into the $qg$ collinear limit. 
However, in the $gg$ collinear limit, while there is good agreement (and therefore successful subtraction) in the azimuthally averaged case, this is not the case for the subtraction point-by-point. 
In this case, the azimuthal terms in the matrix element are not subtracted and the agreement does not get better deeper into the limit. 
These plots demonstrate the validity of combining phase-space points that are correlated by angular rotations in the antenna-subtraction scheme. 
They also show that we do not need azimuthal correlations in the idealised antenna functions. 

We define the soft down-projector by its action on invariants as
\begin{equation}
    \PSdown_j: \begin{cases} 
    s_{ij} \mapsto \lambda s_{ij}, 
    s_{jk} \mapsto \lambda s_{jk}, \\
    s_{ijk} \mapsto s_{ik},
    \end{cases}
\end{equation}
and keep only the terms proportional to $\lambda^{-2}$.
For the corresponding up-projector $\PSup_j$ we choose a trivial mapping which leaves all variables unchanged.
The collinear down-projector we define as,
\begin{equation}
    \label{eq:PCdown}
    \PCdown_{ij}: \begin{cases}
    s_{ij} \mapsto \lambda s_{ij}, \\
    s_{ik} \mapsto \omxj (s_{ik}+s_{jk}),
    s_{jk} \mapsto \xj (s_{ik}+s_{jk}), 
    s_{ijk} \mapsto s_{ik} + s_{jk}.
    \end{cases}
\end{equation}
and keep only terms proportional to $\lambda^{-1}$, while its corresponding up-projector we define as,
\begin{equation}
    \label{eq:PCup}
    \PCup_{ij}: \begin{cases}
    \xj \mapsto s_{jk}/s_{ijk},
    \omxj \mapsto s_{ik}/s_{ijk} \\
    s_{ik}+s_{jk} \mapsto s_{ijk}
    \end{cases}
\end{equation}

For convenience, we define a general single-real radiation tree-level antenna function in terms of the contributions produced by the algorithm in Eq.~\eqref{eq:algorithm} as,
\begin{equation}
X_3^0(i^h,j,k^h) = \Ssoft(i^h,j,k^h) + \Scol(i^h,j;k^h) + \Scol(k^h,j;i^h) \, ,
\label{eq:X30def}
\end{equation}
where the individual pieces are given by
\begin{align}
  \Ssoft(i^h,j,k^h) &= \PSup_{j} L_1(i^h,j,k^h) = L_1(i^h,j,k^h), \\
  \Scol(i^h,j;k^h) &= \PCup_{ij}\left(L_2(i^h,j;k^h) - \PCdown_{ij} \Ssoft(i^h,j,k^h)\right), \\
  \Scol(k^h,j;i^h) &= \PCup_{kj}\left(L_3(k^h,j;i^h) - \PCdown_{kj} \left(\Ssoft(i^h,j,k^h) + \Scol(i^h,j;k^h)\right)\right) \nonumber, \\
                   &\equiv \PCup_{kj}\left(L_3(k^h,j;i^h) - \PCdown_{kj} \Ssoft(i^h,j,k^h)\right),
\end{align}
and we have used the fact that the overlap between the two collinear contributions is contained entirely in the projection of $\Ssoft(i^h,j,k^h)$ such that
\begin{equation}
    \PCdown_{kj} \Scol(i^h,j;k^h) = 0 \, .
\end{equation}
This algorithm guarantees that
\begin{align}
\PSdown_j X_3^0(i^h,j,k^h) &= L_1(i^h,j,k^h),\\
\PCdown_{ij} X_3^0(i^h,j,k^h) &= L_2(i^h,j,k^h),\\
\PCdown_{kj} X_3^0(i^h,j,k^h) &= L_3(i^h,j,k^h).
\end{align}

In the following subsections, we derive the single-real radiation antennae for pairs of quark-antiquark, quark-gluon and gluon-gluon parents. As a check, we also give the analytic form of the three-particle antennae integrated over the fully-inclusive $d$-dimensional antenna phase space, 
\begin{equation}
\label{eq:phijkFF}
{\cal X}_{3}^0(s_{ijk}) =
\left(8\pi^2\left(4\pi\right)^{-\e} e^{\e\gamma_E}\right)
\int {\rm d}\, \Phi_{X_{ijk}} X_{3}^0,
\end{equation}
with $d=4-2\e$.
As in Ref.~\cite{Gehrmann-DeRidder:2005btv}, we have included a normalisation factor to account for powers of the QCD coupling constant. 

\subsection{Quark-Antiquark Antennae}

Building the antenna iteratively according to Eq.~\eqref{eq:algorithm} using the list of limits in Eq.~\eqref{eq:NLOL}, we find that
the three-parton tree-level antenna function with quark-antiquark parents is given (to all orders in $\e$) by 
\begin{equation}
\label{eq:A30}
A_3^0(i_q^h,j_g,k_{\bar{q}}^h) =
\frac{2s_{ik}}{s_{ij}s_{jk}} 
+ \frac{\ome s_{jk}}{s_{ijk}s_{ij}}
+ \frac{\ome s_{ij}}{s_{ijk}s_{jk}}.
\end{equation}
Eq.~\eqref{eq:A30} differs from $\Xold{A}$, given in Eq.~(5.5) of Ref.~\cite{Gehrmann-DeRidder:2005btv} that was derived directly from the squared matrix element of $\gamma^* \to qg\bar{q}$, at $\order{\e}$. We note that in general $X_3^0$ can differ from $\Xold{X}$ at $\order{\e^0}$ as we only require that they have the same unresolved limits. Only in this specific case is the difference $\order{\e}$. 

Integrating over the final-final antenna phase space, Eq.~\eqref{eq:phijkFF}, yields 
\begin{eqnarray}
\label{eq:A30int}
{\cal A}_3^0(s_{ijk}) &=&
S_{ijk}
\left[ \frac{1}{\e^2} + \frac{3}{2\e}
+\frac{19}{4} -\frac{7\pi^2}{12}
+\left(\frac{113}{8}-\frac{7\pi^2}{8}-\frac{25\zeta_3}{3}\right)\e
\right. \nonumber \\
&& \left. \hspace{1cm}
+\left(\frac{675}{16}-\frac{133\pi^2}{48}-\frac{71\pi^4}{1440}-\frac{25\zeta_3}{2} \right)\e^2
+ \order{\e^3} \right], 
\end{eqnarray}
which, as expected, differs from the final-final integral of $\Xold{A}$, in Eq.~(5.6) of Ref.~\cite{Gehrmann-DeRidder:2005btv}, starting from $\order{\e}$. We also use the convenient notation, 
\begin{equation}
    S_{ijk} = \left(\frac{s_{ijk}}{\mu^2}\right)^{-\e}.
\end{equation}
Integrals of the universal soft- and collinear-remainder terms in Eq.~\eqref{eq:X30def} (and in this case Eq.~\eqref{eq:A30}) are given in Appendix~\ref{app:X30regionintegrations} for increased clarity. 

\subsection{Quark-Gluon Antennae}
Building the antenna iteratively according to Eq.~\eqref{eq:algorithm} and using the list of soft and collinear limits given in Eq.~\eqref{eq:NLOL} adapted to the particle content, we find that
\begin{equation}
\label{eq:D30}
D_3^0(i_q^h,j_g,k_g^h) =
\frac{2s_{ik}}{s_{ij}s_{jk}} 
+ \frac{\ome s_{jk}}{s_{ijk}s_{ij}}
+ \frac{s_{ij}s_{ik}}{s_{ijk}^2 s_{jk}}.
\end{equation}
Note that, owing to its origin in the neutralino-decay process, the $\Xold{D}$ antenna function given in Eqs.~(4.3) and (4.9) of Ref.~\cite{Gehrmann-DeRidder:2005svg} contains two antennae, corresponding to the configurations: $j_g$ radiated between $i_q$ and $k_g$, and $k_g$ radiated between $i_q$ and $j_g$.
In Eq.~(6.13) of Ref.~\cite{Gehrmann-DeRidder:2005btv}, the full $\Xold{D}$ was decomposed into two sub-antennae, $\Xold{d}(i_q^h,j_g,k_g^h)$ and $\Xold{d}(i_q^h,k_g,j_g^h)$, each of which contains the soft limit of one of the gluons and part of the collinear limit between the gluons.  $D_3^0(i_q^h,j_g,k_g^h)$ in Eq.~\eqref{eq:D30} is therefore to be compared with $\Xold{d}(i_q^h,j_g,k_g^h)$ given in Eq.~(6.13) of Ref.~\cite{Gehrmann-DeRidder:2005btv}. Eq.~\eqref{eq:D30} is more compact because it only  contains terms that contribute to the soft and collinear limits. The difference starts at $\order{\e^0}$ in the dimensional regularisation parameter $\e$. 
Integrating over the antenna phase space yields,
\begin{eqnarray}
\label{eq:D30int}
{\cal D}_3^0(s_{ijk}) &=&
S_{ijk}
\left[ \frac{1}{\e^2} + \frac{5}{3\e}
+\frac{61}{12} -\frac{7\pi^2}{12}
+\left(\frac{121}{8}-\frac{35\pi^2}{36}-\frac{25\zeta_3}{3}\right)\e
\right. \nonumber \\
&& \left. \hspace{1cm}
+\left(\frac{723}{16}-\frac{427\pi^2}{144}-\frac{71\pi^4}{1440}-\frac{125\zeta_3}{9} \right)\e^2
+ \order{\e^3} \right].
\end{eqnarray}
This differs from the final-final integral of $\Xold{d}(i_q^h,j_g,k_g^h)$, which is a half of Eq.~(6.9) in Ref.~\cite{Gehrmann-DeRidder:2005btv}, starting at $\order{\e^0}$.

Similarly, for the three-quark antenna, we find
\begin{equation}
\label{eq:E30}
E_3^0(i_q^h,j_{\bar{Q}},k_Q^h) =
\frac{1}{s_{jk}} 
-\frac{2s_{ij}s_{ik}}{s_{ijk}^2 s_{jk} \ome},
\end{equation}
which is to be compared with $\Xold{E}$ given in Eq.~(4.9) of Ref.~\cite{Gehrmann-DeRidder:2005svg}, which contains terms that do not contribute to the quark-antiquark collinear limit starting at $\order{\e^0}$.
This has the consequence that the integrated antenna,
\begin{eqnarray}
\label{eq:E30int}
{\cal E}_3^0(s_{ijk}) &=&
S_{ijk}
\left[ - \frac{1}{3\e}
 -\frac{3}{4}
+\left(-\frac{15}{8}+\frac{7\pi^2}{36}\right)\e
\right. \nonumber \\
&& \left. \hspace{1cm}
+\left(-\frac{81}{16}+\frac{7\pi^2}{16}+\frac{25\zeta_3}{9} \right)\e^2
+ \order{\e^3} \right] ,
\end{eqnarray}
differs from the final-final integral of $\Xold{E}$given in Eq.~(6.15) of Ref.~\cite{Gehrmann-DeRidder:2005btv} starting at $\order{\e^0}$.

\subsection{Gluon-Gluon Antennae}
Directly constructing the antenna using Eq.~\eqref{eq:algorithm} with the list of soft and collinear limits given in Eq.~\eqref{eq:NLOL} adapted to the particle content, we find that
\begin{equation}
\label{eq:F30}
F_3^0(i_g^h,j_g,k_g^h) =
\frac{2s_{ik}}{s_{ij}s_{jk}} 
+ \frac{s_{ik}s_{jk}}{s_{ijk}^2 s_{ij}}
+ \frac{s_{ij}s_{ik}}{s_{ijk}^2 s_{jk}}.
\end{equation}
The $\Xold{F}$ antenna function, obtained from Higgs boson decay (Eq.~(4.3) of Ref.~\cite{Gehrmann-DeRidder:2005alt}), has limits when any of the three gluons are soft. 
For this reason, it was split into three permutations of the sub-antenna function $\Xold{f}(i_g^h,j_g,k_g^h)$, given by Eq.~(7.13) of Ref.~\cite{Gehrmann-DeRidder:2005btv}, which differs from Eq.~\eqref{eq:F30} by terms that do not contribute in any of the unresolved limits.
Therefore, the integrated form, 
\begin{eqnarray}
\label{eq:F30int}
{\cal F}_3^0(s_{ijk}) &=&
S_{ijk}
\left[ \frac{1}{\e^2} + \frac{11}{6\e}
+\frac{65}{12} -\frac{7\pi^2}{12}
+\left(\frac{129}{8}-\frac{77\pi^2}{72}-\frac{25\zeta_3}{3}\right)\e
\right. \nonumber \\
&& \left. \hspace{1cm}
+\left(\frac{771}{16}-\frac{455\pi^2}{144}-\frac{71\pi^4}{1440}-\frac{275\zeta_3}{18} \right)\e^2
+ \order{\e^3} \right] ,
\end{eqnarray}
differs from the integrated form of $\Xold{f}(i_g^h,j_g,k_g^h)$, which is a third of Eq.~(7.9) in Ref.~\cite{Gehrmann-DeRidder:2005btv}, starting at $\order{\e^0}$.

The $g\bar{Q}Q$ antenna function is given by
\begin{equation}
\label{eq:G30}
G_3^0(i_g^h,j_{\bar{Q}},k_Q^h) =
\frac{1}{s_{jk}} 
-\frac{2s_{ij}s_{ik}}{s_{ijk}^2 s_{jk} \ome}.
\end{equation}
Note that $G_3^0(i_g^h,j_{\bar{Q}},k_Q^h)$ is identical to $E_3^0(i_q^h,j_{\bar{Q}},k_Q^h)$ because in both cases, the only limit that is required is the $g \to \bar{Q}Q$ collinear limit which is independent of the particle type of the other hard radiator.
The antenna derived from the Higgs-decay matrix element $\Xold{G}$ is given in Eq.~(7.14) of Ref.~\cite{Gehrmann-DeRidder:2005btv} and contains terms that do not contribute in the collinear limit. Integrating over the phase space, we find
\newpage
\begin{eqnarray}
\label{eq:G30int}
{\cal G}_3^0(s_{ijk}) &\equiv& {\cal E}_3^0(s_{ijk})\\
&=&
S_{ijk}
\left[ - \frac{1}{3\e}
 -\frac{3}{4}
+\left(-\frac{15}{8}+\frac{7\pi^2}{36}\right)\e
\right. \nonumber \\
&& \left. \hspace{1cm}
+\left(-\frac{81}{16}+\frac{7\pi^2}{16}+\frac{25\zeta_3}{9} \right)\e^2
+ \order{\e^3} \right] ,
\end{eqnarray}
which differs from the integrated form of $\Xold{G}$, given in Eq.~(7.15) of Ref.~\cite{Gehrmann-DeRidder:2005btv}, at $\order{\e^0}$.

\section{Double-Real Radiation Antennae}
\label{sec:double-real-antennae}
At NNLO, we want to construct four-particle antenna functions $\X(i^h_a,j_b,k_c,l^h_d)$, where the particle types are denoted by $a$, $b$, $c$ and $d$, which carry four-momenta $i$, $j$, $k$ and $l$ respectively. Particles $a$ and $d$ should be hard and the antenna functions must have the correct limits when particles $b$ and $c$ are unresolved. Frequently, we again drop explicit reference to the particle labels in favour of a specific choice of $X$ according to Table~\ref{tab:X40}.

\begin{table}[t]
\centering
\begin{tabular}{ccc}
\underline{Quark-antiquark} & & \\
$qgg\bar{q}$ & $\X(i_q^h,j_g,k_g,l_{\bar{q}}^h)$ & $\A(i^h,j,k,l^h)$ \\
$q\gamma\gamma\bar{q}$ & $\Xt(i_q^h,j_\gamma,k_\gamma,l_{\bar{q}}^h)$ & $\At(i^h,j,k,l^h)$ \\
$q\bar{Q}{Q}\bar{q}$ & $\X(i_q^h,j_{\bar{Q}},k_Q,l_{\bar{q}}^h)$ & $\B(i^h,j,k,l^h)$ \\
$q\bar{q}{q}\bar{q}$ & $\X(i_q^h,j_{\bar{q}},k_q,l_{\bar{q}}^h)$ & $\C(i^h,j,k,l^h)$ \\
\underline{Quark-gluon} & &  \\
$qggg$ & $\X(i_q^h,j_g,k_g,l_g^h)$ & $\D(i^h,j,k,l^h)$  \\
       & $\Xt(i_q^h,j_g,k_g,l_g^h)$ & $\Dt(i^h,j,k,l^h)$  \\
$q\bar{Q}Qg$ & $\X(i_q^h,j_{\bar{Q}},k_Q,l_g^h)$  & $\Ea(i^h,j,k,l^h)$ \\ 
$qg\bar{Q}Q$ & $\X(i_q^h,j_g,k_{\bar{Q}},l_Q^h)$  & $\Eb(i^h,j,k,l^h)$ \\ 
$q\bar{Q}gQ$ & $\Xt(i_q^h,j_{\bar{Q}},k_g,l_Q^h)$  & $\Et(i^h,j,k,l^h)$ \\ 
\underline{Gluon-gluon} & & \\
$gggg$ & $\X(i_g^h,j_g,k_g,l_g^h)$ & $\F(i^h,j,k,l^h)$  \\
       & $\Xt(i_g^h,j_g,k_g,l_g^h)$ & $\Ft(i^h,j,k,l^h)$  \\
$g\bar{Q}Qg$ & $\X(i_g^h,j_{\bar{Q}},k_Q,l_g^h)$  & $\Ga(i^h,j,k,l^h)$ \\ 
$gg\bar{Q}Q$ & $\X(i_g^h,j_g,k_{\bar{Q}},l_Q^h)$  & $\Gb(i^h,j,k,l^h)$ \\ 
$g\bar{Q}gQ$ & $\Xt(i_g^h,j_{\bar{Q}},k_g,l_Q^h)$  & $\Gt(i^h,j,k,l^h)$ \\ 
$\bar{q}q\bar{Q}Q$ & $\X(i_{\bar{q}}^h,j_q,k_{\bar{Q}},l_Q^h)$ & $\H(i^h,j,k,l^h)$  \\
\end{tabular}
\caption{Identification of $\X$ antennae according to particle type and colour-structure. These antennae only contain singular limits when one or both of particles $b$ and $c$ (or equivalently momenta $j$ and $k$) are unresolved. Antennae are classified as quark-antiquark, quark-gluon and gluon-gluon according to the particle type of the parents (i.e. after the antenna mapping). }
\label{tab:X40}
\end{table}

For double-real-radiation antenna functions, we have to distinguish between the case where the two unresolved particles are colour connected (which we denote by 
$\X(i^h_a,j_b,k_c,l^h_d)$) and the case where they are not (which we denote by $\Xt(i^h_a,j_b,k_c,l^h_d)$).
The list of limits included in each case is different, due to different possible double- and simple-collinear limits. Specifically, in $\Xt$ there are no $bc$-collinear limits but there are $ac$- and $bd$-collinear limits, which are absent in $\X$.

As at NLO, we systematically start from the most singular limit and build the list of target functions using double- and single unresolved limits. The list of limits for the 
$\X(i^h_a,j_b,k_c,l^h_d)$ double-real antenna function, from most singular to least singular, is given by
\begin{equation}
\begin{split}
    L_1(i^h,j,k,l^h) &= S_{bc}^{(0)}(i^h,j,k,l^h) \, , \\
    L_2(i^h,j,k;l^h) &= P_{abc}^{(0)}(i^h,j,k) \, , \\
    L_3(i^h,j,k,l^h) &= P_{dcb}^{(0)}(l^h,k,j) \, , \\
    L_4(i^h,j,k,l^h) &= P_{ab}^{(0)}(i^h,j) P_{dc}^{(0)}(l^h,k) \, ,  \\
    L_5(i^h,j,k,l^h) &= S_{b}^{(0)}(i^h,j,k^h) \, X_{3}^{0}(i^h,k,l^h) \, , \\
    L_6(i^h,j,k,l^h) &= S_{c}^{(0)}(j^h,k,l^h) \, X_{3}^{0}(i^h,j,l^h) \, , \\
    L_7(i^h,j,k,l^h) &= P_{ab}^{(0)}(i^h,j) \, X_{3}^{0}((i+j)^h,k,l^h) \, , \\
    L_8(i^h,j,k,l^h) &= P_{bc}^{(0)}(j,k) \, X_{3}^{0}(i^h,(j+k),l^h) \, , \\
    L_9(i^h,j,k,l^h) &= P_{dc}^{(0)}(l^h,k) \, X_{3}^{0}(i^h,j,(l+k)^h) \, ,
\end{split}
\label{eq:listX40}
\end{equation}
where for readability, we have suppressed the labels for the particle types in $X_3^0$.
Here, $L_1$ contains the double soft contribution; $L_2$ and $L_3$ the triple-collinear contributions; $L_4$ the double-collinear contribution; $L_5$ and $L_6$ the single-soft limits; and $L_7$, $L_8$ and $L_9$ the simple-collinear limits. 
The specific choices of $X_3^0$ in $L_5$ -- $L_9$ are fixed by the flavour structure of the relevant single unresolved limits.
Because the iterative procedure is organised such that there are no overlaps between contributions of the same level, the ordering of limits of the same type is not important -- e.g. between $L_2$ and $L_3$.
For the 
$\Xt(i^h_a,j_b,k_c,l^h_d)$, the list of limits is given by
\begin{equation}
\begin{split}
    \tilde{L}_1(i^h,j,k,l^h) &= S_{bc}^{(0)}(i^h,j,k,l^h) \, , \\
    \tilde{L}_2(i^h,j,k;l^h) &= P_{abc}^{(0)}(i^h,j,k) \, , \\
    \tilde{L}_3(i^h,j,k,l^h) &= P_{dcb}^{(0)}(l^h,k,j) \, , \\
    \tilde{L}_4(i^h,j,k,l^h) &= P_{ab}^{(0)}(i^h,j) P_{dc}^{(0)}(l^h,k) \, , \\
    \tilde{L}_5(i^h,j,k,l^h) &= P_{ac}^{(0)}(i^h,k) P_{db}^{(0)}(l^h,j) \, , \\
    \tilde{L}_6(i^h,j,k,l^h) &= S_{b}^{(0)}(i^h,j,l^h) \, X_{3}^{0}(i^h,k,l^h) \, , \\
    \tilde{L}_7(i^h,j,k,l^h) &= S_{c}^{(0)}(i^h,k,l^h) \, X_{3}^{0}(i^h,j,l^h) \, , \\
    \tilde{L}_8(i^h,j,k,l^h) &= P_{ab}^{(0)}(i^h,j) X_{3}^{0}((i+j)^h,k,l^h) \, , \\
    \tilde{L}_9(i^h,j,k,l^h) &= P_{ac}^{(0)}(i^h,k) X_{3}^{0}((i+k)^h,j,l^h) \, , \\
    \tilde{L}_{10}(i^h,j,k,l^h) &= P_{dc}^{(0)}(l^h,k) X_{3}^{0}(i^h,j,(l+k)^h) \, , \\
    \tilde{L}_{11}(i^h,j,k,l^h) &= P_{db}^{(0)}(l^h,j) X_{3}^{0}(i^h,k,(l+j)^h) \, .
\end{split}
\label{eq:listX40t}
\end{equation}
The tree-level double-soft factors are given in Eqs.~\eqref{eq:Sggtwo}-\eqref{eq:Sqq} and zero otherwise.
The triple-collinear splitting functions $P_{abc}^{(0)}(i^h,j,k)$, when particle $a$ is hard, are given in their decomposed form in Chapter~\ref{chapter:paper1}. 
It may be useful for the reader to note that $P_{g \gamma \gamma}^{(0)} (i^h,j,k) \equiv P_{ggg}^{(0)} (k,i^h,j)$. 
We exploit the decomposition into strongly-ordered iterated contributions (which are products of the usual spin-averaged two-particle splitting functions) and a remainder function $R_{abc}^{(0)} (i,j,k)$ that is finite when any pair of $\{i,j,k\}$ are collinear. 
Additionally, the projections of the double-soft factors in the triple-collinear phase space are given in Appendix~\ref{app:TCprojectionsofDS} and they indicate the overlap between the two limits.
The three-particle antennae appearing in Eqs.~\eqref{eq:listX40} and \eqref{eq:listX40t} are those discussed in Section~\ref{sec:single-real-antennae}.  For example, $X_3^0((i+j)^h,k,l^h)$ denotes the antenna with particle types $(ab)$, $c$ and $d$ according to Table~\ref{tab:X30} carrying momenta $(i+j)$, $k$ and $l$ respectively. In Appendix~\ref{app:X40limits}, we list in full the limits for each $\X$, for convenient reference. 

We highlight two observations here. 
Firstly, neither Eq.~\eqref{eq:listX40} nor Eq.~\eqref{eq:listX40t} include an explicit soft-collinear limit. 
These limits are present (and are verified after $\X$ construction) but arise naturally from the combination of double-soft and triple-collinear limits. 
Secondly, because the antenna functions are built from spin-averaged splitting functions, there are no azimuthal correlations in the constructed $\X$. The effect of azimuthal-averaging has been discussed in Section~\ref{sec:single-real-antennae}. 

Together with the list of limits, we need to have a procedure for mapping the antenna into the particular limit subspace and then returning to the full phase space.
We define the double-soft down-projector for particles $j$ and $k$ soft as
\begin{equation}
    \PDSdown_{jk}: \begin{cases}
        s_{jk} \mapsto \lambda^2 s_{jk}, \\
        s_{ij} \mapsto \lambda s_{ij}, \,
        s_{ik} \mapsto \lambda s_{ik}, \,
        s_{jl} \mapsto \lambda s_{jl}, \,
        s_{kl} \mapsto \lambda s_{kl}, \\
        s_{ijk} \mapsto \lambda s_{ijk}, \,
        s_{jkl} \mapsto \lambda s_{jkl}, \\
        s_{ijl} \mapsto s_{il}, \,
        s_{ikl} \mapsto s_{il}, \,
        s_{ijkl} \mapsto s_{il},
    \end{cases}
\end{equation}
and we keep only terms proportional to $\lambda^{-4}$. As at NLO, for the corresponding up-projector $\PDSup_{jk}$ we choose a trivial mapping which leaves all variables unchanged.

We define the triple-collinear down-projector for collinear particles $i$, $j$ and $k$ as
\begin{equation}
    \PTCdown_{ijk}: \begin{cases}
        s_{ij} \mapsto \lambda s_{ij}, \,
        s_{ik} \mapsto \lambda s_{ik}, \,
        s_{jk} \mapsto \lambda s_{jk}, \,
        s_{ijk} \mapsto \lambda s_{ijk}, \\
        s_{ijkl} \mapsto s_{il}+s_{jl}+s_{kl}, \\
        s_{il} \mapsto \xi \left (s_{il}+s_{jl}+s_{kl}\right ),\,
        s_{jkl} \mapsto \omxi \left (s_{il}+s_{jl}+s_{kl}\right ), \\
        s_{jl} \mapsto \xj \left (s_{il}+s_{jl}+s_{kl}\right ),\,
        s_{ikl} \mapsto \omxj\left (s_{il}+s_{jl}+s_{kl}\right ), \\
        s_{kl} \mapsto \xk \left (s_{il}+s_{jl}+s_{kl}\right ),
        s_{ijl} \mapsto \omxk\left (s_{il}+s_{jl}+s_{kl}\right ),
    \end{cases}
\end{equation}
with $\xi+\xj+\xk = 1$ and we keep only terms proportional to $\lambda^{-2}$. Note that when $\omxi$ appears in the numerator, expressions are simplified according to $\omxi = \xj + \xk$ and so on. 
The corresponding up-projector we define as
\begin{equation}
    \PTCup_{ijk}: \begin{cases}
        \xi  \mapsto s_{il}/(s_{il}+s_{jl}+s_{kl}),\\ 
        \omxi  \mapsto s_{jkl}/(s_{il}+s_{jl}+s_{kl}),\\
        \xj  \mapsto s_{jl}/(s_{il}+s_{jl}+s_{kl}),\\ 
        \omxj  \mapsto s_{ikl}/(s_{il}+s_{jl}+s_{kl}),\\
        \xk  \mapsto s_{kl}/(s_{il}+s_{jl}+s_{kl}),\\ 
        \omxk  \mapsto s_{ijl}/(s_{il}+s_{jl}+s_{kl}),\\
        s_{il}+s_{jl}+s_{kl} \mapsto s_{ijkl}. \\
    \end{cases}
\end{equation}

The double-collinear down-projector for particles $i||j$ and $k||l$ we choose as
\begin{equation}
    \PDCdown_{ij;kl}: \begin{cases}
        s_{ij} \mapsto \lambda s_{ij}, \,
        s_{kl} \mapsto \mu s_{kl}, \\
        s_{il} \mapsto \omxj\omyk s_{ijkl}, \,
        s_{jl} \mapsto \xj\omyk s_{ijkl}, \\
        s_{ik} \mapsto \omxj\yk s_{ijkl}, \,
        s_{jk} \mapsto \xj\yk s_{ijkl}, \\
        s_{ijk} \mapsto \yk s_{ijkl} , \,
        s_{ijl} \mapsto \omyk s_{ijkl} , \\
        s_{ikl} \mapsto \omxj s_{ijkl}, \,
        s_{jkl} \mapsto \xj s_{ijkl}, \\
        s_{ijkl} \mapsto s_{ik}+s_{jk}+s_{il}+s_{jl}, \\
    \end{cases}
\end{equation}
and we keep only terms proportional to $\lambda^{-1}\mu^{-1}$ in order to count the divergences of both collinear limits separately.
The corresponding up projector is
\begin{equation}
    \PDCup_{ij;kl}: \begin{cases}
        \xj\yk \mapsto  s_{jk}/s_{ijkl}, \\
        \xj \mapsto (s_{jk}+s_{jl})/s_{ijkl}, \\
        \yk \mapsto (s_{ik}+s_{jk})/s_{ijkl}, \\
        1/\yk \mapsto  s_{ijkl}/s_{ijk} , \,
        1/\omyk \mapsto  s_{ijkl}/s_{ijl}  , \\
        1/\xj \mapsto s_{ijkl}/s_{jkl} , \,
        1/\omxj \mapsto  s_{ijkl}/s_{ikl}.
    \end{cases}
\end{equation}

The single-soft down-projector acts on invariants as
\begin{equation}
    \PSdown_j: \begin{cases} 
    s_{ij} \mapsto \lambda s_{ij}, \,
    s_{jk} \mapsto \lambda s_{jk}, \,
    s_{jl} \mapsto \lambda s_{jl}, \\
    s_{ijk} \mapsto s_{ik}, \,
    s_{ijl} \mapsto s_{il}, \,
    s_{jkl} \mapsto s_{kl}, \\
    s_{ijkl} \mapsto s_{ikl},
    \end{cases}
\end{equation}
and keeps only terms proportional to $\lambda^{-2}$.
For the corresponding up-projector $\PSup_j$, we again choose a trivial mapping which leaves all variables unchanged, in line with the choice for the single-soft up-projector in Section~\ref{sec:single-real-antennae}.

The simple-collinear down-projector we choose as,
\begin{equation}
    \PCdown_{ij}: \begin{cases}
    s_{ij} \mapsto \lambda s_{ij}, \\
    s_{ik} \mapsto \omxj s_{ijk}, \,
    s_{jk} \mapsto \xj s_{ijk}, \\
    s_{il} \mapsto \omxj s_{ijl}, \,
    s_{jl} \mapsto \xj s_{ijl}, \\    
    s_{ijk} \mapsto s_{ik} + s_{jk}, \\
    s_{ijl} \mapsto s_{il} + s_{jl}, \\
    \end{cases}
\end{equation}
and keep only terms proportional to $\lambda^{-1}$, while its corresponding up-projector we define as,
\begin{equation}
    \PCup_{ij}: \begin{cases}
    \xj (s_{ik}+s_{jk}) \mapsto s_{jk}, \,\,
    \omxj (s_{ik}+s_{jk}) \mapsto s_{ik} \\
    \xj (s_{il}+s_{jl}) \mapsto s_{jl}, \,\,
    \omxj (s_{il}+s_{jl}) \mapsto s_{il} \\
    s_{ik}+s_{jk} \mapsto s_{ijk}, \\
    s_{il}+s_{jl} \mapsto s_{ijl}, \\
    \xj \omxj \mapsto s_{jk}s_{il}/(s_{ijk} s_{ijl}), \\
    \xj \mapsto s_{jk}/s_{ijk}.
    \end{cases}
\end{equation}
Note that there is an ambiguity on how to write $\xj$ and $\omxj$ in terms of invariants.  In particular, we use the following identities,
\begin{eqnarray}
\xj^2 &=& \xj -\xj\omxj, \\
\omxj^2 &=& 1-\xj -\xj\omxj,\\
\xj\omxj &\mapsto& \frac{s_{jk}s_{il}}{s_{ijk}s_{ijl}} \text{ or } \frac{s_{ik}s_{jl}}{s_{ijk}s_{ijl}},
\end{eqnarray}
to avoid repeated powers of triple invariants.  Additional identities are imposed to preserve the symmetry of the antenna. 

Note that even after defining all projectors, some ambiguity remains.
For example, if we consider what happens to a term like $s_{ijl}/s_{ijl}$ in the $ijk$ triple-collinear subspace, then 
\begin{alignat*}{2}
1 \equiv \frac{s_{il}+s_{jl}+s_{ij}}{s_{ijl}} 
&\stackrel{\PTCdown_{ijk}}{\longrightarrow} \frac{\xi+\xj}{\omxk} \equiv 1 &&\stackrel{\PTCup_{ijk}}{\longrightarrow} 1,
\end{alignat*}
while applying the process to the individual terms does not give back unity upon summation in the full phase space,
\begin{alignat*}{2}
\frac{s_{il}}{s_{ijl}} 
&\stackrel{\PTCdown_{ijk}}{\longrightarrow} \frac{\xi}{\omxk} &&\stackrel{\PTCup_{ijk}}{\longrightarrow} \frac{s_{il}}{s_{ijl}},\\
\frac{s_{jl}}{s_{ijl}} 
&\stackrel{\PTCdown_{ijk}}{\longrightarrow} \frac{\xj}{\omxk} &&\stackrel{\PTCup_{ijk}}{\longrightarrow} \frac{s_{jl}}{s_{ijl}},\\
\frac{s_{ij}}{s_{ijl}} 
&\stackrel{\PTCdown_{ijk}}{\longrightarrow} 0 
&&\stackrel{\PTCup_{ijk}}{\longrightarrow} 0.
\end{alignat*}
It is clear that the sum of the last three lines is equivalent to unity in the triple-collinear subspace. 
Exactly how one returns to the full space can introduce differences there.   
These differences do not affect the limit in question but do influence less singular limits. 
These are corrected by the iterative process which systematically moves from a more singular limit to a less singular limit, guaranteeing that each limit is correctly described.  
However, after the iterative construction of the antenna function is complete, there can still be differences in finite terms that do not contribute to any limit.

For antennae where $i,j,k$ and $l$ are colour connected to adjacent particles, i.e. $i$ to $j$ to $k$ to $l$, we define the general antenna function as
\newpage
\begin{eqnarray}
\X(i^h,j,k,l^h) &=& \Dsoft(i^h,j,k,l^h) \nonumber \\
    && + \Tcol(i^h,j,k;l^h) + \Tcol(l^h,k,j;i^h) \nonumber \\
    && + \Dcol(i^h,j; k,l^h) \nonumber \\
    && + \Ssoft(i^h,j,k;l^h) + \Ssoft(l^h,k,j;i^h) \nonumber \\
    && + \Scol(i^h,j;k,l^h) + \Scol(j,k;i^h,l^h) + \Scol(l^h,k;j,i^h), \hspace{1cm} 
\label{eq:X40def}
\end{eqnarray}
where the individual pieces are given by
\begin{align}
\label{eq:Xpieces}
  \Dsoft(i^h,j,k,l^h) &= \PSup_{jk} L_1(i^h,j,k,l^h) = L_1(i^h,j,k,l^h) \, , \\
  \Tcol(i^h,j,k;l^h) &= \PTCup_{ijk}\left(L_2(i^h,j,k,l^h) 
  - \PTCdown_{ijk} \Dsoft(i^h,j,k,l^h)\right) \, , \\
  \Tcol(l^h,k,j;i^h) &= \PTCup_{lkj}\left(L_3(i^h,j,k,l^h) 
  - \PTCdown_{lkj} \left(\Dsoft(i^h,j,k,l^h) + \Tcol(i^h,j,k;l^h)\right)\right) \nonumber \\
                   &\equiv \PTCup_{lkj}\left(L_3(i^h,j,k,l^h) - \PTCdown_{lkj} \Dsoft(i^h,j,k,l^h)\right) \, , \\
  \Dcol(i^h,j; k,l^h) &= \PDCup_{ij;kl}\left(L_4(i^h,j,k,l^h) - \PDCdown_{ij;kl} X_{4;3}^{0}(i^h,j,k,l^h)\right) \, , \\
  \Ssoft(i^h,j,k;l^h) &= \PSup_{j}\left(L_5(i^h,j,k,l^h) - \PSdown_{j} X_{4;4}^{0}(i^h,j,k,l^h))\right) \, , \\
  \Ssoft(j,k,l^h;i^h) &= \PSup_{k}\left(L_6(i^h,j,k,l^h) - \PSdown_{k} \left(X_{4;4}^{0}(i^h,j,k,l^h)+\Ssoft(i^h,j,k;l^h)\right)\right) \nonumber \\
  &\equiv \PSup_{k}\left(L_6(i^h,j,k,l^h) - \PSdown_{k} X_{4;4}^{0}(i^h,j,k,l^h)\right) \, , \\
  \Scol(i^h,j;k,l^h) &= \PCup_{ij}\left(L_7(i^h,j,k,l^h) - \PCdown_{ij} X_{4;6}^{0}(i^h,j,k,l^h)\right) \, , \\
  \Scol(j,k;i^h,l^h) &= \PCup_{jk}\left(L_8(i^h,j,k,l^h) - \PCdown_{jk} \left (X_{4;6}^{0}(i^h,j,k,l^h)+ \Scol(i^h,j;k,l^h)\right)\right) \nonumber \\
  &\equiv \PCup_{jk}\left(L_8(i^h,j,k,l^h) - \PCdown_{jk} X_{4;6}^{0}(i^h,j,k,l^h)\right) \, , \\
  \Scol(l^h,k;j,i^h) &= \PCup_{kl}\left(L_9(i^h,j,k,l^h) - \PCdown_{kl} \left( X_{4;6}^{0}(i^h,j,k,l^h)+\Scol(i^h,j;k,l^h)\right.\right. \nonumber \\
  & \hspace{5cm}\left.\left. + \Scol(j,k;i^h,l^h)\right)\right) \nonumber \\
  &\equiv \PCup_{kl}\left(L_9(i^h,j,k,l^h) - \PCdown_{kl} X_{4;6}^{0}(i^h,j,k,l^h)\right) \, ,
\end{align}
where we have used
\begin{eqnarray}
\label{eq:Xpiecerelations1}
\PTCdown_{lkj} \Tcol(i^h,j,k;l^h) &=&0  , \\
\PSdown_{k} \Ssoft(i^h,j,k;l^h) &=&0  , \\
\PCdown_{jk} \Scol(i^h,j;k,l^h) &=&0  , \\
\PCdown_{kl} \Scol(i^h,j;k,l^h) &=&0  , \\
\label{eq:Xpiecerelations11}
\PCdown_{kl} \Scol(j,k;i^h,l^h) &=&0  ,
\end{eqnarray}
together with
\begin{eqnarray}
\label{eq:Xpiecerelations2}
X_{4;3}^{0}(i^h,j,k,l^h) &=& \Dsoft(i^h,j,k,l^h) + \Tcol(i^h,j,k;l^h) + \Tcol(l^h,k,j;i^h)  , \hspace{2cm} \\
X_{4;4}^{0}(i^h,j,k,l^h) &=& X_{4:3}^{0}(i^h,j,k,l^h) + \Dcol(i^h,j; k,l^h),   \\
X_{4;6}^{0}(i^h,j,k,l^h) &=& X_{4:4}^{0}(i^h,j,k,l^h) + \Ssoft(i^h,j,k;l^h) + 
\Ssoft(j,k,l^h;i^h). \hspace{2cm} 
\end{eqnarray}
An example of this type of antenna would be the leading-colour antenna with a quark and an antiquark as hard radiators, emitting two gluons. 
In the same way as for the $X_3^0$ antennae, the algorithm ensures that for all limits
\begin{equation}
    \PPdown_i \X(i^h,j,k,l^h) = L_i(i^h,j,k,l^h).
\end{equation}
In addition, $\X(i^h,j,k,l^h)$ has the correct soft-collinear limits. 

For sub-leading-colour antennae, where $j$ and $k$ are each colour connected to the hard radiators and not to each other, we define a general antenna function $\Xt$ as
\begin{eqnarray}
\Xt(i^h,j,k,l^h) & =& \Dsoft(i^h,j,k,l^h) \nonumber \\
    && + \Tcol(i^h,{j},k;l^h) + \Tcol(l^h,k,j;i^h) \nonumber \\
    && + \Dcol(i^h,{j};k,l^h) + \Dcol(i^h,k; j,l^h) \nonumber \\
    && + \Ssoft(i^h,{j},l^h;k) + \Ssoft(i^h,k,l^h;j) \nonumber \\
    && + \Scol(i^h,{j};k,l^h) + \Scol(i^h,k;j,l^h) \nonumber \\
    && + \Scol(l^h,j;k,i^h) + \Scol(l^h,k;j,i^h),
\label{eq:X40tdef}
\end{eqnarray}
where $\Dsoft$, $\Tcol$, $\Dcol$, $\Ssoft$ and $\Scol$ are defined analogously to Eq.~\eqref{eq:Xpieces} with a similar absence of overlaps between terms at the same level as in Eqs.~\eqref{eq:Xpiecerelations1} -~\eqref{eq:Xpiecerelations11}.
An example of this type of antenna would be one with quark and antiquark hard radiators, emitting two photons/abelianised gluons. 
Again, the algorithm ensures that $\Xt$ satisfies
\begin{equation}
    \PPdown_i \Xt(i^h,j,k,l^h) = \tilde{L}_i(i^h,j,k,l^h),
\end{equation}
for all limits and has the correct soft-collinear limits.

We also give the analytic form of the four-particle antennae integrated over the fully inclusive $d$-dimensional antenna phase space~\cite{Gehrmann-DeRidder:2005btv,Gehrmann-DeRidder:2003pne}, 
\begin{equation}
\label{eq:phijklFF}
\calX(s_{ijkl}) =
\left(8\pi^2\left(4\pi\right)^{-\e} e^{\e\gamma_E}\right)^2
\int {\rm d}\, \Phi_{X_{ijkl}} \X,
\end{equation}
including a normalisation factor to account for powers of the QCD coupling constant.

\subsection{Quark-Antiquark Antennae}
As shown in Table~\ref{tab:X40}, there are four tree-level four-parton antennae with quark-antiquark parents that describe the emission of
\begin{enumerate}
    \item two colour-connected gluons ($\A$),
    \item two photons, or equivalently two gluons that are not colour connected ($\At$),
    \item the emission of a quark-antiquark pair of different flavour to the radiators ($\B$),
    \item the emission of a quark-antiquark pair of the same flavour as the radiators ($\C$).
\end{enumerate}
These antenna functions can straightforwardly be obtained from matrix elements for $\gamma^*\to 4~\text{partons}$~\cite{Gehrmann-DeRidder:2005btv} and the antenna functions constructed here are directly related to the antenna functions given in Ref.~\cite{Gehrmann-DeRidder:2005btv} by
\begin{align}
\A(i_q^h, j_g, k_g, l_{\bar{q}}^h) &\sim
\Aold(i_q, j_g, k_g, l_{\bar{q}}),\\
\At(i_q^h, j_g, k_g, l_{\bar{q}}^h) &\sim
\Atold(i_q, j_g, k_g, l_{\bar{q}}),\\
\B(i_q^h, j_{\bar{Q}}, k_Q, l_{\bar{q}}^h) &\sim
\Bold(i_q, j_{\bar{Q}}, k_Q, l_{\bar{q}}),\\
\C(i_q^h, j_{\bar{q}}, k_q, l_{\bar{q}}^h) &\sim
\Cold(i_q, j_{\bar{q}}, k_q, l_{\bar{q}}),
\end{align}
in the sense that the new antennae should have the same double unresolved limits as the original antenna (as indicated by the $\sim$ symbol).

\begin{figure}[p]
    \centering
    \includegraphics[height=0.22\textheight,page=2]{A40n.pdf}
    \includegraphics[height=0.22\textheight,page=10]{A40n.pdf} 
    \includegraphics[height=0.22\textheight,page=6]{A40n.pdf} 
    \includegraphics[height=0.22\textheight,page=1]{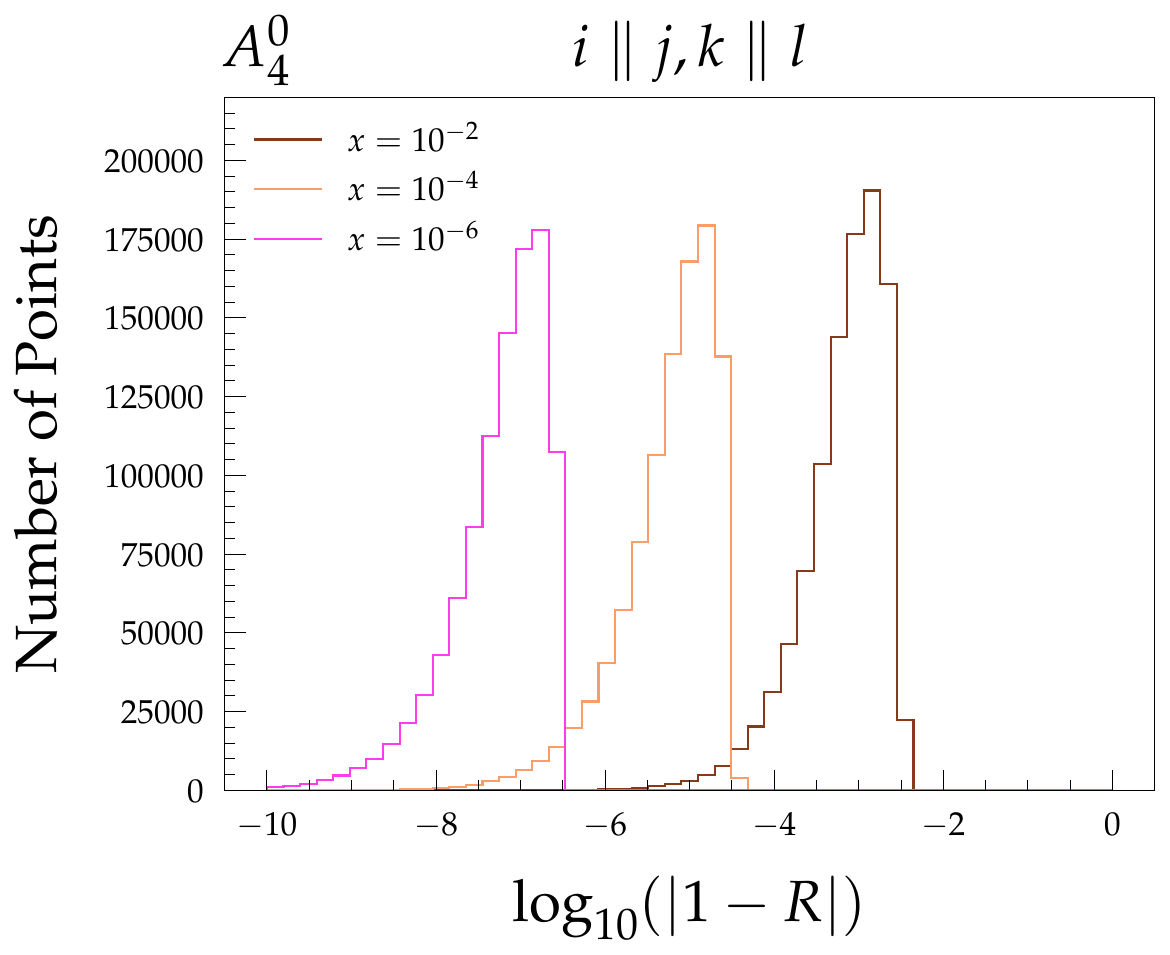}  
    \includegraphics[height=0.22\textheight,page=3]{A40n.pdf} 
    \includegraphics[height=0.22\textheight,page=4]{A40n.pdf} 
    \includegraphics[height=0.22\textheight,page=8]{A40n.pdf}    
    \caption{Numerical tests of the new $\A$ against $\Aold$ in all relevant singular limits. For three different values of the scaling parameter $x$, the relative agreement of the ratio $R = \A/\Aold$ with $1$ is shown on a logarithmic axis.}
    \label{fig:spiketests_A40}
\end{figure}

To build the $\A$ antenna using the algorithm described in Section~\ref{sec:algorithm1}, we simply identify the particles in the list of required limits given in Eq.~\eqref{eq:listX40} -- $b$ and $c$ are gluons, while $a$ ($d$) are the quark (antiquark) hard radiators. 
In this case, the three-particle antennae appearing in the required limits are of type $A_3^0$.
The resulting expression for $\A$ is included in the auxiliary materials of Ref.~\cite{paper2}.  It satisfies the line reversal property,
\begin{equation}
\A(i^h,j,k,l^h) = \A(l^h,k,j,i^h).
\end{equation}
As an example of the numerical behaviour of the new antenna functions, in Fig.~\ref{fig:spiketests_A40} we show numerical tests of the the new $\A$ against the original $\Aold$. 
We follow Ref. \cite{Pires:2010jv} and build trajectories into unresolved limits by scaling the relevant invariants by a fraction $x$ relative to the antenna invariant mass, $s_{ij\ldots} = x s_{ijkl}$, similarly to Fig.~\ref{fig:azimuthal}. Due to the absence of azimuthal terms in our antenna functions, we combine phase-space points that are correlated by angular rotations about the collinear direction in every (multi-)collinear and soft-collinear limit.
Each histogram shows the relative agreement of $\A$ and $\Aold$ in digits,
\begin{equation}
    \log_{10}\left(\left\vert 1-R\right\vert\right) \text{ with } R = \frac{\A}{\Aold} \, .
\end{equation}
We wish to point out that due to the explicit line-reversal symmetry of $\A$, we only show representative examples for each limit. 
In all unresolved limits, we find percent-level agreement already for $x=10^{-2}$ and increasing agreement for smaller values of $x$.
Taking into account the different scaling behaviour of the double- and single unresolved limits, it is evident that the antenna function develops quantitatively similar behaviour in approaching the singularity.

Integrating over the antenna phase space, we find 
\newpage
\begin{eqnarray}
\label{eq:A40int}
\calA (s_{ijkl}) &=& S_{ijkl}^2 \Biggl [
+\frac{3}{4\e^4}
+\frac{65}{24\e^3}
+\frac{1}{\e^2} \left(
\frac{217}{18}
-\frac{13}{12}\pi^2
\right)
+\frac{1}{\e} \left(
\frac{44087}{864}
-\frac{589}{144}\pi^2
-\frac{71}{4}\zeta_3
\right)
\nonumber \\&& \hspace{1cm}
 + \left(
\frac{1134551}{5184}
-\frac{8117}{432}\pi^2
-\frac{1327}{18}\zeta_3
+\frac{373}{1440}\pi^4
\right)
 + \order{\e}\Biggr],
\end{eqnarray} 
which differs from $\calAold$ in Eq.~(5.31) of Ref.~\cite{Gehrmann-DeRidder:2005btv}, starting from the rational part at $\order{1/\e}$. This is completely understood and is simply because the $A_3^0$ given in Eq.~\eqref{eq:A30} differs at $\order{\e}$ from $\Xold{A}$ given in Eq.~(5.5) of Ref.~\cite{Gehrmann-DeRidder:2005btv}. 
The choice of $A_3^0$ impacts the algorithm at the stage of the single-soft limits, $L_{5,6} \sim \Sg A_3^0$. Integrating the  difference $2 \Sg(\Xold{A}-A_3^0)$ over the antenna phase space yields precisely the observed discrepancy of $\order{1/\e}$. Integrated forms of the four-particle antennae constructed by using the original set of three-particle antennae, $\Xold{X}$, derived directly from the squared matrix elements are listed in Appendix~\ref{app:intX4oldX3}. The pole structure of Eq.~\eqref{eq:A40intoldX3} agrees precisely with Eq.~(5.31) of Ref.~\cite{Gehrmann-DeRidder:2005btv} and differs only at $\order{\e^0}$. We also use the convenient notation, 
\begin{equation}
    S_{ijkl} = \left(\frac{s_{ijkl}}{\mu^2}\right)^{-\e}.
\end{equation}
Integrals of the universal double unresolved contributions in Eq.~\eqref{eq:X40def} for all $\X$ are given in Appendix~\ref{app:X40regionintegrations} for increased clarity. 

To build the $\At$ antenna, we identify $a$ ($d$) as the quark (antiquark) hard radiators and $b$ and $c$ as photons (or abelianised gluons). 
The resulting expression for $\At$ is included in the auxiliary materials of Ref.~\cite{paper2}. It is symmetric under both line reversal
\begin{equation}
\At(i^h,j,k,l^h) = \At(l^h,k,j,i^h)
\end{equation}
and exchange of the two gluons,
\begin{equation}
\At(i^h,j,k,l^h) = \At(i^h,k,j,l^h).
\end{equation}
Integrating over the antenna phase space, we find
\begin{eqnarray}
\label{eq:A40tint}
\calAt (s_{ijkl}) &=& S_{ijkl}^2 \Biggl [
+\frac{1}{\e^4}
+\frac{3}{\e^3}
+\frac{1}{\e^2} \left(
13
-\frac{3}{2}\pi^2
\right)
+\frac{1}{\e} \left(
\frac{861}{16}
-\frac{9}{2}\pi^2
-\frac{80}{3}\zeta_3
\right)
\nonumber \\&& \hspace{2cm}
 + \left(
\frac{7105}{32}
-\frac{39}{2}\pi^2
-80\zeta_3
+\frac{29}{120}\pi^4
\right)
 + \order{\e}\Biggr],
\end{eqnarray}
which differs from $\calAtold$ in Eq.~(5.32) of Ref.~\cite{Gehrmann-DeRidder:2005btv}, starting from the rational part at $\order{1/\e}$. The integrated form of $\At$ using the original set of three-particle antenna functions, $\Xold{X}$, is given in Eq.~\eqref{eq:A40tintoldX3}, where the pole structure agrees precisely with $\calAtold$ in Eq.~(5.32) of Ref.~\cite{Gehrmann-DeRidder:2005btv} and differs only at $\order{\e^0}$.

To construct $\B$ we identify $b$ and $c$ as a quark-antiquark pair (that is of different flavour to the hard radiators $a$ and $d$).  
The resulting expression for $\B$ is included in the auxiliary materials of Ref.~\cite{paper2}. It is symmetric under line reversal, 
\begin{equation}
\B(i^h,j,k,l^h) = \B(l^h,k,j,i^h),
\end{equation}
and exchange of the quark-antiquark pair,
\begin{equation}
\B(i^h,j,k,l^h) = \B(i^h,k,j,l^h).
\end{equation}
Integrating over the antenna phase space, we find
\begin{eqnarray}
\label{eq:B40int}
\calB (s_{ijkl}) &=& S_{ijkl}^2 \Biggl [
-\frac{1}{12\e^3}
-\frac{7}{18\e^2}
+\frac{1}{\e} \left(
-\frac{407}{216}
+\frac{11}{72}\pi^2
\right)
\nonumber \\&& \hspace{2cm}
+ \left(
-\frac{5917}{648}
+\frac{145}{216}\pi^2
+\frac{67}{18}\zeta_3
\right)
+ \order{\e}\Biggr],
\end{eqnarray}
which differs from Eq.~(5.39) of Ref.~\cite{Gehrmann-DeRidder:2005btv}, starting from the rational part of $\order{\e^0}$.

The $\C$ antenna has no double-soft or single-soft limits. It is therefore considerably simpler than the other quark-antiquark antennae. By construction, the only limit it contains is when particles $b$, $c$ and $d$ are collinear -- the $\bar{q}q\bar{q}$ triple-collinear limit for identical quarks.  
The resulting expression for $\C$ is included in the auxiliary materials of Ref.~\cite{paper2}. It is symmetric under the exchange of the two identical antiquarks in the triple-collinear group, 
\begin{equation}
\C(i^h,j,k,l^h) = \C(i^h,l,k,j^h).
\end{equation}
This is as a direct result of the symmetry in the $\Pqqq$ splitting function. 
Integrating over the antenna phase space, we find
\begin{eqnarray}
\label{eq:C40int}
\calC (s_{ijkl}) &=& S_{ijkl}^2 \Biggl [
+\frac{1}{\e} \left(
-\frac{13}{32}
+\frac{1}{16}\pi^2
-\frac{1}{4}\zeta_3
\right)
\nonumber \\&& \hspace{2cm}
 + \left(
-\frac{73}{16}
+\frac{23}{96}\pi^2
+\frac{23}{8}\zeta_3
-\frac{1}{45}\pi^4
\right)
 + \order{\e}\Biggr],
\end{eqnarray}
which differs from $\calCold$ in Eq.~(5.44) of Ref.~\cite{Gehrmann-DeRidder:2005btv}, starting from $\order{\e^0}$.

\subsection{Quark-Gluon Antennae}
The antenna functions for quark-gluon parents have been systematically derived from an effective Lagrangian describing heavy neutralino decay~\cite{Gehrmann-DeRidder:2005svg}.  
There are three different configurations corresponding to the tree level processes $\widetilde{\chi} \to \widetilde{g}ggg$ (labelled $\Dold$) and $\widetilde{\chi} \to \widetilde{g}q\bar{q}g$ (with leading-colour and sub-leading colour antennae labelled $\Eold$ and $\Etold$ respectively).  
Since these were based on matrix elements, the $\Dold$ and $\Eold$ antennae did not strictly follow the design principles laid out in Section~\ref{sec:design-principles}.  
In particular, the antennae did not clearly specify which particles should be hard radiators and over-included unresolved limits that are not desirable. 
In Ref.~\cite{Gehrmann-DeRidder:2007foh}, work was done to divide both $\D$ and $\E$ into sub-antennae with better properties, however this yielded functions that were not analytically integrable. 
Here we derive antennae that contain specific limits using the algorithm.
As indicated in Table~\ref{tab:X40}, there are five antennae in total:
\begin{itemize}
\item Two types of antennae describe the $qggg$ system:
$\D(i_q^h,j_g, k_g, l_g^h)$ and \\ $\Dt(i_q^h,j_g, k_g, l_g^h)$.
In $\D$, the two unresolved gluons are colour-connected, while in $\Dt$ they are disconnected.  
In terms of the antennae of Ref.~\cite{Gehrmann-DeRidder:2005btv}, 
\begin{equation}
\Dold(i_q,j_g,k_g,l_g) \sim
  \D(i^h,j,k,l^h)
+ \D(i^h,l,k,j^h)
+ \Dt(i^h,j,l,k^h) \, .
\label{eq:Doldtonew}
\end{equation}
\item Three types of antennae describe the $q\bar{Q}{Q}g$ system:
$\Ea(i_q^h,j_{\bar{Q}},k_{Q},l_g^h)$, \\
$\Eb(i_q^h,j_g,k_{\bar{Q}},l_{Q}^h)$
and
$\Et(i_q^h,j_{\bar{Q}},k_g,l_{Q}^h)$.
At leading colour, two configurations are necessary: $\Ea$ in which the $Q\bar{Q}$ pair can be soft and the gluon is a hard radiator, and $\Eb$ where the gluon can be soft.
The soft-gluon singularities are therefore all contained in $\Eb$ but the triple-collinear $gQ\bar{Q}$ singularities are distributed between $\Ea$ and $\Eb$ according to Eq.~\eqref{eq:Pgqbq}.
At sub-leading colour, only one antenna is needed. 
These antennae are related to the antennae of Ref.~\cite{Gehrmann-DeRidder:2005btv} by,
\begin{align}
\Eold(i_q,j_{\bar Q},k_Q,l_g) &\sim
  \Ea(i^h,j,k,l^h)
+ \Eb(i^h,l,k,j^h) \\
\Etold(i_q,j_{\bar Q},k_Q,l_g) & \sim
  \Et(i^h,j,l,k^h).
\end{align}
\end{itemize}
To build the antennae using the algorithm, we simply identify the particles in the list of required limits given in Eq.~\eqref{eq:listX40}. 
For the $\D$ and $\Dt$ antennae, $a$ is a quark and $b$, $c$ and $d$ are gluons. 
For $\D$ the double-soft limit is $\Sgg$, while for $\Dt$ the double-soft limit is $\Spp$.
The colour-connected triple-gluon collinear limit is shared between the three antennae in Eq.~\eqref{eq:Doldtonew} according to the decomposition in Eq.~\eqref{eq:Pggg}. 
The three-particle antennae that appear in the single unresolved limits can be either of type $A_3^0$ or type $D_3^0$.

The resulting expressions for $\D$ and $\Dt$ are included in the auxiliary materials of Ref.~\cite{paper2}. $\D$ has no symmetries, while $\Dt$ is symmetric under exchange of the unresolved abelianised gluons,
\begin{equation}
\Dt(i^h,j,k,l^h) = \Dt(i^h,k,j,l^h).
\end{equation}
Note, in particular, that $\Dt(i^h,j,k,l^h)$ encapsulates the triple-collinear limits $\Pqpp(i^h,j,k)$ and $\Pggg(j,l^h,k)$, as given in Eq.~\eqref{eq:Pqpp} and Eq.~\eqref{eq:Pgpp} respectively.
Integrating over the antenna phase space, we find
\begin{eqnarray}
\label{eq:D40int}
\calD (s_{ijkl}) &=& S_{ijkl}^2 \Biggl [
+\frac{3}{4\e^4}
+\frac{71}{24\e^3}
+\frac{1}{\e^2} \left(
\frac{118}{9}
-\frac{13}{12}\pi^2
\right)
+\frac{1}{\e} \left(
\frac{11849}{216}
-\frac{35}{8}\pi^2
-\frac{35}{2}\zeta_3
\right)
\nonumber \\&& \hspace{2cm}
 + \left(
\frac{74369}{324}
-\frac{8579}{432}\pi^2
-\frac{5473}{72}\zeta_3
+\frac{9}{32}\pi^4
\right)
 + \order{\e}\Biggr], \\
\label{eq:D40tint}
\calDt (s_{ijkl}) &=& S_{ijkl}^2 \Biggl [
+\frac{1}{\e^4}
+\frac{10}{3\e^3}
+\frac{1}{\e^2} \left(
\frac{29}{2}
-\frac{3}{2}\pi^2
\right)
+\frac{1}{\e} \left(
\frac{26749}{432}
-5\pi^2
-\frac{83}{3}\zeta_3
\right)
\nonumber \\&& \hspace{2cm}
 + \left(
\frac{113227}{432}
-\frac{1045}{48}\pi^2
-\frac{818}{9}\zeta_3
+\frac{19}{120}\pi^4
\right)
 + \order{\e}\Biggr].
\end{eqnarray}
The combination $2 \calD + \calDt$ agrees with the pole structure for $\calDold$ (given in Eq.~(6.45) of Ref.~\cite{Gehrmann-DeRidder:2005btv}), up to the rational part at $\order{1/\e^2}$.  
This is because the $D_3^0$ antenna differs at $\order{\e^0}$ from the three-particle antenna, $\Xold{d}$, used in Ref.~\cite{Gehrmann-DeRidder:2005btv}.  
The integrated forms of $\D$ and $\Dt$ using the original set of three-particle antenna, given in Appendix~\ref{app:intX4oldX3}, restore the agreement with the pole structure of Eq.~(6.45) of Ref.~\cite{Gehrmann-DeRidder:2005btv} through to $\order{\e^0}$.

\begingroup
The expressions for $\Ea$ and $\Eb$ are given in the auxiliary materials of Ref.~\cite{paper2} and have no symmetries. 
Integrating over the antenna phase space yields
\begin{eqnarray}
\label{eq:E40aint}
\calEa (s_{ijkl}) &=& S_{ijkl}^2 \Biggl [
-\frac{1}{12\e^3}
-\frac{5}{12\e^2}
+\frac{1}{\e} \left(
-\frac{1463}{864}
+\frac{1}{8}\pi^2
\right)
\nonumber \\&& \hspace{2cm}
+ \left(
-\frac{38401}{5184}
+\frac{77}{108}\pi^2
+\frac{20}{9}\zeta_3
\right)
 + \order{\e}\Biggr], \\
\label{eq:E40bint}
\calEb (s_{ijkl}) &=& S_{ijkl}^2 \Biggl [
-\frac{1}{3\e^3}
-\frac{35}{24\e^2}
+\frac{1}{\e} \left(
-\frac{5537}{864}
+\frac{1}{2}\pi^2
\right)
\nonumber \\&& \hspace{2cm}
+ \left(
-\frac{47345}{1728}
+\frac{35}{16}\pi^2
+\frac{80}{9}\zeta_3
\right)
 + \order{\e}\Biggr].
\end{eqnarray}
The combination $\calEa + \calEb$ agrees with the pole structure of $\calEold$ (given in Eq.~(6.51) of Ref.~\cite{Gehrmann-DeRidder:2005btv}), up to the rational part at $\order{1/\e^2}$. 
This is because the $E_3^0$ antenna differs at $\order{\e^0}$ from the three-particle antenna, $\Xold{E}$, used in Ref.~\cite{Gehrmann-DeRidder:2005btv}. 
Using the original set of three-particle antenna functions leads to the integrated four-particle antennae listed in Appendix~\ref{app:intX4oldX3}, which restores the agreement with the pole structure of $\calEold$ in Eq.~(6.51) of Ref.~\cite{Gehrmann-DeRidder:2005btv} through to $\order{\e^0}$.
\endgroup

The sub-leading-colour antenna $\Et (i_q^h,j_{\bar{Q}},k_g,l_{Q}^h)$ only contains singularities associated with the $\bar{Q}gQ$ cluster -- namely the triple-collinear limit, the soft-gluon limit and the collinear limits $\bar{Q}g$ and $gQ$.  
These limits are independent of the particle type of the first hard radiator $a$. 
The expression for $\Et$ is included in the auxiliary materials of Ref.~\cite{paper2}. 
$\Et$ is symmetric under exchange of the quark-antiquark pair,
\begin{equation}
\Et(i^h,j,k,l^h) = \Et(i^h,l,k,j^h).
\end{equation}
Integrating over the antenna phase space, we find
\begin{eqnarray}
\label{eq:E40tint}
\calEt (s_{ijkl}) &=& S_{ijkl}^2 \Biggl [
-\frac{1}{6\e^3}
-\frac{13}{18\e^2}
+\frac{1}{\e} \left(
-\frac{80}{27}
+\frac{1}{4}\pi^2
\right)
\nonumber \\&& \hspace{2cm}
+ \left(
-\frac{7501}{648}
+\frac{13}{12}\pi^2
+\frac{40}{9}\zeta_3
\right)
 + \order{\e}\Biggr].
\end{eqnarray}
The pole structure for $\calEt$ agrees with $\calEtold$ (given in Eq.~(6.52) of Ref.~\cite{Gehrmann-DeRidder:2005btv}), up to the rational part at $\order{1/\e^2}$. 
This is because the $E_3^0$ antenna differs at $\order{\e^0}$ from the three-particle antenna, $\Xold{E}$, used in Ref.~\cite{Gehrmann-DeRidder:2005btv}. 
Using as input the original three-particle antenna, $\Xold{E}$, restores the agreement through to $\order{\e^0}$.

\subsection{Gluon-Gluon Antennae}
Antenna functions for gluon-gluon parents have been systematically derived from the effective Lagrangian describing Higgs boson decay into gluons~\cite{Gehrmann-DeRidder:2005alt}. There are four possibilities, 
$ H \to gggg$ (labelled $\Fold$),
$H \to gg Q\bar{Q}$ (with leading-colour and sub-leading colour antennae labelled $\Gold$ and $\Gtold$ respectively)
and 
$H \to q\bar q Q \bar Q$ (labelled $\Hold$). 
The $\Fold$ and $\Gold$ antennae also did not follow the design principles laid out in Section~\ref{sec:design-principles}, as they did not clearly specify which particles should be hard radiators and/or over-included unresolved limits that are not desirable. 
Ref.~\cite{NigelGlover:2010kwr} reorganised the $\Fold$ antenna into $F_{4, \,a}^{0,\text{OLD}}$ and $F_{4, \,b}^{0,\text{OLD}}$ sub-antennae that had better properties but at the cost of introducing composite denominators. 
Similar work was also carried out for the $\Gold$ antenna. 

Using our algorithm, we can build gluon-gluon antennae that do satisfy the ideal design principles. 
As shown in Table~\ref{tab:X40} there are six antennae in total:
\begin{itemize}
\item Two types of antenna describe the $gggg$ system: $\F(i_g^h,j_g,k_g,l_g^h)$ and \\ $\Ft(i_g^h,j_g,k_g,l_g^h)$. 
They are related to the antenna of Ref.~\cite{Gehrmann-DeRidder:2005btv} by,
\begin{align}
\label{eq:Foldtonew}
\Fold(i,j,k,l) &\sim
\F(i^h,j,k,l^h)
+ 3~{\rm cyclic~permutations} \nonumber \\
&+\Ft(i^h,j,l,k^h) 
+\Ft(l^h,i,k,j^h).
\end{align}
In $\F$ the two unresolved gluons are colour-connected, while in $\Ft$ they are disconnected. 
\item Three types of antenna describe the $g Q\bar{Q}g$ system:
$\Ga(i_g^h,j_Q,k_{\bar{Q}},l_g^h)$, \\
$\Gb(i_g^h,j_g,k_Q,l_{\bar{Q}}^h)$ and
$\Gt(i_g^h,j_Q,k_g,l_{\bar{Q}}^h)$. 
At leading colour, two configurations are necessary: $\Ga$ in which the $Q\bar{Q}$ pair can be soft and both gluons are hard radiators, and $\Gb$ where one of the gluons can be soft (and the other is a hard radiator).
The soft-gluon singularities are therefore all contained in $\Gb$ but the triple-collinear $gQ\bar{Q}$ singularities are distributed between $\Ga$ and $\Gb$. 
They are related to the antenna of Ref.~\cite{Gehrmann-DeRidder:2005btv} by,  
\begin{align}
\Gold(i_g,j_Q,k_{\bar{Q}},l_g) &\sim
\Ga(i^h,j,k,l^h) \nonumber \\
& + \Gb(l^h,i,j,k^h)
+ \Gb(i^h,l,k,j^h) \, .
\end{align}
At sub-leading colour, only one antenna, $\Gt (i_g^h,j_{\bar{Q}},k_g,l_Q^h)$,  is needed. There are no double-soft limits and only one triple-collinear limit describing the $\bar{Q} g Q$ cluster. 
$\Gt$ is related to the antenna of Ref.~\cite{Gehrmann-DeRidder:2005btv} by,  
\begin{equation}
\Gtold(i_g,j_Q,k_{\bar{Q}},l_g) \sim
  \Gt(i^h,j,l,k^h)
+ \Gt(l^h,k,i,j^h) \, .
\end{equation}
\item Finally, one gluon-gluon antenna is needed to describe the $q\bar q Q\bar{Q}$ final state, called $\H(i_{\bar{q}}^h,j_{q},k_{\bar{Q}},l_{Q}^h)$, which is directly related to the analogous antenna in Ref.~\cite{Gehrmann-DeRidder:2005btv},
\begin{equation}
\Hold(i_{\bar{q}},j_{q},k_{\bar{Q}},l_{Q}) \sim
\H(i^h,j,k,l^h) \, .
\end{equation}
Note that only different quark flavours need to be considered, since the identical-flavour contribution to this final state is finite.
\end{itemize}
As for the previous antennae, we simply identify the particles in the list of required limits given in Eq.~\eqref{eq:listX40}. 
For the $\F$ and $\Ft$ antennae, all particles are gluons. 
For $\F$ the double-soft limit is $\Sgg$, while for $\Ft$ the double-soft limit is $\Spp$. 
The assignment of the triple-gluon splitting function to $\F$ and $\Ft$ exactly parallels the division for $\D$ and $\Dt$ and the triple-gluon collinear limit is shared between the two antennae according to the decomposition in Eq.~\eqref{eq:Pggg}. 
The three-particle antennae that appear in the single unresolved limits are all of type $F_3^0$.

The resulting expressions for $\F$ and $\Ft$ are included in the auxiliary materials of Ref.~\cite{paper2}. Both $\F$ and $\Ft$ satisfy the line reversal property,
\begin{align}
\F(i^h,j,k,l^h)  &= \F(l^h,k,j,i^h) \, ,\\
\Ft(i^h,j,k,l^h) &= \Ft(l^h,k,j,i^h) \, ,
\end{align}
while $\Ft$ is symmetric under the exchange of the two unresolved gluons,
\begin{equation}
\Ft(i^h,j,k,l^h) = \Ft(i^h,k,j,l^h) \, .
\end{equation}
After integration over the final-final antenna phase space, Eq.~\eqref{eq:phijkFF}, we find the following infrared pole structure, 
\begin{eqnarray}
\label{eq:F40int}
\calF (s_{ijkl}) &=& S_{ijkl}^2 \Biggl [
+\frac{3}{4\e^4}
+\frac{77}{24\e^3}
+\frac{1}{\e^2} \left(
\frac{511}{36}
-\frac{13}{12}\pi^2
\right)
+\frac{1}{\e} \left(
\frac{50801}{864}
-\frac{671}{144}\pi^2
-\frac{69}{4}\zeta_3
\right)
\nonumber \\&& \hspace{1cm}
 + \left(
\frac{415927}{1728}
-\frac{9059}{432}\pi^2
-\frac{2819}{36}\zeta_3
+\frac{437}{1440}\pi^4
\right)
 + \order{\e}\Biggr],\\
\label{eq:F40tint}
\calFt (s_{ijkl}) &=& S_{ijkl}^2 \Biggl [
+\frac{1}{\e^4}
+\frac{11}{3\e^3}
+\frac{1}{\e^2} \left(
\frac{289}{18}
-\frac{3}{2}\pi^2
\right)
+\frac{1}{\e} \left(
\frac{30347}{432}
-\frac{11}{2}\pi^2
-\frac{86}{3}\zeta_3
\right)
\nonumber \\&& \hspace{2cm}
 + \left(
\frac{785743}{2592}
-\frac{193}{8}\pi^2
-\frac{916}{9}\zeta_3
+\frac{3}{40}\pi^4
\right)
 + \order{\e}\Biggr].
\end{eqnarray}
The combination $4 \calF + 2\calFt$ agrees with the pole structure for $\calFold$ (given in Eq.~(7.45) of Ref.~\cite{Gehrmann-DeRidder:2005btv}), up to the rational part at $\order{1/\e^2}$. 
This is because the $F_3^0$ antenna differs at $\order{\e^0}$ from the three-particle antenna, $\Xold{f}$, used in Ref.~\cite{Gehrmann-DeRidder:2005btv}. 
Using as input the original three-particle antenna, $\Xold{f}$, restores the agreement through to $\order{\e^0}$.

The resulting expressions for $\G$ and $\Gb$ are included in the auxiliary materials of Ref.~\cite{paper2}. 
$\Gb$ has no symmetries, while $\Ga$ satisfies the line reversal property
\begin{eqnarray}
\Ga(i^h,j,k,l^h) &=& \Ga(l^h,k,j,i^h).
\end{eqnarray}
After integration over the antenna phase space, we find
\begin{eqnarray}
\label{eq:G40aint}
\calGa (s_{ijkl}) &=& S_{ijkl}^2 \Biggl [
-\frac{1}{12\e^3}
-\frac{4}{9\e^2}
+\frac{1}{\e} \left(
-\frac{649}{432}
+\frac{7}{72}\pi^2
\right)
\nonumber \\&& \hspace{2cm}
+ \left(
-\frac{1637}{288}
+\frac{163}{216}\pi^2
+\frac{13}{18}\zeta_3
\right)
 + \order{\e}\Biggr],\\
\label{eq:G40bint}
\calGb (s_{ijkl}) &=& S_{ijkl}^2 \Biggl [
-\frac{1}{3\e^3}
-\frac{109}{72\e^2}
+\frac{1}{\e} \left(
-\frac{5741}{864}
+\frac{1}{2}\pi^2
\right)
\nonumber \\&& \hspace{2cm}
+ \left(
-\frac{146651}{5184}
+\frac{109}{48}\pi^2
+\frac{80}{9}\zeta_3
\right)
 + \order{\e}\Biggr].
\end{eqnarray}
The combination $ \calGa + 2\calGb$ agrees with the pole structure for $\calGold$ (given in Eq.~(7.52) of Ref.~\cite{Gehrmann-DeRidder:2005btv}), up to the rational part at $\order{1/\e^2}$. 
This is because the $G_3^0$ antenna differs at $\order{\e^0}$ from the three-particle antenna, $\Xold{G}$, used in Ref.~\cite{Gehrmann-DeRidder:2005btv}. 
Using as input the original three-particle antenna restores the agreement through to $\order{\e^0}$.

As for the $\Et (i_q^h,j_{\bar{Q}},k_g,l_{Q}^h)$ antenna, the sub-leading colour antenna $\Gt (i_g^h,j_{\bar{Q}},k_g,l_Q^h)$ only contains singularities associated with the $\bar{Q}gQ$ cluster. 
These are the triple-collinear limit, the soft-gluon limit and the collinear limits for $\bar{Q}g$ and $gQ$. 
These limits are independent of the particle type of the first hard radiator $a$. The expression for $\Gt$ is therefore the same as for $\Et$ and is included in the auxiliary materials of Ref.~\cite{paper2}. 
$\Gt$ is symmetric under exchange of the quark-antiquark pair,
\begin{equation}
\Gt(i^h,j,k,l^h) = \Gt(i^h,l,k,j^h).
\end{equation}
Integrating over the antenna phase space, we find
\begin{eqnarray}
\label{eq:G40tint}
\calGt (s_{ijkl}) &=& S_{ijkl}^2 \Biggl [
-\frac{1}{6\e^3}
-\frac{13}{18\e^2}
+\frac{1}{\e} \left(
-\frac{80}{27}
+\frac{1}{4}\pi^2
\right)
\nonumber \\&& \hspace{2cm}
+ \left(
-\frac{7501}{648}
+\frac{13}{12}\pi^2
+\frac{40}{9}\zeta_3
\right)
 + \order{\e}\Biggr].
\end{eqnarray}
The combination $2\calGt$ agrees with the pole structure for $\calGtold$ (given in Eq.~(7.52) of Ref.~\cite{Gehrmann-DeRidder:2005btv}), up to the rational part at $\order {1/\e^2} $. 
This is as expected because the $G_3^0$ antenna differs at $\order{\e^0}$ from the three-particle antenna, $\Xold{G}$, used in Ref.~\cite{Gehrmann-DeRidder:2005btv}. 
Using as input the original three-particle antenna restores the agreement through to $\order{\e^0}$.

The $\bar{q}q\bar{Q}Q$ antenna, $\H$, contains no double-soft, no triple-collinear and no single-soft limits.
It is composed entirely from the limits where $q$ and $\bar{q}$ and/or $Q$ and $\bar{Q}$ are collinear. 
$\H$ is symmetric under the exchange of $q$ and $\qb$ and/or $Q$ and $\Qb$, as well as the interchange of the quark pairs,
\begin{align}
&\H(i,j,k,l) = \H(j,i,k,l) = \H(i,j,l,k) = \H(j,i,l,k)\nonumber \\
=&\H(k,l,i,j) = \H(k,l,j,i) = \H(l,k,i,j) = \H(l,k,j,i).
\end{align}

The resulting expression for $\H$ is given by
\begin{equation}
\begin{split}
\label{eq:H40}
\H(i_{\qb},j_{q},k_{\Qb},l_{Q}) &=
\frac{1}{s_{ij}s_{k l}} 
+ \frac{2\left(s_{ik}s_{jl}+s_{il}s_{jk}\right)}{s_{ij}s_{kl}s_{ijkl}^2\ome^2} \\
&
- \frac{2 ((s_{ik}+s_{jl})(s_{il}+s_{jk}) + 2s_{ik}s_{jl} + 2
s_{il}s_{jk})}{s_{ij}s_{kl}s_{ijkl}^2\ome} \\
& -\frac{\left(s_{ijkl}+s_{ik}+s_{jk}+s_{il}+s_{jl}\right)}
{ s_{ij}s_{ijkl}^2}
\times
\frac{\left(s_{ik}s_{jl}+s_{il}s_{jk}\right)}{s_{ijk}s_{ijl}\ome} \\
& - \frac{\left(s_{ijkl}+s_{ik}+s_{jk}+s_{il}+s_{jl}\right)}
{s_{kl}s_{ijkl}^2}
\times
\frac{\left(s_{ik}s_{jl}+s_{il}s_{jk}\right)}{s_{jkl}s_{ikl}\ome }.
\end{split}
\end{equation}
Integrating over the antenna phase space, we find
\begin{eqnarray}
\label{eq:H40int}
\calH (s_{ijkl}) &=& S_{ijkl}^2 \Biggl [
+\frac{1}{9\e^2}
+\frac{1}{2\e}
+ \left(
\frac{283}{162}
-\frac{17}{108}\pi^2
\right)
 + \order{\e}\Biggr].
\end{eqnarray}
This agrees with the pole structure for $\calHold$ (given in Eq.~(7.59) of Ref.~\cite{Gehrmann-DeRidder:2005btv}), up to the rational part at $\order {1/\e}$. 
It is instructive to compare the antenna generated by our algorithm, Eq.~\eqref{eq:H40}, with the result given in Ref.~\cite{Gehrmann-DeRidder:2005btv} derived from matrix elements, 
\begin{align}
\label{eq:H40old}
\Hold(i_{\bar{q}},j_{q},k_{\bar{Q}},l_{Q}) &= 
\frac{1}{s_{ijkl}^2} \Bigg[ 
\frac{2\left(s_{ik}s_{j l}-s_{i l}s_{jk}\right)^2}{s_{ij}^2s_{k l}^2 \ome} + \frac{\left(s_{ik}+s_{il}+s_{jk}+s_{jl}\right)^2}{s_{ij}s_{k l}} + \frac{2}{\ome} \nonumber\\
&\hspace{1.5cm} 
- \frac{2 ((s_{ik}+s_{j l})(s_{i l}+s_{jk}) + 2s_{ik}s_{j l} + 2
s_{i l}s_{jk})}{s_{ij}s_{k l}\ome}\Bigg ] \, .
\end{align}
Comparing Eq.~\eqref{eq:H40} with Eq.~\eqref{eq:H40old}, we make the following observations:
\begin{itemize}
    \item The absence of double poles in $s_{ij}$ and $s_{k l}$ in Eq.~\eqref{eq:H40}.  
    This is because angular-averaged splitting functions were used to construct $\H$, thereby ensuring that azimuthal correlations are not present in the antenna.
    \item The presence of triple invariants in $s_{ijk}$ and $s_{jkl}$ in the denominator.  
    This is a consequence of the $\PCup$ projector, which defines the momentum fraction with respect to one of the other two particles in the antenna.    
    For example, for the $\PCup_{ij}$ projector, $\xj\omxj \to s_{il}s_{jk}/(s_{ijk}s_{ijl})$. 
    For most antennae, this is a natural choice and generates singular structures that are already present in the triple-collinear limit. 
    In this particular case however, there is no triple-collinear limit and this looks unnatural.

\end{itemize}

To demonstrate how the constructed antenna is affected by the choice of $X_3^0$ antenna in describing the single unresolved limits, Eq.~\eqref{eq:H40OLDX3} shows the expression generated by the algorithm for $\H$ using the original $\Xold{G}$ antenna for the single unresolved limits,
\newpage
\begin{eqnarray}
\H(i_{\qb},j_{q},k_{\Qb},l_{Q})[\Xold{G}] &=&
\frac{1}{s_{ij}s_{k l}} 
+ \frac{2\left(s_{ik}s_{jl}+s_{il}s_{jk}\right)}{s_{ij}s_{kl}s_{ijkl}^2\ome^2} \nonumber \\
&&
- \frac{2 ((s_{ik}+s_{jl})(s_{il}+s_{jk}) + 2s_{ik}s_{jl} + 2
s_{il}s_{jk})}{s_{ij}s_{kl}s_{ijkl}^2\ome} \nonumber \\
&& -\frac{\left(s_{ijkl}+s_{ik}+s_{jk}+s_{il}+s_{jl}\right)}
{s_{ij}s_{ijkl}^2} \nonumber \\
&&
  -\frac{\left(s_{ijkl}+s_{ik}+s_{jk}+s_{il}+s_{jl}\right)}
{s_{kl}s_{ijkl}^2}  .
\label{eq:H40OLDX3}
\end{eqnarray}
We see that the double unresolved contributions in the first two lines of Eq.~\eqref{eq:H40OLDX3} are exactly the same as the double unresolved contribution in Eq.~\eqref{eq:H40}.  
The single unresolved contributions (third and fourth lines) however are different because of the choice of single-real antenna.
In the $i||j$ limit, the difference between Eq.~\eqref{eq:H40OLDX3} and Eq.~\eqref{eq:H40}  is proportional to
\begin{equation}
    \frac{1}{s_{ij}} \Pqq(\xj) \left( \Xold{G}((i+j),k,l)-G_3^{0}((i+j)^h,k,l^h)\right)
\end{equation}
with
\begin{equation}
\label{eq:G3diff}
    \left( \Xold{G}((i+j),k,l)-G_3^{0}((i+j)^h,k,l^h)\right) \sim 
    -\frac{\left(s_{ijkl}+s_{ik}+s_{jk}+s_{il}+s_{jl}\right)}
{s_{ijkl}^2}
\end{equation} 
and
\begin{equation}
    \Pqq(x_j) = 1 - \frac{2\xj\omxj}{\ome} \stackrel{\PCup_{ij}}{\longrightarrow} 1 - \frac{\left(s_{ik}s_{jl}+s_{il}s_{jk}\right)}{s_{ijk}s_{ijl}\ome}.
\end{equation}
The dependence on the momentum fraction $\xj$ in Eq.~\eqref{eq:H40} is precisely cancelled, leading to the absence of triple invariants in the denominator of Eq.~\eqref{eq:H40OLDX3}.

\begin{figure}[t]
    \centering
    \includegraphics[width=0.49\textwidth,page=1]{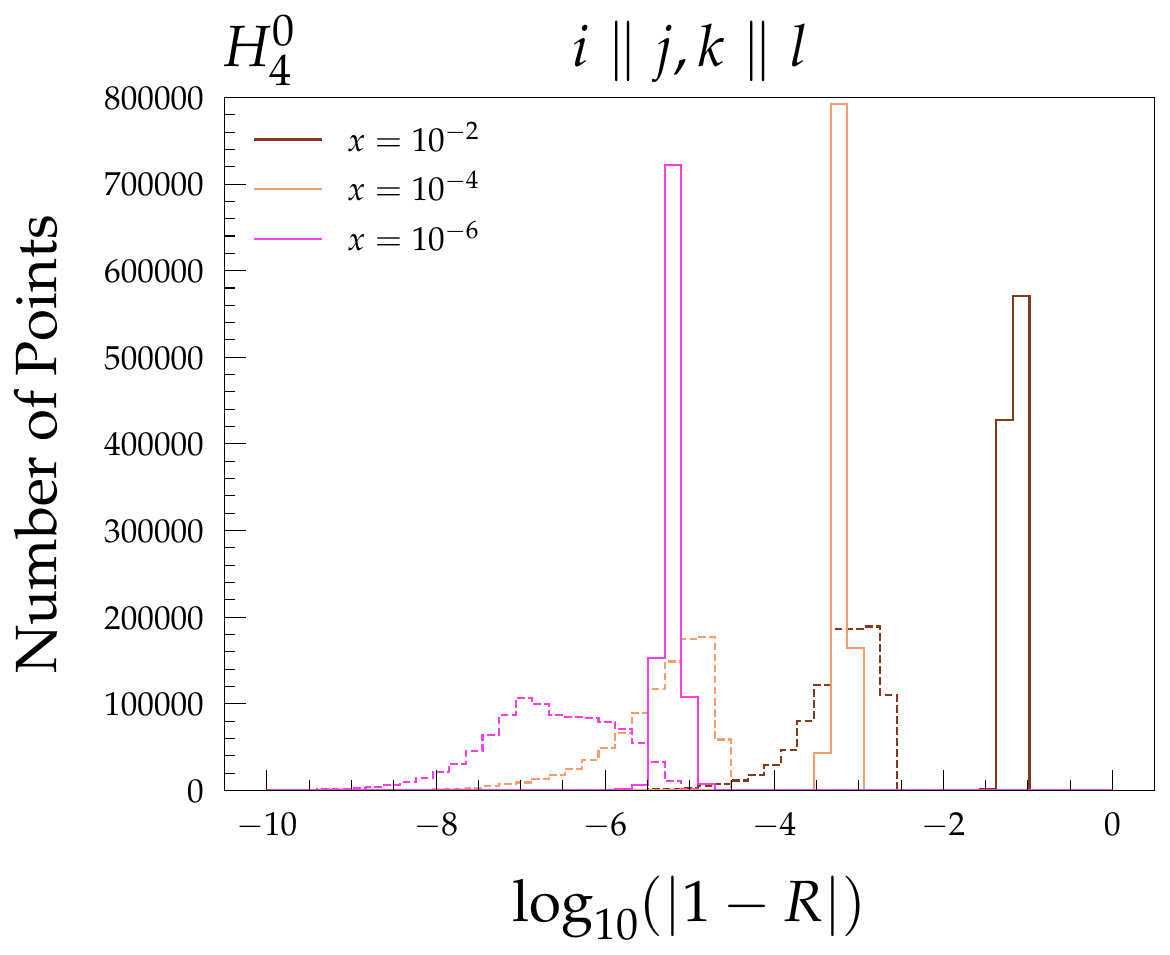}
    \includegraphics[width=0.49\textwidth,page=2]{H40n.pdf}
    \caption{Numerical tests of the new $\H$ against $\Hold$ in all relevant singular limits. Solid lines represent tests of $\H$ constructed with the new $G_3^0$ antennae and dashed lines represent tests with $\H$ constructed with $\Xold{G}$. For three different values of the scaling parameter $x$, the relative agreement of the ratio $R = \H/\Hold$ with $1$ is shown on a logarithmic axis.}
    \label{fig:spiketests_H40}
\end{figure}

In Fig.~\ref{fig:spiketests_H40}, we show numerical tests of $\H$ against $\Hold$ in the relevant double unresolved and single unresolved limits. 
The procedure is analogous to the one for Fig.~\ref{fig:spiketests_A40} and we show both versions of $\H$: the one constructed using the new form for $G_3^0$ in Eq.~\eqref{eq:H40} and the one using the original  $\Xold{G}$ antenna function in Eq.~\eqref{eq:H40OLDX3}.

In the double-collinear limit, both versions of $\H$ agree well with $\Hold$; as $x$ decreases they both describe the double-collinear limit increasingly correctly. 
We observe that the version constructed using the new $G_3^0$ antenna produces much sharper peaks at slightly larger values of $\log(|1-R|)$ compared to that based on the $\Xold{G}$ antenna. 

In the single unresolved limits, we see much bigger differences; for the $\H$ using the $\Xold{G}$ antenna, there is good agreement, while there is very poor agreement for the $\H$ using the new $G_3^0$ antenna.  
This is as expected!  In the single unresolved region, the $\Hold$ antenna function, which is based on the $H \to q\qb Q\Qb$ matrix element behaves as,
\begin{align}
    \Hold \sim &M_4^0(i_{\qb},j_{q},k_{\Qb},l_{Q}) \stackrel{i ||j}{\longrightarrow} \frac{1}{s_{ij}} \Pqq(\xj) M_3^0((i+j)_g,k_{\Qb},l_{Q}),
\end{align}
while by construction, the new $\H $ antenna behaves as,
\begin{align}
    \H \stackrel{i ||j}{\longrightarrow} \frac{1}{s_{ij}} \Pqq(\xj) G_3^0((i+j)^h_g,k_{\Qb},l_{Q}).
\end{align}
Since $\Xold{G}$ is constructed from the three particle matrix element, $M_3^0((i+j)_g,k_{Q},l_{\Qb})$, the $\H$ antenna based on $\Xold{G}$ is guaranteed to have the correct limit. 
However, the finite $\order{\e^0}$ differences between $G_3^{0}$ and $\Xold{G}$ (see Eq.~\eqref{eq:G3diff}) are multiplied by $1/s_{ij}$ and therefore lead to a different single unresolved limit for the $\H$ antenna using $G_3^{0}$. 
In fact, although these differences are sub-leading in the double unresolved region, they are responsible for the small numerical differences observed there.

Note that we do not require that the $\X$ antenna correctly describes the single unresolved region; its role is to correctly describe the double unresolved limits.  
The single unresolved limits are always correctly described by other subtraction terms. 
As a result, the single unresolved limits introduced in the $X_4^0$ must be subtracted away. 
This leads to groups of subtraction terms like,
\begin{eqnarray}
\label{eq:H4subtraction}
   \H(i^h,j,k,l^h) - E_3^0(k^h,j,i^h)G_3^0(I^h,K,l^h)
        - E_3^0(j^h,k,l^h) G_3^0(L^h,J,i^h),
\end{eqnarray} 
where we use antenna momentum mappings $(k,j,i) \to (K,I)$ and $(j,k,l) \to (J,L)$ for the two iterated terms respectively. 
Provided that the same $G_3^0$ is used for both the construction of $\H$ and in the analogues of Eq.~\eqref{eq:H4subtraction}, the single unresolved limits will be correctly described in the wider application of the subtraction scheme. 

Additionally, we summarise the symmetry properties of the $\X$ constructed in this chapter in Table~\ref{tab:X40sym}. 

\begin{table}[t]
\centering
\begin{tabular}{ccc}
$\X (1^h,2,3,4^h)$ & Symmetries \\
\hline
 $\A(1^h,2,3,4^h)$ & $(1,2,3,4) \leftrightarrow (4,3,2,1)$ \\
 $\At(1^h,2,3,4^h)$ & $(1,2,3,4) \leftrightarrow (4,3,2,1)$ and $(1,2,3,4) \leftrightarrow (1,3,2,4)$\\
 $\B(1^h,2,3,4^h)$ & $(1,2,3,4) \leftrightarrow (4,3,2,1)$ and $(1,2,3,4) \leftrightarrow (1,3,2,4)$ \\
 $\C(1^h,2,3,4^h)$ & $(1,2,3,4) \leftrightarrow (1,4,3,2)$ \\
$\D(1^h,2,3,4^h)$ & None  \\
 $\Dt(1^h,2,3,4^h)$ & $(1,2,3,4) \leftrightarrow (1,3,2,4)$   \\
$\Ea(1^h,2,3,4^h)$ & None  \\ 
 $\Eb(1^h,2,3,4^h)$ & None\\ 
 $\Et(1^h,2,3,4^h)$ & $(1,2,3,4) \leftrightarrow (1,4,3,2)$ \\ 
$\F(1^h,2,3,4^h)$ & $(1,2,3,4) \leftrightarrow (4,3,2,1)$   \\
$\Ft(1^h,2,3,4^h)$ & $(1,2,3,4) \leftrightarrow (4,3,2,1)$ and $(1,2,3,4) \leftrightarrow (1,3,2,4)$ \\
$\Ga(1^h,2,3,4^h)$ & $(1,2,3,4) \leftrightarrow (4,3,2,1)$ \\ 
 $\Gb(1^h,2,3,4^h)$ & None \\ 
 $\Gt(1^h,2,3,4^h)$ & $(1,2,3,4) \leftrightarrow (1,4,3,2)$\\ 
$\H(1^h,2,3,4^h)$ & $(1,2,3,4) \leftrightarrow (2,1,3,4)$ and $(1,2,3,4) \leftrightarrow (1,2,3,4)$ \\
& and $(1,2,3,4) \leftrightarrow (3,4,1,2)$ \\
\end{tabular}
\caption{Symmetries present for each $\X (1^h,2,3,4^h)$ antenna.}
\label{tab:X40sym}
\end{table}

\section{Summary and Outlook}
\label{sec:outlook1}
We have proposed a general algorithm for the construction of real-radiation antenna functions directly from their desired unresolved limits.
The technique makes use of an iterative procedure to remove overlaps between different singular factors that are subsequently projected into the full phase space.
As the technique produces only denominators that match physical propagators, all antenna functions can straightforwardly be integrated analytically, which is a cornerstone of the antenna-subtraction method.

We have implemented the algorithm in an automated framework for the construction of antenna functions and explicitly demonstrated that our technique can be used for single-real and double-real radiation antenna functions relevant to NLO and NNLO calculations, respectively.
In particular, we presented a full set of single-real and double-real tree-level antenna functions.
All of the idealised antenna functions we have presented here have been checked both analytically and numerically against the respective singular factors. 
As another strong validation, we have confirmed the correctness of their pole structure explicitly on the integrated level.
While the single-real antenna functions differ from previously used ones only in finite pieces, a residual dependence on the choice of single-real antennae is left in the construction of double-real antenna functions. 
This ambiguity is reflected in a difference in sub-leading poles starting from $1/\e^2$, while the deeper $1/\e^4$ and $1/\e^3$ poles correspond to the universal double-soft and triple-collinear behaviour. 
This is understood and poses no issues in application to the antenna-subtraction scheme when used with the idealised single-real antenna functions.

This work marks the first step towards a refined antenna-subtraction framework, as for the first time we have calculated a full set of double-real antenna functions that correspond to true antennae, meaning that they consist of exactly two hard radiators and contain no spurious singularities.
These features are vital for the subtraction scheme, as they eliminate the need for auxiliary subtraction terms to remove such spurious singularities.

Another avenue for future work might include the introduction of antenna functions with azimuthal correlations.  
Here the main issue is matching the azimuthal correlations with the process-specific matrix elements under consideration.  
For this reason, in this chapter we removed the azimuthal correlations from the antennae by using spin-averaged limits. 
In principle the azimuthal terms can be directly removed from the matrix elements by using pairs (or multiple pairs) of correlated phase-space points.  
Mass effects could also be included.  
The unresolved limits are known, see Refs.~\cite{Catani:2000ef,Keller:1998tf,Dhani:2023uxu,Craft:2023aew}, however the integration of the antennae over the massive phase space is more involved.

To build a complete NNLO subtraction scheme, this work needs to be supplemented by an equivalent construction of one-loop single-real antenna functions that serve the purpose of real-virtual subtraction terms. 
This requires additional manipulation of explicit $\e$-poles and hypergeometric functions, alongside the single-real unresolved radiation. 
We leave this, as well as the description of a refined antenna-subtraction framework, to Chapter~\ref{chapter:paper3}.

The algorithm presented here can also be used to construct idealised real-radiation antennae in the IF and II configurations, since these limits are also known. 
This will be necessary for calculations involving initial coloured radiation, like in $pp$ and $ep$ colliders. 
The $X_3^0$ for the IF and II configurations have already been constructed in Ref.~\cite{Fox:2023bma}. Construction of the NNLO idealised antennae for the IF and II configurations is left to future work. 

While we have focused on the application to single-real and double-real antenna functions, we wish to emphasise that the method can, in the future, be applied to build antenna functions for triple-real radiation as well, given that all singular factors with three unresolved particles are known~\cite{Catani:2019nqv,DelDuca:1999iql,DelDuca:2019ggv,DelDuca:2020vst,DelDuca:2022noh}.
This will form a substantial contribution to enable fully-differential calculations at N$^3$LO, where triple-real antenna functions can be used to subtract triple-unresolved singularities.

\chapter{Constructing Idealised Antenna Functions with Real and Virtual Radiation}
\label{chapter:paper3}

\section{Introduction}

As remarked earlier, in the antenna subtraction scheme, antenna functions are used to subtract specific sets of unresolved singularities, so that a typical subtraction term for a matrix element $M_m^L$ has the form
\begin{equation} 
    X_{n+2}^\ell(i_1^h,i_3,\ldots,i_{n+2},i_2^h) 
    M_{m-n}^{L-\ell}(\ldots,I_1^h,I_2^h,\ldots) \, ,
\end{equation}
where $X_{n+2}^\ell$ represents an $\ell$-loop, $(n+2)$-particle antenna,
$i_1^h$ and $i_2^h$ represent the hard radiators, and $i_3$ to $i_{n+2}$ denote the $n$ unresolved particles.   
As the hard radiators may either be in the initial or in the final state, final-final (FF), initial-final (IF) and initial-initial (II) configurations need to be considered in general. 
$M_{m-n}^{L-\ell}$ is the reduced matrix element, with $n$ fewer final-state particles, $\ell$ fewer loops and where $I_1^h$ and $I_2^h$ represent the particles obtained through an appropriate antenna mapping,
\begin{align}
\{ p_{i_1},p_{i_3},\ldots,p_{i_{n+2}},p_{i_2} \} \mapsto \{ p_{I_1}, p_{I_2} \}  
\end{align}
with $p_i^{\mu}$ representing the four-momentum of particle $i$. At NLO antennae have $n=1$ and $\ell=0$, at NNLO one needs antennae with $n=2,~\ell = 0$ and with $n=1,~\ell=1$, while at \NthreeLO, one needs antennae with $n=3,~\ell = 0$, with $n=2,~\ell = 1$ and with $n=1,~\ell = 2$. 
In the original formulation of the antenna scheme, the antennae were based on matrix elements describing radiation from processes with two coloured particles: $\gamma^* \to q\bar{q}$, $\tilde{\chi} \to \tilde{g}g$ and $H \to gg$, covering the cases where the coloured particles are massless quarks and gluons.  
The corresponding $X_4^0$, $X_3^1$ and $X_3^0$ antennae are therefore perfect subtraction terms for the NNLO contributions to processes with two coloured particles.  
It was straightforward to utilise these matrix-element-based antennae for processes with three coloured particles, such as $e^+e^- \to 3$~jets, $pp \to V$+jet, $pp \to H$+jet and for the leading colour contributions to four coloured particle processes like $pp \to 2$~jets. 
Pushing to the next step, the full colour $pp \to 2$~jets required significant additional work~\cite{Chen:2022clm}. 
Going beyond the current state of the art with the matrix-element-based antenna approach is a formidable task. 
This is because the complexity associated with the subtraction terms becomes increasingly challenging as the particle multiplicity grows.  
This complexity stems from two primary reasons.

Firstly, the double-real-radiation antenna functions obtained from matrix elements do not always indicate which particles act as the hard radiators. 
This is particularly the case for antennae involving gluons. 
To address this issue, sub-antenna functions are introduced. 
However, constructing these sub-antenna functions at NNLO is an arduous task and often involves introducing unphysical denominators that complicate the analytic integration of the subtraction term. 
Additionally, analytic integrals are usually known only for the complete antenna functions. 
As a result, the assembly of antenna-subtraction terms requires careful manipulation to ensure that the sub-antenna functions combine appropriately to form the full antenna functions before integration.

Secondly, NNLO antenna functions can exhibit spurious limits that need to be eliminated through explicit counterterms. 
However, these counterterms can introduce further spurious limits themselves. 
Consequently, this can initiate a complex chain of interdependent subtraction terms that do not necessarily reflect the actual singularity structure of the underlying process.

Both of these issues must be addressed in order to facilitate higher-multiplicity calculations at NNLO. 
Additionally, the same issues will be present at \NthreeLO to a much greater degree. 
In Chapter~\ref{chapter:paper2} we addressed these issues in the construction of idealised real-radiation antennae. 
Specifically, we introduced a general algorithm for building real-radiation antenna functions directly from a specified set of infrared limits with a uniform template, in a way that simplifies the construction of subtraction terms in general, while still being straightforwardly analytically integrable. 
We then applied it to the case of single-real and double-real radiation, required for NLO and NNLO calculations. 
The technique makes use of an iterative procedure to remove overlaps between different singular factors that are subsequently projected into the full phase space. 
As the technique produces only denominators that match physical propagators, all antenna functions could straightforwardly be integrated analytically, which is a cornerstone of the antenna-subtraction method. 

In this chapter, we extend the general algorithm to the construction of antennae with $\ell \neq 0$. 
Unlike in the solely real-radiation case, the mixed real and virtual antenna functions contain both explicit and implicit singularities. 
To illustrate the algorithm, we construct the real-virtual antennae $(n=1,~\ell=1)$ explicitly. 
The real-virtual antenna functions are built directly from the relevant one-loop limits, properly accounting for the overlap between different limits. 
The universal factorisation properties of multi-particle loop matrix elements, when one or more particles are unresolved, have been well studied in the literature~\cite{Bern:1994zx,Bern:1998sc,Kosower:1999rx,Bern:1999ry} and serve as an input to the algorithm. 

In addition to building a full set of idealised $X_3^1$ antennae with both hard radiators in the final state, we demonstrate that the idealised antenna functions (along with the $X_3^0$ and $\X$ of Chapter~\ref{chapter:paper2}) form a complete NNLO subtraction scheme in which the subtraction terms cancel the explicit singularities in the one- and two-loop matrix elements, without leftover infrared singularities hiding in the matrix elements (either by under-counting or over-counting). 
This means that the idealised antenna functions have to satisfy particular constraints. 
First, the cancellation of poles at the real-virtual level means that the explicit poles in the $X_3^1$ antenna have to cancel against other real-virtual subtraction terms.  
In the antenna scheme, these explicit poles are proportional to $X_3^0$ antennae. 
Therefore, the $X_3^1$ must have a particular pole structure multiplying an $X_3^0$ antenna function. 
At the double-virtual level, the combinations of integrated antennae coming from the double-real and real-virtual levels must match the explicit pole structure of the two-loop matrix elements.  
In the antenna scheme, this is encoded through a combination of the $\J{2}$ and $\J{1}$ dipole operators in colour space~\cite{Currie:2013vh}. 
Provided that the pole structure from the relevant combination of $\J{2}$ and $\J{1}$ is unchanged, the subtraction terms will cancel the explicit poles in the two-loop matrix elements.  

The current approach to automation of antenna subtraction~\cite{Chen:2022ktf,Gehrmann:2023dxm} involves a reformulation of the colour-ordered antenna subtraction technique in colour space. 
This method, known as `colourful antenna subtraction', offers a systematic way to construct antenna subtraction terms by working downwards from the most virtual layer, rather than starting from the maximally real layer and working up. 
By translating infrared poles of virtual corrections captured by $\J{2}$ and $\J{1}$ into real-radiation dipole insertions in colour space, the method efficiently constitutes subtraction terms for single-real radiation up to one-loop level and for double-real radiation at the tree level. 
One of the key advantages of this approach is the avoidance of directly handling the divergent behaviour of real-emission corrections. This feature represents a significant simplification at NNLO. 
The double-real subtraction term can be obtained as the final step of a fully automatable procedure, eliminating the need to deal with the involved infrared structure of double-real radiation matrix elements. 
The completion of a consistent set of idealised antenna functions for (double-)real and real-virtual radiation presented here will further reduce the complexity of the subtraction terms, because they avoid the need to subtract spurious limits and therefore reduce the computational overhead associated with precision calculations.

The chapter is structured as follows. We outline the design principles for constructing general $X_{n+2}^\ell$ antenna functions in Section~\ref{sec:principles} as well as the principles for matching to the other elements of an antenna-subtraction scheme. 
We describe the general construction algorithm in Section~\ref{sec:algorithm2} and give the specific details for the construction of final-final $X_3^1$ in Section~\ref{sec:X31construction}. 
Using the previous sections, we illustrate the algorithm by explicitly constructing a full set of $X_3^1$ real-virtual antenna functions for hard radiators in the final state in Section~\ref{sec:X31}. 
Finally, we define the $\J{2}$ and $\J{1}$ operators in this NNLO antenna-subtraction scheme (out of the idealised $\{X_3^0,\X,X_3^1\}$) and compare their pole structure to the generic double-virtual pole structures in Section~\ref{sec:J22}. 
This demonstrates that the idealised subtraction terms will cancel the explicit poles in the two-loop matrix elements and form a complete NNLO subtraction scheme. 
We summarise the chapter in Section~\ref{sec:outlook2}. 

\section{Design Principles}
\label{sec:principles}

Within the antenna-subtraction framework, subtraction terms are constructed using antenna functions that describe the unresolved partonic radiation (both soft and collinear) emitted from a pair of hard radiator partons. The construction of an antenna-subtraction term typically involves the following elements:
\begin{itemize}
\item antennae composed of two hard radiators that accurately capture the infrared singularities arising from the emission of $n$ unresolved partons;
\item an on-shell momentum mapping that ensures that the invariant mass of the antenna is preserved while producing the on-shell momenta that appear in the ``reduced'' matrix element; and
\item a	colour factor associated with the specific process and antenna.
\end{itemize}
The latter two items on this list have been solved for general processes, while the first is subject of this thesis and Refs.~\cite{paper2,paper3}.

In the following, we will describe the design principles we impose upon a general idealised $X_{n+2}^\ell$ antenna function, with at least one loop. 
As opposed to the $\ell=0$ case, antenna functions with additional virtual elements contain explicit poles in the dimensional-regularisation parameter $\e$. 
We therefore impose two different sets of design principles: the generic design principles discussed in Section~\ref{sec:generalprinciples} and the antenna-scheme-dependent design principles discussed in Section~\ref{sec:matching}.
The former principles ensure that the antenna function has the correct infrared limits but does not fix these unambiguously. 
This ambiguity is resolved by the latter principles which
match the explicit singularity structure of the idealised antenna functions onto a specific antenna-subtraction scheme.

\subsection{Generic Design Principles}
\label{sec:generalprinciples}

The generic design principles outlined in Chapter~\ref{chapter:paper2} are sufficient to ensure that the antenna has the correct infrared limits.   
Specifically, we impose the following requirements:
\begin{enumerate}
\item each antenna function has exactly two hard particles (``radiators'') which cannot become unresolved;
\item each antenna function captures all (multi-)soft limits of its unresolved particles;
\item where appropriate (multi-)collinear and mixed soft and collinear limits are decomposed over ``neighbouring'' antennae;
\item antenna functions do not contain any spurious (unphysical) limits;
\item antenna functions only contain singular factors corresponding to physical propagators; and
\item where appropriate, antenna functions obey physical symmetry relations (such as line reversal).
\end{enumerate}
As mentioned earlier, the original NNLO antenna functions derived in Refs.~\cite{Gehrmann-DeRidder:2005svg,Gehrmann-DeRidder:2005alt,Gehrmann-DeRidder:2005btv} do not obey all of these requirements, as they typically violate (some of) these principles. 
This is particularly the case for quark-gluon or gluon-gluon antennae because the matrix elements they are derived from will inevitably have a divergent limit when one of the gluonic radiators becomes soft (thereby violating principle 1). 

These principles will form the core of the algorithm for constructing  $X_{n+2}^\ell$ antennae with the desired infrared limits.

\subsection{Antenna-scheme-dependent Design Principles}
\label{sec:matching}

The generic principles are sufficient to produce compact analytic expressions that correctly capture the unresolved behaviour of $\ell$-loop matrix elements in the (multi-)soft and (multi-)collinear limits. 
Unlike the $\ell=0$ case, these unresolved limits have explicit singularities and therefore the $X_{n+2}^\ell$ antennae constructed from them will also carry explicit $\e$-poles.  

However, it is straightforward to find combinations of terms that contain explicit singularities but which do not contribute in any of the unresolved regions.  
Such terms can be added to $X_{n+2}^\ell$ without violating any of the generic principles.  
However, doing so will clearly change the explicit pole structure.  
This means that the generic principles alone lead to an inherent ambiguity in defining the $X_{n+2}^\ell$ antennae.

If one wishes to design a full subtraction scheme, the real-virtual antennae must {\bf both} have the correct unresolved limits {\bf and} have explicit pole structures of the correct form that cancels against other terms in the subtraction scheme.  
Therefore, we need to resolve the ambiguity in the explicit $\e$ singularities by {\bf matching} onto a set of {\bf target} $\e$-pole structures that ensure that the subtraction terms in each multiplicity layer (a) correctly describe the unresolved limits of the matrix elements and (b) precisely cancel the $\e$ singularities of the matrix elements.

To match onto a particular antenna-subtraction scheme, we therefore introduce one further principle: 
\begin{enumerate}
\item[7.] where appropriate, combinations of terms that are not singular in any unresolved region can be added to match onto ``target poles'', $T(i_1^h,i_3,...,i_{n+2},i_2^h)$.
\end{enumerate}
To illustrate this principle for the case of the $X_3^1$, ``target poles'' $T(i^h,j,k^h)$ take the following schematic form within the NNLO antenna-subtraction scheme~\cite{Gehrmann-DeRidder:2005btv}:
\begin{equation}
    T(i^h,j,k^h) = \frac{1}{\e^2} \left( \sum_{s} \left(\frac{s}{\mu^2}\right)^{-\e} \right) X_3^0(i^h,j,k^h).
\end{equation}
In order to match onto such ``target poles'', we are free to add certain combinations of terms. An example of such a combination of terms is 
\begin{equation}
    \label{eq:Tpoleseg}
    \frac{1}{\e^2}\mu^{2\e}\left( s_{ik}^{-\e} + s_{ijk}^{-\e} - (s_{ij}+s_{ik})^{-\e} - (s_{ik}+s_{jk})^{-\e}\right) X_3^0 (i^h,j,k^h),
\end{equation}
which is not divergent in any unresolved limit. The combination of bracketed terms suppresses the singular behaviour of $A_3^0$ in the soft $j$, collinear $ij$ and collinear $jk$ regions. 
Terms like those in Eq.~\eqref{eq:Tpoleseg} therefore do not affect the unresolved behaviour of the $X_3^1$, if added in a second stage.
However, adding such terms clearly affects the explicit poles in the $X_3^1$ antenna as can be seen from the expansion in $\e$,
\begin{equation}
 \frac{1}{\e}\ln\left(1+\frac{s_{ij}s_{jk}}{s_{ijk}s_{ik}}\right) X_3^0 (i^h,j,k^h) +\order{\e^0}.
\end{equation}
This allows us to match the pole structure of the $X_3^1$ antenna to the other subtraction terms in a way that cancels the explicit poles at the real-virtual level.

These seven principles are sufficient to devise an algorithm for constructing a general $X_{n+2}^\ell$ antenna function and here we will apply it to the construction of $X_3^1$ antenna functions with final-final kinematics. 
We will build the $X_3^1$ antenna functions from the infrared limits and match them to the NNLO antenna-subtraction scheme. 
The idealised $X_3^1$ antenna functions form the final ingredients for improved final-final antenna-subtraction at NNLO (along with the results of Chapter~\ref{chapter:paper2}). 
To test for the consistency of these ingredients, one has to integrate the real-virtual antennae over the antenna phase space and combine all the various integrated implicit singularities to cancel the explicit singularities of the two-loop matrix elements. 
This is detailed in full in Section~\ref{sec:J22}.

\section{The Algorithm}
\label{sec:algorithm2}

In Chapter~\ref{chapter:paper2}, we proposed a general algorithm to build (multiple-)real radiation antenna functions at tree-level. 
In this chapter (and originally Ref.~\cite{paper3}), we extend this algorithm to the construction of $X_{n+2}^\ell$ antenna functions, where $\ell \neq 0$.

Unlike the algorithm for real-radiation antenna functions, the algorithm for $X_{n+2}^\ell$ antenna functions has two distinct stages:
\begin{description}
    \item[Stage 1.] In this step we ensure that the antenna function has the correct infrared singular limits.
    This stage closely follows the algorithm for real-radiation antennae in Chapter~\ref{chapter:paper2}.
    We systematically start from the most singular limit and build the list of target functions, $\{L_i\}$, from relevant (multi-)soft, (multi-)collinear and soft-collinear limits. 
    
    As in Chapter~\ref{chapter:paper2}, we define a down- ($\PPdown_i$) and up-projector ($\PPup_i$) for each unresolved limit ($L_i$) to be included. A down-projector $\PPdown_i$ maps the invariants of the full phase space to the relevant subspace. 
    An associated up-projector $\PPup_i$ restores the full phase space by re-expressing all variables valid in the sub-space in terms of invariants valid in the full phase space. 
    It is to be emphasised that down-projectors $\PPdown_i$ and up-projectors $\PPup_i$ are typically not inverse to each other, as down-projectors destroy information about less-singular and finite pieces.

    The down-projectors are necessary to identify the overlapping region between the antenna function developed so far and the target function associated with the unresolved limit under consideration. 
    Conversely, up-projectors express the argument in terms of antenna invariants. Furthermore, through careful selection of the up-projectors, the antenna function can be exclusively represented using invariants corresponding to physical propagators.

    The set of target functions provides a clear definition of the antenna function's behaviour in all unresolved limits specific to the particular antenna being considered. 
    In each unresolved limit, the antenna function must approach the corresponding target function to accurately capture the singular behaviour exhibited by the squared matrix element. 
    Additionally, the antenna function must remain finite in all limits not explicitly described by a target function. 
    This crucial aspect guarantees the absence of spurious singularities (unlike antenna functions extracted directly from physical matrix elements).

    As explained in Chapter~\ref{chapter:paper2}, the algorithm, which ensures the above characteristics and meets the generic design principles, can be written as 
    \begin{equation}
        \label{eq:algorithm2}
    \begin{split}
    X^\ell_{n+2;1} &= \PPup_1 L_1 \, , \\
    X^\ell_{n+2;2} &= X^\ell_{n+2;1} + \PPup_2 (L_2 - \PPdown_2 X^\ell_{n+2;1}) \, ,\\
    & \vdots \\
    X^\ell_{n+2;N} &= X^\ell_{n+2;N-1} + \PPup_N (L_N-\PPdown_N X^\ell_{n+2;N-1}) \, ,
    \end{split}
    \end{equation}
    where $X^\ell_{n+2;N}$ is the output of {\bf Stage 1}. 

    \item[Stage 2.] The output of {\bf Stage 1} guarantees that $X^\ell_{n+2;N}$ has the chosen unresolved limits $\{L_i\}$.
    However, as discussed above this does not uniquely determine the mixed real-virtual antenna since one can construct a term which contains poles in $\e$ but does not contribute in any of the unresolved limits (which we will denote by $\FinPoles$). One is therefore at liberty to define different antenna-subtraction schemes that differ by explicit $\e$-singular terms that do not affect the unresolved singular limits of the antenna.  
    
    We therefore add an antenna-scheme-dependent {\bf Stage 2} that ensures that the $X_{n+2}^\ell$ antenna has the correct explicit poles to {\bf match} onto the other types of subtraction terms in the desired antenna-subtraction scheme. We {\bf fix}  the scheme by 
    specifying that the explicit $\e$-poles match certain defined ``target poles'', 
    \begin{equation}
        T = T(i_1^h,i_3,...,i_{n+2},i_2^h).
    \end{equation} 
    These target poles must be selected such that the constructed $X_{n+2}^\ell$ is more convenient for use in a wider $\text{N}^{n+\ell} \text{LO}$ subtraction scheme. Different schemes would entail different choices for $T$.  
   
    As in {\bf Stage 1}, we introduce certain projectors $\PPdown_T,\PPup_T$ (at the relevant perturbative order) to identify these additional $\e$-singular contributions which meet all the design principles. Schematically, we can write this final step of the algorithm as
    \begin{equation}
    \label{eq:Xfinal}
        X^\ell_{n+2} \equiv X^\ell_{n+2;N} + \PPup_T (T-\PPdown_T X^\ell_{n+2;N}).
    \end{equation}
    and we require
    \begin{equation}
        \PPdown_i \left[ \PPup_T (T-\PPdown_T X^\ell_{n+2;N}) \right] \equiv 0 \qquad \forall \hspace{0.2cm} i=1,..,N.
    \end{equation}
    For later convenience we define the contribution from {\bf Stage 2} to be,
    \begin{equation}
    \FinPoles \equiv \PPup_T (T-\PPdown_T X^\ell_{n+2;N}).
    \end{equation}
    
\end{description}

Taking into account both {\bf Stage 1} and {\bf Stage 2}, the constructed mixed real-virtual antenna for a given set of infrared limits $\{L_i\}$ and matched to a scheme in which the required $\e$-poles are defined by $T$,  will satisfy
    \begin{align}
    \PPdown_i X^\ell_{n+2} &\equiv  L_i \qquad \forall  \hspace{0.2cm} i=1,..,N \, , \\
    \PPdown_T X^\ell_{n+2} &\equiv T.
    \end{align}
    
\section{Construction of Real-Virtual Antenna Functions}
\label{sec:X31construction}

The above design principles and algorithm have been set-out for the construction of a general $X^\ell_{n+2}$ antenna function. 
Now we specialise to the case of constructing real-virtual $X_3^1$ antenna functions. 
Together with the idealised $X_3^0$ and $\X$ of Chapter~\ref{chapter:paper2}, the $X_3^1$ functions complete the re-formulation of all antenna functions necessary for NNLO calculations, which now meet the design principles. 

We demonstrate the construction of real-virtual antenna functions, $X_3^1(i^h_a,j_b,k^h_c)$, where the particle types are denoted by $a$, $b$ and $c$, which carry four-momenta $i$, $j$ and $k$, respectively. 
Particles $a$ and $c$ should be hard and the antenna functions must have the correct limits when particle $b$ is unresolved. 
Frequently, we drop explicit reference to the particle labels in favour of a specific choice of $X$ according to Table~\ref{tab:X31}.

\begin{table}[t]
\centering
\begin{tabular}{ccc}
\underline{Quark-antiquark} & & \\
$qg\bar{q}$ & $X_3^1(i_q^h,j_g,k_{\bar{q}}^h)$ & $\Arv(i^h,j,k^h)$ \\
 & $\widetilde{X}_3^1(i_q^h,j_g,k_{\bar{q}}^h)$ & $\Atrv(i^h,j,k^h)$ \\
 & $\widehat{X}_3^1(i_q^h,j_g,k_{\bar{q}}^h)$ & $\Ahrv(i^h,j,k^h)$ \\
\underline{Quark-gluon} & &  \\
$qgg$ & $X_3^1(i_q^h,j_g,k_g^h)$ & $\Drv(i^h,j,k^h)$  \\
 & $\widetilde{X}_3^1(i_q^h,j_g,k_g^h)$ & $\Dtrv(i^h,j,k^h)$  \\
 & $\widehat{X}_3^1(i_q^h,j_g,k_g^h)$ & $\Dhrv(i^h,j,k^h)$  \\
$q\bar{Q}Q$ & $X_3^1(i_q^h,j_{\bar{Q}},k_Q^h)$  & $\Erv(i^h,j,k^h)$ \\ 
 & $\widetilde{X}_3^1(i_q^h,j_{\bar{Q}},k_Q^h)$  & $\Etrv(i^h,j,k^h)$ \\ 
  & $\widehat{X}_3^1(i_q^h,j_{\bar{Q}},k_Q^h)$  & $\Ehrv(i^h,j,k^h)$ \\ 
\underline{Gluon-gluon} & & \\
$ggg$ & $X_3^1(i_g^h,j_g,k_g^h)$ & $\Frv(i^h,j,k^h)$  \\
 & $\widehat{X}_3^1(i_g^h,j_g,k_g^h)$ & $\Fhrv(i^h,j,k^h)$  \\
$g\bar{Q}Q$ & $X_3^1(i_g^h,j_{\bar{Q}},k_Q^h)$  & $\Grv(i^h,j,k^h)$ \\ 
 & $\widetilde{X}_3^1(i_g^h,j_{\bar{Q}},k_Q^h)$  & $\Gtrv(i^h,j,k^h)$ \\ 
  & $\widehat{X}_3^1(i_g^h,j_{\bar{Q}},k_Q^h)$  & $\Ghrv(i^h,j,k^h)$ \\ 
\end{tabular}
\caption{Identification of $X_3^1$ antennae according to particle type and colour-structure. These antennae only contain singular limits when particle $b$ (or equivalently momentum $j$) is unresolved, in addition to explicit $\e$ poles. Antennae are classified as quark-antiquark, quark-gluon and gluon-gluon according to the particle type of the parents (i.e. after the antenna mapping). }
\label{tab:X31}
\end{table}

For the specific case of $X_3^1(i^h_a,j_b,k^h_c)$ there are three such limits (meeting the generic design principles), corresponding to particle $b$ becoming soft, particles $a$ and $b$ becoming collinear, or particles $c$ and $b$ becoming collinear, so that the list of target functions is the following,
\begin{eqnarray}
    \label{eq:RVL}
    L_1(i^h,j,k^h) &=& S_b^{(1)}(i^h,j,k^h)  , \\
    L_2(i^h,j,k^h) &=& P^{(1)}_{ab}(i^h,j)  , \\
    L_3(i^h,j,k^h) &=& P^{(1)}_{cb}(k^h,j)  .
\end{eqnarray}
The precise definitions of the one-loop soft factor $S_b^{(1)}$ and the one-loop splitting functions $P^{(1)}_{ab}$ are well known and we reproduce them in our notation in Section~\ref{sec:limits}. 

In order to match onto a particular antenna-subtraction scheme, we require a target pole structure, $T = T(i^h,j,k^h)$.  
We want to match the constructed $X_3^1$ to the full NNLO antenna-subtraction scheme and so we require the $\e$-poles to have a similar $\e$-pole structure to the $X_3^{1,\text{OLD}}$ of Ref.~\cite{Gehrmann-DeRidder:2005btv}. This means a collection of $\e$-poles multiplying $X_3^0$ antennae. 
By removing the contribution to the poles from the renormalisation term, we write the full set of target poles, $T(i^h,j,k^h)$, for the unrenormalised $X_3^1$ as, 
\begin{eqnarray}
\label{eq:targetA}
    T(i_q^h,j_g,k_{\bar{q}}^h)  &=&  -\frac{\Af}{\e^2} \left(S_{ij} + S_{jk} - S_{ijk}  \right) A_3^0 (i^h,j,k^h), \\ 
    \widetilde{T}(i_q^h,j_g,k_{\bar{q}}^h) &=&  
    -\frac{\Af}{\e^2} \left(S_{ijk} - S_{ik} \right) A_3^0 (i^h,j,k^h), \\ 
     \widehat{T}(i_q^h,j_g,k_{\bar{q}}^h) &=&  0, \\ 
     T(i_q^h,j_g,k_{g}^h) &=& -\frac{\Af}{\e^2} \left(S_{ij} + S_{[ik+jk]} + S_{jk} -2 S_{ijk} \right) D_3^{0} (i^h,j,k^h), \\ 
     \widetilde{T}(i_q^h,j_g,k_{g}^h) &=& -\frac{\Af}{\e^2} \left(S_{ik} - S_{[ik+jk]}  \right) D_3^{0} (i^h,j,k^h), \\ 
     \widehat{T}(i_q^h,j_g,k_{g}^h) &=&   0, \\ 
     T(i_q^h,j_{\bar{Q}},k_{Q}^h) &=&  -\Af\left[\frac{1}{\e^2} \left(S_{ij} + S_{ik} - 2 S_{ijk}   \right) - \frac{13}{6\e}S_{jk} \right] E_3^0 (i^h,j,k^h), \\ 
     \widetilde{T}(i_q^h,j_{\bar{Q}},k_{Q}^h) &=&  - \Af \left( \frac{1}{\e^2} + \frac{3}{2 \e} \right) S_{jk} E_3^0 (i^h,j,k^h), \\ 
     \widehat{T}(i_q^h,j_{\bar{Q}},k_{Q}^h) &=& - \Af \frac{2}{3 \e} S_{jk} E_3^0 (i^h,j,k^h), \\ 
     T(i_g^h,j_g,k_{g}^h) &=&  -\frac{\Af}{\e^2} \left(S_{ij} + S_{ik} + S_{jk} -2 S_{ijk} \right) F_3^0 (i^h,j,k^h), \\ 
     \widehat{T}(i_g^h,j_g,k_{g}^h) &=&  0 , \\ 
     T(i_g^h,j_{\bar{Q}},k_{Q}^h) &=&  -\Af\left[\frac{1}{\e^2} \left(S_{ij} + S_{ik} - 2 S_{ijk}   \right) - \frac{13}{6\e}S_{jk} \right] G_3^0 (i^h,j,k^h), \\ 
     \widetilde{T}(i_g^h,j_{\bar{Q}},k_{Q}^h) &=&  - \Af \left( \frac{1}{\e^2} + \frac{3}{2 \e} \right) S_{jk} G_3^0 (i^h,j,k^h), \\
\label{eq:targetG}
\widehat{T}(i_g^h,j_{\bar{Q}},k_{Q}^h) &=& - \Af \frac{2}{3 \e} S_{jk} G_3^0 (i^h,j,k^h).
\end{eqnarray}
A reminder that $\Af$ is an overall factor defined in Eq.~\eqref{eq:Afdef}.
This factor ensures that the $X_3^1$ antennae derived here have the same overall normalisation as those in Ref.~\cite{Gehrmann-DeRidder:2005btv}. We have also used the convenient notation, $S_{ij}$, to separate the loop-type structures from the unresolved-type structures:
\begin{equation}
\label{eq:Sdefs}
    S_{ij} = \left(\frac{s_{ij}}{\mu^2}\right)^{-\e},\qquad
    S_{ijk} = \left(\frac{s_{ijk}}{\mu^2}\right)^{-\e}, \qquad 
    S_{[ik+jk]} = \left(\frac{s_{ik}+s_{jk}}{\mu^2}\right)^{-\e}.
\end{equation}

The $X_3^0 (i^h,j,k^h)$ antennae appearing in Eqs.~\eqref{eq:targetA}--\eqref{eq:targetG} are those constructed in Chapter~\ref{chapter:paper2}. Therefore, there are some differences compared to the pole structure 
in Ref.~\cite{Gehrmann-DeRidder:2005btv} precisely due to differences between the definitions of the $X_3^0 (i^h,j,k^h)$ of Chapter~\ref{chapter:paper2} and the $X_3^{0,\text{OLD}} (i^h,j,k^h)$ of Ref.~\cite{Gehrmann-DeRidder:2005btv} (see Ref.~\cite{paper2} for a full discussion). 
This is particularly the case for the $D_3^0$ and $F_3^0$ antennae where the unresolved singularities of $F_3^{0,\text{OLD}}$ are assigned to three
$F_3^{0}$ antennae, while the unresolved singularities present in 
$D_3^{0,\text{OLD}}$ are found by combining two $D_3^{0}$ antennae.   

Finally, we note that the $\e$-singularities of 
$D_3^{1,\text{OLD}}$ of Ref.~\cite{Gehrmann-DeRidder:2005btv} are split between $\Drv$ and a new type of antenna, $\Dtrv$,
\begin{equation}
   \PPdown_T D_3^{1,\text{OLD}} \sim T(i_q^h,j_g,k_{g}^h) +\widetilde{T}(i_q^h,j_g,k_{g}^h) + ( j \leftrightarrow k) ,
\end{equation}
where $\sim$ reflects the fact that the LHS multiplies $D_3^{0,\text{OLD}}$ while the RHS multiplies $D_3^{0}$. The reasons for the definition of $\Dtrv$ will become clear later in this chapter.

\subsection{Template Antennae}
For convenience, we define a general unrenormalised real-virtual antenna function in terms of the contributions produced by the various steps of the algorithm.  At leading-colour, we have
\begin{eqnarray}
\label{eq:X31def}
X_3^1(i^h,j,k^h) &=& \Ssoft^{(1)}(i^h,j,k^h) + \Scol^{(1)}(i^h,j;k^h) + \Scol^{(1)}(k^h,j;i^h) \nonumber \\  
&+&   \FinPoles(i^h,j,k^h) \, ,
\end{eqnarray}
while the corresponding sub-leading-colour expression is
\begin{eqnarray}
\label{eq:X31tdef}
\widetilde{X}_3^1(i^h,j,k^h) &=& \Scolt^{(1)}(i^h,j;k^h) + \Scolt^{(1)}(k^h,j;i^h) \nonumber
\\ 
&+&  \FinPolest(i^h,j,k^h) \, ,
\end{eqnarray}
and the quark-loop contribution is
\begin{eqnarray}
\label{eq:X31hdef}
\widehat{X}_3^1(i^h,j,k^h) &=& \Scolh^{(1)}(i^h,j;k^h) + \Scolh^{(1)}(k^h,j;i^h) \nonumber \\ 
&+&  \FinPolesh(i^h,j,k^h) \, ,
\end{eqnarray}
since the one-loop soft factor is only non-zero at leading-colour.
The meanings of the individual terms in Eqs.~\eqref{eq:X31def}--\eqref{eq:X31hdef} will be made clear in the following subsections, however, we note that in each equation the first line is produced by {\bf Stage 1} of the algorithm and the second line is added in {\bf Stage 2}.  Therefore, we expect that 
\begin{align}
    \PSdown_j \FinPoles(i^h,j,k^h) &= 0 \, , \\
    \PCdown_{ij} \FinPoles(i^h,j,k^h) &= 0 \, , \\
    \PCdown_{jk} \FinPoles(i^h,j,k^h) &= 0 \, ,   
\end{align}
and similarly for $\FinPolest(i^h,j,k^h)$ and $\FinPolesh(i^h,j,k^h)$.

\subsection{Stage 1}

All $X_3^1 (i^h,j,k^h)$ antenna functions are defined over the full three-particle phase space, whereas each unresolved limit lives on a restricted part of phase space: the $j$ soft limit, the $ij$ collinear limit and the $jk$ collinear limit. 

We define the soft down-projector by its action on integer powers of invariants as
\begin{equation}
    \PSdown_j: \begin{cases} 
    s_{ij} \mapsto \lambda s_{ij}, 
    s_{jk} \mapsto \lambda s_{jk}, \\
    s_{ijk} \mapsto s_{ik},
    \end{cases}
\end{equation}
and keep only the terms proportional to $\lambda^{-2}$. 

For the corresponding up-projector $\PSup_j$ we choose a trivial mapping which leaves all variables unchanged. The collinear down-projector acts on integer powers of invariants and is defined in analogy to Eq.~\eqref{eq:PCdown},
\begin{equation}
    \PCdown_{ij}: \begin{cases}
    s_{ij} \mapsto \lambda s_{ij}, \\
    s_{ik} \mapsto \omxj (s_{ik}+s_{jk}),
    s_{jk} \mapsto \xj (s_{ik}+s_{jk}), 
    s_{ijk} \mapsto s_{ik} + s_{jk},
    \end{cases}
\end{equation}
but keeps only terms of order $\lambda^{-1}$.  The corresponding up-projector is the same as in Eq.~\eqref{eq:PCup},
\begin{equation}
    \PCup_{ij}: \begin{cases}
    \xj \mapsto s_{jk}/s_{ijk},
    \omxj \mapsto s_{ik}/s_{ijk}, \\
    s_{ik}+s_{jk} \mapsto s_{ijk}.
    \end{cases} \, 
\end{equation}
This up-projector ensures the presence of $s_{ijk}$ denominators, which are present in matrix elements corresponding to physical propagators and means that the same integration tools for one-loop matrix elements can be used in the integration of the constructed $X_3^1$ over its Lorentz-invariant antenna phase space.

The subtracted single unresolved one-loop factors are built from unrenormalised colour-ordered limits and are given by
\begin{align}
    \Ssoft^{(1)}(i^h,j,k^h) &= \PSup_j S_b^{(1)}(i^h,j,k^h) \, ,\\
    \Scol^{(1)}(i^h,j;k^h) &= \PCup_{ij}\left(P_{ab}^{(1)}(i^h,j) - \PCdown_{ij}\Ssoft^{(1)}(i^h,j,k^h) \right) \, , \\
    \Scol^{(1)}(k^h,j;i^h) &= \PCup_{kj}\left(P_{cb}^{(1)}(k^h,j) - \PCdown_{kj}\left(\Ssoft^{(1)}(i^h,j,k^h) + \Scol^{(1)}(i^h,j;k^h) \right) \right) \, , \nonumber \\
    &\equiv \PCup_{kj}\left(P_{cb}^{(1)}(k^h,j) - \PCdown_{kj}\Ssoft^{(1)}(i^h,j,k^h)  \right) \, ,
\end{align}
and analogously for the sub-leading colour and quark-flavour contributions. The subscripts $a,b,c$ represent the particle types which carry momenta $i,j,k$ respectively. The unrenormalised one-loop single unresolved limits are listed in full in Section~\ref{sec:limits}. We have used the feature that the only overlap between the $ij$- and $jk$-collinear limits occurs in $\Ssoft^{(1)}$ so that
\begin{equation}
    \PCdown_{kj}\Scol^{(1)}(i^h,j;k^h) = 0,
\end{equation}
which was also observed in Chapter~\ref{chapter:paper2}. 

At this point, we have iteratively constructed the quantity 
\begin{equation}
\label{eq:X33}
    X_{3;3}^1(i^h,j,k^h) = 
    \Ssoft^{(1)}(i^h,j,k^h)+
    \Scol^{(1)}(i^h,j;k^h)+
    \Scol^{(1)}(k^h,j;i^h),
\end{equation}
such that,
\begin{align}
    \PSdown_j X_{3;3}^1(i^h,j,k^h) &= S_b^{(1)}(i^h,j,k^h) \, , \\
    \PCdown_{ij} X_{3;3}^1(i^h,j,k^h) &= P_{ab}^{(1)}(i^h,j) \, , \\
    \PCdown_{jk} X_{3;3}^1(i^h,j,k^h) &= P_{cb}^{(1)}(k^h,j) \, ,   
\end{align}
which carries all of the desired unresolved limits.

\subsection{Stage 2}
\label{subsec:stage2}

We now turn to the construction of $\FinPoles(i^h,j,k^h)$, which
does not contribute to any unresolved limit but does contain explicit $\e$ poles. In the language of Section~\ref{sec:algorithm2}, then schematically,
\begin{equation}
\FinPoles(i^h,j,k^h) 
\equiv \PPup_T \left(T(i^h,j,k^h)-\PPdown_T X^1_{3;3(i^h,j,k^h)}\right).
\end{equation}

We observe that each of the target pole structures in Eqs.~\eqref{eq:targetA}--\eqref{eq:targetG} is of the form,
\begin{equation}
    \Poles \times \XLO (i^h,j,k^h),
\end{equation}
where $\Poles$ is combination of $\e$-poles and factors like $s_{ij}^{-\e}$. 
We therefore choose to achieve {\bf Stage 2} through two iterative steps (rather than one), adding a projector for each step:
\begin{description}
\item[Step 1.] We introduce a projector $\PXup$ (and the trivial projector $\PXdown$) to ensure that the $\e$-poles are proportional to $\XLO(i^h,j,k^h)$ and
\item[Step 2.] we introduce projectors $\Peup$ and $\Pedown$ to adjust the pole structure multiplying $\XLO$ to match $T(i^h,j,k^h)$.
\end{description}

\subsection*{Step 1}
We define the projector $\PXup$ such that,
\begin{equation}
    \PXup: \begin{cases}
    P_{ab}^{(0)}(i^h,j) \mapsto X_3^0 (i^h,j,k^h),\\
    P_{cb}^{(0)}(k^h,j) \mapsto X_3^0 (i^h,j,k^h).
    \end{cases} 
\end{equation}
The inverse projector $\PXdown$ is simply unity.  

We define $\FinPoles_X(i^h,j,k^h)$ to be the contribution arising from the action of $\PXup$ such that,
\begin{align}
\label{eq:finpolesX}
    \FinPoles_X(i^h,j,k^h) &= (\PXup - 1) ( \Ssoft^{(1)}(i^h,j,k^h) + \Scol^{(1)}(i^h,j;k^h) + \Scol^{(1)}(k^h,j;i^h) ), \nonumber \\
    &= (\PXup - 1) X_{3;3}^1(i^h,j,k^h).
\end{align}
$\Ssoft^{(1)}(i^h,j,k^h)$ does not contain splitting functions, so the action of $\PXup$ on $\Ssoft$ is trivial,
\begin{equation}
   \PXup \Ssoft^{(1)}(i^h,j,k^h) \equiv \Ssoft^{(1)}(i^h,j,k^h), 
\end{equation}
which guarantees,
\begin{align}
\label{eq:noIRfinpolesX1}
    \PSdown_j \FinPoles_X(i^h,j,k^h) &= 0 \, .  
\end{align}
Furthermore, because of the structure of the one-loop splitting functions, we also have
\begin{eqnarray}
    \PCdown_{ij} \FinPoles_X(i^h,j,k^h) &=& 0  , \\
    \label{eq:noIRfinpolesX2}
    \PCdown_{jk} \FinPoles_X(i^h,j,k^h) &=& 0  .  
\end{eqnarray}
We will explain more clearly how this is achieved in the specific example of $A_3^1$ in Section~\ref{sec:A31construct}.
We also note that following the iterative structure of Eq.~\eqref{eq:algorithm2}, we define the fourth step of the algorithm to be
\begin{align}
\label{eq:X34}
    X_{3;4}^1(i^h,j,k^h) &= X_{3;3}^1(i^h,j,k^h) + (\PXup-1) X_{3;3}^1(i^h,j,k^h), \\
              &\equiv \PXup X_{3;3}^1(i^h,j,k^h),
\end{align}
where $X_{3;3}^1(i^h,j,k^h)$ is given in Eq.~\eqref{eq:X33}.

\subsection*{Step 2}

The operators $\Pedown$ and $\Peup$ are defined as follows.

$\Pedown$ is defined by Laurent-expanding the argument in $\e$ and discarding terms of $\order{\e^0}$ and higher. 
    
$\Peup$ is defined by extending the argument to an all-orders  expression in $\e$, which agrees with the argument up to $\order{\e^0}$. This is not a unique action. For the case of the $X_3^1$ we choose, where possible, for $\Peup$ to result in linear combinations of $\{ s_{ik}^{-\e}, s_{ijk}^{-\e}, (s_{ik} + s_{jk})^{-\e} \}$ (which are simply-integrable objects) multiplied by simple $\e$-poles and $X_3^0$.
Only two structures are required for the construction of the $X_3^1$:
\begin{equation}
   \Peup \frac{1}{\e} \ln \left(1 + \frac{s_{ij} s_{jk} }{s_{ik} s_{ijk}} \right)  = \frac{1}{\e^2} \Lambda_1(i^h,j,k^h) , 
\end{equation}
\begin{equation}
   \Peup \frac{2}{\e} \ln \left( 1- \frac{s_{jk} }{s_{ijk}} \right)  = \frac{2}{\e^2} \Lambda_2(i^h,j,k^h) , 
\end{equation}
where 
\begin{eqnarray}
    \Lambda_1(i^h,j,k^h) &=&  S_{ik} + S_{ijk} - S_{[ik+jk]} - S_{[ik+ij]} , \\
    \Lambda_2(i^h,j,k^h) &=& S_{ijk} - S_{[ik+ij]} ,
\end{eqnarray}
with $S_{ik}$ etc. defined as in Eq.~\eqref{eq:Sdefs}.

We define $\FinPoles_{\e}(i^h,j,k^h)$ to be the contribution arising from the action of $\Peup$ and $\Pedown$ such that,
\begin{eqnarray}
\label{eq:finpolese}
    \FinPoles_{\e}(i^h,j,k^h) &=& \Peup \bigg( \Pedown T(i^h,j,k^h) - \Pedown \bigg[ \Ssoft^{(1)}(i^h,j,k^h) \\ \nonumber
    && + \Scol^{(1)}(i^h,j;k^h)  +  \Scol^{(1)}(k^h,j;i^h) +  \FinPoles_X (i^h,j,k^h) \bigg] \bigg).
\end{eqnarray}
These contributions typically contain a factor which suppresses all the unresolved limits in the $X_3^0$ to which it multiplies. $\Lambda_1$ suppresses any contributions to the soft-$j$ limit or the collinear-$ij$ or collinear-$jk$ limits. $\Lambda_2$ suppresses contributions to the soft-$j$ or collinear-$jk$ limits so that
\begin{eqnarray}
\label{eq:noIRfinpolese}
    \PSdown_j \FinPoles_{\e}(i^h,j,k^h) &=& 0  , \\
    \PCdown_{ij} \FinPoles_{\e}(i^h,j,k^h) &=& 0  , \\
    \label{eq:noIRfinpolese2}
    \PCdown_{jk} \FinPoles_{\e}(i^h,j,k^h) &=& 0  .  
\end{eqnarray}
We will explain more clearly how this works in detail in the specific example of $A_3^1$ in Section~\ref{sec:A31construct}.

In the iterative language of Eq.~\eqref{eq:algorithm2} the fifth and final step of the algorithm is
\begin{align}
    X_{3;5}^1(i^h,j,k^h) &=   X_{3;4}^1(i^h,j,k^h) +  \Peup \left(\Pedown T(i^h,j,k^h) - \Pedown X_{3;4}^1(i^h,j,k^h) \right),
\end{align}
where we define the complete constructed antenna function,
\begin{equation}
     X_3^1(i^h,j,k^h) \equiv X_{3;5}^1(i^h,j,k^h) .
\end{equation}

It is convenient to combine the contributions from Eqs.~\eqref{eq:finpolesX} and \eqref{eq:finpolese} to obtain a single contribution (as in the antenna templates of Eqs.~\eqref{eq:X31def}, \eqref{eq:X31tdef} and \eqref{eq:X31hdef}) and we define
\begin{equation}
    \FinPoles(i^h,j,k^h) \equiv \FinPoles_X(i^h,j,k^h) + \FinPoles_{\e}(i^h,j,k^h).
\end{equation}
It is to be emphasised again that $\FinPoles(i^h,j,k^h)$ does not contribute in any unresolved limit but does carry explicit poles in $\e$.
Indeed, using Eqs.~\eqref{eq:noIRfinpolesX1} -~\eqref{eq:noIRfinpolesX2} and Eqs.~\eqref{eq:noIRfinpolese} -~\eqref{eq:noIRfinpolese2}, it is straightforward to see that
\begin{align}
\label{eq:noIRfinpoles}
    \PSdown_j \FinPoles(i^h,j,k^h) &= 0 \, , \\
    \PCdown_{ij} \FinPoles(i^h,j,k^h) &= 0 \, , \\
    \PCdown_{jk} \FinPoles(i^h,j,k^h) &= 0 \, .  
\end{align}

Finally, the algorithm presented here ensures that
\begin{align}
    \PSdown_j X_3^1(i^h,j,k^h) &= S_b^{(1)}(i^h,j,k^h) \, , \\
    \PCdown_{ij} X_3^1(i^h,j,k^h) &= P_{ab}^{(1)}(i^h,j) \, , \\
    \PCdown_{jk} X_3^1(i^h,j,k^h) &= P_{cb}^{(1)}(k^h,j) \, ,   
\end{align}
and 
\begin{align}
    \Pedown X_3^1(i^h,j,k^h) &= \Pedown T(i^h,j,k^h)\, .
\end{align}

\subsection{Renormalisation}

As a final step, we renormalise the antennae at scale $\mu^2$,
\begin{eqnarray}
X_3^1(i^h,j,k^h) &\mapsto & X_3^1(i^h,j,k^h) -  \frac{ \bz}{\e} X_3^0(i^h,j,k^h) \, , \\
\label{eq:X31defR}
\widetilde{X}_3^1(i^h,j,k^h) &\mapsto& \widetilde{X}_3^1(i^h,j,k^h) \, , \\
\label{eq:X31tdefR}
\widehat{X}_3^1(i^h,j,k^h) &\mapsto& \widehat{X}_3^1(i^h,j,k^h)-  \frac{ \bzf}{\e}  X_3^0(i^h,j,k^h) \, .
\label{eq:X31hdefR}
\end{eqnarray}
We use the colour decomposition of $\Bz$,
\begin{equation}
    \Bz = N_c \bz + N_F \bzf ,
\end{equation}
where $\bz = 11/6$ and $\bzf=-1/3$.

\subsection{$\Arv$ Construction in Full Detail}
\label{sec:A31construct}

To make the construction explicit, we work through the construction of $A_3^1$ as an example before describing the full set of idealised real-virtual antenna functions in Section~\ref{sec:X31}. 
$\Arv (i_q^h,j_g,k_{\bar{q}}^h)$ is the leading-colour antenna function with quark and antiquark hard radiators, which encapsulates the one-loop limits when the gluon becomes unresolved. 
The relevant unresolved limits are
\begin{equation}
    \Sgone(i^h,j,k^h), \qquad\Pqgone(i_q^h,j_g), \qquad\Pqgone(k_{\bar{q}},j_g) ,
\end{equation}
which are given in Section~\ref{sec:limits}. 
Additionally, we choose a target for the $\e$-poles before renormalisation, which is consistent with the above limits but matches the $\e$-pole structures appearing in the antenna-subtraction scheme. 
The target pole structure for $\Arv$ is given by
\begin{equation}
\label{eq:Atarget}
T(i_q^h,j_g,k_{\bar{q}}^h)  =  \frac{\Af}{\e^2} \left(S_{ij} + S_{jk} - S_{ijk}  \right) A_3^0 (i^h,j,k^h).
\end{equation}
We want to match the constructed $X_3^1$ to the full NNLO antenna-subtraction scheme; we therefore require the $\e$-poles to have a similar $\e$-pole structure to the $X_3^{1, \text{OLD}}$ in Ref.~\cite{Gehrmann-DeRidder:2005btv}.

We choose to simplify our notation by using the following structure 
\begin{eqnarray}
G(w,\e) &=&  {}_2F_1\left(1,\e,1+\e,-w\right) -1  , \nonumber \\
&=& - \sum_{n=1}^{\infty} (-\e)^n \Li_{n}\left(-w\right), \nonumber \\
&\equiv& (1+w)^{-\e} {}_2F_1\left(\e,\e,1+\e,\frac{w}{1+w}\right) -1  ,
\end{eqnarray}
where $w = s_{jk}/s_{ik}$. 
Note that in the $w \to 0$ limit, $G(w,\e)$ vanishes. 

Before renormalisation, $\Arv$ is built iteratively in the following order,
\begin{eqnarray}
\Arv(i_q^h,j_g,k_{\bar{q}}^h) &=& \Ssoft^{(1)}(i_q^h,j_g,k_{\bar{q}}^h) \\ \nonumber
&&+ \Scol^{(1)}(i_q^h,j_g;k_{\bar{q}}^h) + \Scol^{(1)}(k_{\bar{q}}^h,j_g;i_q^h) 
+  \FinPoles(i_q^h,j_g,k_{\bar{q}}^h) \, ,
\label{eq:A31def}
\end{eqnarray}
with the $\FinPoles$ contribution constructed in two steps as in Section~\ref{subsec:stage2} according to Eqs.~\eqref{eq:finpolesX} and \eqref{eq:finpolese},
\begin{equation}
\FinPoles(i_q^h,j_g,k_{\bar{q}}^h) 
= \FinPoles_X(i_q^h,j_g,k_{\bar{q}}^h) 
+  \FinPoles_{\e}(i_q^h,j_g,k_{\bar{q}}^h). 
\end{equation}

The first contribution is simply the one-loop soft factor,
\begin{equation}
\label{eq:SsoftA31}
    \Ssoft^{(1)}(i_q^h,j_g,k_{\bar{q}}^h) = \PSup_j \Sgone(i^h,j,k^h)  = - \Af\frac{\Gamma(1-\e)\Gamma(1+\e)}{\e^2} \frac{S_{ij}S_{jk}}{S_{ik}}  \Sgzero (i^h,j,k^h), 
\end{equation}
where $\Sgzero$ is the tree-level eikonal factor given in Section~\ref{sec:tree-single-lims}. 

The second piece is given by the overlap of the one-loop splitting function $\Pqgone(i^h,j)$ and the one-loop soft factor $\Ssoft^{(1)}$ in the $ij$ collinear limit, up-projected into the full phase-space,
\begin{eqnarray}
\lefteqn{     \Scol^{(1)}(i_q^h,j_g;k_{\bar{q}}^h) } \nonumber \\
&=& \PCup_{ij}\left(\Pqgone(i^h,j) - \PCdown_{ij}\Ssoft^{(1)}(i_q^h,j_g,k_{\bar{q}}^h) \right), \nonumber \\ 
     &=&  \frac{\Af}{\e^2} \bigg[ -\Gamma(1-\e)\Gamma(1+\e) \frac{S_{ij}S_{jk}}{S_{ik}} \frac{\ome s_{jk}}{s_{ij}s_{ijk}} + S_{ij} G\left(\frac{s_{jk}}{s_{ik}},\e \right) \Pqgzero(i_q^h,j_g;k_{\bar{q}}^h)  \bigg]  \nonumber \\ 
    \label{eq:ScolA31a}
     &&+ S_{ij} \Af \frac{(s_{ijk}-\e s_{jk} )}{s_{ij}s_{ijk}} \frac{1}{2 (1-2\e)} . 
\end{eqnarray}
Here we use the short-hand notation 
\begin{equation}
  P_{ab}^{(n)} (i^h,j;k) = \PCup_{ij} P_{ab}^{(n)} (i^h,j), 
\end{equation}
to indicate an $n$-loop splitting function up-projected into the full phase space of the antenna and the tree-level splitting functions, $P_{ab}^{(0)}$, are given in Section~\ref{sec:tree-single-lims}. 

The third contribution is given by
\begin{eqnarray}
\lefteqn{\Scol^{(1)}(k_{\bar{q}}^h,j_g;i_q^h)} \nonumber \\
&=& \PCup_{kj}\left(\Pqgone(k_q^h,j_g) - \PCdown_{kj}\left(\Ssoft^{(1)}(i_q^h,j_g,k_{\bar{q}}^h) + \Scol^{(1)}(i_q^h,j_g;k_{\bar{q}}^h) \right) \right), \nonumber \\
&=& \PCup_{kj}\left(\Pqgone(k_q^h,j_g) - \PCdown_{kj}\Ssoft^{(1)}(i_q^h,j_g,k_{\bar{q}}^h)  \right) , \nonumber \\ 
    &=& \frac{\Af}{\e^2} \bigg[ -\Gamma(1-\e)\Gamma(1+\e) \frac{S_{ij}S_{jk}}{S_{ik}} \frac{\ome s_{ij}}{s_{jk}s_{ijk}} + S_{jk} G\left(\frac{s_{ij}}{s_{ik}},\e \right) \Pqgzero(k_{\bar{q}}^h,j_g;i_{q}^h)  \bigg] \nonumber \\ 
    \label{eq:ScolA31b}
     &&+  S_{jk} \Af \frac{(s_{ijk}-\e s_{ij} )}{s_{jk}s_{ijk}}\frac{1}{2 (1-2\e)}  . 
\end{eqnarray}
Recalling from Eq.~\eqref{eq:A30} that 
\begin{eqnarray}
    A_3^0 (i_q^h,j_g,k_{\bar{q}}^h) &\equiv& \Sgzero (i^h,j,k^h) + \frac{\ome s_{jk}}{s_{ij}s_{ijk}} + \frac{\ome s_{ij}}{s_{jk}s_{ijk}}, \nonumber \\ 
    &\equiv& \Pqgzero(i_q^h,j_g;k_{\bar{q}}^h) + \frac{\ome s_{ij}}{s_{jk}s_{ijk}} , \nonumber \\ 
    &\equiv& \Pqgzero(k_{\bar{q}}^h,j_g;i_{q}^h) + \frac{\ome s_{jk}}{s_{ij}s_{ijk}} ,
\end{eqnarray}
it is straightforward to see that the terms proportional to $S_{ij}S_{jk}/S_{ik}$ in Eqs.~\eqref{eq:SsoftA31},~\eqref{eq:ScolA31a} and \eqref{eq:ScolA31b}, combine to give a term which factorises onto $A_3^0(i_q^h,j_g,k_{\bar{q}}^h)$ such that 
\newpage
\begin{eqnarray}
\label{eqn:runningA31}
\lefteqn{   \Ssoft^{(1)}(i_q^h,j_g,k_{\bar{q}}^h) + \Scol^{(1)}(i_q^h,j_g;k_{\bar{q}}^h) + \Scol^{(1)}(k_{\bar{q}}^h,j_g;i_q^h) } \nonumber \\ 
   &=&+ \frac{\Af}{\e^2}\bigg[ -\Gamma(1-\e)\Gamma(1+\e) \frac{S_{ij}S_{jk}}{S_{ik}} A_3^0(i^h,j,k^h) \nonumber \\ 
   &&+S_{ij} G\left(\frac{s_{jk}}{s_{ik}},\e \right) \Pqgzero(i_q^h,j_g;k_{\bar{q}}^h) + S_{jk}G\left(\frac{s_{ij}}{s_{ik}},\e \right) \Pqgzero(k_{\bar{q}}^h,j_g;i_{q}^h) \bigg] \nonumber \\ 
    &&  + S_{ij} \Af \frac{(s_{ijk}-\e s_{jk} )}{s_{ij}s_{ijk}} \frac{1}{2 (1-2\e)} 
    +  S_{jk} \Af \frac{(s_{ijk}-\e s_{ij} )}{s_{jk}s_{ijk}}\frac{1}{2 (1-2\e)}  .
\end{eqnarray}
This combination completes Stage 1 of the algorithm and is to some extent a complete construction of $\Arv$. 
It is complete in the sense that it encapsulates the fundamental one-loop unresolved limits we require.

{\bf Stage 2} of the algorithm preserves the unresolved limits but includes explicit poles that do not contribute in any limit.  
The next piece $\FinPoles_X$ ensures that the explicit pole structure of the $\Arv$ factors onto $A_3^0$:
\begin{align}
    \FinPoles_X(i_q^h,j_g,k_{\bar{q}}^h) &= (\PXup - 1) \left( \Ssoft^{(1)}(i_q^h,j_g,k_{\bar{q}}^h) + \Scol^{(1)}(i_q^h,j_g;k_{\bar{q}}^h) + \Scol^{(1)}(k_{\bar{q}}^h,j_g;i_q^h) \right), \nonumber \\ 
    &= \frac{\Af}{\e^2} \bigg[\hspace{0.3cm} S_{ij} G\left(\frac{s_{jk}}{s_{ik}},\e \right) \left(A_3^0 (i_q^h,j_g,k_{\bar{q}}^h) - \Pqgzero(i_q^h,j_g;k_{\bar{q}}^h) \right) \nonumber \\ 
    &\hspace{1cm} + S_{jk}G\left(\frac{s_{ij}}{s_{ik}},\e \right) \left( A_3^0 (i_q^h,j_g,k_{\bar{q}}^h) - \Pqgzero(k_{\bar{q}}^h,j_g;i_{q}^h) \right) \bigg], \nonumber \\ 
    &= \frac{\Af}{\e^2} \bigg[ S_{ij} G\left(\frac{s_{jk}}{s_{ik}},\e \right) \frac{\ome s_{ij}}{s_{jk}s_{ijk}} + S_{jk}G\left(\frac{s_{ij}}{s_{ik}},\e \right) \frac{\ome s_{jk}}{s_{ij}s_{ijk}} \bigg].
    \label{eq:FinPolesA31A}
\end{align}
This term vanishes in the unresolved region for the following reason.  
The first term in the final line appears to have a singularity in the $jk$ collinear limit due to the $1/s_{jk}$ factor.  However, in this limit the hypergeometric function $G(s_{jk}/s_{ik},\e)$ approaches zero  and this behaviour therefore suppresses the singularity due to the $1/s_{jk}$ factor.  A similar argument holds for the second term. As such, neither term in Eq.~\eqref{eq:FinPolesA31A} contributes to any unresolved limit, although they evidently do contribute explicit $\e$ poles. 
In summary,
\begin{align}
    \PSdown_j \left( \frac{1}{s_{ij}s_{jk}} \times G\left(\frac{s_{jk}}{s_{ik}},\e \right) 
    \right) 
     &=  0 \, , \\
    \PSdown_j  \left( \frac{1}{s_{ij}s_{jk}} \times G\left(\frac{s_{ij}}{s_{ik}},\e \right)\right)  &=  0 \, , \\
    \PCdown_{ij}  \left( \frac{1}{s_{ij}} \times G\left(\frac{s_{ij}}{s_{ik}},\e \right) \right) &=  0 \, , \\
    \PCdown_{jk} \left( \frac{1}{s_{jk}} \times  G\left(\frac{s_{jk}}{s_{ik}},\e \right) \right) &=  0 \, . 
\end{align}

The running total for $\Arv$ is given by 
\begin{eqnarray}
&& \Ssoft^{(1)}(i_q^h,j_g,k_{\bar{q}}^h) + \Scol^{(1)}(i_q^h,j_g;k_{\bar{q}}^h) + \Scol^{(1)}(k_{\bar{q}}^h,j_g;i_q^h) + \FinPoles_X(i_q^h,j_g,k_{\bar{q}}^h)  \nonumber \\ 
   && \hspace{1cm} = + \frac{\Af}{\e^2}\bigg[ -\Gamma(1-\e)\Gamma(1+\e) \frac{S_{ij}S_{jk}}{S_{ik}} \nonumber  \\
   && \hspace{1cm} +S_{ij} G\left(\frac{s_{jk}}{s_{ik}},\e \right) + S_{jk}G\left(\frac{s_{ij}}{s_{ik}},\e \right) \bigg] A_3^0 (i_q^h,j_g,k_{\bar{q}}^h) , \nonumber \\ 
    && \hspace{1cm} + S_{ij} \Af \frac{(s_{ijk}-\e s_{jk} )}{s_{ij}s_{ijk}} \frac{1}{2 (1-2\e)} 
    +  S_{jk} \Af \frac{(s_{ijk}-\e s_{ij} )}{s_{jk}s_{ijk}}\frac{1}{2 (1-2\e)}  .
\end{eqnarray}
Effectively, the tree-level splitting functions in Eq.~\eqref{eqn:runningA31} have been promoted to full $A_3^0$ antenna functions.

The next contribution, $\FinPoles_{\e}$, is also part of antenna-scheme matching, for which we have the target pole structure proportional to $A_3^0$, given in Eq.~\eqref{eq:Atarget}. The resulting expression is given by
\begin{eqnarray}
    \FinPoles_{\e}(i_q^h,j_g,k_{\bar{q}}^h) &=& \Peup \bigg( \Pedown T(i_q^h,j_g,k_{\bar{q}}^h) - \Pedown \bigg[ \Ssoft^{(1)}(i_q^h,j_g,k_{\bar{q}}^h) \nonumber \\ 
    && + \Scol^{(1)}(i_q^h,j_g;k_{\bar{q}}^h) + \Scol^{(1)}(k_{\bar{q}}^h,j_g;i_q^h) +  \FinPoles_X (i_q^h,j_g,k_{\bar{q}}^h) \bigg] \bigg) , \nonumber \\ 
    &=& \Peup \bigg( \frac{1}{\e}  \ln \left(1 + \frac{s_{ij} s_{jk} }{s_{ik} s_{ijk}} \right)  A_3^0(i^h,j,k^h)  \bigg) , \nonumber  \\ 
    &=& \frac{\Af}{\e^2} \Lambda_1(i^h,j,k^h) A_3^0(i^h,j,k^h) .
\end{eqnarray}
As discussed earlier, the logarithmic structure of $\Lambda_1$ suppresses all the unresolved limits present in the $A_3^0$ antenna at every order in $\e$. This structure also carries a $1/\e^2$ factor, so $\FinPoles_{\e}(i_q^h,j_g,k_{\bar{q}}^h)$ contains explicit $\e$ poles (which are important for antenna-scheme matching) but does not contribute in the unresolved limits. 
In summary,
\begin{align}
    \PSdown_j \left( \Lambda_1(i^h,j,k^h) A_3^0(i^h,j,k^h) \right)  &=  0 \, , \\
    \PCdown_{ij} \left(\Lambda_1(i^h,j,k^h) A_3^0(i^h,j,k^h) \right)  &=  0 \, , \\
    \PCdown_{jk} \left(\Lambda_1(i^h,j,k^h) A_3^0(i^h,j,k^h) \right)  &=  0 \, .
\end{align}

Finally, including the renormalisation term and combining terms together, we find a compact expression for $\Arv$ given by 
\begin{equation}
\begin{split}
    \Arv(i^h,j,k^h) &= \frac{\Af}{\e^2}\bigg[ -\Gamma(1-\e)\Gamma(1+\e) \frac{S_{ij}S_{jk}}{S_{ik}} +S_{ij} G\left(\frac{s_{jk}}{s_{ik}},\e \right)  + S_{jk}G\left(\frac{s_{ij}}{s_{ik}},\e \right) \\
    & \hspace{0.5cm}  + \Lambda_1(i^h,j,k^h) \bigg] A_3^0(i^h,j,k^h) 
     - \frac{\bz}{\e} A_3^0(i^h,j,k^h)\\
    & \hspace{0.5cm} + S_{ij} \Af \frac{(s_{ijk}-\e s_{jk} )}{s_{ij}s_{ijk}} \frac{1}{2 (1-2\e)} 
    +  S_{jk} \Af \frac{(s_{ijk}-\e s_{ij} )}{s_{jk}s_{ijk}}\frac{1}{2 (1-2\e)} \, .
\end{split}
\end{equation}

We verify that
\begin{align}
    \PSdown_j \Arv(i^h,j,k^h) &= S_g^{(1)}(i^h,j,k^h) \, , \\
    \PCdown_{ij} \Arv(i^h,j,k^h) &= P_{qg}^{(1)}(i^h,j) \, , \\
    \PCdown_{jk} \Arv(i^h,j,k^h) &= P_{qg}^{(1)}(k^h,j) \, ,   
\end{align}
and 
\begin{align}
    \Pedown \Arv(i^h,j,k^h) &= \Pedown T(i_q^h,j_g,k_{\bar{q}}^h)\, .
\end{align}

\section{Real-Virtual Antenna Functions}
\label{sec:X31}

In this section, we give compact expressions for the full set of real-virtual antennae and their integrals.  
In deriving and integrating these antennae, we have made use of MAPLE, hypexp~\cite{Huber:2005yg,Huber:2007dx} and FORM~\cite{Vermaseren:2000nd,Kuipers:2012rf}. 

\subsection{Quark-Antiquark Antennae}

As shown in Table~\ref{tab:X31}, there are three one-loop three-parton antennae with quark-antiquark parents that describe the emission of a gluon, organised by colour structure: $\Arv$, $\Atrv$ and $\Ahrv$. The antenna functions constructed here are directly related to the antenna functions given in Ref.~\cite{Gehrmann-DeRidder:2005btv} by 
\begin{align}
    \Arvold (i_q,j_g,k_{\bar{q}}) &\sim \Arv (i_q^h,j_g,k_{\bar{q}}^h) \, , \\
    \Atrvold (i_q,j_g,k_{\bar{q}}) &\sim \Atrv (i_q^h,j_g,k_{\bar{q}}^h) \, , \\
    \Ahrvold (i_q,j_g,k_{\bar{q}}) &\sim \Ahrv (i_q^h,j_g,k_{\bar{q}}^h) \, ,
\end{align}
where $\sim$ means that they contain the same limits as $j_g$ becomes unresolved, although they may contain different $\e$ poles. 

In order to build these antennae using the algorithm in Section~\ref{sec:algorithm2}, we identify the particles, included in the antenna, to specify the limits encapsulated by the $X_3^1$ and identify the target poles for the $X_3^1$, factorising on to the respective $X_3^0$ (here $A_3^0$). The resulting formula from the algorithm (copied from above) is given by
\begin{equation}
\begin{split}
    \Arv(i^h,j,k^h) &= \frac{\Af}{\e^2}\bigg[ -\Gamma(1-\e)\Gamma(1+\e) \frac{S_{ij}S_{jk}}{S_{ik}} +S_{ij} G\left(\frac{s_{jk}}{s_{ik}},\e \right)  + S_{jk}G\left(\frac{s_{ij}}{s_{ik}},\e \right) \\
    & \hspace{0.5cm}  + \Lambda_1(i^h,j,k^h) \bigg] A_3^0(i^h,j,k^h) 
    - \frac{\bz}{\e} A_3^0(i^h,j,k^h) 
    \\
    & \hspace{0.5cm} + S_{ij} \Af \frac{(s_{ijk}-\e s_{jk} )}{s_{ij}s_{ijk}} \frac{1}{2 (1-2\e)} 
    +  S_{jk} \Af \frac{(s_{ijk}-\e s_{ij} )}{s_{jk}s_{ijk}}\frac{1}{2 (1-2\e)} \, .
\end{split}
\label{eqn:A31}
\end{equation}

Integrating over the single unresolved antenna phase space (more details are given in Appendix~\ref{app:integration}), we yield the integrated antenna
\begin{align}
{\cal A}_3^1 (s_{ijk}) &= S_{ijk}^2\Biggl [
-\frac{1}{4\e^4}
-\frac{31}{12\e^3}
+\frac{1}{\e^2} \left(
-\frac{53}{8}
+\frac{11}{24}\pi^2
\right)
+\frac{1}{\e} \left(
-\frac{659}{24}
+\frac{22}{9}\pi^2
+\frac{23}{3}\zeta_3
\right) \nonumber \\
&\hspace{2.5cm}
+ \left(
-\frac{1345}{12}
+\frac{199}{24}\pi^2
+\frac{635}{18}\zeta_3
+\frac{13}{1440}\pi^4
\right)
+ \order{\e}\Biggr] \, .
\label{eq:A31int}
\end{align}
This expansion differs from $\calArvold$ in Eq.~(5.18) of Ref.~\cite{Gehrmann-DeRidder:2005btv}, starting from the rational part at $\order{1/\e}$. In a similar way to the constructed $\A$ in Chapter~\ref{chapter:paper2}, this is simply because the $A_3^0$ given in Eq.~\eqref{eq:A30} differs at $\order{\e}$ from $\Xold{A}$ of Ref.~\cite{Gehrmann-DeRidder:2005btv}. The choice of $A_3^0$ impacts the $\e$ poles of $\Arv$ at both the unintegrated and integrated levels, because $A_3^0$ factorises onto explicit $1/\e^2$ poles in Eq.~\eqref{eqn:A31}. If instead the original $\Xold{A}$, of Ref.~\cite{Gehrmann-DeRidder:2005btv}, is used in Eq.~\eqref{eqn:A31}, the integrated antenna in Eq.~\eqref{eq:A31intoldX3} contains exactly the same poles as $\calArvold$ in Eq.~(5.18) of Ref.~\cite{Gehrmann-DeRidder:2005btv} and differs only at $\order{\e^0}$. 

Similarly, for the sub-leading-colour $q \bar{q}$ antenna,
\begin{align}
    \Atrv(i^h,j,k^h) &= -\frac{\Af}{\e^2} \bigg[ S_{ij} G\left(\frac{s_{jk}}{s_{ik}},-\e \right)  + S_{jk}G\left(\frac{s_{ij}}{s_{ik}},-\e \right)  - \Lambda_1(i^h,j,k^h) \bigg] A_3^0(i^h,j,k^h) \nonumber \\
    &\hspace{1cm} - S_{ij} \Af \frac{(s_{ijk}-\e s_{jk} )}{s_{ij}s_{ijk}} \frac{1}{2 (1-2\e)} 
    - S_{jk} \Af \frac{(s_{ijk}-\e s_{ij} )}{s_{jk}s_{ijk}}\frac{1}{2 (1-2\e)} \, ,
\label{eqn:At31}
\end{align}
and after integration we find the expression,
\begin{align}
\widetilde{{\cal A}}_3^1 (s_{ijk}) &= S_{ijk}^2\Biggl [
+\frac{1}{\e^2} \left(
-\frac{5}{8}
+\frac{1}{6}\pi^2
\right)
+\frac{1}{\e} \left(
-\frac{19}{4}
+\frac{1}{4}\pi^2
+7\zeta_3
\right) \nonumber \\
&\hspace{2.5cm}
+ \left(
-\frac{447}{16}
+\frac{29}{16}\pi^2
+\frac{21}{2}\zeta_3
+\frac{7}{60}\pi^4
\right)
+ \order{\e}\Biggr] \, .
\label{eq:A31tint}
\end{align}
This expansion only differs from $\calAtrvold$ in Eq.~(5.19) of Ref.~\cite{Gehrmann-DeRidder:2005btv} at $\order{\e^0}$. In this case, the choice of $A_3^0$ does not impact the $\e$ poles of $\Atrv$ because they are at most $1/\e$ and the $A_3^0$ given in Eq.~\eqref{eq:A30} differs only at $\order{\e}$ from $\Xold{A}$ of Ref.~\cite{Gehrmann-DeRidder:2005btv}. 

For the quark-loop $q \bar{q}$ antenna, there are no unrenormalised unresolved limits and so the antenna is simply a renormalisation term:
\begin{equation}
    \Ahrv(i^h,j,k^h) = - \frac{\bzf}{\e}   A_3^0(i^h,j,k^h) \, .
\label{eqn:Ah31}
\end{equation}
The integrated version is given by
\begin{align}
\widehat{{\cal A}}_3^1 (s_{ijk}) &= S_{ijk}^2\Biggl [
+\frac{1}{3\e^3}
+\frac{1}{2\e^2}
+\frac{1}{\e} \left(
\frac{19}{12}
-\frac{7}{36}\pi^2
\right) \nonumber \\
&\hspace{2.5cm}
+ \left(
\frac{113}{24}
-\frac{7}{24}\pi^2
-\frac{25}{9}\zeta_3
\right)
+ \order{\e}\Biggr] \, ,
\label{eq:A31hint}
\end{align}
which only differs from $\calAhrvold$ in Eq.~(5.20) of Ref.~\cite{Gehrmann-DeRidder:2005btv} at $\order{\e^0}$. In this case, the choice of $A_3^0$ does not impact the $\e$ poles of $\Ahrv$ because they are at most $1/\e$ and the $A_3^0$ given in Eq.~\eqref{eq:A30} differs only at $\order{\e}$ from $\Xold{A}$ of Ref.~\cite{Gehrmann-DeRidder:2005btv}.

\subsection{Quark-Gluon Antennae}

As shown in Table~\ref{tab:X31}, there are six one-loop three-parton antennae with quark-gluon parents organised by colour structure: $\Drv$, $\Dtrv$, $\Dhrv$, $\Erv$, $\Etrv$ and $\Ehrv$. The antenna functions constructed here are directly related to the antenna functions given in Ref.~\cite{Gehrmann-DeRidder:2005btv} by 
\begin{align}
    \Drvold (i_q,j_g,k_{g}) &\sim \Drv(i_q^h,j_g,k_{g}^h)
    +\Dtrv (i_q^h,j_g,k_{g}^h) + (j \leftrightarrow k) \, , \label{eq:Doldnew} \\
    \Dhrvold (i_q,j_g,k_{\bar{q}}) &\sim \Dhrv (i_q^h,j_g,k_{g}^h) + \Dhrv (i_q^h,k_g,j_{g}^h) \, , \\
    \Ervold (i_q,j_{\bar{Q}},k_{Q}) &\sim \Erv(i_q^h,j_{\bar{Q}},k_{Q}^h) \, , \\
    \Etrvold (i_q,j_{\bar{Q}},k_{Q}) &\sim \Etrv(i_q^h,j_{\bar{Q}},k_{Q}^h) \, , \\ 
    \Ehrvold (i_q,j_{\bar{Q}},k_{Q}) &\sim \Ehrv(i_q^h,j_{\bar{Q}},k_{Q}^h) \, .
\end{align}
Note that $\Drvold$ was extracted from an effective Lagrangian describing heavy neutralino decay into a gluino-gluon pair, where the gluino plays the role of the quark~\cite{Gehrmann-DeRidder:2005svg}. 
Firstly, $\Drvold$ contains unresolved configurations where either of the gluons can be soft, so this is decomposed here such that only one gluon can be soft.
Secondly, the extracted antennae $\Drvold$ (and $\Dold$) contain both leading-colour and sub-leading colour limits and they receive special treatment in the antenna scheme. 
In Chapter~\ref{chapter:paper2} and Ref.~\cite{paper2}, we effectively split $\Dold$ into a combination of $\D$ and $\Dt$ antennae and we perform a similar decomposition here of $\Drvold$ into $\Drv$ and $\Dtrv$. 
Due to the absence of a sub-leading colour $\Dtrvold$ antenna, we only have target poles, $T(i_q^h,j_g,k_g^h)$, for the combination of $\Drv(i_q^h,j_g,k_g^h)+\Dtrv(i_q^h,j_g,k_g^h)$. We choose to place the resulting $\FinPoles_{\e}$ term in the formula for $\Drv$. 
To recap, the combination of $\Drv(i_q^h,j_g,k_g^h)+\Dtrv(i_q^h,j_g,k_g^h)$ have been used to match $\e$ poles in the existing antenna-subtraction-scheme, while $\Drv(i_q^h,j_g,k_g^h)$ contains the leading-colour limits when $j$ is unresolved and $\Dtrv(i_q^h,j_g,k_g^h)$ contains the sub-leading-colour limits when $j$ is unresolved. 
This means the two antennae, $\Drv$ and $\Dtrv$, could in principle be used independently in subtraction terms to cancel relevant one-loop unresolved limits but the $\e$-pole cancellation may require specific attention. 

%
 %
%
%
%
  %

%
%
%

The $\Drv$ formula is given by 
\begin{eqnarray}
\label{eqn:D31}
    \Drv(i^h,j,k^h) &=& \frac{\Af}{\e^2}\bigg[ -\Gamma(1-\e)\Gamma(1+\e) \frac{S_{ij}S_{jk}}{S_{ik}} \nonumber \\ 
    &&+S_{ij} G\left(\frac{s_{jk}}{s_{ik}},\e \right) +S_{jk} G\left(\frac{s_{ij}}{s_{ik}},\e \right)  - S_{jk}G\left(\frac{s_{ij}}{s_{ik}},-\e \right) \nonumber \\ 
    && + 2 \Lambda_1(i^h,j,k^h) \bigg] D_3^0 (i^h,j,k^h) 
    - \frac{\bz}{\e} D_3^0 (i^h,j,k^h) \nonumber
    \\ 
    && + S_{ij} \Af \frac{(s_{ijk}-\e s_{jk} )}{s_{ij}s_{ijk}} \frac{1}{2 (1-2\e)} \nonumber \\ 
    && + \frac{\Af}{2(1-\e)(1-2\e)(3-2\e)} \frac{S_{jk}}{s_{jk}} \left( 1-2 \e \frac{s_{ij}s_{ik}}{s_{ijk}^2} \right)  ,
\end{eqnarray}
and after integration we find the expression
\begin{eqnarray}
\label{eq:d31int}
{\cal D}_3^1 (s_{ijk}) &=& S_{ijk}^2\Biggl [
-\frac{1}{4\e^4}
-\frac{8}{3\e^3}
+\frac{1}{\e^2} \left(
-\frac{1109}{144}
+\frac{13}{24}\pi^2
\right)
+\frac{1}{\e} \left(
-\frac{14603}{432}
+\frac{49}{18}\pi^2
+\frac{73}{6}\zeta_3
\right)\nonumber \\
&& \hspace{2cm}
 + \left(
-\frac{7985}{54}
+\frac{8561}{864}\pi^2
+\frac{535}{12}\zeta_3
+\frac{79}{480}\pi^4
\right)
 + \order{\e}\Biggr].
\end{eqnarray}
	 %
The $\Dtrv$ formula is given by 
\begin{eqnarray}
\label{eqn:Dt31}
    \Dtrv(i^h,j,k^h) &=& -\frac{\Af}{\e^2} S_{ij} G\left(\frac{s_{jk}}{s_{ik}},-\e \right) D_3^0(i^h,j,k^h)  \nonumber \\ 
    && - S_{ij} \Af \frac{(s_{ijk}-\e s_{jk} )}{s_{ij}s_{ijk}} \frac{1}{2 (1-2\e)}  ,
\end{eqnarray}
and after integration we find the expression
\begin{eqnarray}
\label{eq:d31tint}
\widetilde{{\cal D}}_3^1 (s_{ijk}) &=& S_{ijk}^2\Biggl [
+\frac{1}{\e^2} \left(
-\frac{5}{16}
+\frac{1}{12}\pi^2
\right)
+\frac{1}{\e} \left(
-\frac{77}{48}
+\frac{11}{72}\pi^2
+\frac{5}{2}\zeta_3
\right)\nonumber \\
&& \hspace{2cm}
 + \left(
-\frac{983}{144}
+\frac{941}{864}\pi^2
+\frac{55}{12}\zeta_3
-\frac{7}{180}\pi^4
\right)
 + \order{\e}\Biggr]. \hspace{1cm}
\end{eqnarray}
The combination of $2(\calDrv + \calDtrv)$ differs from $\calDrvold$ in Eq.~(6.22) of Ref.~\cite{Gehrmann-DeRidder:2005btv}, starting from $\order{1/\e^2}$. 
In a similar way to the constructed $\D$ and $\Dt$ in Chapter~\ref{chapter:paper2}, this is simply because the $D_3^0$ given in Eq.~\eqref{eq:D30} differs at $\order{\e^0}$ from $\Xold{d}$ of Ref.~\cite{Gehrmann-DeRidder:2005btv}. 
The choice of $D_3^0$ impacts the $\e$ poles of $\Drv$ and $\Dtrv$ at both the unintegrated and integrated levels because $D_3^0$ factorises onto explicit $1/\e^2$ poles in Eq.~\eqref{eqn:D31} and Eq.~\eqref{eqn:Dt31}. 

The quark-loop $qg$ antenna function is given by
\begin{align}
    \Dhrv(i^h,j,k^h) &= -\Af \frac{S_{jk}}{s_{jk}} \frac{1}{2(1-\e)^2(1-2\e)(3-2\e)} \left(1-2 \e \frac{s_{ij}s_{ik}}{s_{ijk}^2} \right) \nonumber \\
    &\hspace{0.5cm} - \frac{\bzf}{\e}    D_3^0(i^h,j,k^h),
\label{eqn:Dh31}
\end{align}
and after integration we find the expression
\begin{align}
\widehat{{\cal D}}_3^1 (s_{ijk}) &= S_{ijk}^2\Biggl [
+\frac{1}{3\e^3}
+\frac{5}{9\e^2}
+\frac{1}{\e} \left(
\frac{125}{72}
-\frac{7}{36}\pi^2
\right)  + \left(
\frac{97}{18}
-\frac{35}{108}\pi^2
-\frac{25}{9}\zeta_3
\right) 
+ \order{\e}\Biggr] \, .
\label{eq:d31hint}
\end{align}
This expansion differs from $\calDhrvold/2$ in Eq.~(6.23) of Ref.~\cite{Gehrmann-DeRidder:2005btv}, starting from the rational part at $\order{1/\e}$. In the $\Dhrv$ formula, the poles are at most $1/\e$ and they only appear in the renormalisation term. The finite difference between $D_3^0$ and $\Xold{d}$ from Ref.~\cite{Gehrmann-DeRidder:2005btv} therefore only impacts the $\order{1/\e}$ poles. 

%

The $\Erv$-type antennae contain contributions to only one limit -- the $jk$ collinear limit, when the $\bar{Q}Q$ pair become collinear and as such they are simpler expressions than the others. The first antenna is given by
\begin{align}
    \Erv(i^h,j,k^h) &= - \frac{\Af}{\e^2} \bigg[ S_{jk} G\left(\frac{s_{ij}}{s_{ik}},-\e \right) + S_{jk} G\left(\frac{s_{ik}}{s_{ij}},-\e \right) \nonumber \\
    &\hspace{0.5cm} - S_{jk} \frac{\e (13 - 8 \e)}{2(3-2\e)(1-2\e)} - 2 \Lambda_2(i^h,j,k^h) \bigg] E_3^0 (i^h,j,k^h) 
    - \frac{\bz}{\e} E_3^0 (i^h,j,k^h)
    \, , 
\label{eqn:E31}
\end{align}
where the $\Lambda_2/\e^2$ term suppresses the only limit in the $E_3^0$ to which it factorises (the $jk$ collinear limit) and thus only affects the $\e$ pole structure of $\Erv$. 
After integration we find the expression
\begin{align}
{\cal E}_3^1 (s_{ijk}) &= S_{ijk}^2\Biggl [
+\frac{11}{18\e^2}
+\frac{1}{\e} \left(
\frac{56}{27}
-\frac{1}{9}\pi^2
\right) + \left(
\frac{4111}{432}
-\frac{131}{216}\pi^2
-4\zeta_3
\right)
+ \order{\e}\Biggr] \, ,
\label{eq:E31int}
\end{align}
which differs from $\calErvold$ in Eq.~(6.34) of Ref.~\cite{Gehrmann-DeRidder:2005btv}, starting from the rational part at $\order{1/\e}$. 
In the $\Erv$ formula, the poles are at most $1/\e$. 
The finite difference between $E_3^0$ and $\Xold{E}$ from Ref.~\cite{Gehrmann-DeRidder:2005btv} therefore only impacts the $\order{1/\e}$ poles. 

%

The sub-leading-colour antenna is given by
\begin{equation}
    \Etrv(i^h,j,k^h) = -\Af S_{jk} \bigg[ \frac{1}{\e^2} + \frac{(3+2\e)}{2 \e (1-2 \e)} \bigg] E_3^0 (i^h,j,k^h) \, , 
\label{eqn:Et31}
\end{equation}
and after integration we find the expression
\begin{align}
\widetilde{{\cal E}}_3^1 (s_{ijk}) &= S_{ijk}^2\Biggl [
+\frac{1}{6\e^3}
+\frac{13}{18\e^2}
+\frac{1}{\e} \left(
\frac{613}{216}
-\frac{1}{4}\pi^2
\right) + \left(
\frac{3359}{324}
-\frac{13}{12}\pi^2
-\frac{31}{9}\zeta_3
\right)
+ \order{\e}\Biggr],
\label{eq:E31tint}
\end{align}
which differs from $\calEtrvold$ in Eq.~(6.35) of Ref.~\cite{Gehrmann-DeRidder:2005btv}, starting from the rational part at $\order{1/\e^2}$. 
In the $\Etrv$ formula, the poles are at most $1/\e^2$. 
The finite difference between $E_3^0$ and $\Xold{E}$ from Ref.~\cite{Gehrmann-DeRidder:2005btv} therefore impacts the $\order{1/\e^2}$ poles. 

The quark-loop antenna is given by
\begin{equation}
\label{eqn:Eh31}
    \Ehrv(i^h,j,k^h) = - \Af \bigg[ S_{jk} \frac{2(1-\e)}{\e (3-2 \e)(1-2\e)} \bigg] E_3^0 (i^h,j,k^h)  
    - \frac{\bzf}{\e}  E_3^0 (i^h,j,k^h),
\end{equation}
and after integration we find the expression
\begin{eqnarray}
\label{eq:E31hint}
\widehat{{\cal E}}_3^1 (s_{ijk}) &=& S_{ijk}^2\Biggl [
+\frac{1}{4\e}
+ \left(
\frac{791}{648}
-\frac{11}{108}\pi^2
\right)
 + \order{\e}\Biggr],
\end{eqnarray}
which differs from $\calEhrvold$ in Eq.~(6.36) of Ref.~\cite{Gehrmann-DeRidder:2005btv}, starting from the rational part at $\order{1/\e}$. 
In the $\Ehrv$ formula, the poles are at most $1/\e$. 
The finite difference between $E_3^0$ and $\Xold{E}$ from Ref.~\cite{Gehrmann-DeRidder:2005btv} therefore impacts the $\order{1/\e}$ poles of $\calEhrv$, although these are the deepest poles in this case. 
This is because of cancellations at $\order{1/\e^2}$ between the integrals of the first and second terms in Eq.~\eqref{eqn:Eh31}. 

\subsection{Gluon-Gluon Antennae}

As shown in Table~\ref{tab:X31}, there are five one-loop three-parton antennae with gluon-gluon parents organised by colour structure: $\Frv$, $\Fhrv$, $\Grv$, $\Gtrv$ and $\Ghrv$. 
The antenna functions constructed here are directly related to the antenna functions given in Ref.~\cite{Gehrmann-DeRidder:2005btv} by 
\begin{align}
    \Frvold (i_g,j_g,k_{g}) &\sim \Frv (i_g^h,j_g,k_{g}^h) + \Frv (j_g^h,k_g,i_{g}^h) + \Frv (k_g^h,i_g,j_{g}^h) \, , \label{eq:Foldnew} \\
    \Fhrvold (i_g,j_g,k_{g}) &\sim \Fhrv (i_g^h,j_g,k_{g}^h) + \Fhrv (j_g^h,k_g,i_{g}^h) + \Fhrv (k_g^h,i_g,j_{g}^h) \, , \\ 
    \Grvold (i_g,j_{\bar{Q}},k_{Q}) &\sim \Grv (i_g^h,j_{\bar{Q}},k_{Q}^h) \, , \\
    \Gtrvold (i_g,j_{\bar{Q}},k_{Q}) &\sim \Gtrv (i_g^h,j_{\bar{Q}},k_{Q}^h) \, , \\
    \Ghrvold (i_g,j_{\bar{Q}},k_{Q}) &\sim \Ghrv (i_g^h,j_{\bar{Q}},k_{Q}^h) \, .
\end{align}
Note that $\Frvold$ was extracted from an effective Lagrangian describing Higgs boson decay into gluons~\cite{Gehrmann-DeRidder:2005alt}. 
This means that $\Frvold$ contains unresolved configurations where any one of the three gluons can be soft, so this is decomposed here such that only one gluon can be soft. 
The same discussion can be applied to $\Fhrv$. 

%
%

The resulting formula for the three-gluon one-loop antenna function at leading-colour is given by
\begin{align}
    \Frv(i^h,j,k^h) &= \frac{\Af}{\e^2}\bigg[ -\Gamma(1-\e)\Gamma(1+\e) \frac{S_{ij}S_{jk}}{S_{ik}} \nonumber \\
    &\hspace{1cm} + S_{ij} G\left(\frac{s_{jk}}{s_{ik}},\e \right) - S_{ij}G\left(\frac{s_{jk}}{s_{ik}},-\e \right) \nonumber \\
    &\hspace{1cm} + S_{jk} G\left(\frac{s_{ij}}{s_{ik}},\e \right) - S_{jk}G\left(\frac{s_{ij}}{s_{ik}},-\e \right) \label{eqn:F31}  \nonumber \\
    &\hspace{1cm} + 2 \Lambda_1(i^h,j,k^h) \bigg] F_3^0 (i^h,j,k^h) 
    - \frac{\bz}{\e} F_3^0 (i^h,j,k^h)
    \nonumber \\
    &\hspace{0.5cm} + \frac{\Af}{2(1-\e)(1-2\e)(3-2\e)} \bigg[ \frac{S_{ij}}{s_{ij}} \left(1-2 \e \frac{s_{jk}s_{ik}}{s_{ijk}^2} \right) + \frac{S_{jk}}{s_{jk}} \left( 1-2 \e \frac{s_{ij}s_{ik}}{s_{ijk}^2} \right) \bigg] \, , 
\end{align}
and after integration we find the expression
\begin{align}
{\cal F}_3^1 (s_{ijk}) &= S_{ijk}^2\Biggl [
-\frac{1}{4\e^4}
-\frac{11}{4\e^3}
+\frac{1}{\e^2} \left(
-\frac{79}{9}
+\frac{5}{8}\pi^2
\right)
+\frac{1}{\e} \left(
-\frac{8339}{216}
+\frac{55}{18}\pi^2
+\frac{44}{3}\zeta_3
\right) \nonumber \\
&\hspace{2.5cm}
+ \left(
-\frac{73169}{432}
+\frac{5137}{432}\pi^2
+\frac{473}{9}\zeta_3
+\frac{181}{1440}\pi^4
\right)
+ \order{\e}\Biggr] \, .
\label{eq:f31int}
\end{align}
This expansion differs from $\calFrvold/3$ in Eq.~(7.22) of Ref.~\cite{Gehrmann-DeRidder:2005btv}, starting from the rational part at $\order{1/\e^2}$. In the $\Frv$ formula, the poles are at most $1/\e^2$. The finite difference between $F_3^0$ and $\Xold{f}$ from Ref.~\cite{Gehrmann-DeRidder:2005btv} therefore impacts the $\order{1/\e^2}$ poles. 


The quark-loop antenna function is given by
\begin{align}
    \Fhrv(i^h,j,k^h) &= \frac{\Af}{2(1-\e)^2(1-2\e)(3-2\e)} \bigg[ \frac{S_{ij}}{s_{ij}} \left( 1-2 \e \frac{s_{jk}s_{ik}}{s_{ijk}^2} \right) \nonumber \\
    &\hspace{0.5cm} + \frac{S_{jk}}{s_{jk}} \left( 1-2 \e \frac{s_{ij}s_{ik}}{s_{ijk}^2} \right) \bigg] - \frac{\bzf}{\e}  F_3^0(i^h,j,k^h) \, ,
\label{eqn:Fh31}
\end{align}
and after integration we find the expression
\begin{align}
\widehat{{\cal F}}_3^1 (s_{ijk}) &= S_{ijk}^2\Biggl [
+\frac{1}{3\e^3}
+\frac{11}{18\e^2}
+\frac{1}{\e} \left(
\frac{17}{9}
-\frac{7}{36}\pi^2
\right) 
+ \left(
\frac{437}{72}
-\frac{77}{216}\pi^2
-\frac{25}{9}\zeta_3
\right)
+ \order{\e}\Biggr] \, .
\label{eq:f31hint}
\end{align}
This expansion differs from $\calFhrvold/3$ in Eq.~(7.23) of Ref.~\cite{Gehrmann-DeRidder:2005btv}, starting from the rational part at $\order{1/\e}$. In the $\Fhrv$ formula, the poles are at most $1/\e$ and they only appear in the renormalisation term. The finite difference between $F_3^0$ and $\Xold{f}$ from Ref.~\cite{Gehrmann-DeRidder:2005btv} therefore only impacts the $\order{1/\e}$ poles. 

%

The formula for the one-loop gluon-splitting $gg$ antenna function at leading-colour is given by
\begin{align}
    \Grv(i^h,j,k^h) &= - \frac{\Af}{\e^2} \bigg[ S_{jk} G\left(\frac{s_{ij}}{s_{ik}},-\e \right) + S_{jk} G\left(\frac{s_{ik}}{s_{ij}},-\e \right) \label{eqn:G31} \nonumber  \\
    &\hspace{1.3cm} - S_{jk} \frac{\e (13 - 8 \e)}{2(3-2\e)(1-2\e)} - 2 \Lambda_2(i^h,j,k^h) \bigg] G_3^0 (i^h,j,k^h) 
    - \frac{\bz}{\e} G_3^0 (i^h,j,k^h)\, , \nonumber \\
\end{align}
and after integration we find the expression
\begin{equation}
\label{eq:G31int}
{\cal G}_3^1 (s_{ijk}) = S_{ijk}^2\Biggl [
+\frac{11}{18\e^2}
+\frac{1}{\e} \left(
\frac{56}{27}
-\frac{1}{9}\pi^2
\right)
+ \left(
\frac{4111}{432}
-\frac{131}{216}\pi^2
-4\zeta_3
\right)
 + \order{\e}\Biggr].
\end{equation}
Firstly, given that $E_3^0 = G_3^0$ and that $\Erv$ and $\Grv$ encapsulate the same limits, these formulae (unintegrated and integrated) are identical for the $\Erv$- and $\Grv$- type antennae:
\begin{align}
    \Grv(i^h,j,k^h) &= \Erv(i^h,j,k^h) \, , \\ 
   \Gtrv(i^h,j,k^h) &= \Etrv(i^h,j,k^h) \, , \\ 
   \Ghrv(i^h,j,k^h) &= \Ehrv(i^h,j,k^h) \, .
\end{align}
Therefore the discussion for the $\Grv$-type antennae is the same as below Eq.~\eqref{eq:E31int}, Eq.~\eqref{eq:E31tint} and Eq.~\eqref{eq:E31hint}, respectively. 
When the  $X_3^{0,\text{OLD}}$ from Ref.~\cite{Gehrmann-DeRidder:2005btv} are used, the $\Grv$- and $\Erv$- type antennae have a different pole structure but the same collinear limits. 

%

The sub-leading-colour antenna function is given by
\begin{equation}
    \Gtrv(i^h,j,k^h) = -\Af S_{jk} \bigg[ \frac{1}{\e^2} + \frac{(3+2\e)}{2 \e (1-2 \e)} \bigg] G_3^0 (i^h,j,k^h) \, ,
\label{eqn:Gt31}
\end{equation}
and after integration we find the expression
\begin{align}
\widetilde{{\cal G}}_3^1 (s_{ijk}) &= S_{ijk}^2\Biggl [
+\frac{1}{6\e^3}
+\frac{13}{18\e^2}
+\frac{1}{\e} \left(
\frac{613}{216}
-\frac{1}{4}\pi^2
\right)
+ \left(
\frac{3359}{324}
-\frac{13}{12}\pi^2
-\frac{31}{9}\zeta_3
\right)
+ \order{\e}\Biggr]  \, .
\label{eq:G31tint}
\end{align}
See the discussion for Eq.~\eqref{eq:E31tint}, which also applies to Eq.~\eqref{eq:G31tint}. 


The quark-loop antenna function is given by
\begin{equation}
    \Ghrv(i^h,j,k^h) = - \Af \bigg[ S_{jk} \frac{2(1-\e)}{\e (3-2 \e)(1-2\e)} \bigg] G_3^0 (i^h,j,k^h)
    - \frac{\bzf}{\e}  G_3^0 (i^h,j,k^h) \, ,
\label{eqn:Gh31}
\end{equation}
and after integration we find the expression
\begin{equation}
\widehat{{\cal G}}_3^1 (s_{ijk}) = S_{ijk}^2\Biggl [
+\frac{1}{4\e}
+ \left(
\frac{791}{648}
-\frac{11}{108}\pi^2
\right)
+ \order{\e}\Biggr] \, .
\label{eq:G31hint}
\end{equation}
See the discussion for Eq.~\eqref{eq:E31hint}, which also applies to Eq.~\eqref{eq:G31hint}. 

\section{Antenna-Subtraction Scheme Consistency Checks}
\label{sec:J22}

In the antenna-subtraction scheme, the virtual (NLO) and double-virtual (NNLO) subtraction terms can be written in terms of integrated dipoles denoted by $\J{1}$ and $\J{2}$ respectively, see Ref.~\cite{Currie:2013vh}.  
These integrated dipoles are formed by systematically combining integrated antenna-function contributions from the real and real-virtual levels (together with appropriate mass factorisation terms in general). 
The NNLO integrated dipoles $\J{2}$ naturally emerge from groups of integrated antenna functions with the same parents at a given colour-level (and mass factorisation kernels) and, together with combinations of $\J{1}$, reproduce and properly subtract the explicit poles of the double-virtual contribution to the NNLO cross section. 
The integrated dipoles are therefore intimately related to Catani’s IR singularity operators (see Ref.~\cite{Catani:1998bh}) which describe the singularities of virtual matrix elements.  
It is a non-trivial check of an antenna scheme constructed directly from unresolved limits that the integrated dipoles cancel the explicit poles of the double-virtual contribution. 
In this section, we write down expressions for $\J{2}$ (and $\J{1}$) and show that they produce the correct pole structure.

We start from the expressions for the integrated dipoles in colour space (see Chapter~\ref{chapter:qcd} and Ref.~\cite{Chen:2022ktf}). 
In order to cancel the singularities of one- and two-loop matrix elements, 
$\Jtot{1}$ and $\Jtot{2}$ ($\Jtotb{2}$) must be related to
$\JtotOLD{1}$ and $\JtotOLD{2}$ ($\JtotbOLD{2}$), given in Ref.~\cite{Chen:2022clm}.
In particular, they must satisfy the following identities, which ensure that they match the known singularity structures at one and two loops.
At NLO,
\begin{equation}
\label{eq:J21I1}
    \Jtot{1} (i,j) = \JtotOLD{1} (i,j)  + \order{\e^0} ,
\end{equation}
and at NNLO,    
\begin{align}
\label{eq:J22I2}
    &\Jtot{2}\left(q,\bar{q}\right) - \frac{\Bz}{\e} \Jtot{1}\left(q,\bar{q}\right) = 
    \JtotOLD{2}\left(q,\bar{q}\right) - \frac{\Bz}{\e} \JtotOLD{1}\left(q,\bar{q}\right) + \mathcal{O}\left(\epsilon^0\right) \, , \\ 
    &\Jtot{2}\left(g,g\right) - \frac{\Bz}{\e}\Jtot{1}\left(g,g\right) =
    \JtotOLD{2}\left(g,g\right) - \frac{\Bz}{\e}\JtotOLD{1}\left(g,g\right)
    +\mathcal{O}\left(\epsilon^0\right) \, , \\
    &\Jtot{2}\left(q,g\right) + \Jtot{2}\left(g,\bar{q}\right) - 2\Jtotb{2}\left(q,\bar{q}\right) - \frac{\Bz}{\e}\left(\Jtot{1}\left(q,g\right) + \Jtot{1}\left(g,\bar{q}\right)\right) \nonumber\\
    &\hspace{3.075cm} = \JtotOLD{2}\left(q,g\right) + \JtotOLD{2}\left(g,\bar{q}\right) - 2\JtotbOLD{2}\left(q,\bar{q}\right) \nonumber \\
    \label{eq:J22I22}
    &\hspace{3.8cm} - \frac{\Bz}{\e}\left(\JtotOLD{1}\left(q,g\right) + \JtotOLD{1}\left(g,\bar{q}\right)\right)
    + \mathcal{O}\left(\epsilon^0\right) \, . 
\end{align}

Here we give the new definitions for the $\J{1}$ and $\J{2}$ pertaining to final-final configurations, 
which satisfy the above identities and are constructed from the integrated versions of $X_3^0$ and $X_4^0$ presented in Chapter~\ref{chapter:paper2}
and the integrated $X_3^1$ constructed in this chapter, 
thus completing the set of antenna functions required for the complete final-final NNLO subtraction scheme. 
The $\J{1}$ are defined by
\begin{eqnarray}
\label{eq:J21QQ}
\J{1} \left(1_{q},2_{\bar{q}}\right) &=& \cala(s_{12}), \\
\label{eq:J21QG}
\J{1}\left(1_{q},2_{g}\right) &=& \cald(s_{12}), \\
\label{eq:J21hQG}
\Jh{1}\left(1_{q},2_{g}\right) &=& \frac{1}{2}\cale(s_{12}), \\
\label{eq:J21GG}
\J{1}\left(1_{g},2_{g}\right) &=& \calf(s_{12}), \\
\label{eq:J21hGG}
\Jh{1}\left(1_{g},2_{g}\right) &=& \calg(s_{12}) ,
\end{eqnarray}
while the $\J{2}$ are defined below. 
For $q\bar{q}$ antennae, the integrated dipoles read
\begin{align}
\J{2} \left(1_{q},2_{\bar{q}}\right) &=
\calA(s_{12}) + \calArv(s_{12}) + \frac{\bz}{\e} \left(\frac{s_{12}}{\mu^2}\right)^{-\e} \cala(s_{12}) - \frac{1}{2} \left[ \cala \otimes \cala \right](s_{12}) \, ,
\label{eq:J22QQ} \\
\Jt{2} \left(1_{q},2_{\bar{q}}\right) &=
\frac{1}{2} \calAt(s_{12}) + 2\calC(s_{12}) + \calAtrv(s_{12}) - \frac{1}{2}\left[ \cala \otimes \cala \right](s_{12}) \, , \label{eq:J22tQQ} \\
\Jh{2} \left(1_{q},2_{\bar{q}}\right) &=
\calB(s_{12}) + \calAhrv(s_{12})+ \frac{\bzf}{\e} \left(\frac{s_{12}}{\mu^2}\right)^{-\e} \cala(s_{12}) \, , \label{eq:J22hQQ} \\
\Jb{2} \left(1_{q},2_{\bar{q}}\right) &=
\frac{1}{2}\calAt(s_{12}) + \calAtrv(s_{12}) - \frac{1}{2}\left[ \cala \otimes \cala \right] (s_{12})\, . \label{eq:J22barQQ}
\end{align}
For $qg$ antennae, the integrated dipoles are given by
\begin{align}
\J{2} \left(1_{q},2_{g}\right) &=
\calD(s_{12}) + \frac{1}{2} \calDt(s_{12}) + \calDrv(s_{12}) +\calDtrv(s_{12}) + \frac{\bz}{\e} \left(\frac{s_{12}}{\mu^2}\right)^{-\e} \cald (s_{12}) 
\nonumber \\
&\hspace{0.5cm}
- \left[ \cald \otimes \cald \right](s_{12}) \, , \label{eq:J22QG} \\
\Jh{2} \left(1_{q},2_{g}\right) &=
\calEa (s_{12})+ \calEb(s_{12}) + \calDhrv(s_{12}) + \frac{1}{2}\calErv (s_{12})
+ \frac{\bzf}{\e} \left(\frac{s_{12}}{\mu^2}\right)^{-\e} \cald (s_{12})\nonumber \\
&\hspace{0.5cm}
+ \frac{1}{2}\frac{\bz}{\e} \left(\frac{s_{12}}{\mu^2}\right)^{-\e} \cale(s_{12}) - \left[ \cald \otimes \cale \right](s_{12}) \, , \label{eq:J22hQG} \\
\Jhh{2}\left(1_{q},2_{g}\right) &= 
\frac{1}{2}\calEhrv(s_{12}) + 
\frac{1}{2} \frac{\bzf}{\e} \left(\frac{s_{12}}{\mu^2}\right)^{-\e} \cale(s_{12}) - \frac{1}{4}\left[ \cale \otimes \cale \right](s_{12}) \, , \label{eq:J22hhQG} \\
\Jht{2} \left(1_{q},2_{g}\right) &=
\frac{1}{2} \calEt(s_{12}) + \frac{1}{2} \calEtrv (s_{12})\, . \label{eq:J22htQG}
\end{align}
\newpage
Finally, for $gg$ antennae, the integrated dipoles are
\begin{align}
\J{2} \left(1_{g},2_{g}\right) &=
\calF(s_{12}) + \frac{1}{2} \calFt(s_{12}) + \calFrv(s_{12}) + \frac{\bz}{\e} \left(\frac{s_{12}}{\mu^2}\right)^{-\e} \calf(s_{12}) \nonumber \\
&\hspace{0.5cm}- \left[ \calf \otimes \calf \right] (s_{12})\, , \label{eq:J22GG} \\
\Jh{2} \left(1_{g},2_{g}\right) &=
\calGa(s_{12}) + 2\calGb(s_{12}) + \calFhrv(s_{12}) + \calGrv(s_{12}) 
+ \frac{\bzf}{\e} \left(\frac{s_{12}}{\mu^2}\right)^{-\e} \calf(s_{12}) \nonumber \\
&\hspace{0.5cm}
+ \frac{\bz}{\e} \left(\frac{s_{12}}{\mu^2}\right)^{-\e} \calg
(s_{12})
- 2\left[ \calf \otimes \calg \right] (s_{12})\, , \label{eq:J22hGG} \\
\Jhh{2} \left(1_{g},2_{g}\right) &=
\frac{1}{2}\calH (s_{12})+ \calGhrv (s_{12})+
 \frac{\bzf}{\e} \left(\frac{s_{12}}{\mu^2}\right)^{-\e} \calg (s_{12}) - \left[ \calg \otimes \calg \right] (s_{12})\, , \label{eq:J22hhGG} \\
\Jht{2}\left(1_{g},2_{g}\right) &=
\calGt(s_{12}) + \calGtrv (s_{12})\, . \label{eq:J22htGG}
\end{align}
Note that, apart from certain well understood rescalings (such as $(1/3){\cal F}_3^0 \mapsto {\cal F}_3^0$, $(1/2){\cal D}_3^0 \mapsto {\cal D}_3^0$ and so on), Eqs.~\eqref{eq:J21QQ}--\eqref{eq:J22htGG} have a very similar structure to those appearing in Ref.~\cite{Currie:2013vh}. 

We observe that Eqs.~\eqref{eq:J21QQ}--\eqref{eq:J21hGG} satisfy Eq.~\eqref{eq:J21I1}.  
This is no surprise since the integrated single-real antenna functions differ from those in Ref.~\cite{Gehrmann-DeRidder:2005svg} only in finite pieces.  

However, a residual dependence on the choice of single-real antennae is left in the construction of double-real antenna functions and the real-virtual antenna functions constructed in this chapter. 
The deepest $1/\e^4$ and $1/\e^3$ poles correspond to the universal unresolved behaviour and are identical but the $1/\e^2$ and $1/\e$ poles are potentially different.  This is understood as finite differences in the single-real antennae and should pose no issues in application to the antenna-subtraction scheme, when used with a consistent set of $X_3^0$, $X_4^0$ and $X_3^1$ antenna functions.
Indeed, this is the case and we find that Eqs.~\eqref{eq:J22QQ}--\eqref{eq:J22htGG} satisfy Eqs.~\eqref{eq:J22I2}--\eqref{eq:J22I22}, thereby demonstrating the consistency of the NNLO antenna subtraction scheme based on the antennae derived directly from the desired singular limits presented here and in Refs.~\cite{paper2,paper3}.

\section{Summary}
\label{sec:outlook2}

In this chapter, we have extended the algorithm to build real-radiation antenna functions presented in Chapter~\ref{chapter:paper2} to the case where explicit poles in $\e$ are present, pertaining to mixed real and virtual corrections in higher-order calculations.
As a proof of the applicability of our new method, we have explicitly derived all $X_3^1$ antenna functions describing real-virtual radiation. Together with the real- and double-real antenna functions, $X_3^0$ and $X_4^0$, derived in Chapter~\ref{chapter:paper2}, this completes the derivation of a consistent set of idealised antenna functions for NNLO calculations of processes with massless partons in the final state.

We have identified two complementary sets of design principles relating to mixed real and virtual antenna functions, which we dubbed generic and subtraction-scheme-dependent design principles. 
While the former ensures that each antenna function has judicious physical properties and obeys all unresolved limits, the latter matches the mixed real-virtual antennae onto the full antenna-subtraction scheme. 
We have explicitly verified the consistency of the method at NNLO by recalculating the integrated dipoles $\J{2}$ and $\J{1}$ in the new scheme. 
These have a direct relation to the general Catani IR singularity operators and therefore the pole structure of one- and two-loop matrix elements.

\chapter{Conclusions and Outlook}
\label{chapter:conc}

In this thesis, we have presented the research and background pertaining to the idealisation of antenna functions. 
Primarily, this involved detailed study of NNLO unresolved limits, in particular the triple-collinear splitting functions, and development of an algorithm to build antenna functions directly from the limits and $\e$-poles we want them to contain. 
This algorithm was explicitly realised for the construction of $X_3^0$, $X_4^0$ and $X_3^1$ antenna functions in the final-final configuration, which underpin a consistent antenna-subtraction framework at NNLO. 

In Chapter~\ref{chapter:qcd}, we introduced the foundational aspects of QCD necessary for the research presented here. 
In Chapter~\ref{chapter:pheno}, we highlighted the phenomenological context for the antenna-subtraction framework. 

Chapter~\ref{chapter:paper1} concerns the decomposition of the triple-collinear splitting functions into products of two $1 \to 2$ splitting functions and a truly $1 \to 3$ remainder function. 
This was achieved by a basis change, which organises terms in the triple-collinear splitting functions by suppressing internal simple-collinear limits with a trace structure. 
The decomposition enabled a full study of all the internal and external singularities encoded within the triple-collinear splitting functions. 
This exposed the intricate interplay between overlapping limits at NNLO. 
The most immediate benefit of this work was in identifying triple-collinear splitting functions with one hard particle; this formed a crucial input for the construction of idealised $X_4^0$ antenna functions, using the algorithm detailed in Chapter~\ref{chapter:paper2}. 
One could also envisage using the remainder functions to isolate the solely double unresolved limits in an NNLO subtraction scheme. 
For the purposes of improving antenna subtraction, it was not favourable to construct solely double unresolved antennae. 
The reason for this is related to the structure of the $d \hat{\sigma}^{T,b_2}$ term, which includes terms with the $X_3^1$ antenna functions. 
This contribution must be $\e$-finite and as such there are counterterms, labelled $J_X$ terms, which must be integrated counterparts of subtraction terms at the double-real level. 
These unintegrated terms have the structure of iterated products of colour-connected $X_3^0$ antenna functions and their only role is to remove iterated colour-connected single unresolved limits from the $X_4^0$ antenna functions. 
Thus the $X_4^0$ antenna functions must include single and double unresolved limits. An alternative reasoning is that the separation between the double unresolved and the iterated single unresolved parts of $X_4^0$ is not unique. 
Instead it is scheme dependent and we define the generic single unresolved structures to be $X_3^0$ in the antenna scheme. 
As such, the scheme-appropriate way to subtract iterated single unresolved limits from $X_4^0$ is by a product of two $X_3^0$ antenna functions.  

In Chapter~\ref{chapter:antennasub}, we discussed the antenna-subtraction method, both at NLO and NNLO, for the final-final configuration. 
While at NLO the subtraction terms take a generic simple form, at NNLO we require many types of subtraction term. 
For the double-real level, we need contributions to handle both the colour-connected, almost-colour-connected and colour-unconnected double unresolved limits, as well as the single unresolved limits. 
For the real-virtual level, we need contributions to handle both the single unresolved limits and the explicit $\e$-poles due to the virtual emission. 
For the double-virtual level, we need contributions to handle the various two-loop IR-divergent structures. 
Within the subtraction terms at the double-real level and the real-virtual level, we require counter-subtraction of spurious limits or $\e$-poles introduced by other terms. 
Since the subtraction procedure at NNLO is innately complex, we must seek to minimise the introduction of spurious limits and optimise the formulation of NNLO subtraction terms. 
Both of these missions are addressed by the colourful antenna-subtraction method and the construction of idealised antenna functions. 

The construction of idealised real-radiation antenna functions was the focus of Chapter~\ref{chapter:paper2}; this was extended to include virtual radiation in Chapter~\ref{chapter:paper3}. 
The algorithm presented here has been used to build a complete set of idealised $X_3^0$, $X_4^0$ and $X_3^1$ antenna functions for the final-final configuration.
These idealised antenna functions encompass the unresolved limits between two hard radiators, making them ideal candidates for generating antenna-subtraction terms. 
The extension of the algorithm, for constructing antenna functions with virtual radiation, consists in a prescription to fix the $\e$-poles at a second stage after fixing the unresolved limits. 
Some of the old antenna functions were already in an idealised form, however many of the $X_4^0$ have improved structures compared to their earlier forms. 
This is where we anticipate simplifications to the double-real subtraction terms for higher-multiplicity processes. 
Improvements to the antenna-subtraction architecture will be essential for going beyond the current state-of-the-art and addressing complicated processes such as $e^+e^- \to 4$ jets at NNLO. 
We believe that, apart from reducing the complexity of subtraction terms, the idealised antenna functions will reduce the computational overhead associated with precision calculations. 
Our assessment is based on the fact that we have chosen our design principles in such a way that they avoid the need for spurious subtraction terms as much as possible. 
The idealised antenna functions could also find their application in parton showers and their matching to NNLO calculations. 
From a more general point of view, the algorithm set-out here is general and could be used to construct various singular structures, including sector antenna functions, fragmentation antenna functions and massive antenna functions. 

In order to address precision phenomenology at hadron colliders such as the LHC, NNLO antenna functions for initial-final and initial-initial configurations are required. 
These can be constructed using the same algorithm set-out in Chapters~\ref{chapter:paper2} and~\ref{chapter:paper3}, using the known initial-final and initial-initial unresolved limits. 
For the cancellation of poles in virtual and double-virtual matrix elements, those antenna functions also need to be integrated over the respective initial-final and initial-initial antenna phase spaces. 
We note that a first step in this direction has recently been taken and the $X_3^0$ antennae for the initial-final and initial-initial configurations have been derived using this approach in Ref.~\cite{Fox:2023bma}. 

We plan for the idealised NNLO antennae to be integrated into the automated antenna framework, in conjunction with the colourful antenna-subtraction method. 
This improved antenna-subtraction scheme would streamline the NNLO QCD calculation of full-colour $pp \to 3$~jets as well as bring other processes such as $pp \to V$+2 jets, $pp \to H$+2 jets and $e^+e^- \to $ 4 jets into scope (as and when the two loop matrix elements become available). 
We can also anticipate interest in calculations for other colliders than the LHC, with simpler configurations, such as the Electron-Ion Collider (EIC), the International Linear Collider (ILC), the Circular Electron-Positron Collider (CEPC) and the initial runs of the Future Circular Collider (FCCee). 
The improved NNLO antenna-subtraction scheme will be equally valuable for providing phenomenological comparison during the runs of these experiments. 

Given the high complexity of the generic antenna-subtraction scheme at NNLO, it is not surprising that developments for an \NthreeLO antenna-subtraction scheme are in their infancy. 
Such a scheme would have to follow a formula like Eq.~\eqref{eq:N3LOsubtract}. 
$S_{RRR}$ would require building out of new $X_5^0$ antenna functions with three unresolved partons at tree-level, while also utilising $X_4^0$ and $X_3^0$ antenna functions. 
$S_{RRV}$ would require building out of new $X_4^1$ antenna functions with two unresolved partons at one-loop, while also utilising $X_4^0$, $X_3^1$ and $X_3^0$ antenna functions. 
$S_{RVV}$ would require building out of new $X_3^2$ antenna functions with one unresolved parton at two-loops, while also utilising $X_3^1$ and $X_3^0$ antenna functions. 
The IR structure of three-loop squared matrix elements is known, see Refs.~\cite{Sterman:2002qn,Becher:2009qa}, however it will be convenient to recast this structure as a set of $\J{3}$ integrated dipoles at three-loops. 
These will depend on the new antenna functions at \NthreeLO as well as the NNLO and NLO antennae.
Recently, these new \NthreeLO antenna functions and their integrals have been computed with the traditional approach, by using decays of a virtual photon~\cite{Jakubcik:2022zdi}, a heavy neutralino in the MSSM~\cite{Chen:2023egx} and a Higgs boson~\cite{Chen:2023fba}. 
Antenna subtraction at \NthreeLO, using these functions, should be viable for low multiplicity processes, since the antenna functions are extracted from the matrix elements they subtract against and are the perfect subtraction terms. 
However at higher multiplicities, formulating subtraction terms may become exceedingly complex, since the \NthreeLO antenna functions contain many overlapping unresolved limits. 
The current limitations of the traditionally extracted $X_4^{0, \text{OLD}}$, explored in Chapter~\ref{chapter:paper2}, will be greatly multiplied in $X_5^0$ and $X_4^1$ extracted from matrix elements. 

We therefore wish to stress that the algorithm presented in this thesis can straightforwardly be promoted to \NthreeLO calculations, provided that the appropriate unresolved limits are known analytically. 
For the case of constructing idealised $X_5^0$ antenna functions, we require a decomposition of the tree-level quadruple-collinear splitting functions, with one identified hard parton. 
However, the challenge of this task is not to be underestimated due to the great complexity in the quadruple-collinear limits, which include products of three iterated $1 \to 2$ splittings, products of one iterated $1 \to 3$ splitting with a $1 \to 2$ splitting and a truly $1 \to 4$ splitting. 
For the case of constructing idealised $X_4^1$ antenna functions, we require a decomposition of the one-loop triple-collinear splitting functions, with one identified hard parton. 
This is likely to require new techniques due to the handling of both implicit and explicit divergences. 
For the case of constructing idealised $X_3^2$ antenna functions, we require a decomposition of the two-loop simple-collinear splitting functions, with one identified hard parton. 
This can be approached in a similar way to the decomposition of the one-loop simple-collinear splitting functions, presented in Section~\ref{sec:limits}. 
There is also the question of where to fix the $\e$-poles of the $X_4^1$ and $X_3^2$ antenna functions. 
For the $X_3^1$, this consisted in matching them to other known elements in the NNLO subtraction scheme. 
We will require a larger balancing act at \NthreeLO to build a fully consistent subtraction scheme. 
As such, there is much work still to be done on improving the antenna-subtraction scheme but it will continue to play a key role in precision calculations for many years.

\appendix
\chapter{$X_4^0$ Appendix}

\section{$\X$ Limits}
\label{app:X40limits}

In this appendix we list, for convenience, the limits of all $\X$ constructed using the algorithm of Chapter~\ref{chapter:paper2}. 
If the result of an NNLO down-projector is not given for a certain $\X$, then its result is $0$.

\subsection*{$\A (1_q^h,2_g,3_g,4^h_{\bar{q}} )$}
\begin{eqnarray} 
\PSdown_{23} A_4^0 (1^h,2,3,4^h) &=& \Sgg(1^h,2,3,4^h) \\
\PTCdown_{123} A_4^0 (1^h,2,3,4^h) &=& \Pqgg(1^h,2,3) \\
\PTCdown_{234} A_4^0 (1^h,2,3,4^h) &=& \Pqgg(4^h,3,2) \\
\PDCdown_{1234} A_4^0 (1^h,2,3,4^h) &=&  \Pqg(1^h,2) \Pqg(4^h,3)\\
\PSdown_{2} A_4^0 (1^h,2,3,4^h) &=& \frac{2 s_{13}}{s_{12} s_{23}} A_3^0(1^h,3,4^h)\\
\PSdown_{3} A_4^0 (1^h,2,3,4^h) &=& \frac{2 s_{24}}{s_{23} s_{34}} A_3^0(1^h,2,4^h)\\
\PCdown_{12} A_4^0 (1^h,2,3,4^h) &=&  \Pqg(1^h,2) A_3^0((1+2)^h,3,4^h)\\
\PCdown_{23} A_4^0 (1^h,2,3,4^h) &=&  \Pgg(2,3) A_3^0(1^h,(2+3),4^h)\\
\PCdown_{34} A_4^0 (1^h,2,3,4^h) &=&  \Pqg(4^h,3) A_3^0(1^h,2,(3+4)^h)\\
\PSCdown_{2;34} A_4^0 (1^h,2,3,4^h) &=& \frac{2 s_{134}}{s_{12} s_{234}}  \Pqg(4^h,3)\\
\PSCdown_{3;12} A_4^0 (1^h,2,3,4^h) &=& \frac{2 s_{124}}{s_{123} s_{34}}  \Pqg(1^h,2)
\end{eqnarray}

\subsection*{$\At (1_q^h,2_\gamma,3_\gamma,4_{\bar{q}}^h )$}
\begin{eqnarray}       
\PSdown_{23} \At (1^h,2,3,4^h) &=&  \Spp(1^h,2,3,4^h)\\
\PTCdown_{123} \At (1^h,2,3,4^h) &=& \Pqpp(1^h,2,3)\\
\PTCdown_{234} \At (1^h,2,3,4^h) &=& \Pqpp(4^h,3,2)\\
\PDCdown_{1234} \At (1^h,2,3,4^h) &=&  \Pqg(1^h,2) \Pqg(4^h,3)\\
\PDCdown_{1324} \At (1^h,2,3,4^h) &=&  \Pqg(1^h,3) \Pqg(4^,2)\\
\PSdown_{2} \At (1^h,2,3,4^h) &=& \frac{2 s_{14}}{s_{12} s_{24}} A_3^0(1^h,3,4^h)\\
\PSdown_{3} \At (1^h,2,3,4^h) &=& \frac{2 s_{14}}{s_{13} s_{34}} A_3^0(1^h,2,4^h)\\
\PCdown_{12} \At (1^h,2,3,4^h) &=&  \Pqg(1^h,2) A_3^0((1+2)^h,3,4^h)\\
\PCdown_{13} \At (1^h,2,3,4^h) &=&  \Pqg(1^h,3) A_3^0((1+3)^h,2,4^h)\\
\PCdown_{24} \At (1^h,2,3,4^h) &=&  \Pqg(4^h,2) A_3^0(1^h,3,(2+4)^h)\\
\PCdown_{34} \At (1^h,2,3,4^h) &=&  \Pqg(4^h,3) A_3^0(1^h,2,(3+4)^h)\\
\PSCdown_{2;13} \At (1^h,2,3,4^h) &=& \frac{2 s_{14}}{s_{12} s_{24}}  \Pqg(1^h,3)\\
\PSCdown_{2;34} \At (1^h,2,3,4^h) &=& \frac{2 s_{14}}{s_{12} s_{24}}  \Pqg(4^h,3)\\
\PSCdown_{3;12} \At (1^h,2,3,4^h) &=& \frac{2 s_{14}}{s_{13} s_{34}}  \Pqg(1^h,2)\\
\PSCdown_{3;24} \At (1^h,2,3,4^h) &=& \frac{2 s_{14}}{s_{13} s_{34}}  \Pqg(4^h,2)
\end{eqnarray}

\subsection*{$\B (1_q^h,2_{\bar{Q}},3_Q,4_{\bar{q}}^h)$}
\begin{eqnarray}
\PSdown_{23} \B (1^h,2,3,4^h) &=& \Sqq(1^h,2,3,4^h)\\
\PTCdown_{123} \B (1^h,2,3,4^h) &=& \PqQQ(1^h,2,3)\\
\PTCdown_{234} \B (1^h,2,3,4^h) &=& \PqQQ(4^h,3,2)\\
\PCdown_{23} \B (1^h,2,3,4^h) &=&  \Pqq(2,3) A_3^0(1^h,(2+3),4^h)
\end{eqnarray}

\subsection*{$\C (1_q^h,2_{\bar{q}},3_q,4_{\bar{q}}^h)$}
\begin{eqnarray}
\PTCdown_{234} \C (1^h,2,3,4^h) &=& \frac{1}{2} \Pqqq(2,3,4) 
\end{eqnarray}

\subsection*{$\D (1_q^h,2_g,3_g,4_g^h)$}
\begin{eqnarray}
\PSdown_{23} \D (1^h,2,3,4^h) &=& \Sgg(1^h,2,3,4^h)\\
\PTCdown_{123} \D (1^h,2,3,4^h) &=& \Pqgg(1^h,2,3)\\
\PTCdown_{234} \D (1^h,2,3,4^h) &=& \Pggg(4^h,3,2)\\
\PDCdown_{1234} \D (1^h,2,3,4^h) &=&  \Pqg(1^h,2) \Pgg(4^h,3)\\
\PSdown_{2} \D (1^h,2,3,4^h) &=&   \frac{2 s_{13}}{s_{12} s_{23}} D_3^0(1^h,3,4^h)\\
\PSdown_{3} \D (1^h,2,3,4^h) &=&   \frac{2 s_{24}}{s_{23} s_{34}} D_3^0(1^h,2,4^h)\\
\PCdown_{12} \D (1^h,2,3,4^h) &=&  \Pqg(1^h,2) D_3^0((1+2)^h,3,4^h)\\
\PCdown_{23} \D (1^h,2,3,4^h) &=&  \Pgg(2,3) D_3^0(1^h,(2+3),4^h)\\ 
\PCdown_{34} \D (1^h,2,3,4^h) &=&  \Pgg(4^h,3) D_3^0(1^h,2,(3+4)^h)\\
\PSCdown_{2;34} \D (1^h,2,3,4^h) &=& \frac{2 s_{134}}{s_{12} s_{234}}  \Pgg(4^h,3)\\
\PSCdown_{3;12} \D (1^h,2,3,4^h) &=& \frac{2 s_{124}}{s_{123} s_{34}}  \Pqg(1^h,2)
\end{eqnarray}

\subsection*{$\Dt (1_q^h,2_g,3_g,4_g^h)$}
\begin{eqnarray}  
\PSdown_{23} \Dt (1^h,2,3,4^h) &=& \Spp(1^h,2,3,4^h)\\
\PTCdown_{123} \Dt (1^h,2,3,4^h) &=& \Pqpp(1^h,2,3)\\
\PTCdown_{234} \Dt (1^h,2,3,4^h) &=& \Pggg(3,4^h,2)\\
\PDCdown_{1234} \Dt (1^h,2,3,4^h) &=&  \Pqg(1^h,2)  \Pgg(4^h,3)\\
\PDCdown_{1324} \Dt (1^h,2,3,4^h) &=&  \Pqg(1^h,2)  \Pgg(4^h,2)\\
\PSdown_{2} \Dt (1^h,2,3,4^h) &=&   \frac{2 s_{14}}{s_{12} s_{24}} D_3^0(1^h,3,4^h)\\
\PSdown_{3} \Dt (1^h,2,3,4^h) &=&   \frac{2 s_{14}}{s_{13} s_{34}} D_3^0(1^h,2,4^h)\\
\PCdown_{12} \Dt (1^h,2,3,4^h) &=&   \Pqg(1^h,2) D_3^0((1+2)^h,3,4^h)\\
\PCdown_{13} \Dt (1^h,2,3,4^h) &=&   \Pqg(1^h,3) D_3^0((1+3)^h,2,4^h)\\
\PCdown_{24} \Dt (1^h,2,3,4^h) &=&   \Pgg(4^h,2) D_3^0(1^h,3,(2+4)^h)\\ 
\PCdown_{34} \Dt (1^h,2,3,4^h) &=&   \Pgg(4^h,3) D_3^0(1^h,2,(3+4)^h)\\
\PSCdown_{2;34} \Dt (1^h,2,3,4^h) &=& \frac{2 s_{134}}{s_{12} s_{234}}  \Pgg(4^h,3)\\
\PSCdown_{3;24} \Dt (1^h,2,3,4^h) &=& \frac{2 s_{124}}{s_{13} s_{234}}  \Pgg(4^h,2)\\  
\PSCdown_{3;12} \Dt (1^h,2,3,4^h) &=& \frac{2 s_{124}}{s_{34} s_{123}}  \Pqg(1^h,2)\\
\PSCdown_{2;13} \Dt (1^h,2,3,4^h) &=& \frac{2 s_{134}}{s_{24} s_{123}}  \Pqg(1^h,3)
\end{eqnarray}

\subsection*{$\Ea (1_q^h,2_{\bar{Q}},3_Q,4_g^h)$}     
\begin{eqnarray}  
\PSdown_{23} \Ea (1^h,2,3,4^h) &=& \Sqq(1^h,2,3,4^h)\\
\PTCdown_{123} \Ea (1^h,2,3,4^h) &=& \PqQQ(1^h,2,3)\\
\PTCdown_{234} \Ea (1^h,2,3,4^h) &=& \Pgqbq(4^h,3,2)\\   
\PCdown_{23} \Ea (1^h,2,3,4^h) &=&  \Pqq(2,3) D_3^0(1^h,(2+3),4^h)
\end{eqnarray}
  
\subsection*{$\Eb (1_q^h,2_g,3_{\bar{Q}},4_Q^h)$}    
\begin{eqnarray}       
\PTCdown_{234} \Eb (1^h,2,3,4^h) &=& \Pgqbq(2,3,4^h)\\      
\PDCdown_{1234} \Eb (1^h,2,3,4^h) &=&  \Pqq(4^h,3) \Pqg(1^h,2)\\            
\PSdown_{2} \Eb (1^h,2,3,4^h) &=& \frac{2 s_{13}}{s_{12} s_{23}} E_3^0(1^h,3,4^h)\\       
\PCdown_{12} \Eb (1^h,2,3,4^h) &=&  \Pqg(1^h,2) E_3^0((1+2)^h,3,4^h)\\
\PCdown_{23} \Eb (1^h,2,3,4^h) &=&  \Pqg(3,2) E_3^0(1^h,(2+3),4^h)\\
\PCdown_{34} \Eb (1^h,2,3,4^h) &=&  \Pqq(4^h,3) D_3^0(1^h,2,(3+4)^h)\\  
\PSCdown_{2;34} \Eb (1^h,2,3,4^h) &=& \frac{2 s_{13}}{s_{12} s_{23}}  \Pqq(4^h,3)
\end{eqnarray}
         
\subsection*{$\Et (1_q^h,2_{\bar{Q}},3_g,4_Q^h)$} 
\begin{eqnarray}       
\PTCdown_{234} \Et (1^h,2,3,4^h) &=& \Ppqbq(4^h,3,2) \\     
\PSdown_{3} \Et (1^h,2,3,4^h) &=& \frac{2 s_{24}}{s_{23} s_{34}} E_3^0(1^h,2,4^h) \\
\PCdown_{23} \Et (1^h,2,3,4^h) &=&  \Pqg(2,3) E_3^0(1^h,(2+3),4^h) \\
\PCdown_{34} \Et (1^h,2,3,4^h) &=&  \Pqg(4^h,3) E_3^0(1^h,2,(3+4)^h) 
\end{eqnarray}    
         
\subsection*{$\F (1_g^h,2_g,3_g,4_g^h)$}      
\begin{eqnarray}  
\PSdown_{23} \F (1^h,2,3,4^h) &=& \Sgg(1^h,2,3,4^h)\\
\PTCdown_{123} \F (1^h,2,3,4^h) &=& \Pggg(1^h,2,3)\\
\PTCdown_{234} \F (1^h,2,3,4^h) &=& \Pggg(4^h,3,2)\\
\PDCdown_{1234} \F (1^h,2,3,4^h) &=&  \Pgg(1^h,2) \Pgg(4^h,3)\\  
\PSdown_{2} \F (1^h,2,3,4^h) &=& \frac{2 s_{13}}{s_{12} s_{23}} F_3^0(1^h,3,4^h)\\
\PSdown_{3} \F (1^h,2,3,4^h) &=& \frac{2 s_{24}}{s_{23} s_{34}} F_3^0(1^h,2,4^h)\\   
\PCdown_{12} \F (1^h,2,3,4^h) &=&  \Pgg(1^h,2) F_3^0((1+2)^h,3,4^h)\\
\PCdown_{23} \F (1^h,2,3,4^h) &=&  \Pgg(2,3) F_3^0(1^h,(2+3),4^h)\\
\PCdown_{34} \F (1^h,2,3,4^h) &=&  \Pgg(4^h,3) F_3^0(1^h,2,(3+4)^h)\\    
\PSCdown_{2;34} \F (1^h,2,3,4^h) &=& \frac{2 s_{134}}{s_{12} s_{234}}  \Pgg(4^h,3)\\
\PSCdown_{3;12} \F (1^h,2,3,4^h) &=& \frac{2 s_{124}}{s_{123} s_{34}}  \Pgg(1^h,2)
\end{eqnarray}
       
\subsection*{$\Ft (1_g^h,2_g,3_g,4_g^h)$}
\begin{eqnarray}  
\PSdown_{23} \Ft (1^h,2,3,4^h) &=& \Spp(1^h,2,3,4^h)\\ 
\PTCdown_{123} \Ft (1^h,2,3,4^h) &=& \Pggg(2,1^h,3)\\
\PTCdown_{234} \Ft (1^h,2,3,4^h) &=& \Pggg(3,4^h,2)\\
\PDCdown_{1234} \Ft (1^h,2,3,4^h) &=&  \Pgg(1^h,2)  \Pgg(4^h,3)\\
\PDCdown_{1324} \Ft (1^h,2,3,4^h) &=&  \Pgg(1^h,3)  \Pgg(4^h,2)\\
\PSdown_{2} \Ft (1^h,2,3,4^h) &=&   \frac{2 s_{14}}{s_{12} s_{24}} F_3^0(1^h,3,4^h)\\
\PSdown_{3} \Ft (1^h,2,3,4^h) &=&   \frac{2 s_{14}}{s_{13} s_{34}} F_3^0(1^h,2,4^h)\\ 
\PCdown_{12} \Ft (1^h,2,3,4^h) &=&   \Pgg(1^h,2) F_3^0((1+2)^h,3,4^h)\\
\PCdown_{13} \Ft (1^h,2,3,4^h) &=&   \Pgg(1^h,3) F_3^0((1+3)^h,2,4^h)\\
\PCdown_{24} \Ft (1^h,2,3,4^h) &=&   \Pgg(4^h,2) F_3^0(1^h,3,(2+4)^h)\\ 
\PCdown_{34} \Ft (1^h,2,3,4^h) &=&   \Pgg(4^h,3) F_3^0(1^h,2,(3+4)^h)\\
\PSCdown_{2;34} \Ft (1^h,2,3,4^h) &=& \frac{2 s_{134}}{s_{12} s_{234}}  \Pgg(4^h,3)\\
\PSCdown_{3;24} \Ft (1^h,2,3,4^h) &=& \frac{2 s_{124}}{s_{13} s_{234}}  \Pgg(4^h,2)\\
\PSCdown_{3;12} \Ft (1^h,2,3,4^h) &=& \frac{2 s_{124}}{s_{34} s_{123}}  \Pgg(1^h,2)\\
\PSCdown_{2;13} \Ft (1^h,2,3,4^h) &=& \frac{2 s_{134}}{s_{24} s_{123}}  \Pgg(1^h,3)
\end{eqnarray}

\subsection*{$\Ga (1_g^h,2_{\bar{Q}},3_Q,4_g^h)$} 
\begin{eqnarray}  
\PSdown_{23} \Ga (1^h,2,3,4^h) &=& \Sqq(1^h,2,3,4^h)\\
\PTCdown_{123} \Ga (1^h,2,3,4^h) &=& \Pgqbq(1^h,2,3)\\
\PTCdown_{234} \Ga (1^h,2,3,4^h) &=& \Pgqbq(4^h,3,2)\\
\PCdown_{23} \Ga (1^h,2,3,4^h) &=&  \Pqq(2,3) F_3^0(1^h,(2+3),4^h)
\end{eqnarray}
 
\subsection*{$\Gb (1_g^h,2_g,3_{\bar{Q}},4_Q^h)$}       
\begin{eqnarray}  
\PTCdown_{234} \Gb (1^h,2,3,4^h) &=& \Pgqbq(2,3,4^h)\\
\PDCdown_{1234} \Gb (1^h,2,3,4^h) &=&  \Pgg(1^h,2) \Pqq(4^h,3)\\ 
\PSdown_{2} \Gb (1^h,2,3,4^h) &=& \frac{2 s_{13}}{s_{12} s_{23}} G_3^0(1^h,3,4^h)\\ 
\PCdown_{12} \Gb (1^h,2,3,4^h) &=&  \Pgg(1^h,2) G_3^0((1+2)^h,3,4^h)\\
\PCdown_{23} \Gb (1^h,2,3,4^h) &=&  \Pqg(3,2) G_3^0(1^h,(2+3),4^h)\\
\PCdown_{34} \Gb (1^h,2,3,4^h) &=&  \Pqq(4^h,3) F_3^0(1^h,2,(3+4)^h)\\  
\PSCdown_{2;34} \Gb (1^h,2,3,4^h) &=& \frac{2 s_{13}}{s_{12} s_{23}}  \Pqq(4^h,3)
\end{eqnarray}

\subsection*{$\Gt (1_g^h,2_{\bar{Q}},3_g,4_Q^h)$}
\begin{eqnarray}
\PTCdown_{234} \Gt (1^h,2,3,4^h) &=& \Ppqbq(4^h,3,2) \\          
\PSdown_{3} \Gt (1^h,2,3,4^h) &=& \frac{2 s_{24}}{s_{23} s_{34}} G_3^0(1^h,2,4^h)\\    
\PCdown_{23} \Gt (1^h,2,3,4^h) &=&  \Pqg(2,3) G_3^0(1^h,(2+3),4^h) \\
\PCdown_{34} \Gt (1^h,2,3,4^h) &=&  \Pqg(4^h,3) G_3^0(1^h,2,(3+4)^h) 
\end{eqnarray}

\subsection*{$\H (1_{\bar{q}}^h,2_q,3_{\bar{Q}},4_Q^h)$}
\begin{eqnarray} 
\PDCdown_{1234} \H (1^h,2,3,4^h) &=&  \Pqq(1^h,2) \Pqq(4^h,3) \\
\PCdown_{12} \H (1^h,2,3,4^h) &=&  \Pqq(1^h,2) G_3^0((1+2)^h,3,4^h) \\
\PCdown_{34} \H (1^h,2,3,4^h) &=&  \Pqq(4^h,3) G_3^0((3+4)^h,1,2^h) 
\end{eqnarray}  

\section{Integrals of Single-Unresolved Contributions}
\label{app:X30regionintegrations}
In this appendix we list the integration over the single-unresolved final-final phase space for the different contributions of the $X_3^0$ antennae.

The integration of the soft contribution $\Ssoft$ is simply the integral of the soft eikonal $\Sg$ and is the same for all antennae that contain a soft limit,
\begin{equation}
\begin{split}
\calSsoft(i^h,j_g,k^h)
&= S_{ijk} \left(
\frac{1}{\e^2}
+\frac{2}{\e}
+6 - \frac{7}{12}\pi^2
+\e\left(18-\frac{25}{3}\zeta_3-\frac{7}{6}\pi^2\right) \right. \\
&\left. \hspace{2.2cm}
+\e^2\left(54-\frac{50}{3}\zeta_3-\frac{7}{2}\pi^2-\frac{71}{1440}\pi^4
\right) 
+{\cal O}(\e^3) 
\right) \, .
\end{split}
\end{equation}
The integrals of the three different collinear remainders $\Scol$ (with the soft contribution subtracted) are,
\begin{align}
\calScol(i_q^h,j_g;k^h)
&= S_{ijk} \left(
-\frac{1}{4\e}
-\frac{5}{8}
+\e\left(-\frac{31}{16}+\frac{7}{48}\pi^2\right) \right.\\
& \left. \hspace{2.2cm}
+ \e^2\left(-
\frac{189}{32}+\frac{25}{12}\zeta_3+\frac{35}{96}\pi^2\right) 
+{\cal O}(\e^3) \right) \, , \nonumber\\
\calScol(i_g^h,j_g;k^h)
&= S_{ijk} \left(
-\frac{1}{12\e}
-\frac{7}{24}
+\e\left(-\frac{15}{16}+\frac{7}{144}\pi^2\right) \right.\\
& \left. \hspace{2.2cm}
+ \e^2\left(-
\frac{93}{32}+\frac{25}{36}\zeta_3+\frac{49}{288}\pi^2\right) 
+{\cal O}(\e^3) \right) \, , \nonumber \\
\calScol(i_{Q}^h,j_{\Qb};k^h)
&=
S_{ijk}
\left( - \frac{1}{3\e}
 -\frac{3}{4}
+\left(-\frac{15}{8}+\frac{7\pi^2}{36}\right)\e
\right.  \\
& \left. \hspace{2.2cm}
+\left(-\frac{81}{16}+\frac{7\pi^2}{16}+\frac{25\zeta_3}{9} \right)\e^2
+ \order{\e^3} \right) \, . \nonumber
\end{align}

Note that in all of the above formulae arguments merely serve the purpose of identifying the particle content and the hard radiators. There is no dependence on the particle momenta in the integrated contributions besides the overall normalisation to the invariant mass $s_{ijk}$.

\section{Integrals of Double-Unresolved Contributions}
\label{app:X40regionintegrations}
In this appendix we list the integration over the double-unresolved final-final phase space for the different contributions to the $X_4^0$ antennae. These contributions do not depend on the form chosen for the single-real antenna functions.

\begingroup
\allowdisplaybreaks
The integrals of the double-soft contribution $\Dsoft$ are simply given by the integration of the double-soft factors $\Sgg$, $\Spp$ and $\Sqq$,
\begin{align}
\calDsoft(i^h,j_g,k_g,l^h) &= 
S_{ijkl}^2
\left[+\frac{3}{4\e^4}
+\frac{89}{24\e^3}
+\frac{1}{\e^2}\left(\frac{599}{36} - \pi^2\right)
\right.\nonumber\\
&\left.\hspace{2.2cm}
+ \frac{1}{\e}\left(\frac{7705}{108} - \frac{787}{144}\pi^2 - \frac{53}{4}\zeta_3\right)
\right. \\
& \left.\hspace{2.2cm}
+ \left(\frac{195547}{648} - \frac{2705}{108}\pi^2 - \frac{3371}{36}\zeta_3 + \frac{199}{480}\pi^4\right)
+ \order{\e}\right] \, , \nonumber\\
\calDsoft(i^h,j_\gamma,k_\gamma,l^h) &= 
S_{ijkl}^2 
\left[+\frac{1}{\e^4}
+\frac{4}{\e^3}
+\frac{1}{\e^2}\left(18 - \frac{3}{2}\pi^2\right)
+\frac{1}{\e}\left(76 - 6\pi^2 - \frac{74}{3}\zeta_3\right)
\right. \nonumber\\
& \left.\hspace{2.2cm}
+\left(312 - 27\pi^2 - \frac{308}{3}\zeta_3 + \frac{49}{120}\pi^4\right)
+ \order{\e}\right] \, , \\
\calDsoft(i^h,j_q,k_{\qb},l^h) &= 
S_{ijkl}^2 
\left[-\frac{1}{12\e^3}
-\frac{17}{36\e^2}
+\frac{1}{\e}\left(-\frac{277}{108} + \frac{11}{72}\pi^2\right)
\right. \nonumber\\
& \left.\hspace{2.2cm}
+\left(-\frac{4199}{324} + \frac{169}{216}\pi^2 + \frac{67}{18}\zeta_3\right)
+ \order{\e}\right] \, .
\end{align}

The integrals over the nine triple-collinear remainders (with the double-soft contribution subtracted) are,
\begin{align}
\calTcol(i_g^h,j_g,k_g;l^h) &= 
S_{ijkl}^2 
\left[-\frac{1}{4\e^3}
+\frac{1}{\e^2}\left(-\frac{499}{288} - \frac{1}{24}\pi^2\right)
+\frac{1}{\e}\left(-\frac{1757}{192} + \frac{5}{18}\pi^2 - 2\zeta_3\right)
\right. \nonumber\\
& \left.\hspace{2.2cm}
+\left(-\frac{440147}{10368} + \frac{1363}{576}\pi^2 + \frac{11}{12}\zeta_3 - \frac{1}{18}\pi^4\right)
+\order{\e}\right] \, , \\
\calTcol(i_g,j_g^h,k_g;l^h) &= S_{ijkl}^2 \left [-\frac{1}{6\e^3}-\frac{9}{16\e^2}+\frac{1}{\e} \left(\frac{413}{864}+\frac{1}{4}\pi^2-2\zeta_3\right) 
\right. \nonumber\\
& \left.\hspace{2.2cm}
+ \left(\frac{25565}{1728}+\frac{79}{96}\pi^2+\frac{4}{9}\zeta_3-\frac{1}{6}\pi^4\right) + \order{\e}\right] \, , \\
\calTcol(i_q^h,j_g,k_g;l^h) &=
S_{ijkl}^2  
\left[-\frac{1}{2\e^3}
+\frac{1}{\e^2}\left(-\frac{187}{96} - \frac{1}{24}\pi^2\right)
+\frac{1}{\e}\left(-\frac{5185}{576} + \frac{37}{48}\pi^2 - \frac{9}{4}\zeta_3\right)
\right. \nonumber\\
& \left.\hspace{2.2cm}
+\left(-\frac{141871}{3456} + \frac{1685}{576}\pi^2 + \frac{347}{24}\zeta_3 - \frac{7}{90}\pi^4\right)
+\order{\e}\right] \, , \\
\calTcol(i_q^h,j_\gamma,k_\gamma;l^h) &=
S_{ijkl}^2 
\left[-\frac{1}{2\e^3}
-\frac{43}{16\e^2}
+\frac{1}{\e}\left(-\frac{377}{32} + \frac{3}{4}\pi^2 -\zeta_3\right)
\right. \nonumber\\
& \left.\hspace{2.2cm}
+\left(-\frac{3003}{64} + \frac{129}{32}\pi^2 + \frac{34}{3}\zeta_3 - \frac{1}{12}\pi^4\right)
+\order{\e}\right] \, , \\
\calTcol(i_q^h,j_g,k_{\qb};l^h) &= 
S_{ijkl}^2 
\left[-\frac{1}{6\e^3}
-\frac{35}{36\e^2}
+\frac{1}{\e}\left(-\frac{277}{54} + \frac{1}{4}\pi^2\right)
\right. \nonumber\\
& \left.\hspace{2.2cm}
+\left(-\frac{7967}{324} + \frac{35}{24}\pi^2 + \frac{40}{9}\zeta_3\right)
+\order{\e}\right]
\, , \\
\calTcol(i_g^h,j_q,k_{\qb};l^h) &= 
S_{ijkl}^2 
\left[+\frac{1}{72\e^2}
+\frac{1}{\e}\left(-\frac{23}{288} + \frac{1}{18}\pi^2\right)
\right. \nonumber\\
& \left.\hspace{2.2cm}
+\left(-\frac{15857}{5184} + \frac{23}{72}\pi^2+3\zeta_3\right)
+\order{\e}\right]
\, , \\
\calTcol(i_{\qb}^h,j_q,k_{g};l^h) &= 
S_{ijkl}^2 
\left[-\frac{1}{3\e^3}-\frac{9}{8\e^2}
+\frac{1}{\e} \left(-\frac{2927}{864}+\frac{1}{2}\pi^2\right)
\right. \nonumber\\
& \left.\hspace{2.2cm}
+ \left(-\frac{16127}{1728}+\frac{27}{16}\pi^2+\frac{80}{9}\zeta_3\right) 
+ \order{\e}\right]
\, , \\
\calTcol(i_q^h,j_{\Qb},k_{Q};l^h) &= 
S_{ijkl}^2 
\left[+\frac{1}{24\e^2}
+\frac{31}{144\e}
+\left(\frac{395}{432} - \frac{1}{18}\pi^2\right)
+\order{\e}\right] \, ,\\
\calTcol(i_q^h,j_{\qb},k_{q};l^h) &=
S_{ijkl}^2 
\left[+\frac{1}{\e}\left(-\frac{13}{32}
+\frac{1}{16}\pi^2
-\frac{1}{4}\zeta_3\right) 
\right. \nonumber\\
& \left.\hspace{2.2cm}
+\left(-\frac{73}{16}
+\frac{23}{96}\pi^2
+\frac{23}{8}\zeta_3
-\frac{1}{45}\pi^4\right)
+\order{\e}\right] \, .
\end{align}

The integrals over the nine double-collinear remainders (with the double-soft and triple-collinear contributions subtracted) are,
\begin{align}
\calDcol(i_q^h,j_{g};k_{g},l_{\qb}^h) &= 
S_{ijkl}^2 \left [-\frac{3}{16\e^2}-\frac{21}{16\e}+
\left(-\frac{55}{8}+\frac{9}{32}\pi^2\right) + \order{\e}\right],\\
\calDcol(i_q^h,j_{\gamma};k_{\gamma},l_{\qb}^h)  &=
S_{ijkl}^2 \left [+\frac{3}{16\e^2}+\frac{17}{16\e}+
\left(\frac{21}{4}-\frac{9}{32}\pi^2\right) + \order{\e}\right],\\
\calDcol(i_q^h,j_{g};k_{g},l_{g}^h) &= 
S_{ijkl}^2 \left
[+\frac{23}{72\e^2}+\frac{335}{144\e}+ \left(\frac{32573}{2592}-
\frac{23}{48}\pi^2\right) + \order{\e}\right],\\
\calDcol(i_q^h,j_{\tilde{g}};k_{\tilde{g}},l_{g}^h) &=
S_{ijkl}^2 \left
[+\frac{1}{8\e^2}+\frac{13}{16\e}+ \left(\frac{407}{96}-
\frac{3}{16}\pi^2\right) + \order{\e}\right],\\
\calDcol(i_q^h,j_{g};k_{\Qb},l_{Q}^h) &=
S_{ijkl}^2 \left [-\frac{1}{4\e}-\frac{95}{48} +
\order{\e}\right],\\
\calDcol(i_g^h,j_{g};k_{g},l_{g}^h)  &=
S_{ijkl}^2 \left
[+\frac{41}{48\e^2}+\frac{97}{16\e}+ \left(\frac{13997}{432}-
\frac{41}{32}\pi^2\right) + \order{\e}\right],\\
\calDcol(i_g^h,j_{\tilde{g}};k_{\tilde{g}},l_{g}^h)&=
S_{ijkl}^2 \left
[+\frac{13}{144\e^2}+\frac{95}{144\e}+ \left(\frac{4693}{1296}-
\frac{13}{96}\pi^2\right) + \order{\e}\right],\\
\calDcol(i_g^h,j_{g};k_{\Qb},l_{Q}^h) &= 
S_{ijkl}^2 \left [-\frac{1}{18\e^2}-
\frac{1}{2\e}+ \left(-\frac{3905}{1296}+\frac{1}{12}\pi^2\right) +
\order{\e}\right],\\
\calDcol(i_{\qb}^h,j_{q};k_{\Qb},l_{Q}^h) &=
S_{ijkl}^2 \left
[+\frac{1}{9\e^2}+\frac{13}{36\e}+ \left(\frac{139}{324}-
\frac{1}{6}\pi^2\right) + \order{\e}\right].
\end{align}

The remaining single-unresolved terms, $\Ssoft$ and $\Scol$, are not universal, since they depend on the choice of single-real antenna function. We do not list the integrals of these contributions.

Note that in all of the above formulae arguments merely serve the purpose of identifying the particle content and the hard radiators. There is no dependence on the particle momenta in the integrated contributions besides the overall normalisation to the invariant mass.
\endgroup

\section{Overlap of Double-Soft and Triple-Collinear Contributions}
\label{app:TCprojectionsofDS}
The projections into the triple-collinear phase space for each of the three double-soft factors are given by,
\begin{align}
\PTCdown_{ijk} \Sgg(i^h,j,k,l^h) &= \frac{2 \ome W_{jk}}{\sjk^2\sijk^2\omxi^2}
+\frac{4\xi\xj\xk}{\sjk\sijk\omxi^3}
+\frac{2\xi^2}{\sij\sijk\xk\omxi} \nonumber \\
& \qquad
+\frac{2\xi}{\sjk\sijk\xk}
-\frac{8\xi}{\sjk\sijk\omxi}
+\frac{2\xi}{\sij\sjk\xk}
+\frac{2\xi}{\sij\sjk\omxi} \, ,  \\
\PTCdown_{ijk} \Spp(i^h,j,k,l^h) &= \frac{4\sjk\xi^2}{\sij\sik\sijk\xj\xk}
+\frac{4\xi^2}{\sij\sijk\xj\xk}
+\frac{4\xi^2}{\sik\sijk\xj\xk} \, , \\
\PTCdown_{ijk} \Sqq(i^h,j,k,l^h) &= -\frac{2 
W_{jk}}{\sjk^2\sijk^2\omxi^2}
-\frac{4\xi\xj\xk}{\sjk\sijk\omxi^3\ome} \nonumber \\
& \qquad
+\frac{2\xi}{\sjk\sijk\omxi} \, , 
\end{align}
where 
\begin{align}
    W_{ij} &= (x_i s_{jk} - x_j s_{ik})^2 - \frac{2}{\ome}\frac{x_i x_j x_k}{(1-x_k)}s_{ij} s_{ijk} \, . 
\end{align}

\section{Integrals of $\X$ Antennae derived using the $X_3^0$ of Ref.~\cite{Gehrmann-DeRidder:2005btv}}
\label{app:intX4oldX3}
\begingroup
\allowdisplaybreaks
In this appendix, we list the integrals over the antenna phase space of the $\X$ antennae constructed using the $X_3^0$ antennae of Ref.~\cite{Gehrmann-DeRidder:2005btv}:
\begin{eqnarray}
\label{eq:A40intoldX3}
\calA (s_{ijkl}) &=& S_{ijkl}^2 \Biggl [
+\frac{3}{4\e^4}
+\frac{65}{24\e^3}
+\frac{1}{\e^2} \left(
\frac{217}{18}
-\frac{13}{12}\pi^2
\right)
+\frac{1}{\e} \left(
\frac{43223}{864}
-\frac{589}{144}\pi^2
-\frac{71}{4}\zeta_3
\right)
\nonumber \\&& \hspace{2cm}
 + \left(
\frac{1094807}{5184}
-\frac{8117}{432}\pi^2
-\frac{1327}{18}\zeta_3
+\frac{373}{1440}\pi^4
\right)
 + \order{\e}\Biggr], \hspace{1.5cm} \\
\label{eq:A40tintoldX3}
\calAt (s_{ijkl}) &=& S_{ijkl}^2 \Biggl [
+\frac{1}{\e^4}
+\frac{3}{\e^3}
+\frac{1}{\e^2} \left(
13
-\frac{3}{2}\pi^2
\right)
+\frac{1}{\e} \left(
\frac{845}{16}
-\frac{9}{2}\pi^2
-\frac{80}{3}\zeta_3
\right)
\nonumber \\&& \hspace{2cm}
 + \left(
\frac{6865}{32}
-\frac{39}{2}\pi^2
-80\zeta_3
+\frac{29}{120}\pi^4
\right)
 + \order{\e}\Biggr], \\
\label{eq:B40intoldX3}
\calB (s_{ijkl}) &=& S_{ijkl}^2 \Biggl [
-\frac{1}{12\e^3}
-\frac{7}{18\e^2}
+\frac{1}{\e} \left(
-\frac{407}{216}
+\frac{11}{72}\pi^2
\right)
\nonumber \\&& \hspace{2cm}
+ \left(
-\frac{5809}{648}
+\frac{145}{216}\pi^2
+\frac{67}{18}\zeta_3
\right)
 + \order{\e}\Biggr], \\
\label{eq:C40intoldX3}
\calC (s_{ijkl}) &=& S_{ijkl}^2 \Biggl [
+\frac{1}{\e} \left(
-\frac{13}{32}
+\frac{1}{16}\pi^2
-\frac{1}{4}\zeta_3
\right)
\nonumber \\&& \hspace{2cm}
 + \left(
-\frac{73}{16}
+\frac{23}{96}\pi^2
+\frac{23}{8}\zeta_3
-\frac{1}{45}\pi^4
\right)
 + \order{\e}\Biggr], \\
\label{eq:D40intoldX3}
\calD (s_{ijkl}) &=& S_{ijkl}^2 \Biggl [
+\frac{3}{4\e^4}
+\frac{71}{24\e^3}
+\frac{1}{\e^2} \left(
\frac{257}{18}
-\frac{13}{12}\pi^2
\right)
+\frac{1}{\e} \left(
\frac{13661}{216}
-\frac{35}{8}\pi^2
-\frac{35}{2}\zeta_3
\right)
\nonumber \\&& \hspace{2cm}
 + \left(
\frac{22286}{81}
-\frac{9335}{432}\pi^2
-\frac{5473}{72}\zeta_3
+\frac{9}{32}\pi^4
\right)
 + \order{\e}\Biggr], \\
\label{eq:D40tintoldX3}
\calDt (s_{ijkl}) &=& S_{ijkl}^2 \Biggl [
+\frac{1}{\e^4}
+\frac{10}{3\e^3}
+\frac{1}{\e^2} \left(
\frac{47}{3}
-\frac{3}{2}\pi^2
\right)
+\frac{1}{\e} \left(
\frac{30313}{432}
-5\pi^2
-\frac{83}{3}\zeta_3
\right)
\nonumber \\&& \hspace{2cm}
 + \left(
\frac{132451}{432}
-\frac{1129}{48}\pi^2
-\frac{818}{9}\zeta_3
+\frac{19}{120}\pi^4
\right)
 + \order{\e}\Biggr], \\
\label{eq:E40aintoldX3}
\calEa (s_{ijkl}) &=& S_{ijkl}^2 \Biggl [
-\frac{1}{12\e^3}
-\frac{5}{12\e^2}
+\frac{1}{\e} \left(
-\frac{1631}{864}
+\frac{1}{8}\pi^2
\right)
\nonumber \\&& \hspace{2cm}
+ \left(
-\frac{46315}{5184}
+\frac{77}{108}\pi^2
+\frac{20}{9}\zeta_3
\right)
 + \order{\e}\Biggr], \\
\label{eq:E40bintoldX3}
\calEb (s_{ijkl}) &=& S_{ijkl}^2 \Biggl [
-\frac{1}{3\e^3}
-\frac{41}{24\e^2}
+\frac{1}{\e} \left(
-\frac{7325}{864}
+\frac{1}{2}\pi^2
\right)
\nonumber \\&& \hspace{2cm}
+ \left(
-\frac{22745}{576}
+\frac{41}{16}\pi^2
+\frac{80}{9}\zeta_3
\right)
 + \order{\e}\Biggr], \\
\label{eq:E40tintoldX3}
\calEt (s_{ijkl}) &=& S_{ijkl}^2 \Biggl [
-\frac{1}{6\e^3}
-\frac{35}{36\e^2}
+\frac{1}{\e} \left(
-\frac{1045}{216}
+\frac{1}{4}\pi^2
\right)
\nonumber \\&& \hspace{2cm}
+ \left(
-\frac{28529}{1296}
+\frac{35}{24}\pi^2
+\frac{40}{9}\zeta_3
\right)
 + \order{\e}\Biggr], \\
\label{eq:F40intoldX3}
\calF (s_{ijkl}) &=& S_{ijkl}^2 \Biggl [
+\frac{3}{4\e^4}
+\frac{77}{24\e^3}
+\frac{1}{\e^2} \left(
\frac{559}{36}
-\frac{13}{12}\pi^2
\right)
+\frac{1}{\e} \left(
\frac{59249}{864}
-\frac{671}{144}\pi^2
-\frac{69}{4}\zeta_3
\right)
\nonumber \\&& \hspace{2cm}
 + \left(
\frac{508343}{1728}
-\frac{9923}{432}\pi^2
-\frac{2819}{36}\zeta_3
+\frac{437}{1440}\pi^4
\right)
 + \order{\e}\Biggr], \hspace{1.5cm} \\
\label{eq:F40tintoldX3}
\calFt (s_{ijkl}) &=& S_{ijkl}^2 \Biggl [
+\frac{1}{\e^4}
+\frac{11}{3\e^3}
+\frac{1}{\e^2} \left(
\frac{313}{18}
-\frac{3}{2}\pi^2
\right)
+\frac{1}{\e} \left(
\frac{34571}{432}
-\frac{11}{2}\pi^2
-\frac{86}{3}\zeta_3
\right)
\nonumber \\&& \hspace{2cm}
 + \left(
\frac{924559}{2592}
-\frac{209}{8}\pi^2
-\frac{916}{9}\zeta_3
+\frac{3}{40}\pi^4
\right)
 + \order{\e}\Biggr], \\
\label{eq:G40aintoldX3}
\calGa (s_{ijkl}) &=& S_{ijkl}^2 \Biggl [
-\frac{1}{12\e^3}
-\frac{4}{9\e^2}
+\frac{1}{\e} \left(
-\frac{745}{432}
+\frac{7}{72}\pi^2
\right)
\nonumber \\&& \hspace{2cm}
+ \left(
-\frac{6431}{864}
+\frac{163}{216}\pi^2
+\frac{13}{18}\zeta_3
\right)
 + \order{\e}\Biggr], \\
\label{eq:G40bintoldX3}
\calGb (s_{ijkl}) &=& S_{ijkl}^2 \Biggl [
-\frac{1}{3\e^3}
-\frac{139}{72\e^2}
+\frac{1}{\e} \left(
-\frac{8669}{864}
+\frac{1}{2}\pi^2
\right)
\nonumber \\&& \hspace{2cm}
+ \left(
-\frac{248495}{5184}
+\frac{139}{48}\pi^2
+\frac{80}{9}\zeta_3
\right)
 + \order{\e}\Biggr], \\
\label{eq:G40tintoldX3}
\calGt (s_{ijkl}) &=& S_{ijkl}^2 \Biggl [
-\frac{1}{6\e^3}
-\frac{41}{36\e^2}
+\frac{1}{\e} \left(
-\frac{1327}{216}
+\frac{1}{4}\pi^2
\right)
\nonumber \\&& \hspace{2cm}
+ \left(
-\frac{38291}{1296}
+\frac{41}{24}\pi^2
+\frac{40}{9}\zeta_3
\right)
 + \order{\e}\Biggr], \\
\label{eq:H40intoldX3}
\calH (s_{ijkl}) &=& S_{ijkl}^2 \Biggl [
+\frac{1}{9\e^2}
+\frac{7}{9\e}
+ \left(
\frac{1345}{324}
-\frac{1}{6}\pi^2
\right)
 + \order{\e}\Biggr].
\end{eqnarray}
In all cases, we find agreement with the analogous integrated antenna given in Ref.~\cite{Gehrmann-DeRidder:2005btv} through to $\order{\e^0}$.
\endgroup

\chapter{$X_3^1$ Appendix}

\section{Integration of $X_3^1$}
\label{app:integration}

The integrated antenna is obtained by integrating over the antenna phase space,
\begin{equation}
{\cal X}_{3}^1(s_{ijk}) =
\left(8\pi^2\left(4\pi\right)^{-\e} e^{\e\gamma_E}\right)
\int d \Phi_{X_{ijk}} X_{3}^1,
\end{equation}
with $d=4-2\e$.
As in Ref.~\cite{Gehrmann-DeRidder:2005btv}, we have included a normalisation factor to account for powers of the QCD coupling constant. 
The antenna phase space is given by
\begin{equation}
    d \Phi_{X_{ijk}}  = \frac{1}{16\pi^2}\frac{1}{\Gamma(1-\e)} \left(\frac{4\pi}{s_{ijk}}\right)^{\e} s_{ijk} {\rm d} I ,
 \end{equation}
 with
 \begin{align}
{\rm d} I &= {\rm d}y_{ij}  {\rm d}y_{jk}
\left(y_{ij} y_{jk} (1-y_{ij}-y_{jk})\right)^{-\e} ,
\end{align}
where $0 < y_{ij} < 1$, $0 < y_{jk} < 1-y_{ij}$ and $y_{IJ} = s_{IJ}/s_{IJK}$.
Setting $y_{jk} = (1-y_{ij}) z$, then
\begin{align}
{\rm d}I &= {\rm d}y_{ij} \, {\rm d}z \, y_{ij}^{-\e} \, \left(1-y_{ij}\right)^{1-2\e} 
 \, z^{-\e} \, \omz^{-\e} ,
\end{align}
with $0 < y_{ij},z < 1$.

The integrals we encounter are of the form,
\begin{align}
&\int_0^1 {}_2F_1(\pm\e,\pm\e,1\pm\e,z)   z^{\alpha} \omz^{\beta}
{\rm d}z \\
& \hspace{2cm}= 
\frac{\Gamma(\alpha+1)\Gamma(\beta+1)}{\Gamma(\alpha+\beta+2)}{}_3F_2 (\pm\e,\pm\e,\alpha+1,1\pm\e,\alpha+\beta+2,1), \nonumber \\
&\int_0^1 {}_2F_1(\pm\e,\pm\e,1\pm\e,1-z)   z^{\alpha} \omz^{\beta}
{\rm d}z \\
& \hspace{2cm}= 
\frac{\Gamma(\alpha+1)\Gamma(\beta+1)}{\Gamma(\alpha+\beta+2)}{}_3F_2 (\pm\e,\pm\e,\beta+1,1\pm\e,\alpha+\beta+2,1), \nonumber \\
&\int_0^1 z^{\alpha} \omz^{\beta}
{\rm d}z = \frac{\Gamma(\alpha+1)\Gamma(\beta+1)}{\Gamma(\alpha+\beta+2)} .
\end{align}
Also note the definitions for the hypergeometric functions are
\begin{equation}
    {}_2F_1(a,b,c,z) = \frac{\Gamma (c)}{\Gamma(b) \Gamma(c-b)} \int_0^1 d t \text{ } t^{b-1} (1-t)^{c-b-1} (1-zt)^{-a} ,
\end{equation}
and recursively,
\begin{eqnarray}
    &&{}_{n+1}F_{m+1} (a_1,...,a_n,c,b_1,...,b_m,d,z) =  \nonumber \\
    && \hspace{0.8cm} \frac{\Gamma (d)}{\Gamma(c) \Gamma(d-c)} \int_0^1 d t \text{ } t^{c-1} (1-t)^{d-c-1} {}_{n}F_{m} (a_1,...,a_n,b_1,...,b_m,tz), \hspace{1cm}
\end{eqnarray}
for $n\ge 2$ and $m\ge 1$ where
\begin{equation}
    \Gamma(x) = \int_0^1 dy \text{ } e^{-y} y^{x-1}, \hspace{1cm} x>0.
\end{equation}
The reader may also find the following series definitions useful,
\begin{equation}
    {}_2F_1(a,b,c,z) = \sum_{n=0}^\infty \frac{(a)_n (b)_n}{(c)_n \Gamma(n+1)} z^n, \hspace{1cm} (l)_n = \frac{\Gamma(n+l)}{\Gamma(l)},
\end{equation}
and
\begin{equation}
    {}_3F_2(a,b,c,d,e,z) = \sum_{n=0}^\infty \frac{(a)_n (b)_n (c)_n}{(d)_n (e)_n \Gamma(n+1)} z^n,
\end{equation}
in addition to the following identities,
\begin{eqnarray}
    \Gamma(x+1) = x \Gamma(x), \hspace{1cm} \Gamma(1) = 1.
\end{eqnarray}
Many further relations can be found in Ref.~\cite{bateman}. 
Additionally, the polylogarithms we refer to are defined by
\begin{equation}
    \Li_2(x) = - \int_0^1 dy \frac{\ln(1-yx)}{y},
\end{equation}
and recursively,
\begin{equation}
    \Li_{n+1}(x) = \int_0^x dy \frac{\Li_{n} (y)}{y},
\end{equation}
for $n \ge 2$. A fuller list of identities can be found in Ref.~\cite{Devoto:1983tc}. 

\section{Expansions of Hypergeometric Functions}
\label{app:3F2}

In this appendix, we give the expansions around $\e=0$ of the ${}_3F_2$ functions necessary for integrating the $X_3^1$. They were performed using a combination of MAPLE code and hypexp~\cite{Huber:2005yg,Huber:2007dx}. 

\begin{eqnarray}
    &&{}_3F_2\left(\e,\e,2-\e,1+\e,3-\e,1\right) = \nonumber \\ && \hspace{1cm} 1 + \e^2 \left(-3/4 + \left(\pi^2/6\right)\right) + \e^3 \left(-9/4 + \left(\pi^2/6\right)\right)  \nonumber \\ && \hspace{1cm}
    + \e^4 \left(-21/4 + \left(7 \left(\pi^2/6\right)\right)/4 + \left(7 \left(\pi^2/6\right)^2\right)/10 + \zeta_3\right) \nonumber \\
    && \hspace{1cm}+ \e^5 \left(-45/4 + \left(13 \left(\pi^2/6\right)\right)/4 + \left(4 \left(\pi^2/6\right)^2\right)/5 + \left(5 \zeta_3\right)/2\right) \nonumber \\ && \hspace{1cm}+ 
    \e^6 \bigg(-93/4 + \left(25 \left(\pi^2/6\right)\right)/4 + \left(59 \left(\pi^2/6\right)^2\right)/40 + \left(31 \left(\pi^2/6\right)^3\right)/70 \nonumber \\ && \hspace{1cm}+ \left(11 \zeta_3\right)/2 - \left(\pi^2/6\right) \zeta_3 + 
      3 \zeta_5\bigg) + \mathcal{O}\left(\e^7\right), \\
 &&{}_3F_2\left(\e,\e,1-\e,1+\e,2-\e,1\right) = \nonumber \\ && \hspace{1cm} 1 + \e^2 \left(-1 + \pi^2/6\right) + \e^3 \left(\pi^2/3 - 4\right)  + \e^4  \left(-12 + \left(5  \pi^2\right)/6 + \left(7  \pi^4\right)/360 + 2  \zeta_3\right) \nonumber \\
 && \hspace{1cm}+\e^5   \left(-32 + 2  \pi^2 + \left(2  \pi^4\right)/45 + 6  \zeta_3\right) + \mathcal{O}\left(\e^6\right), \\
 &&{}_3F_2\left(-\e,-\e,1-\e,1-\e,2-3 \e,1\right) \nonumber = {}_2F_1\left(-\e,-\e,2-3 \e,1\right) =  \\ && \hspace{1cm} 1 + \e^2 \left(-1 + \pi^2/6\right) + \e^3 \left(4 \zeta_3 - 4\right) + \e^4 \left(-12 - 1/6 \pi^2 + 11/72 \pi^4\right) \nonumber \\
 && \hspace{1cm}+ \e^5 \left(-32 + \left(2 \zeta_3 \pi^2\right)/3 - 4 \zeta_3 - \left(2 \pi^2\right)/3 + 36 \zeta_5\right) \nonumber \\ && \hspace{1cm} + \e^6 \left(-80 - 16 \zeta_3 + 8 \zeta_3^2 - 2 \pi^2 - \left(11 \pi^4\right)/72 + \left(281 \pi^6\right)/2160\right) + \mathcal{O}\left(\e^7\right),  \\
 &&{}_3F_2\left(\e,\e,-\e,1+\e,2-\e,1\right) = \nonumber \\ && \hspace{1cm} 1 + \e^3 \left(-1 + \pi^2/6 - \zeta_3\right) + \e^4 \left(-4 + \zeta_3 + \pi^2/3 - \pi^4/360\right) \nonumber \\
 && \hspace{1cm} + \e^5 \left(-12 + 2 \zeta_3 - 3 \zeta_5 + \left(5 \pi^2\right)/6 + \pi^4/45 + \zeta_3 \pi^2/6\right) \nonumber \\ && \hspace{1cm} + \e^6 \left(-32 + \left(2 \pi^4\right)/45 + 6 \zeta_3 + 2 \pi^2 + \zeta_3^2 - \zeta_3 \pi^2/6 + 3 \zeta_5 - \pi^6/630\right) \nonumber \\ && \hspace{1cm}+ \mathcal{O}\left(\e^7\right), \\
 &&{}_3F_2\left(-\e,-\e,-\e,1-\e,2-3 \e,1\right) = \nonumber \\ && \hspace{1cm} 1 + \e^3 \left(-1 + \left(\pi^2/6\right) - \zeta_3\right) \nonumber \\ && \hspace{1cm} + \e^4 \left(-6 + 2 \left(\pi^2/6\right) - \left(17 \left(\pi^2/6\right)^2\right)/10 + 5 \zeta_3\right) \nonumber \\
 && \hspace{1cm} +  \e^5 \left(-24 + 3 \left(\pi^2/6\right) + \left(36 \left(\pi^2/6\right)^2\right)/5 + 10 \zeta_3 - 7 \left(\pi^2/6\right) \zeta_3 - \zeta_5\right) \nonumber \\ && \hspace{1cm} + 
    \e^6 \bigg(-80 + 2 \left(\pi^2/6\right) + \left(72 \left(\pi^2/6\right)^2\right)/5 - \left(100 \left(\pi^2/6\right)^3\right)/21 \nonumber \\ && \hspace{1cm} + 16 \zeta_3 + 11 \left(\pi^2/6\right) \zeta_3 - 
      17 \zeta_3^2 + 37 \zeta_5\bigg) + \mathcal{O}\left(\e^7\right), \\
 &&{}_3F_2\left(\e,\e,2-\e,1+\e,4-\e,1\right) = \nonumber \\ && \hspace{1cm} 
 1 + \e^2 \left(-37/36 + \left(\pi^2/6\right)\right) + \e^3 \left(-179/54 + \left(5 \left(\pi^2/6\right)\right)/3\right) \nonumber \\ && \hspace{1cm}+ 
    \e^4 \left(-215/27 + \left(101 \left(\pi^2/6\right)\right)/36 + \left(7 \left(\pi^2/6\right)^2\right)/10 + \left(5 \zeta_3\right)/3\right) \nonumber \\
    && \hspace{1cm} + \e^5 \left(-8413/486 + \left(31 \left(\pi^2/6\right)\right)/6 + \left(4 \left(\pi^2/6\right)^2\right)/3 + \left(23 \zeta_3\right)/6\right) \nonumber \\ && \hspace{1cm}+ 
    \e^6 \bigg(-52549/1458 + \left(799 \left(\pi^2/6\right)\right)/81 + \left(169 \left(\pi^2/6\right)^2\right)/72 \nonumber \\ && \hspace{1cm} + \left(31 \left(\pi^2/6\right)^3\right)/70 + 
      \left(229 \zeta_3\right)/27 - \left(5 \left(\pi^2/6\right) \zeta_3\right)/3 + 5 \zeta_5\bigg) + \mathcal{O}\left(\e^7\right), \\
 &&{}_3F_2\left(\e,\e,-\e,1+\e,1-\e,1\right) \nonumber =  \\ && \hspace{1cm} 1 - \e^3 \zeta_3  - \left(\e^4 \left(\pi^2/6\right)^2\right)/10 \nonumber \\
 && \hspace{1cm} + 
 \e^5 \left(\left(\pi^2/6\right) \zeta_3 - 3 \zeta_5\right) + \e^6 \left(\left(-12 \left(\pi^2/6\right)^3\right)/35 + \zeta_3^2\right) + \mathcal{O}\left(\e^7\right), \\
 &&{}_3F_2\left(\e,\e,1-\e,1+\e,3-\e,1\right) = \nonumber \\ && \hspace{1cm} 1 + \e^2 \left(-5/4 + \pi^2/6\right) + \e^3 \left(\pi^2/2 - 11/2\right) \nonumber \\
 && \hspace{1cm} + \e^4 \left(\left(7 \pi^4\right)/360 + \left(29 \pi^2\right)/24 + 3 \zeta_3 - 17\right) +\mathcal{O}\left(\e^5\right), \\
 &&{}_3F_2\left(-\e,-\e,2-\e,1-\e,3-3 \e,1\right) \nonumber =  \\ && \hspace{1cm} 1 + \e^2 \left(-3/4 + \left(\pi^2/6\right)\right) + 
 \e^3 \left(-15/4 + 4 \zeta_3\right)  \nonumber \\
 && \hspace{1cm} + \e^4 \left(-49/4 - \left(3 \left(\pi^2/6\right)\right)/4 + \left(11 \left(\pi^2/6\right)^2\right)/2\right) \nonumber \\ && \hspace{1cm} + 
 \e^5 \left(-135/4 - \left(15 \left(\pi^2/6\right)\right)/4 - 3 \zeta_3 + 4 \left(\pi^2/6\right) \zeta_3 + 36 \zeta_5\right) \nonumber \\ && \hspace{1cm}+ \e^6 \bigg(-341/4 - \left(49 \left(\pi^2/6\right)\right)/4 - \left(33 \left(\pi^2/6\right)^2\right)/8  \nonumber
    \\ && \hspace{1cm} + \left(281 \left(\pi^2/6\right)^3\right)/10 - 15 \zeta_3 + 8 \zeta_3^2\bigg) + \mathcal{O}\left(\e^7\right), \\
 &&{}_3F_2\left(-\e,-\e,2-\e,1-\e,4-3 \e,1\right) = \nonumber \\ && \hspace{1cm} 1 + \e^2 \left(-37/36 + \left(\pi^2/6\right)\right) + \e^3 \left(-227/54 + 4 \zeta_3\right) \nonumber \\ && \hspace{1cm} + \e^4 \left(-337/27 - \left(37 \left(\pi^2/6\right)\right)/36 + \left(11 \left(\pi^2/6\right)^2\right)/2\right) \nonumber \\ 
 && \hspace{1cm}+ 
    \e^5 \left(-15847/486 - \left(227 \left(\pi^2/6\right)\right)/54 - \left(37 \zeta_3\right)/9 + 4 \left(\pi^2/6\right) \zeta_3 + 36 \zeta_5\right) \nonumber \\
 && \hspace{1cm} + \e^6 \bigg(-116293/1458 - \left(337 \left(\pi^2/6\right)\right)/27 - \left(407 \left(\pi^2/6\right)^2\right)/72 \nonumber \\ && \hspace{1cm} + 
      \left(281 \left(\pi^2/6\right)^3\right)/10 - \left(454 \zeta_3\right)/27 + 8 \zeta_3^2\bigg) \nonumber \\  && \hspace{1cm} + \mathcal{O}\left(\e^7\right), \\
 &&{}_3F_2\left(-\e,-\e,-\e,1-\e,1-3 \e,1\right) \nonumber =  \\ && \hspace{1cm} 1 - \e^3 \zeta_3 - \left(17 \e^4 \left(\pi^2/6\right)^2\right)/10  \nonumber \\
 && \hspace{1cm} + 
 \e^5 \left(-7 \left(\pi^2/6\right) \zeta_3 - \zeta_5\right)  + \e^6 \left(\left(-100 \left(\pi^2/6\right)^3\right)/21 - 17 \zeta_3^2\right) + \mathcal{O}\left(\e^7\right), \\
 &&{}_3F_2\left(-\e,-\e,1-\e,1-\e,3-3 \e,1\right) \nonumber = {}_2F_1\left(-\e,-\e,3-3 \e,1\right) = \\ && \hspace{1cm} 1 + \e^2 \left(-5/4 + \pi^2/6\right) + \e^3 \left(4 \zeta_3 - 9/2\right) \nonumber \\
 && \hspace{1cm} + \e^4 \left(-5/24 \pi^2 + 11/72 \pi^4 - 25/2\right) +\mathcal{O}\left(\e^5\right), \\
 &&{}_3F_2\left(-\e,-\e,1-\e,1-\e,4-3 \e,1\right) = {}_2F_1\left(-\e,-\e,4-3 \e,1\right) = \nonumber \\ && \hspace{1cm} 1 + \e^2 \left(\pi^2/6 - 49/36\right) + \e^3 \left(4 \zeta_3 - 251/54\right) \nonumber \\
 && \hspace{1cm} + \e^4 \left(-49/216 \pi^2 + 11/72 \pi^4 - 1351/108\right) +\mathcal{O}\left(\e^5\right), \\
 &&{}_3F_2\left(-\e,-\e,3-\e,1-\e,4-3 \e,1\right) \nonumber =  \\ && \hspace{1cm} 1 + \e^2 \left(-11/18 + \left(\pi^2/6\right)\right) + 
 \e^3 \left(-403/108 + 4 \zeta_3\right) \nonumber \\ && \hspace{1cm}+ \e^4 \left(-2717/216 - \left(11 \left(\pi^2/6\right)\right)/18 + \left(11 \left(\pi^2/6\right)^2\right)/2\right)  \nonumber \\
    && \hspace{1cm}+ 
    \e^5 \left(-134981/3888 - \left(403 \left(\pi^2/6\right)\right)/108 - \left(22 \zeta_3\right)/9 + 4 \left(\pi^2/6\right) \zeta_3 + 36 \zeta_5\right) \nonumber \\
    && \hspace{1cm}+ \e^6 \bigg(-2037343/23328 - \left(2717 \left(\pi^2/6\right)\right)/216 - 
      \left(121 \left(\pi^2/6\right)^2\right)/36 \nonumber \\ && \hspace{1cm}+ \left(281 \left(\pi^2/6\right)^3\right)/10 - \left(403 \zeta_3\right)/27 + 8 \zeta_3^2\bigg) + \mathcal{O}\left(\e^7\right).  
\end{eqnarray}

\section{Integrals of $X_3^1$ Antennae derived using the $X_3^0$ of Ref.~\cite{Gehrmann-DeRidder:2005btv}}
\label{app:X31oldX30}

\begingroup
\allowdisplaybreaks
In this appendix, we list the integrals over the antenna phase space of the renormalised $X_3^1$ antenna constructed using the $X_3^0$ antennae of Ref.~\cite{Gehrmann-DeRidder:2005btv}.

\begin{eqnarray}
\label{eq:A31intoldX3}
{\cal A}_3^1 (s_{ijk}) &=& S_{ijk}^2\Biggl [
-\frac{1}{4\e^4}
-\frac{31}{12\e^3}
+\frac{1}{\e^2} \left(
-\frac{53}{8}
+\frac{11}{24}\pi^2
\right)
+\frac{1}{\e} \left(
-\frac{647}{24}
+\frac{22}{9}\pi^2
+\frac{23}{3}\zeta_3
\right) \nonumber \\
&& \hspace{2cm}
 + \left(
-\frac{1289}{12}
+\frac{199}{24}\pi^2
+\frac{635}{18}\zeta_3
+\frac{13}{1440}\pi^4
\right)
 + \order{\e}\Biggr] \, ,\\
\label{eq:A31tintoldX3}
\widetilde{{\cal A}}_3^1 (s_{ijk}) &=& S_{ijk}^2\Biggl [
+\frac{1}{\e^2} \left(
-\frac{5}{8}
+\frac{1}{6}\pi^2
\right)
+\frac{1}{\e} \left(
-\frac{19}{4}
+\frac{1}{4}\pi^2
+7\zeta_3
\right)
\nonumber \\
&& \hspace{2cm}
 + \left(
-\frac{435}{16}
+\frac{29}{16}\pi^2
+\frac{21}{2}\zeta_3
+\frac{7}{60}\pi^4
\right)
 + \order{\e}\Biggr] \, ,
\\
\label{eq:A31hintoldX3}
\widehat{{\cal A}}_3^1 (s_{ijk}) &=& S_{ijk}^2\Biggl [
+\frac{1}{3\e^3}
+\frac{1}{2\e^2}
+\frac{1}{\e} \left(
\frac{19}{12}
-\frac{7}{36}\pi^2
\right)
+ \left(
\frac{109}{24}
-\frac{7}{24}\pi^2
-\frac{25}{9}\zeta_3
\right)
\nonumber \\
&& \hspace{2cm}
 + \order{\e}\Biggr] \, ,
\\
\label{eq:d31intoldX3}
{\cal D}_3^1 (s_{ijk}) &=& S_{ijk}^2\Biggl [
-\frac{1}{4\e^4}
-\frac{8}{3\e^3}
+\frac{1}{\e^2} \left(
-\frac{1193}{144}
+\frac{13}{24}\pi^2
\right)
+\frac{1}{\e} \left(
-\frac{8473}{216}
+\frac{49}{18}\pi^2
+\frac{73}{6}\zeta_3
\right)
\nonumber \\
&& \hspace{2cm}
 + \left(
-\frac{18937}{108}
+\frac{9485}{864}\pi^2
+\frac{535}{12}\zeta_3
+\frac{79}{480}\pi^4
\right)
 + \order{\e}\Biggr] \, ,
\\
\label{eq:d31tintoldX3}
\widetilde{{\cal D}}_3^1 (s_{ijk}) &=& S_{ijk}^2\Biggl [
+\frac{1}{\e^2} \left(
-\frac{5}{16}
+\frac{1}{12}\pi^2
\right)
+\frac{1}{\e} \left(
-\frac{13}{6}
+\frac{11}{72}\pi^2
+\frac{5}{2}\zeta_3
\right)\nonumber \\
&& \hspace{2cm}
 + \left(
-\frac{395}{36}
+\frac{941}{864}\pi^2
+\frac{55}{12}\zeta_3
-\frac{7}{180}\pi^4
\right)
 + \order{\e}\Biggr] \, ,
\\
\label{eq:d31hintoldX3}
\widehat{{\cal D}}_3^1 (s_{ijk}) &=& S_{ijk}^2\Biggl [
+\frac{1}{3\e^3}
+\frac{5}{9\e^2}
+\frac{1}{\e} \left(
\frac{139}{72}
-\frac{7}{36}\pi^2
\right)
+ \left(
\frac{443}{72}
-\frac{35}{108}\pi^2
-\frac{25}{9}\zeta_3
\right)\nonumber \\
&& \hspace{2cm}
 + \order{\e}\Biggr] \, ,
\\
\label{eq:E31intoldX3}
{\cal E}_3^1 (s_{ijk}) &=& S_{ijk}^2\Biggl [
+\frac{11}{18\e^2}
+\frac{1}{\e} \left(
\frac{74}{27}
-\frac{1}{9}\pi^2
\right)
+ \left(
\frac{1441}{108}
-\frac{149}{216}\pi^2
-4\zeta_3
\right)
 + \order{\e}\Biggr] \, ,
\\
\label{eq:E31tintoldX3}
\widetilde{{\cal E}}_3^1 (s_{ijk}) &=& S_{ijk}^2\Biggl [
+\frac{1}{6\e^3}
+\frac{35}{36\e^2}
+\frac{1}{\e} \left(
\frac{509}{108}
-\frac{1}{4}\pi^2
\right)
+ \left(
\frac{1670}{81}
-\frac{35}{24}\pi^2
-\frac{31}{9}\zeta_3
\right)\nonumber \\
&& \hspace{2cm}
 + \order{\e}\Biggr] \, ,
\\
\label{eq:E31hintoldX3}
\widehat{{\cal E}}_3^1 (s_{ijk}) &=& S_{ijk}^2\Biggl [
+\frac{1}{3\e}
+ \left(
\frac{172}{81}
-\frac{11}{108}\pi^2
\right)
 + \order{\e}\Biggr] \, ,
\\
\label{eq:f31intoldX3}
{\cal F}_3^1 (s_{ijk}) &=& S_{ijk}^2\Biggl [
-\frac{1}{4\e^4}
-\frac{11}{4\e^3}
+\frac{1}{\e^2} \left(
-\frac{85}{9}
+\frac{5}{8}\pi^2
\right)
+\frac{1}{\e} \left(
-\frac{9827}{216}
+\frac{55}{18}\pi^2
+\frac{44}{3}\zeta_3
\right)
\nonumber \\
&& \hspace{2cm}
 + \left(
-\frac{88961}{432}
+\frac{5665}{432}\pi^2
+\frac{473}{9}\zeta_3
+\frac{181}{1440}\pi^4
\right)
 + \order{\e}\Biggr], \hspace{1cm} \, 
\\
\label{eq:f31hintoldX3}
\widehat{{\cal F}}_3^1 (s_{ijk}) &=& S_{ijk}^2\Biggl [
+\frac{1}{3\e^3}
+\frac{11}{18\e^2}
+\frac{1}{\e} \left(
\frac{19}{9}
-\frac{7}{36}\pi^2
\right)
+ \left(
\frac{167}{24}
-\frac{77}{216}\pi^2
-\frac{25}{9}\zeta_3
\right)\nonumber \\
&& \hspace{2cm}
 + \order{\e}\Biggr], \, 
\\
\label{eq:G31intoldX3}
{\cal G}_3^1 (s_{ijk}) &=& S_{ijk}^2\Biggl [
+\frac{11}{18\e^2}
+\frac{1}{\e} \left(
\frac{169}{54}
-\frac{1}{9}\pi^2
\right)
+ \left(
\frac{3355}{216}
-\frac{161}{216}\pi^2
-4\zeta_3
\right)
 + \order{\e}\Biggr] , \hspace{1.2cm} \, 
\\
\label{eq:G31tintoldX3}
\widetilde{{\cal G}}_3^1 (s_{ijk}) &=& S_{ijk}^2\Biggl [
+\frac{1}{6\e^3}
+\frac{41}{36\e^2}
+\frac{1}{\e} \left(
\frac{325}{54}
-\frac{1}{4}\pi^2
\right)
+ \left(
\frac{9053}{324}
-\frac{41}{24}\pi^2
-\frac{31}{9}\zeta_3
\right)\nonumber \\
&& \hspace{2cm}
 + \order{\e}\Biggr] \, ,
\\
\label{eq:G31hintoldX3}
\widehat{{\cal G}}_3^1 (s_{ijk}) &=& S_{ijk}^2\Biggl [
+\frac{7}{18\e}
+ \left(
\frac{895}{324}
-\frac{11}{108}\pi^2
\right)
 + \order{\e}\Biggr] \, .
\end{eqnarray}

\endgroup
For the $A$-type, $E$-type and $G$-type antennae, we find complete agreement with the pole structure of the analogous integrated antennae given in Ref.~\cite{Gehrmann-DeRidder:2005btv}.  For the $D$-type and $F$-type antennae, we have utilised the $X_3^0$ sub-antenna given in Eqs.~(6.13) and (7.13) of Ref.~\cite{Gehrmann-DeRidder:2005btv} respectively and therefore the pole structures of the combinations $2 \left( {{\cal D}}_3^1+\widetilde{{\cal D}}_3^1 \right)$, $2 \widehat{{\cal D}}_3^1$, $3 {{\cal F}}_3^1$ and $3 \widehat{{\cal F}}_3^1$ agrees with the expressions for $\Drvold$, $\Dhrvold$, $\Frvold$ and $\Fhrvold$ respectively,
given by Eqs. (6.22), (6.23), (7.22) and (7.23) of Ref.~\cite{Gehrmann-DeRidder:2005btv} respectively to $\order{\e^0}$.

\bibliographystyle{JHEP}
\bibliography{bibliography}

\end{document}

%% file: preamble.tex
\usepackage[utf8]{inputenc} 
\usepackage[T1]{fontenc} 
\usepackage{lmodern}
\usepackage[british]{babel}
\usepackage[nottoc]{tocbibind} 
\usepackage[figuresright]{rotating} 
\usepackage[hidelinks,pdfusetitle]{hyperref} 
\usepackage{cite} 
\usepackage{epsfig}

\usepackage{ifthen}
\newboolean{printBibInSubfiles}
\setboolean{printBibInSubfiles}{true} 
\def\bib{\ifthenelse{\boolean{printBibInSubfiles}}
        {\bibliographystyle{JHEP}\bibliography{bibliography}}
        {}
    }

\usepackage[shrink=10,stretch=10]{microtype}
\makeatletter
\@ifpackageloaded{microtype}{%
	\providecommand{\disableprotrusion}{\microtypesetup{protrusion=false}}
	\providecommand{\enableprotrusion}{\microtypesetup{protrusion=true}}
}{%
	\providecommand{\disableprotrusion}{}
	\providecommand{\enableprotrusion}{}
}
\makeatother

\usepackage{booktabs} 

\usepackage[margin=15mm,hang]{caption}
\usepackage{subcaption}
\usepackage{perpage} 
\MakePerPage{footnote} 

\graphicspath{{img/}}

\newenvironment{itemize*}%
{\begin{itemize}%
	\setlength{\itemsep}{0pt}%
	\setlength{\parskip}{0pt}}%
{\end{itemize}}
\newenvironment{enumerate*}%
{\begin{enumerate}%
	\setlength{\itemsep}{0pt}%
	\setlength{\parskip}{0pt}}%
{\end{enumerate}}

\usepackage{amsmath,amssymb}
\numberwithin{equation}{section}
\allowdisplaybreaks 
\usepackage{amssymb}
\usepackage{bm} 
\usepackage{bbm} 
\usepackage{xfrac} 
\usepackage{array} 
\newcolumntype{L}{>{$}l<{$}} 

\usepackage{cleveref} 

\usepackage[italic]{hepnames} 
\usepackage{braket} 
\usepackage{slashed} 
\usepackage{siunitx} 
\DeclareSIUnit\fb{\femto\barn}

\usepackage{xcolor}
\usepackage{makecell}
\usepackage{rotating}
\usepackage{pdflscape}
\usepackage{float}

\DeclareMathAlphabet\mathbfcal{OMS}{cmsy}{b}{n}

\usepackage{tikz}
\usetikzlibrary{shapes.geometric, arrows, positioning}
\tikzstyle{fullspace} = [rectangle, minimum width=4cm, minimum height=1.5cm, text centered, text width=3cm, draw=black, fill=white]
\tikzstyle{subspace} = [rectangle, rounded corners, minimum width=4cm, minimum height=1.5cm,text centered, text width=3cm, draw=black, fill=white]
\tikzstyle{arrow} = [thick,->,>=stealth]
\tikzstyle{circlebox} = [circle,text centered, minimum width = 1.5 cm, minimum height = 1.5 cm, draw=black,fill=white]
\tikzstyle{recbox} = [rectangle,text centered, minimum width = 1.5 cm, minimum height = 1.5 cm, draw=black,fill=white]

\tikzstyle{startstop} = [rectangle, rounded corners, 
minimum width=4cm, 
minimum height=1cm,
text centered, 
draw=black, 
fill=red!30]

\tikzstyle{io} = [trapezium, 
trapezium stretches=true, 
trapezium left angle=70, 
trapezium right angle=110, 
minimum width=1.5cm, 
minimum height=1cm, text centered, 
draw=black, fill=blue!30]

\tikzstyle{process} = [rectangle, 
minimum width=1cm, 
minimum height=1cm, 
text centered, 
draw=black, 
fill=orange!30]

\tikzstyle{decision} = [diamond, 
minimum width=3cm, 
minimum height=1cm, 
text centered, 
draw=black, 
fill=green!30]

\usepackage{tikz-feynman}
\tikzfeynmanset{compat=1.1.0}

\newcommand{\Tr}[4]{\mathrm{Tr}(\slashed{#1}\slashed{#2} \slashed{#3} \slashed{#4})}
\newcommand{\ome}{(1-\epsilon)}
\newcommand{\e}{\epsilon}

\newcommand{\order}[1]{\ensuremath{\mathcal{O}}\left( #1 \right)\xspace}

\newcommand{\fa}{a}
\newcommand{\fb}{b}
\newcommand{\fc}{\tilde{b}}
\newcommand{\fd}{c}
\newcommand{\fe}{\tilde{c}}
\newcommand{\ff}{d}
\newcommand{\fg}{\tilde{d}}

\newcommand{\NthreeLO}{N$^3$LO\xspace}

\newcommand{\Pqgzero}{P_{qg}^{(0)}}

\newcommand{\Pggzero}{P_{gg}^{(0)}}
\newcommand{\Pqqzero}{P_{q\bar{q}}^{(0)}}
\newcommand{\PggSzero}{P_{gg}^{\text{sub},(0)}}

\newcommand{\Pqg}{P_{qg}^{(0)}}
\newcommand{\Pgq}{P_{gq}^{(0)}}
\renewcommand{\Pgg}{P_{gg}^{(0)}}
\newcommand{\Pqq}{P_{q\bar{q}}^{(0)}}

\newcommand{\PggS}{P_{gg}^{\text{sub},(0)}}

\newcommand\nameggg{ggg \to g}

\newcommand\nameqgg{qgg \to q}
\newcommand\nameqpp{q\gamma\gamma \to q}
\newcommand\namegqbq{g\bar q q\to g}
\newcommand\namepqbq{q g \bar q \to \gamma}
\newcommand\nameqQQ{q\bar Q Q\to q}
\newcommand\nameqqq{q \bar q q\to q}

\renewcommand\nameggg{ggg}

\renewcommand\nameqgg{qgg}
\renewcommand\nameqpp{q\gamma\gamma}
\renewcommand\namegqbq{g\bar q q}
\renewcommand\namepqbq{q g \bar q }
\renewcommand\nameqQQ{q\bar Q Q}
\renewcommand\nameqqq{q \bar q q}

\newcommand{\Bz}{\beta_0}
\newcommand{\bz}{b_0}
\newcommand{\bzf}{b_{0,F}}

\newcommand{\Pggg}{P_{\nameggg}^{(0)}}
\newcommand{\Rggg}{R_{\nameggg}^{(0)}}

\newcommand{\Rgggsub}{R_{g(gg)}^{(0)}}
\newcommand{\Pqgg}{P_{\nameqgg}^{(0)}}
\newcommand{\Rqgg}{R_{\nameqgg}^{(0)}}
\newcommand{\Pqpp}{P_{\nameqpp}^{(0)}}
\newcommand{\Rqpp}{R_{\nameqpp}^{(0)}}
\newcommand{\Pgqbq}{P_{\namegqbq}^{(0)}}
\newcommand{\Rgqbq}{R_{\namegqbq}^{(0)}}
\newcommand{\Ppqbq}{P_{\namepqbq}^{(0)}}
\newcommand{\Rpqbq}{R_{\namepqbq}^{(0)}}
\newcommand{\PqQQ}{P_{\nameqQQ}^{(0)}}
\newcommand{\RqQQ}{R_{\nameqQQ}^{(0)}}
\newcommand{\Pqqq}{P_{\nameqqq}^{(0)}}
\newcommand{\Rqqq}{R_{\nameqqq}^{(0)}}

\newcommand{\softorder}{(0)}
\newcommand{\Sgg}{S_{gg}^{\softorder}}
\newcommand{\Spp}{S_{\gamma\gamma}^{\softorder}}
\newcommand{\Sqq}{S_{q\qb}^{\softorder}}
\newcommand{\Sg}{S_{g}^{\softorder}}
\newcommand{\Sp}{S_{\gamma}^{\softorder}}

\newcommand{\Sgone}{S_{g}^{(1)}}
\newcommand{\Sgtone}{\widetilde{S}_{g}^{(1)}}
\newcommand{\Sghone}{\widehat{S}_{g}^{(1)}}
\newcommand{\Sgzero}{S_{g}^{(0)}}
\newcommand{\Pqgone}{P_{qg}^{(1)}}
\newcommand{\Pqgtone}{\widetilde{P}_{qg}^{(1)}}
\newcommand{\Pqghone}{\widehat{P}_{qg}^{(1)}}

\newcommand{\Pggone}{P_{gg}^{(1)}}
\newcommand{\Pgghone}{\widehat{P}_{gg}^{(1)}}
\newcommand{\Pggonesub}{P_{gg}^{\mathrm{sub},(1)}}
\newcommand{\Pggtone}{\widetilde{P}_{gg}^{(1)}}
\newcommand{\Pgghonesub}{\widehat{P}_{gg}^{\mathrm{sub},(1)}}
\newcommand{\Pqqone}{P_{q\bar{q}}^{(1)}}
\newcommand{\Pqqtone}{\widetilde{P}_{q\bar{q}}^{(1)}}
\newcommand{\Pqqhone}{\widehat{P}_{q\bar{q}}^{(1)}}

\newcommand{\Sq}{S_{q}^{\softorder}}
\newcommand{\Sqb}{S_{\bar{q}}^{\softorder}}
\newcommand{\Sb}{S_{b}^{\softorder}}

\newcommand{\PDSup}{\mathrm{\mathbf{DS}}^\uparrow\xspace}
\newcommand{\PTCup}{\mathrm{\mathbf{TC}}^\uparrow\xspace}

\newcommand{\PDCup}{\mathrm{\mathbf{DC}}^\uparrow\xspace}
\renewcommand{\PSup}{\mathrm{\mathbf{S}}^\uparrow\xspace}
\newcommand{\PCup}{\mathrm{\mathbf{C}}^\uparrow\xspace}
\newcommand{\PXup}{\mathrm{\mathbf{P}}_X^\uparrow\xspace}
\newcommand{\PXdown}{\mathrm{\mathbf{P}}_X^\downarrow\xspace}
\newcommand{\PDSdown}{\mathrm{\mathbf{DS}}^\downarrow\xspace}
\newcommand{\PTCdown}{\mathrm{\mathbf{TC}}^\downarrow\xspace}
\newcommand{\PSCdown}{\mathrm{\mathbf{SC}}^\downarrow\xspace}
\newcommand{\PDCdown}{\mathrm{\mathbf{DC}}^\downarrow\xspace}
\renewcommand{\PSdown}{\mathrm{\mathbf{S}}^\downarrow\xspace}
\newcommand{\PCdown}{\mathrm{\mathbf{C}}^\downarrow\xspace}

\newcommand{\PPup}{\mathrm{\mathbf{P}}^\uparrow\xspace}
\newcommand{\PPdown}{\mathrm{\mathbf{P}}^\downarrow\xspace}
\newcommand{\Peup}{\mathrm{\mathbf{P}}_\e^\uparrow\xspace}
\newcommand{\Pedown}{\mathrm{\mathbf{P}}_\e^\downarrow\xspace}

\newcommand{\XLO}{X_3^0}
\newcommand{\X}{X_4^0}
\newcommand{\Xt}{\widetilde{X}_4^0}
\newcommand{\A}{A_4^0}
\newcommand{\B}{B_4^0}
\newcommand{\C}{C_4^0}
\newcommand{\D}{D_4^0}
\newcommand{\F}{F_4^0}
\newcommand{\E}{E_4^0}
\newcommand{\Ea}{E_4^{0}}
\newcommand{\Eb}{\overline{E}_4^{0}}
\newcommand{\G}{G_4^0}
\newcommand{\Ga}{G_4^{0}}
\newcommand{\Gb}{\overline{G}_4^{0}}
\renewcommand{\H}{H_4^0}

\newcommand{\Arv}{A_3^1}
\newcommand{\Atrv}{\widetilde{A}_3^1}
\newcommand{\Ahrv}{\widehat{A}_3^1}
\newcommand{\Drv}{D_3^1}

\newcommand{\Dtrv}{\widetilde{D}_3^1}
\newcommand{\Dhrv}{\widehat{D}_3^1}
\newcommand{\Erv}{E_3^1}
\newcommand{\Etrv}{\widetilde{E}_3^1}
\newcommand{\Ehrv}{\widehat{E}_3^1}
\newcommand{\Frv}{F_3^1}
\newcommand{\Fhrv}{\widehat{F}_3^1}
\newcommand{\Grv}{G_3^1}
\newcommand{\Gtrv}{\widetilde{G}_3^1}
\newcommand{\Ghrv}{\widehat{G}_3^1}

\newcommand{\Arvold}{A_3^{1,\oldant}}
\newcommand{\Atrvold}{\widetilde{A}_3^{1,\oldant}}
\newcommand{\Ahrvold}{\widehat{A}_1^{1,\oldant}}
\newcommand{\Drvold}{D_3^{1,\oldant}}
\newcommand{\Dtrvold}{\widetilde{D}_3^{1,\oldant}}
\newcommand{\Dhrvold}{\widehat{D}_3^{1,\oldant}}
\newcommand{\Ehrvold}{\widehat{E}_3^{1,\oldant}}
\newcommand{\Ervold}{E_3^{1,\oldant}}
\newcommand{\Etrvold}{\widetilde{E}_3^{1,\oldant}}
\newcommand{\Frvold}{F_3^{1,\oldant}}
\newcommand{\Gtrvold}{\widetilde{G}_3^{1,\oldant}}
\newcommand{\Fhrvold}{\widehat{F}_3^{1,\oldant}}
\newcommand{\Grvold}{G_3^{1,\oldant}}
\newcommand{\Ghrvold}{\widehat{G}_3^{1,\oldant}}

\newcommand{\calArvold}{{\cal A}_3^{1,\oldant}}
\newcommand{\calAtrvold}{\widetilde{\cal A}_3^{1,\oldant}}
\newcommand{\calAhrvold}{\widehat{\cal A}_1^{1,\oldant}}
\newcommand{\calDrvold}{{\cal D}_3^{1,\oldant}}
\newcommand{\calDhrvold}{\widehat{\cal D}_3^{1,\oldant}}
\newcommand{\calEhrvold}{\widehat{\cal E}_3^{1,\oldant}}
\newcommand{\calErvold}{{\cal E}_3^{1,\oldant}}
\newcommand{\calEtrvold}{\widetilde{\cal E}_3^{1,\oldant}}
\newcommand{\calFrvold}{{\cal F}_3^{1,\oldant}}

\newcommand{\calFhrvold}{\widehat{\cal F}_3^{1,\oldant}}

\newcommand{\calArv}{{\cal A}_3^1}
\newcommand{\calAtrv}{\widetilde{{\cal A}}_3^1}
\newcommand{\calAhrv}{\widehat{{\cal A}}_3^1}
\newcommand{\calDrv}{{\cal D}_3^1}

\newcommand{\calDtrv}{\widetilde{{\cal D}}_3^1}
\newcommand{\calDhrv}{\widehat{{\cal D}}_3^1}
\newcommand{\calErv}{{\cal E}_3^1}
\newcommand{\calEtrv}{\widetilde{{\cal E}}_3^1}
\newcommand{\calEhrv}{\widehat{{\cal E}}_3^1}
\newcommand{\calFrv}{{\cal F}_3^1}
\newcommand{\calFhrv}{\widehat{{\cal F}}_3^1}
\newcommand{\calGrv}{{\cal G}_3^1}
\newcommand{\calGtrv}{\widetilde{{\cal G}}_3^1}
\newcommand{\calGhrv}{\widehat{{\cal G}}_3^1}

\newcommand{\At}{\widetilde{A}_4^0}
\newcommand{\Dt}{\widetilde{D}_4^0}
\newcommand{\Ft}{\widetilde{F}_4^0}
\newcommand{\Et}{\widetilde{E}_4^0}
\newcommand{\Gt}{\widetilde{G}_4^0}

\newcommand{\oldant}{\, \mathrm{OLD}}
\newcommand{\Xold}[1]{#1_3^{0,\oldant}}

\newcommand{\Aold}{A_4^{0,\oldant}}
\newcommand{\Atold}{\widetilde{A}_4^{0,\oldant}}
\newcommand{\Bold}{B_4^{0,\oldant}}
\newcommand{\Cold}{C_4^{0,\oldant}}
\newcommand{\Dold}{D_4^{0,\oldant}}
\newcommand{\Fold}{F_4^{0,\oldant}}
\newcommand{\Eold}{E_4^{0,\oldant}}
\newcommand{\Etold}{\widetilde{E}_4^{0,\oldant}}
\newcommand{\Gold}{G_4^{0,\oldant}}
\newcommand{\Gtold}{\widetilde{G}_4^{0,\oldant}}
\newcommand{\Hold}{H_4^{0,\oldant}}

\newcommand{\calX}{{\cal X}_4^0}

\newcommand{\calA}{{\cal A}_4^0}
\newcommand{\calAold}{{\cal A}_4^{0,\oldant}}
\newcommand{\calB}{{\cal B}_4^0}

\newcommand{\calC}{{\cal C}_4^0}
\newcommand{\calCold}{{\cal C}_4^{0,\oldant}}
\newcommand{\calD}{{\cal D}_4^0}
\newcommand{\calDold}{{\cal D}_4^{0,\oldant}}
\newcommand{\calF}{{\cal F}_4^0}
\newcommand{\calFold}{{\cal F}_4^{0,\oldant}}
\newcommand{\calEold}{{\cal E}_4^{0,\oldant}}
\newcommand{\calEa}{{\cal E}_4^{0}}
\newcommand{\calEb}{\overline{{\cal E}}_4^{0}}

\newcommand{\calGold}{{\cal G}_4^{0,\oldant}}
\newcommand{\calGa}{{\cal G}_4^{0}}
\newcommand{\calGb}{\overline{{\cal G}}_4^{0}}
\newcommand{\calH}{{\cal H}_4^0}
\newcommand{\calHold}{{\cal H}_4^{0,\oldant}}
\newcommand{\calAt}{\widetilde{{\cal A}}_4^0}
\newcommand{\calDt}{\widetilde{{\cal D}}_4^0}
\newcommand{\calFt}{\widetilde{{\cal F}}_4^0}
\newcommand{\calEt}{\widetilde{{\cal E}}_4^0}
\newcommand{\calGt}{\widetilde{{\cal G}}_4^0}
\newcommand{\calAtold}{\widetilde{{\cal A}}_4^{0,\oldant}}

\newcommand{\calEtold}{\widetilde{{\cal E}}_4^{0,\oldant}}
\newcommand{\calGtold}{\widetilde{{\cal G}}_4^{0,\oldant}}

\newcommand{\cala}{{\cal A}_3^0}
\newcommand{\cald}{{\cal D}_3^0}
\newcommand{\calf}{{\cal F}_3^0}
\newcommand{\cale}{{\cal E}_3^0}
\newcommand{\calg}{{\cal G}_3^0}

\newcommand{\Ssoft}{\mathrm{Ssoft}}
\newcommand{\Scol}{\mathrm{Scol}}
\newcommand{\Dsoft}{\mathrm{Dsoft}}
\newcommand{\Tcol}{\mathrm{Tcol}}
\newcommand{\Dcol}{\mathrm{Dcol}}

\newcommand{\Scolt}{\widetilde{\mathrm{Scol}}}

\newcommand{\Scolh}{\widehat{\mathrm{Scol}}}

\newcommand{\FinPoles}{\text{TPoles}}
\newcommand{\FinPolest}{\widetilde{\text{TPoles}}}
\newcommand{\FinPolesh}{\widehat{\text{TPoles}}}
\newcommand{\Poles}{{\mathcal Poles}}

\newcommand{\calSsoft}{\mathcal{S}\kern-0.05em{s\kern-0.05em o\kern-0.05em f\kern-0.05em t}}
\newcommand{\calScol}{\mathcal{S}\kern-0.05em{c\kern-0.05em o\kern-0.05em l}}
\newcommand{\calDsoft}{\mathcal{D}\kern-0.05em{s\kern-0.05em o\kern-0.05em f\kern-0.05em t}}
\newcommand{\calTcol}{\mathcal{T}\kern-0.05em{c\kern-0.05em o\kern-0.05em l}}
\newcommand{\calDcol}{\mathcal{D}\kern-0.05em{c\kern-0.05em o\kern-0.05em l}}

\newcommand{\qb}{\bar{q}}
\newcommand{\Qb}{\bar{Q}}

\newcommand{\omz}{(1-z)}
\newcommand{\Li}{\mathrm{Li}}

\newcommand{\Af}{R_{\e}}

\newcommand{\omxi}{(1-x_i)}
\newcommand{\omxj}{(1-x_j)}
\newcommand{\omxk}{(1-x_k)}

\newcommand{\omyk}{(1-y_k)}
\newcommand{\sij}{s_{ij}}
\newcommand{\sjk}{s_{jk}}
\newcommand{\sik}{s_{ik}}
\newcommand{\sijk}{s_{ijk}}
\renewcommand{\xi}{x_{i}}
\newcommand{\xj}{x_{j}}
\newcommand{\xk}{x_{k}}

\newcommand{\yk}{y_{k}}

\newcommand{\Jcol}[1]{\mathbfcal{J}^{(#1)}}
\newcommand{\Jcolb}[1]{\overline{\mathbfcal{J}}^{(#1)}}

\newcommand{\Jtot}[1]{\mathcal{J}^{(#1)}_2}
\newcommand{\Jtotb}[1]{\overline{\mathcal{J}}^{(#1)}_2}
\newcommand{\JtotOLD}[1]{\mathcal{J}^{(#1),\oldant}_2}
\newcommand{\JtotbOLD}[1]{\overline{\mathcal{J}}^{(#1),\oldant}_2}

\newcommand{\J}[1]{J_{2}^{(#1)}}
\newcommand{\Jh}[1]{\widehat{J_{2}}^{(#1)}}
\newcommand{\Jt}[1]{\widetilde{J_{2}}^{(#1)}}
\newcommand{\Jht}[1]{\widehat{\widetilde{J_{2}}}^{(#1)}}
\newcommand{\Jhh}[1]{\widehat{\widehat{J_{2}}}^{(#1)}}
\newcommand{\Jb}[1]{\overline{J}_{2}^{(#1)}}